\documentclass[12pt]{article}
\usepackage[utf8]{inputenc}
\usepackage{amsmath,amsfonts,amssymb,amsthm}
\usepackage[mathcal]{eucal}
\usepackage[longnamesfirst,sort&compress]{natbib}
\usepackage[margin=1in]{geometry}
\usepackage{indentfirst}
\usepackage{setspace,graphicx,pdflscape,datetime,subfigure,verbatim,float,tabularx,booktabs}
\usepackage{hyperref} \hypersetup{pdfpagemode=FullScreen,colorlinks=true,citecolor=blue,urlcolor=blue}
\usepackage[]{caption}  

\newtheorem{obs}{Theorem}

\newtheorem{observation}[obs]{Observation}
\newcommand{\comments}[1]{}

\newcolumntype{F}{>{\raggedright\arraybackslash}X}
\newcolumntype{G}{>{\raggedleft\arraybackslash}X}
\newcolumntype{N}{>{\centering\arraybackslash}X}
\newcommand\RPtxtsize[1]{\normalsize{#1}}

\newcommand{\RPprep}[5]{%
  \def\ArgI{#1}
  \def\ArgII{#2}
  \def\ArgIII{#3}
  \def\ArgIV{#4}
  \def\ArgV{#5}
  \ifnum \ArgII=1
    \renewcommand\RPtxtsize[1]{\footnotesize{##1}}
  \else
    \renewcommand\RPtxtsize[1]{\normalsize{##1}}
  \fi
}

\newcommand\RPtab[1]{%
		\ifnum \ArgIII=1
			\begin{landscape}
		\fi
		\begin{table}[!hbp]
		\caption{\ArgI}
		\ArgV
		\vfill
		\centering
		\RPtxtsize{
		#1
		}
		\label{tab:\ArgIV}
	\end{table}
	\ifnum \ArgIII=1
		\end{landscape}
	\fi
}	

\newcommand{\RPfig}[1]{
	\ifnum \ArgIII=1
		\begin{landscape}
	\fi
	\begin{figure}[!hbp]
		\centering
		\RPtxtsize{
		#1
		}
		\caption{\ArgI. %
		    \ArgV
		}
		\vfill
		\medskip
		\label{fig:\ArgIV}
	\end{figure}
	\ifnum \ArgIII=1
		\end{landscape}
	\fi
}

\begin{document}

\title{Facts of US Firm Scale and Growth 1970-2019:\protect\\An Illustrated Guide}

\author{
	Robert Parham\footnote{University of Virginia (robertp@virginia.edu). Code to reproduce all figures and tables in this paper is available from the author.}
}
\date{\today}
\renewcommand{\thefootnote}{\fnsymbol{footnote}}
\maketitle

\vspace{-.2in}
\begin{abstract}
\noindent This work analyzes data on all public US firms in the 50 year period 1970-2019, and presents 18 stylized facts of their scale, income, growth, return, investment, and dynamism. Special attention is given to (i) identifying distributional forms; and (ii) scale effects --- systematic difference between firms based on their scale of operations. Notable findings are that the Difference-of-Log-Normals (DLN) distribution has a central role in describing firm data, scale-dependant heteroskedasticity is rampant, and small firms are systematically different from large firms.
\end{abstract}

\medskip
\thispagestyle{empty}
\setcounter{footnote}{0}
\renewcommand{\thefootnote}{\arabic{footnote}}
\setcounter{page}{1}
\doublespacing 

\clearpage

\section{Introduction}

Firm growth is one of the most researched subjects in all of economics. Macroeconomic growth is micro-founded on firm growth --- firms being the productive side of the economy. Asset pricing deals with the growth of one firm size measure in particular --- market value of equity, adjusted for capital dispersions to equity owners --- whose growth is simply equity return. The firm size/scale distribution (whose first difference is firm growth) has seen such intense research interest that it has its own JEL classification: L11.

Nevertheless, the stylized facts related to firm size, scale (defined as the natural log of size) and growth are dispersed over a large body of literature spanning several decades. Much of the ``traditional'' knowledge has been overturned by latter findings, such as prior observations on the shape of the firm scale distribution (rejecting the Pareto and Zipf), or the shape of the return distribution (rejecting the Stable and Laplace). I am not aware of any effort to systematically consider stylized facts which hold true regardless of the specific measure of size being used. Similarly, I am not aware of efforts to systematically consider scale effects --- differences in firm behavior dependant on the scale of operations.

This paper aims to fill this gap, and provide a systematic and fresh view of the empirical data on firms. To that end, it analyzes data on all public US firms in the 50-year period 1970-2019, and presents evidence on the distributions and scaling patterns of firm scale, income, growth, return, investment, and dynamism. In that, it follows a long tradition of ``stylized facts'' papers spurring research interest by highlighting salient facts of the data which have been neglected or gone unnoticed by researchers.\footnote{E.g. \cite{Kaldor1961}, \cite{Little1962}, \cite{Fama1965}, \cite{Lucas1978}, \cite{ChanEtAl2003}, \cite{BuldyrevEtAl2007}, \cite{AngeliniGenerale2008}, and \cite{KondoEtAl2018}.}

In a short companion paper, \cite{Parham2022} presents the difference-of-log-Normals (DLN) distribution and posits that it is a fundamental distribution in nature. The DLN arises when considering phenomena in which two opposing multiplicative forces are at play, as a direct result of the Central Limit Theorems. A clear example of such a phenomenon is firm income, defined to be sales minus expenses, both distributing approximately log-Normally. \cite{Parham2022} posits that firm income should distribute DLN, as should all firm growth distributions. A core finding of this paper is to corroborate this theoretical hypothesis, finding that many firm outcomes distribute DLN. Among them: income (or cashflows); capital dispersions to/from all stakeholders and to/from equity holders; growth in: sales, expenses, physical and total capital, equity and total value (with both adjusted to capital dispersions). 

The paper further shows that returns and excess returns at the daily, monthly, and yearly frequency all distribute DLN as well, while rejecting the previous candidates in the literature, the Stable and Laplace distributions. Firm investment and investment intensity, measured in several alternative ways, are also shown to distribute DLN. The fit between the DLN and equity returns is especially noteworthy, given the voluminous literature on the determinants, fat-tails, and statistical properties of equity returns.

A second notable pattern explored in this paper is scaling effects --- systematic differences between firms based on their scale. Both level and dispersion (i.e. heteroskedasticity) effects are considered, and are shown to be rampant across a wide variety of firm outcomes. Systematic decreasing dispersion is especially notable in the data, yet gained little attention from corporate finance scholars. Furthermore, the paper documents a striking difference between small firms (below the median size in the data) and large firms (above the median size in the data). The two groups of firms exhibit strikingly different behavior.

Section~\ref{sec:ScaleIncome} begins by presenting the data being analyzes. The section describes data sources, variable definitions, data screens, and deflators. It then analyzes the firm scale distribution by considering several measures of firm scale and adjusting for time effects. I show that skew-Normality of the firm scale distribution cannot be rejected, and that scales are highly persistent and cointegrated. Next, the section analyzes the firm income distributions, which are shown to distribute DLN, and to be subject to significant time and scale effects.

Next, Section~\ref{sec:GrowthReturn} analyzes the various facets of firm growth distributions. It begins by presenting the heteroskedasticity of growth w.r.t. scale --- growth dispersion is decreasing with scale --- and discusses the relation between this finding and the internal structure of the firm. Adjusting for time and scale effects, the section shows that growth distributes DLN as well. Equity returns are a special case of firm growth (i.e., growth in equity value, adjusted for dividends and net repurchase). As data on firm returns are available at high frequency, I present similar tests for firm return distributions. DLN is not rejected for raw or excess returns, at the daily, monthly and yearly frequencies.

Section~\ref{sec:GrowthReturn} then goes on to analyze the growth of income. Firm income is sometimes negative (e.g., losses or cash infusions from stake holders for the purpose of investment) and hence traditional measures of growth such as differences in log income between consecutive periods are undefined. The DLN methods paper \cite{Parham2022a} develops a generalized measure of growth for sometimes-negative DLN RVs, and I use the measure to show that income growth distributes DLN as well.

Section~\ref{sec:InvestDyn} first discusses firm investments. It analyzes several measures of investment --- total, physical, and R\&D investment, under various definitions. I show that investments, too, are subject to scale effects and are distributed DLN. The section documents peculiar anomalies in the behavior of R\&D distributions, related to the level and skewness of R\&D w.r.t. scale. To further explore this anomaly, the section then documents several facts related to firm ratios, which propose small firms and large firms are systematically different from each other is important ways. The section concludes with a discussion of firm dynamism (entry and exit). I show that scale is highly persistent in the long-term, analyze entry and exit patterns by scale and age, and document the relation between scale and age more broadly.

\section{Scale and income}
\label{sec:ScaleIncome}

\subsection{Data}
\label{sec:Data}

The data analyzed in this paper cover all public US firms in the 50-year period 1970-2019. Data are predominantly derived from the yearly CRSP/Compustat data set. For some tests related to equity returns I use higher-frequency CRSP data, and for some tests related to firm dynamism I use Compustat data ranging back to 1950. Other minor data sources include the nominal and real GDP series from FRED and factor returns from Ken French's website.

Data variables are identified throughout by two capital letter mnemonics (e.g., SL for firm sales). Table~\ref{tab:DataDef} defines all data variables and provides a mapping to Compustat items used to construct them. I rely on the sources and uses identity,
\begin{equation}
\label{eq:SrcUse}
\underbrace{\text{sales}}_{SL} - \underbrace{\text{expenses}}_{XS} = \underbrace{\text{income}}_{CF} = \underbrace{\text{total net dividends}}_{DI} + \underbrace{\text{total net investment}}_{IT}
\end{equation}
to calculate expenses as dissipated sales (i.e., sales - income). This guarantees all expenses, including cost of goods, selling, general, administrative, taxes, and various ``special'' and ``one-time'' expenses are accounted for.

\RPprep{Data definitions}{0}{0}{DataDef}{%
    This table defines all data items used. The first column is the name of each data item and the second is the mnemonic used throughout. The third column is the mapping to Compustat items or previously defined mnemonics, and the fourth is a short description. The core accounting identity used is the sources and uses equation: income = sales - expenses = total dividends + total investment, with dividends broadly defined below. The last two data items are alternative definitions, used for comparability with previous work. The ``L.'' is the lag operator.
}
\RPtab{%
    \begin{tabularx}{\linewidth}{lllF}
    \toprule
    Name & XX & Definition & Description \\
    \midrule
    Equity value & EQ & mve & market, year end \\
    Debt value & DB & lt  & book total liabilities \\
    Total value & VL & EQ + DB & equity + debt \\
    Equity dividends & DE & dvt + (prstkc - sstk)  & dividends + net repurchase \\
    Debt dividends & DD & xint + (L.DB-DB) & interest paid + decrease in debt \\
    Total dividends & DI & DE + DD & to equity and debt \\
    Physical capital & KP & ppent & PP\&E, net of depreciation \\
    Total capital & KT & at & total assets (tangible) \\
    Depreciation & DP & dp & of physical capital \\
    Physical investment & IP & KP - L.KP + DP &  growth in net physical capital \\
    Total investment & IT & KT - L.KT + DP & growth in net assets \\
    Income & CF & DI + IT & bottom-up free cash flows \\
    Sales & SL & sl & total sales \\
    Expenses & XS & SL - CF & dissipated sales \\
    R\&D & RD & xrd & an expense included in XS \\
    Expenses (alt.) & XA & cogs + xsga + txt & top-down definition \\
    Income (alt.) & CA & SL - XA & top-down definition \\
    Investment (alt.) & IA & capx - sppe & cash-flow statement definition \\
	\bottomrule
    \end{tabularx}
}

I define three subsets of the data: (i) \{All\} - 283K raw firm-year observations on 25K firms, representing the entire Compustat universe; (ii) \{Good\} - 200K filtered observations on 20K firms, with the data filters described momentarily; and (iii) \{Non-Bank\} - the 164K subset of \{Good\} observations on 16K firms, excluding financial and utility firms. The \{All\} subset is used to verify data filters do not introduce substantial bias, and the \{Non-Bank\} subset is used when conducting some analyses related to firm assets, as both excluded industries define assets differently than the rest of the sample. When the subset used is not mentioned, it is implied to be the \{Good\} subset. The filters applied to reach that subset remove 62K observations for missing critical data items, 16K for being too small (with a cutoff of 1M 2019 dollars on value, capital, sales), and 5K for major restructuring (an M\&A larger than 20\% of assets). Panel (a) of Figure~\ref{fig:FSDfacts} presents the number of firms per year in the sample, with a rising trend from less than $2000$ firms in $1970$ to a significant peak of $6000$ around the late 90's, followed by a continued decline to about $3500$ in $2019$.

\RPprep{Scale - Stylized facts}{0}{0}{FSDfacts}{%
    This figure presents stylized facts of the firm scale distribution (FSD). Panel (a) presents the number of firms per year. Panels (b) and (c) show the time trend of the FSD using real and GDP-adjusted dollars, respectively. For each year, the panels graph the $\{25,50,75\}^{th}$ percentiles of the FSD. Panels (d)-(f) present: un-adjusted equity scale FSD(EQ) w/ fitted Normal; time-adjusted equity scale FSD($\widetilde{\text{EQ}}$) w/ fitted Normal; and FSD($\widetilde{\text{EQ}}$) w/ fitted skew-Normal, respectively. Panels (g)-(i) present the respective q-q plots of panels (d)-(f).
}
\RPfig{%
	\begin{tabular}{ccc} 
		\subfigure[Number of firms]{\includegraphics[width=2in]{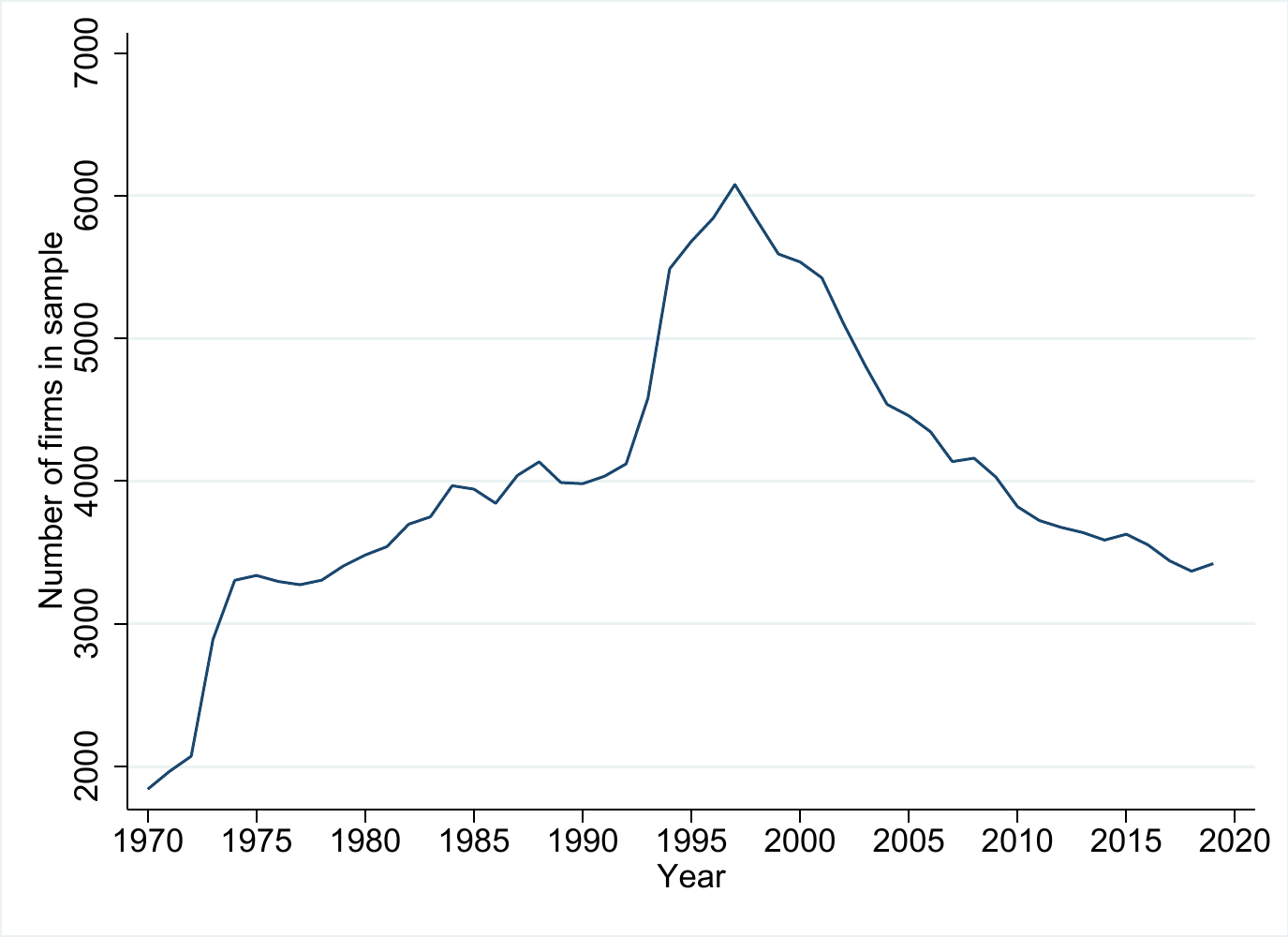}} & \subfigure[Time trend of FSD (real)]{\includegraphics[width=2in]{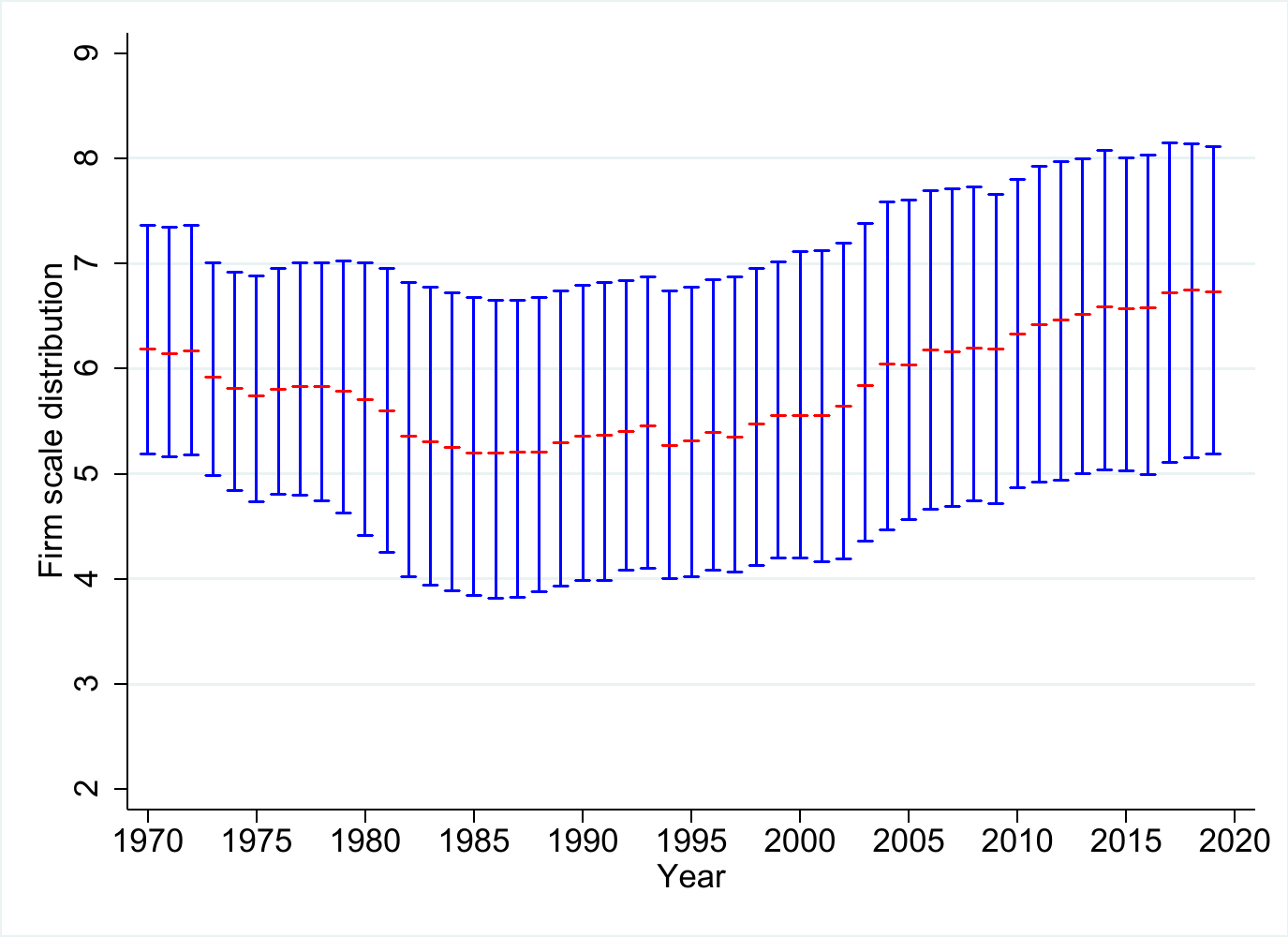}} &
		\subfigure[Time trend of FSD (GDP)]{\includegraphics[width=2in]{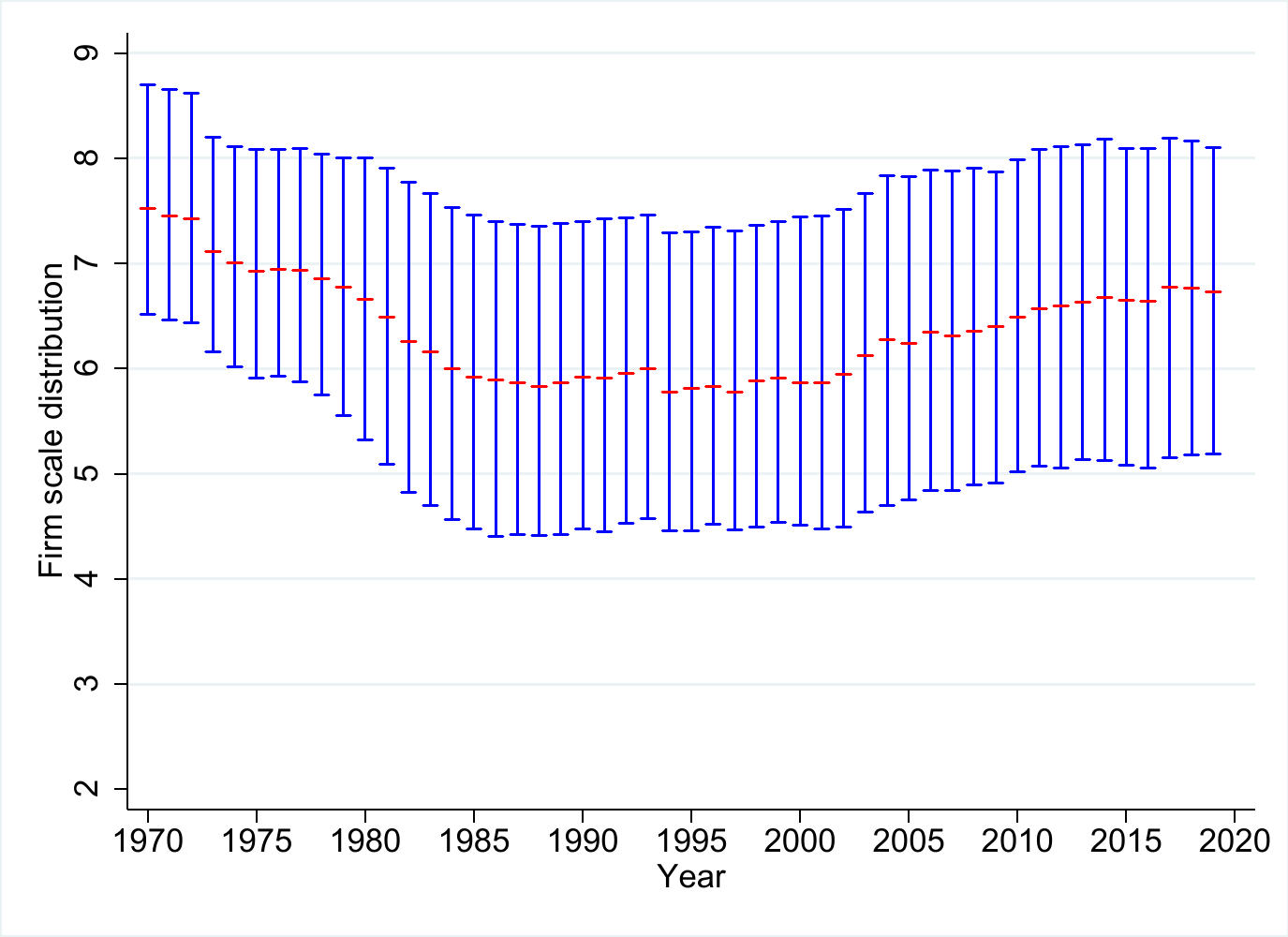}} \\ \\
		\subfigure[FSD(EQ) w/ N]{\includegraphics[width=2in]{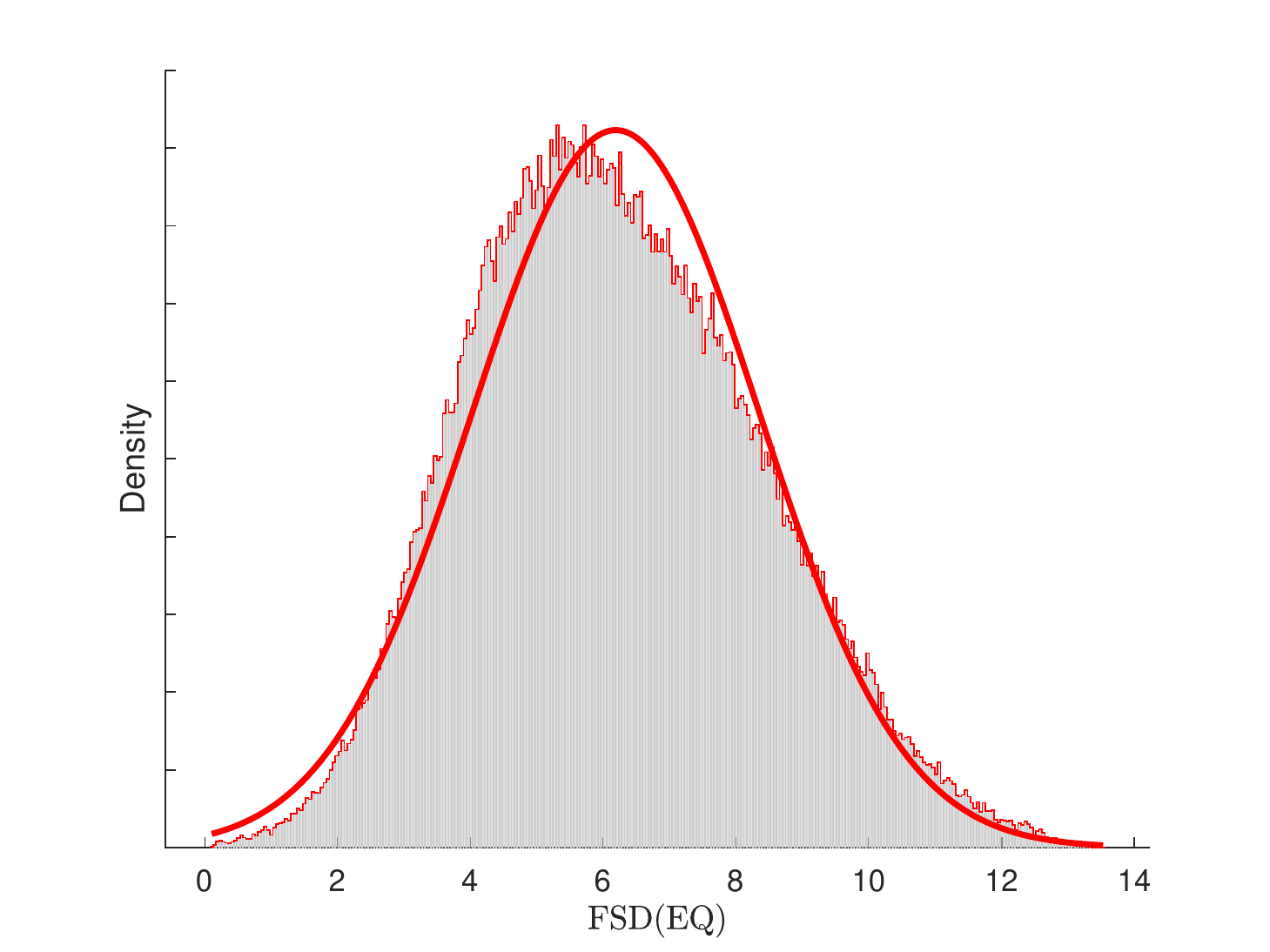}} &
		\subfigure[FSD($\widetilde{\text{EQ}}$) w/ N]{\includegraphics[width=2in]{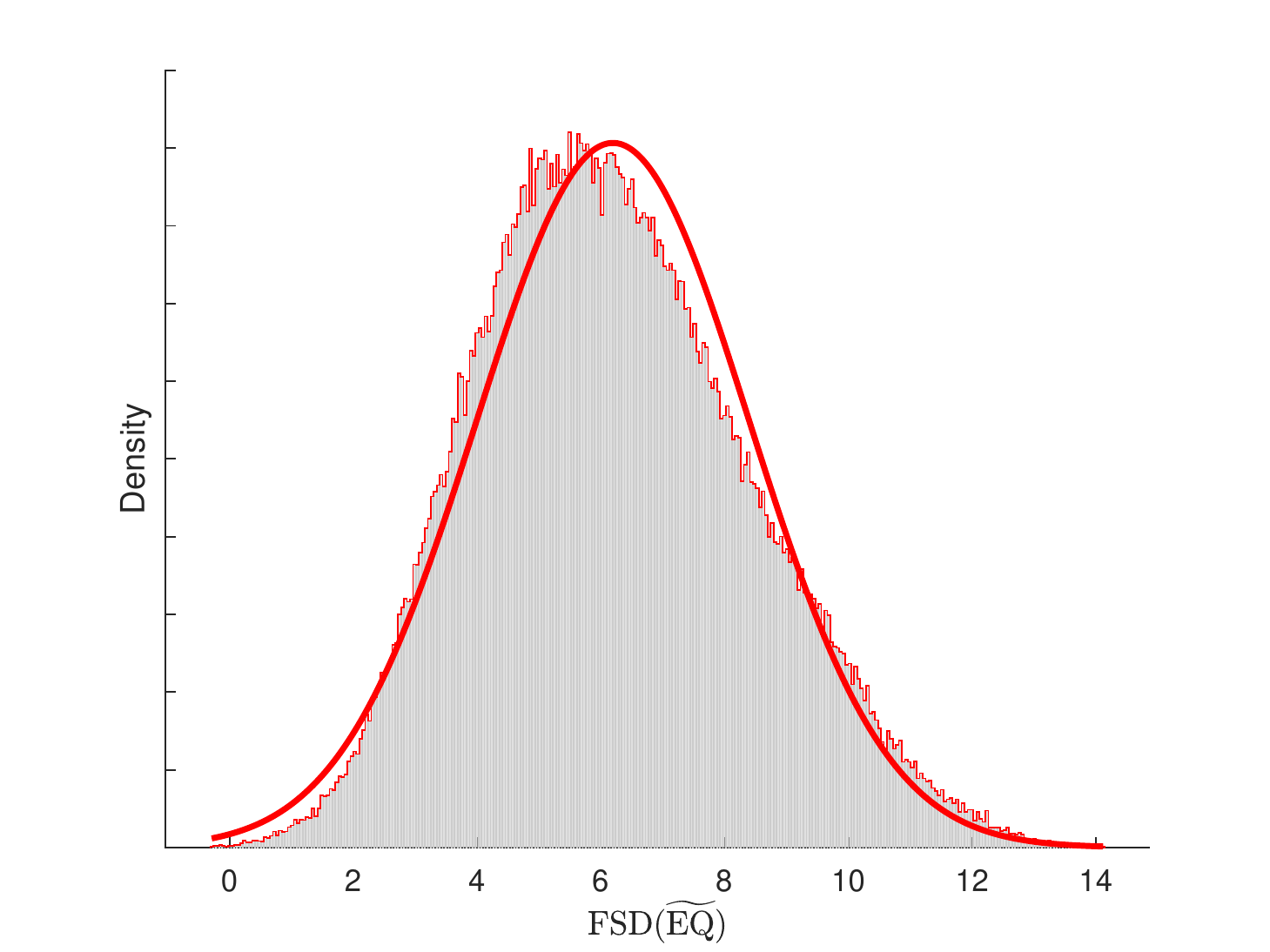}} &
		\subfigure[FSD($\widetilde{\text{EQ}}$) w/ SN]{\includegraphics[width=2in]{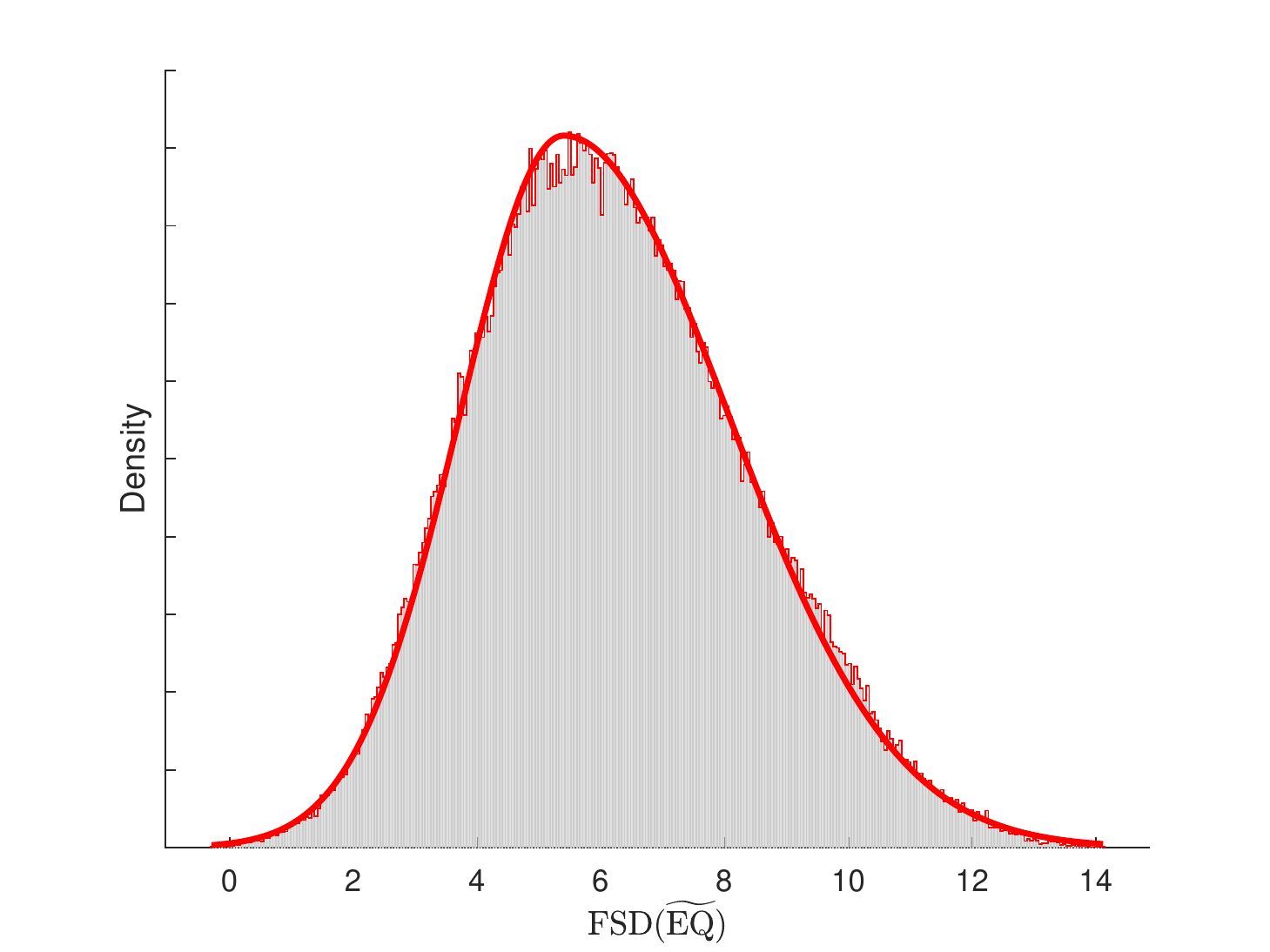}} \\ \\
		\subfigure[q-q FSD(EQ) vs. N]{\includegraphics[width=2in]{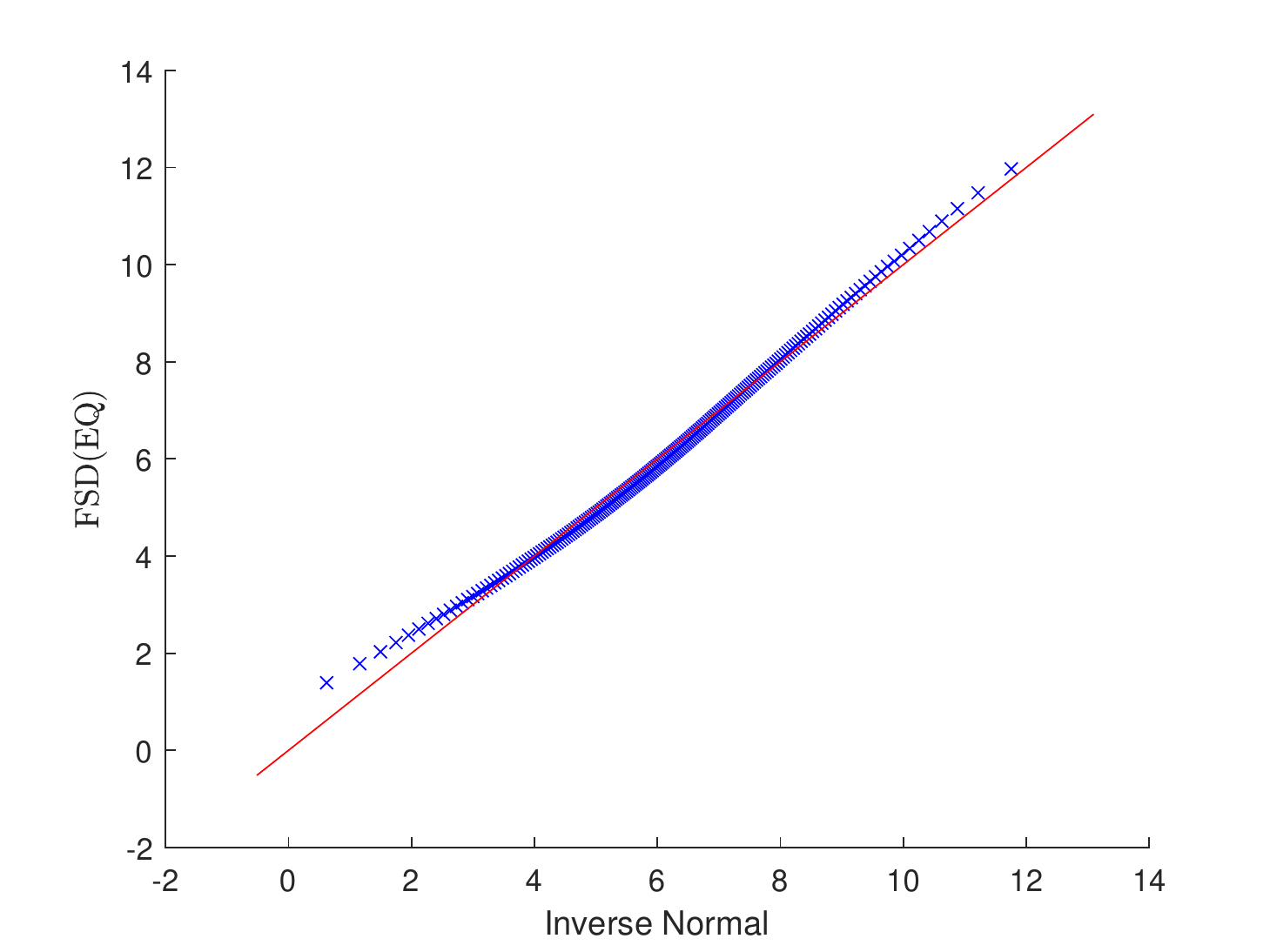}} &
		\subfigure[q-q FSD($\widetilde{\text{EQ}}$) vs. N]{\includegraphics[width=2in]{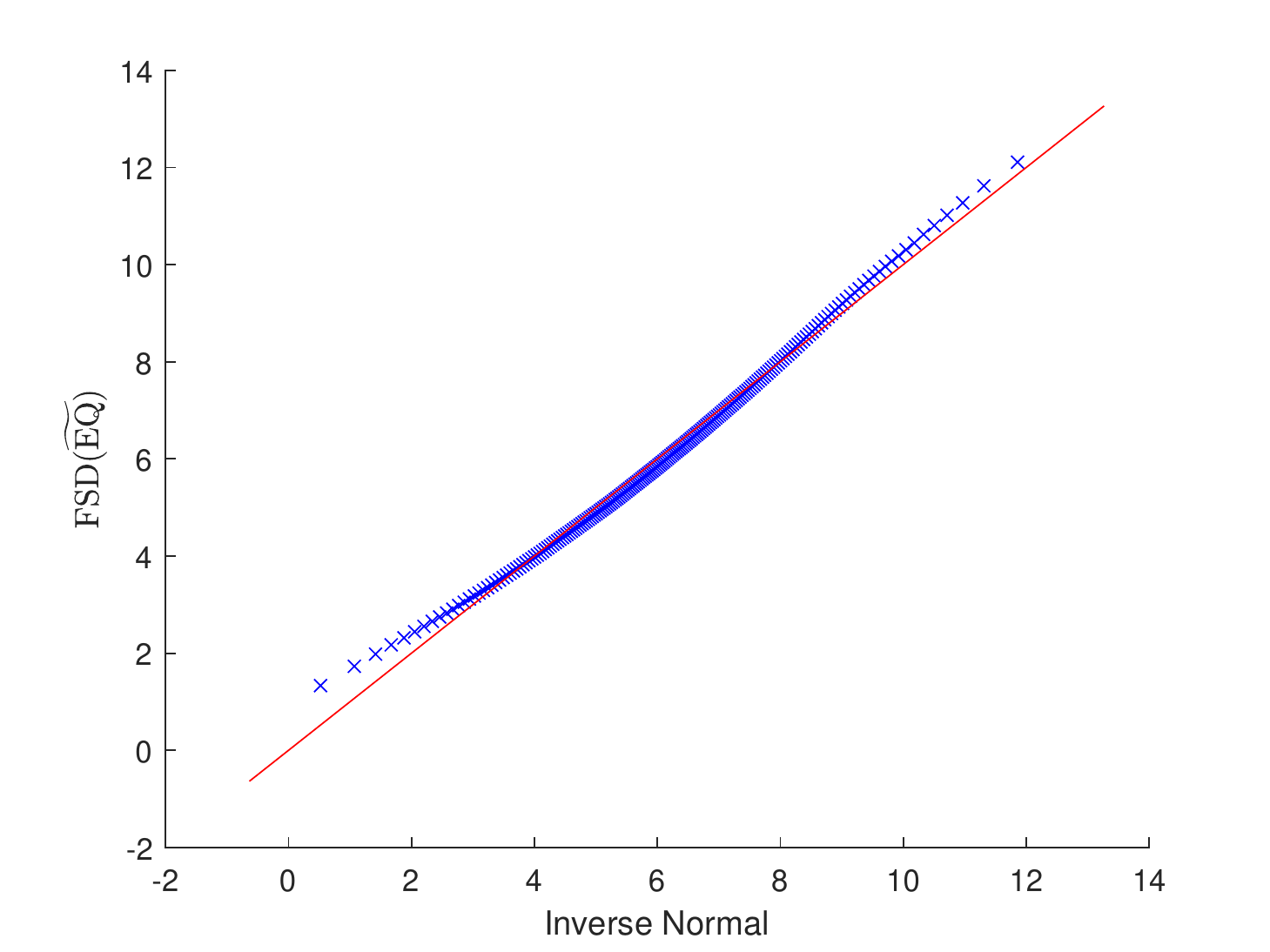}} &
		\subfigure[q-q FSD($\widetilde{\text{EQ}}$) vs. SN]{\includegraphics[width=2in]{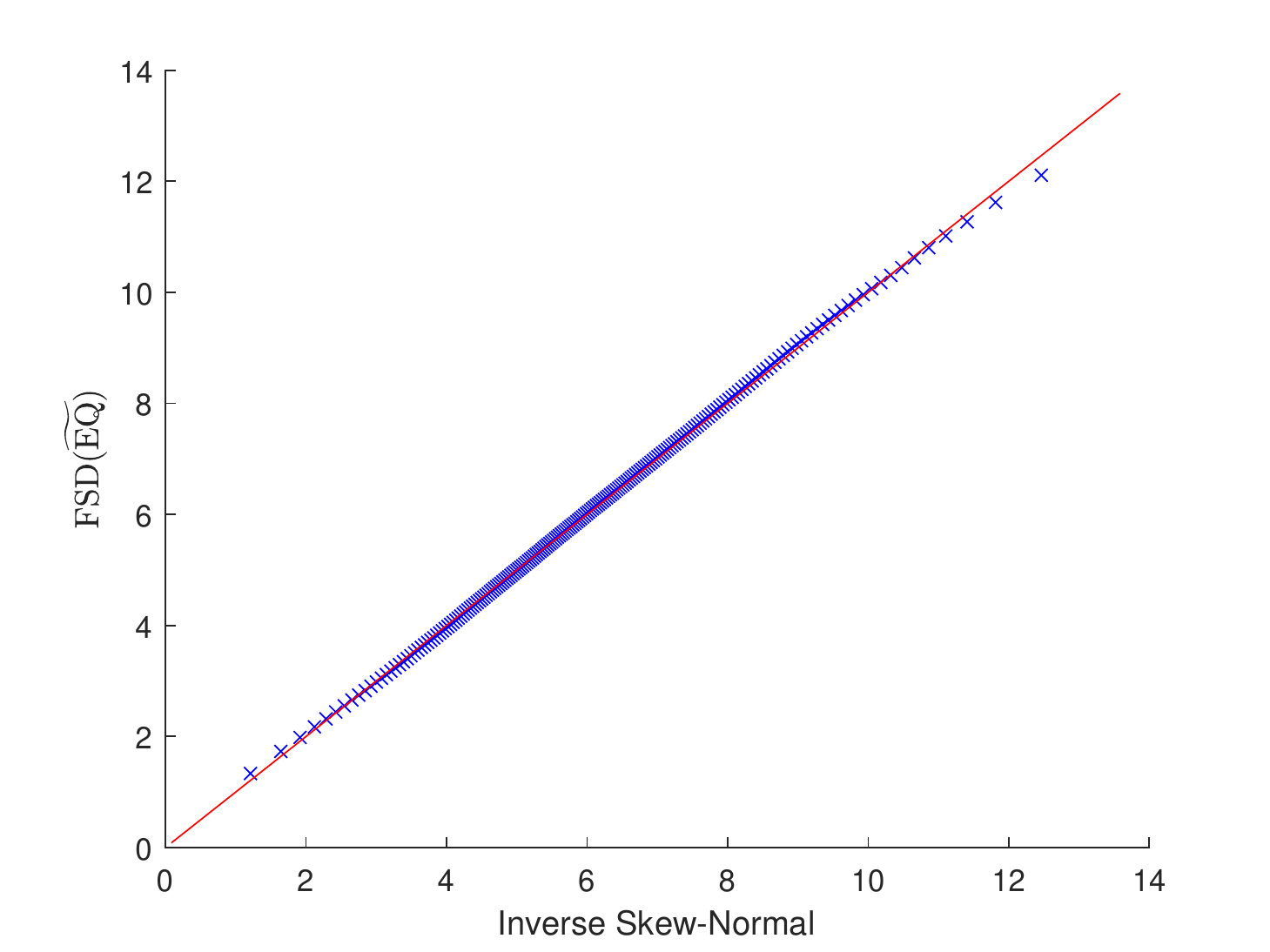}} \\ \\
	\end{tabular}
}

I consider two approaches to adjusting historical dollar values to 2019 dollars: (i) using the (real) GDP-deflator to adjust to real 2019 dollars; and (ii) using nominal GDP as the deflator. The first approach only adjusts for inflation, while the second adjusts for both inflation and the economy-wide ``Romerian'' secular growth trend, thus yielding a stationary scale distribution. To see that, Panels (b) and (c) of Figure~\ref{fig:FSDfacts} both show the time-trend of firm scale distributions over the sample period. Panel (b) uses the real GDP-deflator, while Panel (c) uses the nominal GDP. Panel (c) shows the median firm scale is today at a similar level as in the 70's, after a downward shift during the 80's and 90's, commensurate with the increase in number of firms. It also shows that scale dispersion is fairly stable starting the mid 80's. I will be using nominal GDP as the deflator throughout, which effectively holds the total size of the economy fixed.

\subsection{Scale}
\label{sec:Scale}

I begin the empirical analysis by considering the firm scale distribution (FSD) in the CRSP/Compustat data. There is a large literature in economics on the distribution of firm size and scale, surveyed by \cite{Sutton1997}. Sutton concludes that the FSD is found to be stable over time and approximately Normal. Latter contributions, including \cite{CooleyQuadrini2001}, \cite{CabralMata2003}, \cite{DesaiEtAl2003} and \cite{AngeliniGenerale2008} concentrate on the observed skewness of the FSD, and relate it to financial frictions hampering the growth of younger/smaller firms.

Which economic measure best captures firm size (and its log, firm scale)? The asset pricing literature tends to use market value of equity (EQ) as the preferred measure of size, the firm growth literature tends to use sales (SL), and the corporate finance literature tends to use physical or total capital stock (KP or KT). To these, I add total firm value (VL) and firm expenses (XS). While using expenses to measure firm size is not common, it is in-line with the common approach of using number of employees as a measure of firm size. Using expenses has the benefits of being widely available for all firms and being more holistic, i.e. neutralizing the ``build vs. buy'' decision of firms.

Panel (a) of Table~\ref{tab:FSDDescMom} reports the first four central moments of each scale measure. Most FSDs, with the exception of the physical capital FSD, denoted FSD(KP), have a mean around 6.5 ($\approx$665M 2019\$) and a standard deviation around 2.1. The FSDs have moderate positive skewness around 0.3 and kurtosis very close to 3, i.e., no heavy tails. Results are nearly identical for the \{Non-Bank\} subset, though mean scales are slightly lower, indicating financial and utility firms are larger than the median firm. For the \{All\} subset, mean scales are even lower, indicating filtered firms are significantly smaller than the median firm.

\RPprep{Scale - Descriptive statistics}{0}{0}{FSDDescMom}{%
    Panel (a) of this table presents the first four central moments of the FSD, based on six measures of size. Variable definitions are in Table~\ref{tab:DataDef}. Time-adjusted versions are marked with $\widetilde{XX}$, and time-adjustment procedure is described in Appendix~\ref{sec:AppdxFE}. The panel also presents the persistence of each measure, using the \cite{ArellanoBover1995}/\cite{BlundellBond1998} panel estimator. Panel (b) presents the correlations between the measures, with the upper triangular presenting correlations between unadjusted measures (XX) and the lower triangular correlations between time-adjusted measures ($\widetilde{\text{XX}}$). Panel (c) presents the results of three cointegration tests between all FSD panels, with the first two tests from \cite{Pedroni2004}, and the third from \cite{Westerlund2005}. The first two test the null of no cointegration vs. the alternative that all panels are cointegrated while the third tests vs. the alternative that some panels are cointegrated. Tests are conducted by decade, on the available balanced sample of time-adjusted measures within each decade.
}
\RPtab{%
    \begin{tabularx}{\linewidth}{Frrrrrrrrrrrr}
    \toprule
	\multicolumn{13}{l}{\textit{Panel (a): Scale = log(XX) moments}}\\
	\midrule
	& EQ & VL & KP & KT & SL & XS & $\widetilde{\text{EQ}}$ & $\widetilde{\text{VL}}$ & $\widetilde{\text{KP}}$ & $\widetilde{\text{KT}}$ & $\widetilde{\text{SL}}$ & $\widetilde{\text{XS}}$\\
	\midrule
    $M_{1}$ (mean) & 6.19 & 6.98 & 4.91 & 6.65 & 6.36 & 6.30 & 6.19 & 7.01 & 4.96 & 6.69 & 6.40 & 6.34 \\
    $M_{2}$ (s.d.) & 2.16 & 2.12 & 2.41 & 2.11 & 2.13 & 2.05 & 2.20 & 2.17 & 2.43 & 2.14 & 2.13 & 2.06 \\
    $M_{3}$ (skew) & 0.27 & 0.34 & 0.32 & 0.36 & 0.09 & 0.23 & 0.30 & 0.37 & 0.37 & 0.41 & 0.16 & 0.29 \\
    $M_{4}$ (kurt) & 2.75 & 2.91 & 2.56 & 2.99 & 2.86 & 2.78 & 2.81 & 2.96 & 2.66 & 3.02 & 2.92 & 2.87 \\ \\
    Persistence & 0.87 & 0.93 & 1.18 & 1.15	& 1.01 & 0.76 & 0.97 & 0.98 & 0.98 & 1.01 & 0.92 & 0.84 \\
    s.e.        & .003 & .002 & .003 & .002 & .004 & .004 & .003 & .002 & .002 & .001 & .002 & .003 \\
    \end{tabularx}

    \begin{tabularx}{3.2in}{Frrrrrr}
	\\ \multicolumn{7}{l}{\textit{Panel (b): Scale correlations}}\\
	\midrule
	& EQ & VL & KP & KT & SL & XS \\
	\midrule
	EQ &  --- & .941 & .779 & .864 & .809 & .799 \\
	VL & .938 &  --- & .822 & .967 & .850 & .837 \\
	KP & .804 & .845 &  --- & .854 & .852 & .834 \\
	KT & .864 & .968 & .869 &  --- & .877 & .860 \\
	SL & .830 & .867 & .848 & .885 &  --- & .982 \\
	XS & .819 & .853 & .829 & .867 & .982 &  --- \\
    \end{tabularx}

    \begin{tabularx}{\linewidth}{Fcccccc}
	\\ \multicolumn{7}{l}{\textit{Panel (c): Scale cointegration tests}}\\
	\midrule
	& Phillips-Perron t & p-val & Dicky-Fuller t & p-val & Variance ratio & p-val \\
	70's & 51.12 & $<$0.001 & -65.76 & $<$0.001 & 17.42 & $<$0.001 \\
	80's & 57.38 & $<$0.001 & -57.41 & $<$0.001 & 21.32 & $<$0.001 \\
	90's & 58.97 & $<$0.001 & -65.87 & $<$0.001 & 22.05 & $<$0.001 \\
	00's & 62.51 & $<$0.001 & -75.13 & $<$0.001 & 21.67 & $<$0.001 \\
	10's & 58.97 & $<$0.001 & -60.87 & $<$0.001 & 22.57 & $<$0.001 \\
    \bottomrule
    \end{tabularx}
}

Panel (a) also reports the central moments for time-adjusted versions of each FSD. The process of time-adjustment standardizes each yearly distribution before pooling them together. The standardization sets the location and dispersion of each yearly distribution equal to the sample-wide location and dispersion. This is done to verify we introduce no time-aggregation artifacts into the pooled distribution, and is akin to taking time fixed-effect from the entire FSD. For further details on the distributional fixed-effects transformations I use throughout, see Appendix~\ref{sec:AppdxFE}. The time-adjustment has little impact on the moments, alleviating concerns about aggregation bias.

Visual evidence on the skew-Normality of the FSD is presented in Panels (d)-(i) of Figure~\ref{fig:FSDfacts}. Panel (d) presents the unadjusted FSD(EQ) overlaid with an MLE-fitted Normal distribution, and Panel (e) presents the time-adjusted version,  FSD($\widetilde{\text{EQ}}$). In both cases, the fit is visually close but not exact, as further highlighted by the quantile-quantile (q-q) plots vs. the Normal in Panels (g) and (h). Panels (f) and (i) then present the FSD($\widetilde{\text{EQ}}$) vs. the skew-Normal, with a visually good fit in both the histogram and q-q views. Similar results hold for the other FSD flavors. 

Formal tests for the unadjusted and adjusted versions of all six FSD flavors, against both the Normal and the skew-Normal, are presented in Table~\ref{tab:FSDdist}. The three goodness-of-fit distributional tests I use are the Kolmogorov-Smirnov (K-S), the Chi-square (C-2), and the Anderson-Darling (A-D) tests. The three tests are sensitive to different distributional deviations --- K-S has uniform power throughout, C-2 is more powerful around the center-mass, and A-D is more powerful around the tails --- hence I report results of all three tests. Panel (a) shows that Normality of the FSD is generally rejected across the board (at the 5\% significance level). In contrast, Skew-Normality is not rejected for any FSD flavor (with the exception of physical capital) by either test, as seen in Panel (b) of Table~\ref{tab:FSDdist}.

\RPprep{Scale - Distributional tests}{1}{0}{FSDdist}{%
    This table presents results of tests of distribution equality for the FSD, based on the scale measures described in Table~\ref{tab:DataDef}. K-S is a Kolmogorov–Smirnov test; C-2 is a binned $\chi^2$ test with 50 bins; A-D is an Anderson-Darling test. Panels (a) and (b) report the test statistics and their p-values for the Normal and skew-Normal, respectively. Panel (c) reports the AIC- and BIC-based relative likelihoods for each distribution.
}
\RPtab{%
    \begin{tabularx}{\linewidth}{Frrrrrrrrrrrr}
    \toprule
	& EQ & VL & KP$^{a}$ & KT$^{a}$ & SL & XS & $\widetilde{\text{EQ}}$ & $\widetilde{\text{VL}}$ & $\widetilde{\text{KP}}^{a}$ & $\widetilde{\text{KT}}^{a}$ & $\widetilde{\text{SL}}$ & $\widetilde{\text{XS}}$\\ 
	\midrule
	\\ \multicolumn{13}{l}{\textit{Panel (a): FSD vs. Normal}}\\
	\midrule
    K-S   & 0.029 & 0.026 & 0.023 & 0.028 & 0.010 & 0.022 & 0.029 & 0.031 & 0.028 & 0.034 & 0.017 & 0.028 \\
    p-val & 0.025 & 0.029 & 0.032 & 0.026 & 0.059 & 0.033 & 0.026 & 0.024 & 0.027 & 0.022 & 0.041 & 0.027 \\
    C-2   & 191   & 200   & 207   & 230   & 47    & 160   & 202   & 233   & 261   & 329   & 88    & 221   \\
    p-val & 0.027 & 0.027 & 0.026 & 0.025 & 0.054 & 0.030 & 0.027 & 0.025 & 0.023 & 0.020 & 0.040 & 0.025 \\ 
    A-D   & 17.18 & 16.91 & 16.15 & 19.70 & 2.89  & 11.92 & 18.40 & 21.62 & 20.84 & 28.26 & 7.08  & 18.47 \\
    p-val & 0.026 & 0.027 & 0.027 & 0.025 & 0.052 & 0.031 & 0.026 & 0.024 & 0.024 & 0.021 & 0.038 & 0.026 \\ 
    \\ \multicolumn{13}{l}{\textit{Panel (b): FSD vs. skew-Normal}}\\
	\midrule
    K-S   & 0.011 & 0.009 & 0.015 & 0.008 & 0.006 & 0.010 & 0.006 & 0.005 & 0.014 & 0.007 & 0.005 & 0.007 \\
    p-val & 0.056 & 0.066 & 0.045 & 0.069 & 0.086 & 0.058 & 0.086 & 0.098 & 0.048 & 0.074 & 0.099 & 0.075 \\
    C-2   & 22.47 & 30.51 & 99    & 32    & 22.26 & 41.73 & 16.82 & 17.65 & 67    & 30    & 17.73 & 23.27 \\
    p-val & 0.077 & 0.066 & 0.038 & 0.064 & 0.077 & 0.057 & 0.089 & 0.087 & 0.046 & 0.067 & 0.087 & 0.075 \\ 
    A-D   & 2.163 & 1.639 & 7.660 & 1.410 & 0.681 & 2.252 & 1.061 & 0.599 & 4.410 & 1.160 & 0.534 & 1.106 \\
    p-val & 0.057 & 0.062 & 0.037 & 0.065 & 0.081 & 0.056 & 0.070 & 0.084 & 0.045 & 0.069 & 0.087 & 0.070 \\ 
    \\ \multicolumn{13}{l}{\textit{Panel (c): Distribution comparison}}\\
	\midrule
	\multicolumn{2}{l}{AIC R.L.:} \\
    Normal  & 0.000 & 0.000 & 0.000 & 0.000 & 0.000 & 0.000 & 0.000 & 0.000 & 0.000 & 0.000 & 0.000 & 0.000 \\
    Skew-N & 1.000 & 1.000 & 1.000 & 1.000 & 1.000 & 1.000 & 1.000 & 1.000 & 1.000 & 1.000 & 1.000 & 1.000 \\
	\multicolumn{2}{l}{BIC R.L.:} \\
    Normal  & 0.000 & 0.000 & 0.000 & 0.000 & 0.000 & 0.000 & 0.000 & 0.000 & 0.000 & 0.000 & 0.000 & 0.000 \\
    Skew-N & 1.000 & 1.000 & 1.000 & 1.000 & 1.000 & 1.000 & 1.000 & 1.000 & 1.000 & 1.000 & 1.000 & 1.000 \\
	\bottomrule
    \end{tabularx}
    \begin{flushleft}
    $^a$ Using the \{Non-Bank\} data subset \\
    \end{flushleft}    
}

The better fit of the skew-Normal distribution might not be too surprising, given that it has an extra degree of freedom (i.e., an extra parameter). To account for the degrees of freedom, I use the relative likelihood test, derived from the AIC statistic of \cite{Akaike1973}. The relative likelihood is a non-nested version of the likelihood ratio test, accounting for the number of parameters.\footnote{For a review of the information-theoretic approach to model selection see, e.g., \cite{BurnhamAnderson2002}.} I also report relative likelihood tests using the BIC statistic, which penalizes extra degrees of freedom more heavily. Panel (c) presents the relative likelihood tests of the Normal and the skew-Normal, showing that the skew-Normal yields a significant improvement over the Normal as a statistical model for the FSD, even after penalizing for the extra parameter.

The data suggest skew-Normality has excellent fit to the public-firm FSD. Given the large extant literature regarding the FSD, the most puzzling fact about the FSD in the CRSP/Compustat data is how unpuzzling it is. I conclude:

\begin{observation}
    \label{ob:1}
    The FSD is distributed skew-Normal.
\end{observation}

After exploring the cross-sectional dimension of the FSD data, I turn to time-dimension facts. The last two rows in Panel (a) of Table~\ref{tab:FSDDescMom} present the persistence of each scale measure using the system GMM estimator of \cite{ArellanoBover1995} and \cite{BlundellBond1998}. The persistence of time-adjusted values is of greater interest as it removes spurious effects stemming from the business cycle. The persistence of values and capitals is close to 1, with sales somewhat less persistent and expenses the least persistent. I conclude:

\begin{observation}
    \label{ob:2}
    Value and capital scales are highly persistent. Sales and particularly expenses somewhat less so.
\end{observation}

The relation between the different measures is explored in Panels (b) and (c) of Table~\ref{tab:FSDDescMom}. The simple pooled correlations between each scale measure are presented in Panel (b), for both adjusted and unadjusted scales. The two highest correlations are between (i) sales and expenses; and (ii) firm value and total capital. The lowest correlation is between equity value and physical capital. Panel (c) then presents panel cointegration tests between the different measures. Because the cointegration tests of \cite{Pedroni2004} and \cite{Westerlund2005} require balanced panels, Panel (c) presents the tests by decade and includes all balanced panels available within the decade. The hypothesis of no cointegration is strongly rejected by all tests for all decades. I conclude:

\begin{observation}
    \label{ob:3}
    FSD measures are cointegrated.
\end{observation}

\subsection{Income}
\label{sec:Income}

Firm income is a variable of considerable importance in the theory of firm scale. Income is both the means to scaling (as it provides money for investments) and the ends of scaling (as it provides money for dividends). This section mainly considers the firm net income distribution, denoted FID(CF) and the income of firm owners --- equity and debt holders --- total dividends, whose distribution is denoted FID(DI). It also presents results for the alternative top-down definition of income FID(CA), commonly used in the literature, and for dispensations to equity holders alone, denoted  FID(DE).

Panels (a) and (d) of Figure~\ref{fig:FIDfacts} present truncated views of FID(CF) and FID(DI), respectively. The panels help explain why income measures are not typically used as measures of scale: income exhibits exponential tails in \emph{both the positive and negative directions}. The common way of dealing with exponential tails --- applying a log transform --- hence cannot be used. About 20\% of CF observations and 40\% of DI observations are negative, corresponding to losses and capital infusions from owners, respectively, and so ignoring negative values is untenable.

\RPprep{Income - Stylized facts}{0}{0}{FIDfacts}{%
    This figure presents stylized facts of the firm income distribution (FID). Panel (a) presents the truncated distribution of FID(CF) in linear scale and Panel (b) presents the untruncated distribution in asinh scale, both overlaid with ML-fitted DLN distributions. Panel (c) presents the q-q for the fit of FID(CF) to the DLN distribution. Panels (d)-(f) repeat for owner income, FID(DI). Panel (g) presents the dependence of the FID on firm scale, by presenting the (10,25,50,75,90)$^{th}$ percentiles of FID(CF) conditional on the sign of CF, for 49 KT scale bins. Panel (h) presents the time- and scale-adjusted FID($\widetilde{\text{CF}}$), with scale adjustment done by considering intensity relative to total capital KT. Panel (i) presents the dependence of FID($\widetilde{\text{CF}}$) on firm scale. Panels (g)-(i) use the \{Non-Bank\} subset.
}
\RPfig{%
	\begin{tabular}{ccc} 
		\subfigure[FID(CF) w/ DLN ]{\includegraphics[width=2in]{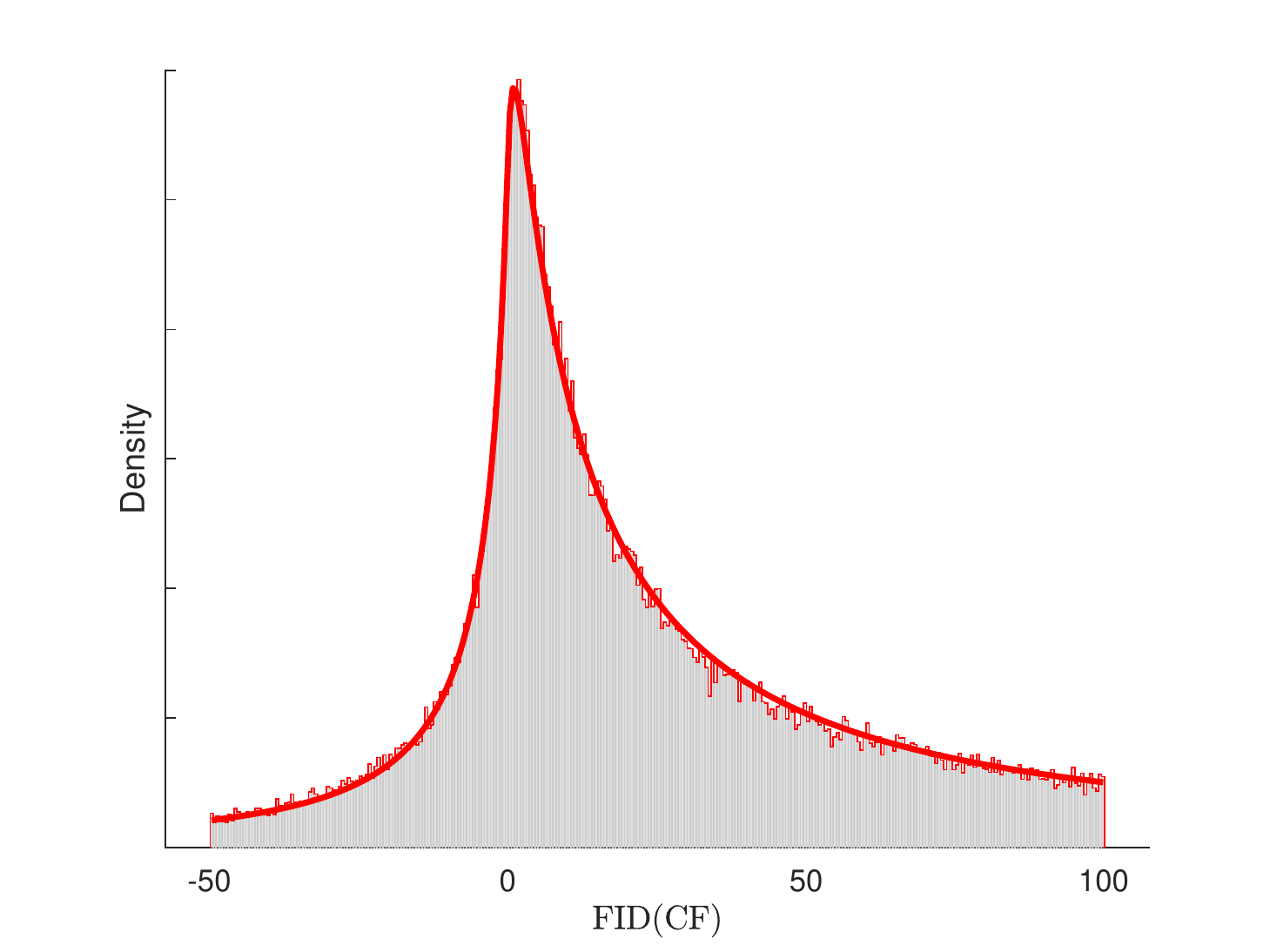}} & \subfigure[FID(CF) w/ DLN ]{\includegraphics[width=2in]{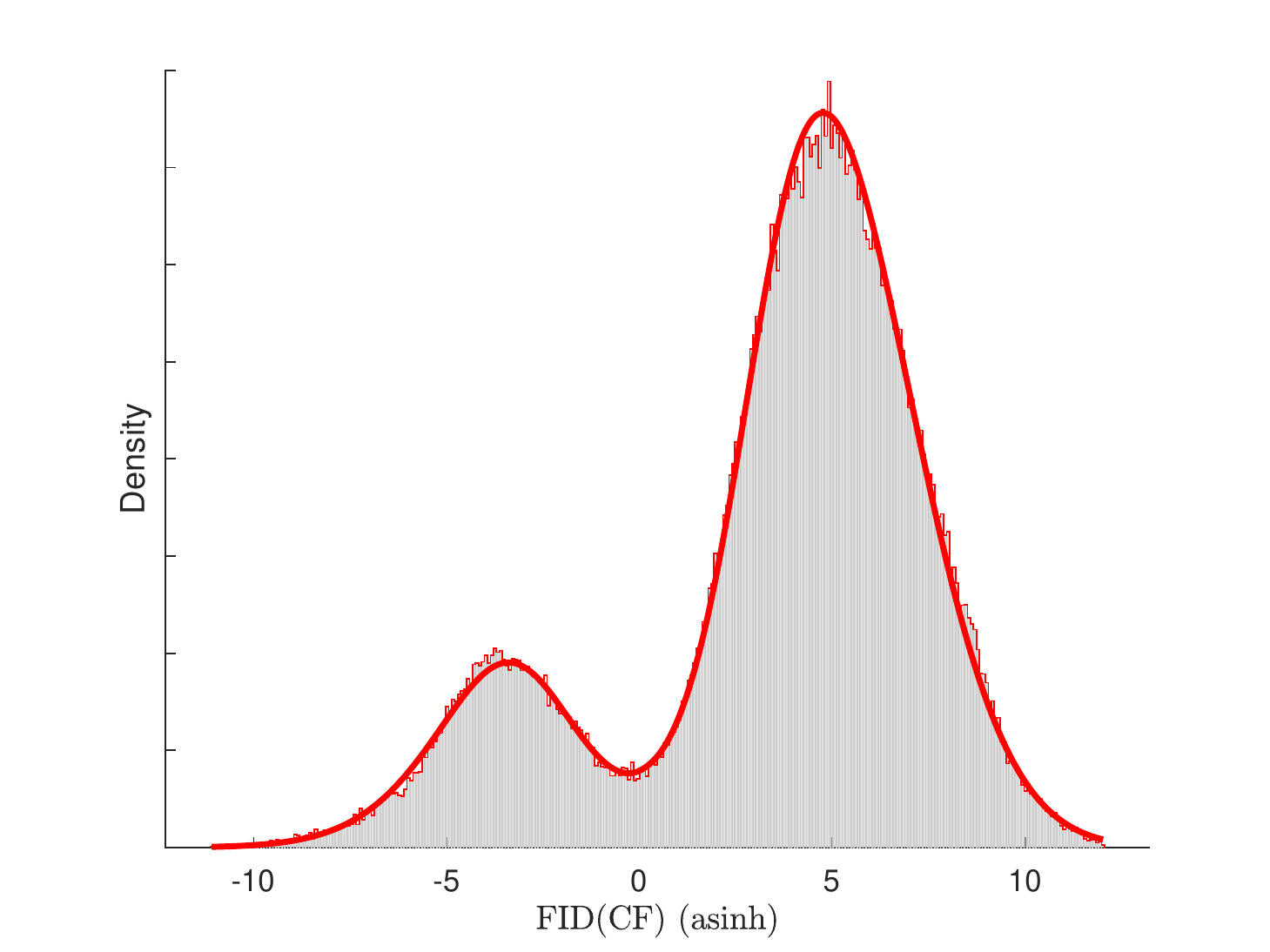}} &
		\subfigure[q-q FID(CF) vs. DLN ]{\includegraphics[width=2in]{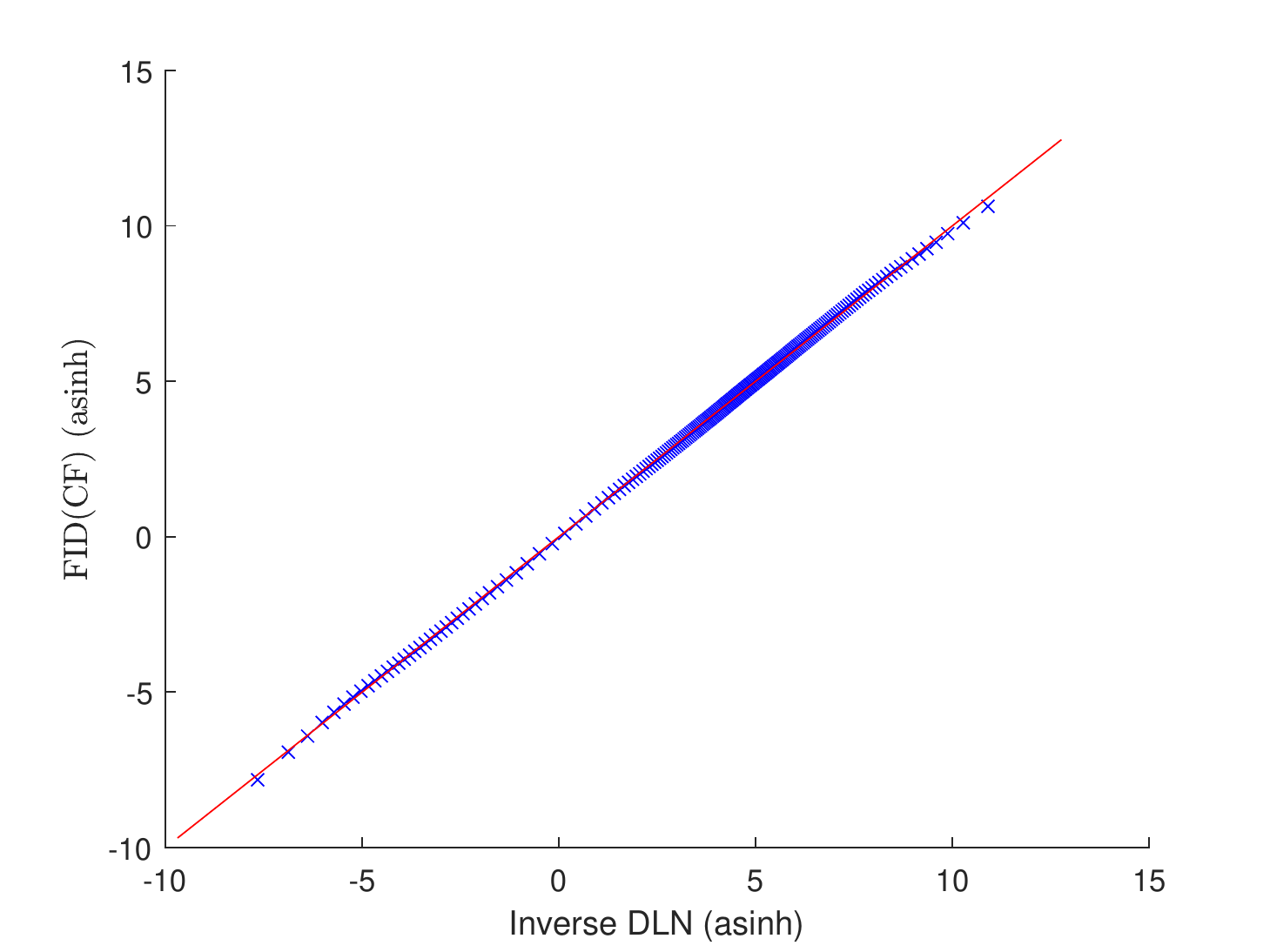}} \\ \\
		\subfigure[FID(DI) w/ DLN ]{\includegraphics[width=2in]{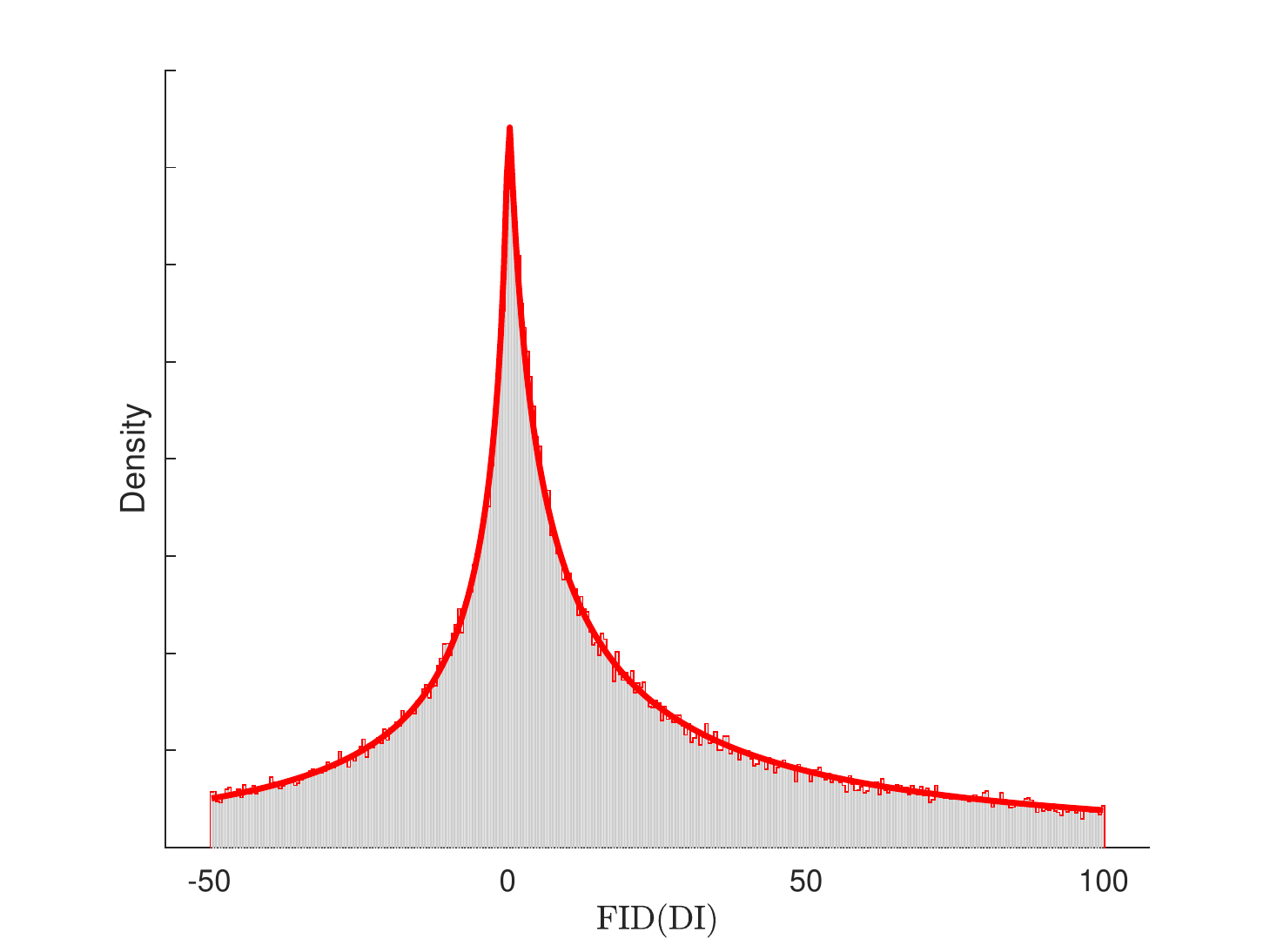}} &
		\subfigure[FID(DI) w/ DLN ]{\includegraphics[width=2in]{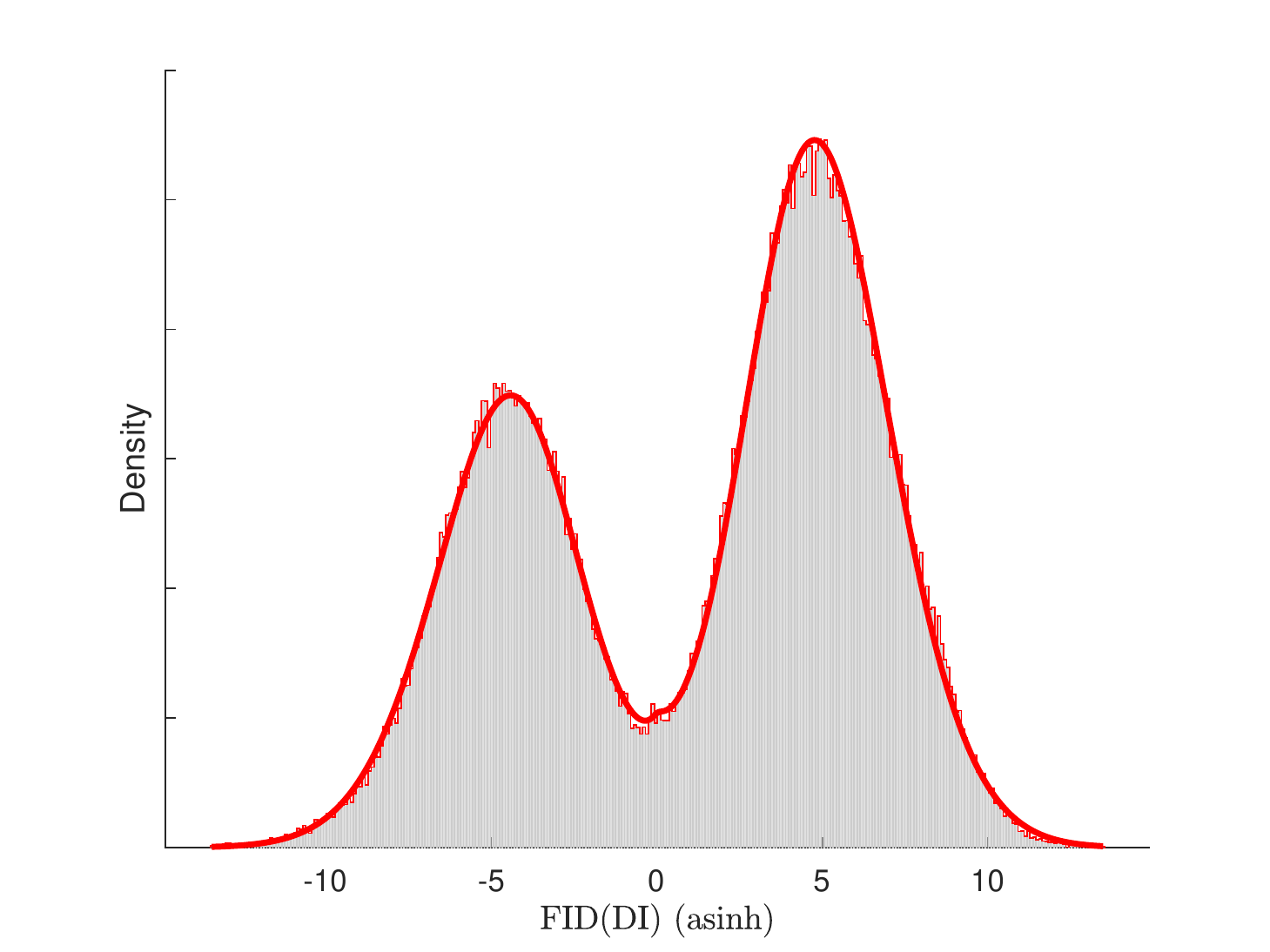}} &
		\subfigure[q-q FID(DI) vs. DLN ]{\includegraphics[width=2in]{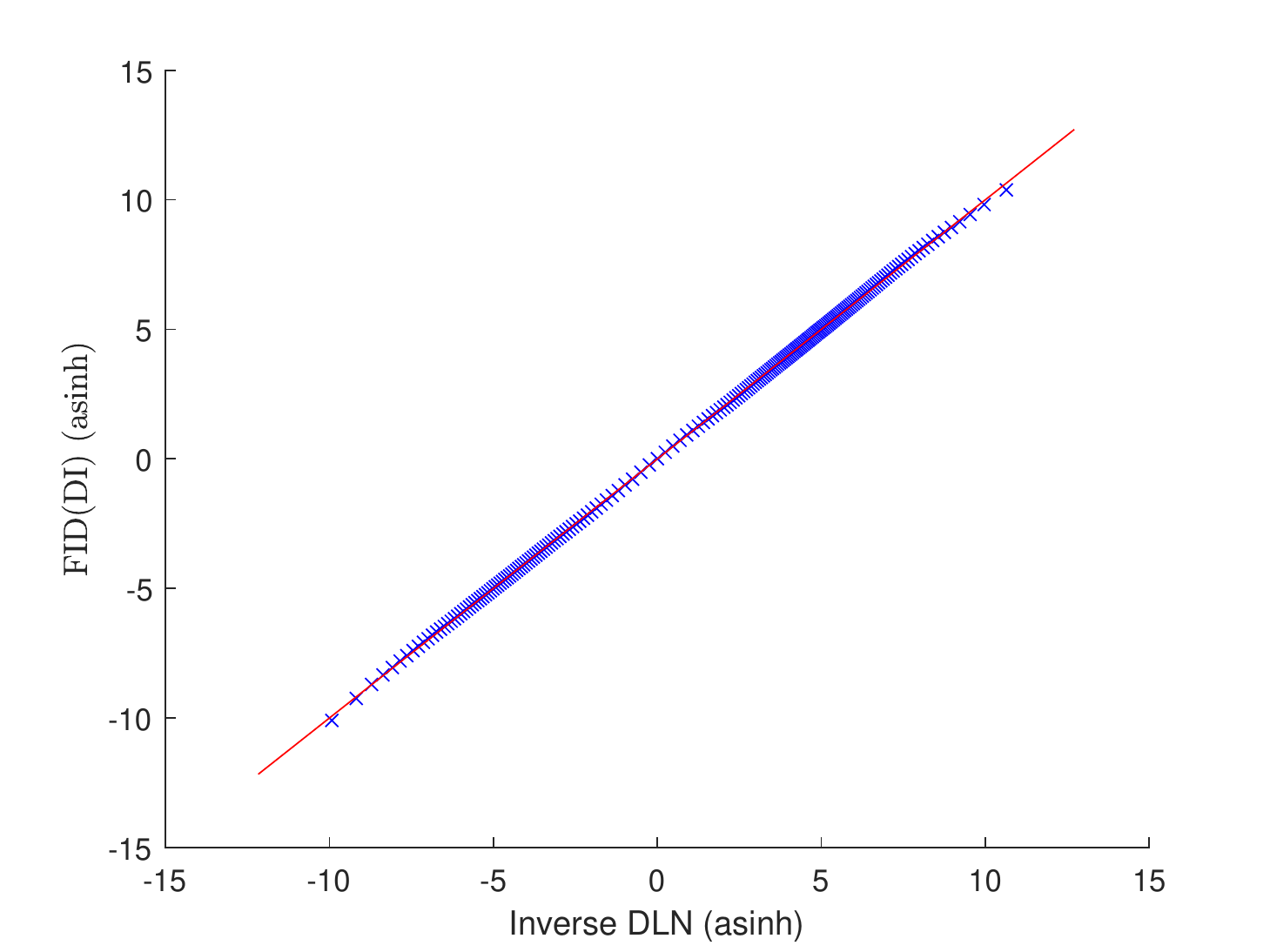}} \\ \\
		\subfigure[FID(CF) by scale ]{\includegraphics[width=2in]{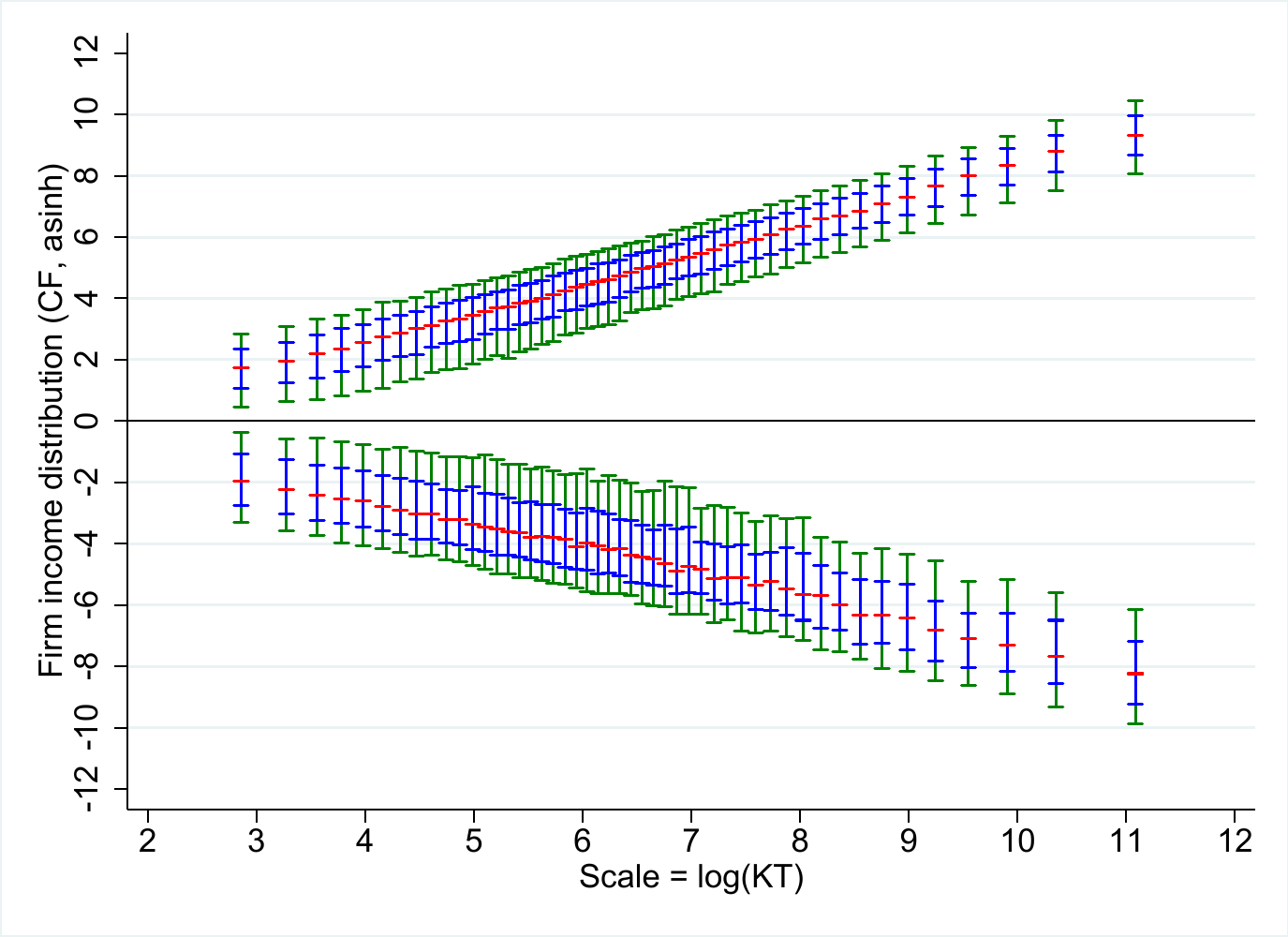}} &
		\subfigure[FID($\widetilde{\text{CF}}$) w/ DLN ]{\includegraphics[width=2in]{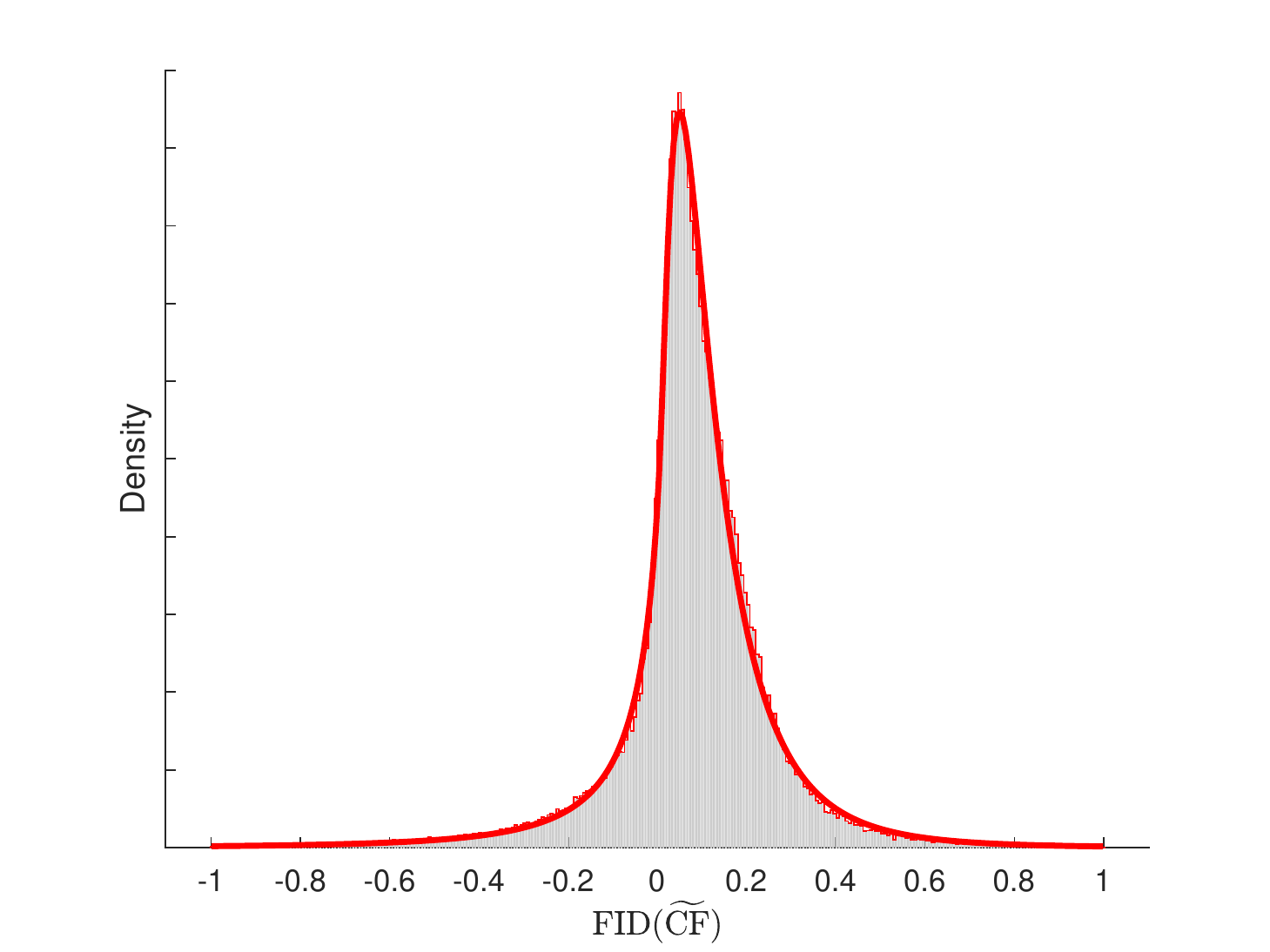}} &
		\subfigure[FID($\widetilde{\text{CF}}$) by scale ]{\includegraphics[width=2in]{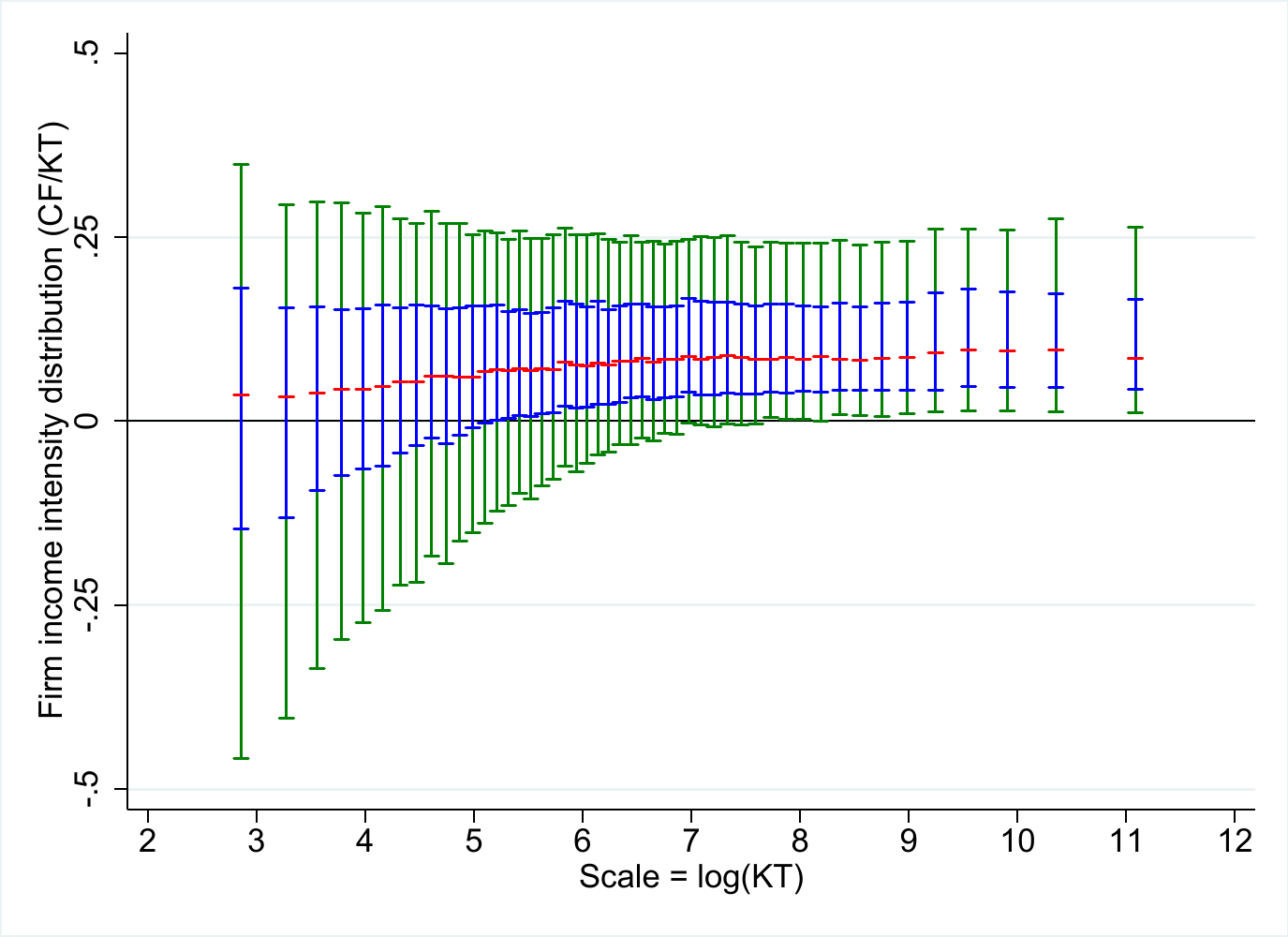}} \\ \\
	\end{tabular}
}

To deal with the double-exponential nature of the tails, Panels (b) and (e) then present the same data, untruncated but with the x-axis transformed to \emph{Inverse Hyperbolic Sine} (asinh) scale. The asinh scale can simply be thought of as a log scale, but in both the positive and negative directions, and allows us to view the entire distribution. Panels (a), (b), (d), (e) are overlaid with MLE-fitted difference-of-log-Normals (DLN) distributions, exhibiting excellent fit. The accompanying q-q plots in Panels (c) and (f) confirm this observation. The companion paper \cite{Parham2022a} provides econometric details, describes the DLN distribution, the sinh and asinh transforms, and the rationale and benefits of using them.

Panel (g) considers the dependence of income on scale (using total capital KT as the measure of scale). I first split the data into $49$ equal bins, based on firm scale, truncating the top and bottom 1\% of the FSD(KT), such that each bin contains 2\% of the observations. For each bin, Panel (g) plots the $(10,25,50,75,90)^{th}$ percentiles of FID(CF). The panel shows income is naturally increasing with scale --- larger firms earn and dispense (or lose and are infused) more money than smaller firms --- so the ``middle'' of the panel hollows as firm scale rises. Panel (h) hence presents the time- and scale-adjusted FID, denoted FID($\widetilde{\text{CF}}$), overlaid with fitted DLN distribution. Adjusting to time-effects is done as before. To account for the ubiquitous scale-effects, I consider income intensity --- income divided by size (KT). The visual fit of the adjusted distribution is again excellent. Panel (i) presents the income intensity distribution by scale, showing scale-dependant systematic patterns in income intensity as well. Income intensity increases with size, and the number of firms reporting losses decreases. Similar results hold for the distributions of CA, DI, and DE, and when adjusting by other measures of scale.

Table~\ref{tab:FIDDescMom} reports descriptive statistics of the FID. Due to the double-exponential nature of the tails, the table presents the statistics separately for positive and negative incomes, in asinh scale. We observe low skewness and kurtosis close to 3 for both halves of the distribution, again indicating approximate (log-)Normality on each side. The 1-lag persistence on the positive side is relatively high, ranging from $0.3$ to $0.8$, while on the negative side persistence is low or negative. This implies firms with negative income revert to positivity quickly. The table reports the slopes of increasing median on the positive and negative sides (e.g., the slopes in panel (g) of Figure~\ref{fig:FIDfacts} for CF). The positive slopes are very close to 1, justifying the use of intensities. Table~\ref{tab:FIDDescMom} also reports the moments for the income intensity distributions --- all are skewed and all are significantly heavy-tailed, with relatively low persistence, and no increasing intensity w.r.t scale.

\RPprep{Income - Descriptive statistics}{0}{0}{FIDDescMom}{%
    The first four columns of this table present statistics for the positive side of the FID, while the next four concentrate on the negative side and the last four on the entire income intensity distribution (i.e., income divided by total capital KT). Variable definitions are in Table~\ref{tab:DataDef}. The positive and negative sides are in asinh scale. $M_1$ to $M_4$ are the first four central moments. Persistence for the positives and negatives is the 1-lag pooled auto-correlation, while for intensities it is the \cite{ArellanoBover1995}/\cite{BlundellBond1998} panel estimate. \%Ob is the percent of total observations in the positive/negative side, and may not sum to 100\% due to zero-valued obs. The scaling coefficient is the slope in a quantile regression of (asinh) income or of intensity on scale.
}
\RPtab{%
    \begin{tabularx}{\linewidth}{Frrrrrrrrrrrr}
    \toprule
	& CF$^{+}$ & CA$^{+}$ & DI$^{+}$ & DE$^{+}$ & CF$^{-}$ & CA$^{-}$ & DI$^{-}$ & DE$^{-}$ & $\widetilde{\text{CF}}$ & $\widetilde{\text{CA}}$ & $\widetilde{\text{DI}}$ & $\widetilde{\text{DE}}$\\
	\midrule
    $M_{1}$ & 5.02 & 5.00 & 4.88 & 3.94 & -3.59 & -3.04 & -4.53 & -2.57 & 0.07  & 0.09  & -0.02 & -0.02  \\
    $M_{2}$ & 2.15 & 2.06 & 2.24 & 2.30 & 1.90  & 1.58  & 2.19  & 2.09  & 0.20  & 0.18  & 0.33  & 0.22   \\
    $M_{3}$ & 0.20 & 0.26 & 0.20 & 0.30 & -0.52 & -0.37 & -0.34 & -0.61 & -0.06 & -3.19 & -4.72 & -9.20  \\
    $M_{4}$ & 2.73 & 2.88 & 2.76 & 2.44 & 3.37  & 3.20  & 3.15  & 2.50  & 47    & 69    & 46    & 135    \\
    Pers.   & 0.59 & 0.80 & 0.33 & 0.62 & -0.10 & 0.20  & 0.01  & -0.02 & 0.16  & 0.48  & 0.04  & 0.06   \\
    \%Ob  & 0.80 & 0.75 & 0.59 & 0.56 & 0.20  & 0.11  & 0.41  & 0.33  & /     & /     & /     & /      \\
    $\beta_{scl}$ & 0.96 & 0.99 & 0.97 & 0.98 & -0.70 & -0.68 & -0.87 & -0.73 & 0.00  & 0.00  & 0.00  & 0.00   \\
    \end{tabularx}
}

The excellent fit of the DLN, suggested in Figure~\ref{fig:FIDfacts}, is confirmed by Table~\ref{tab:FIDdist}, which presents the formal distributional tests. Given the double-exponential nature of income, I concentrate on tests vs. the Stable, Laplace, and DLN distributions --- all yielding double-exponential (i.e. heavy) tails. The tests strongly reject the Stable and Laplace on all income and income intensity variables considered, while the DLN is generally not rejected (with the exception of equity dispensations --- DE). The relative likelihood tests strongly favor the DLN as a statistical model of income, even for the rejected DE. Upon inspection, the DE data include a large mass of observations around zero distorting the shape of the distribution and likely driven by fixed costs leading to an inaction region in firm policy. I conclude:

\begin{observation}
    \label{ob:4}
    The FID is distributed DLN.
\end{observation}

\RPprep{Income - Distributional tests}{1}{0}{FIDdist}{%
    This table presents results of tests of distribution equality for the four measures of income: CF, CA, DI, and DE, described in Table~\ref{tab:DataDef}, as well as their time- and scale-adjusted versions. Scale adjustments use total capital (KT) as the size measure. K-S is a Kolmogorov–Smirnov test; C-2 is a binned $\chi^2$ test with 50 bins; A-D is an Anderson-Darling test. Panels (a)-(c) report the test statistics and their p-values for the Stable, Laplace, and DLN, respectively. Panel (d) reports the relative likelihoods for each distribution. The raw variables (columns 1-4) use the \{Good\} data subset. The scale adjusted versions use the \{Non-Bank\} subset.
}
\RPtab{%
    \begin{tabularx}{\linewidth}{Frrrrrrrr}
    \toprule
	& CF & CA & DI & DE & $\widetilde{\text{CF}}$ & $\widetilde{\text{CA}}$ & $\widetilde{\text{DI}}$ & $\widetilde{\text{DE}}$\\ 
	\midrule
	\\ \multicolumn{9}{l}{\textit{Panel (a): Income vs. Stable}}\\
	\midrule
    K-S   & 0.035 & 0.033 & 0.023 & 0.041 & 0.040 & 0.050 & 0.055 & 0.096 \\
    p-val & 0.021 & 0.022 & 0.032 & 0.017 & 0.018 & 0.013 & 0.011 & 0.001 \\
    C-2   & 698   & $>$999& 364   & $>$999& 366   & 677   & 523   & $>$999\\
    p-val & 0.011 & 0.007 & 0.019 & 0.005 & 0.019 & 0.011 & 0.014 & 0.003 \\ 
    A-D   & 23.66 & 29.03 & 17.27 & 45.30 & 24.57 & 39.41 & 31.26 & 224.9 \\
    p-val & 0.023 & 0.020 & 0.026 & 0.016 & 0.022 & 0.017 & 0.020 & 0.001 \\ 
    \\ \multicolumn{9}{l}{\textit{Panel (b): Income vs. Laplace}}\\
	\midrule
    K-S   & 0.365 & 0.401 & 0.260 & 0.369 & 0.071 & 0.071 & 0.123 & 0.268 \\
    p-val & 0.000 & 0.000 & 0.000 & 0.000 & 0.006 & 0.006 & 0.000 & 0.000 \\
    C-2   & $>$999& $>$999& $>$999& $>$999& 885   & 953   & $>$999& $>$999\\
    p-val & 0.000 & 0.000 & 0.000 & 0.000 & 0.009 & 0.008 & 0.000 & 0.000 \\ 
    A-D   & $>$999& $>$999& $>$999& $>$999& 54.48 & 55.89 & 190.4 & 761.2 \\
    p-val & 0.000 & 0.000 & 0.000 & 0.000 & 0.014 & 0.013 & 0.002 & 0.000 \\
    \\ \multicolumn{9}{l}{\textit{Panel (c): Income vs. DLN}}\\
	\midrule
    K-S   & 0.005 & 0.004 & 0.004 & 0.030 & 0.010 & 0.022 & 0.006 & 0.132 \\
    p-val & 0.092 & 0.108 & 0.120 & 0.024 & 0.061 & 0.033 & 0.088 & 0.000 \\
    C-2   & 16    & 16    & 11    & 183   & 57    & 152   & 20    & $>$999\\
    p-val & 0.092 & 0.091 & 0.117 & 0.028 & 0.049 & 0.031 & 0.082 & 0.000 \\ 
    A-D   & 0.70  & 0.39  & 0.29  & 10.36 & 1.86  & 7.36  & 0.47  & 204.08\\
    p-val & 0.080 & 0.095 & 0.105 & 0.033 & 0.059 & 0.037 & 0.090 & 0.001 \\
    \\ \multicolumn{9}{l}{\textit{Panel (d): Distribution comparison}}\\
	\midrule
	\multicolumn{2}{l}{AIC R.L.:} \\
    Stable  & 0.000 & 0.000 & 0.000 & 0.000 & 0.000 & 0.000 & 0.000 & 0.000 \\
    Laplace & 0.000 & 0.000 & 0.000 & 0.000 & 0.000 & 0.000 & 0.000 & 0.000 \\
    DLN     & 1.000 & 1.000 & 1.000 & 1.000 & 1.000 & 1.000 & 1.000 & 1.000 \\
	\multicolumn{2}{l}{BIC R.L.:} \\
    Stable  & 0.000 & 0.000 & 0.000 & 0.000 & 0.000 & 0.000 & 0.000 & 0.000 \\
    Laplace & 0.000 & 0.000 & 0.000 & 0.000 & 0.000 & 0.000 & 0.000 & 0.000 \\
    DLN     & 1.000 & 1.000 & 1.000 & 1.000 & 1.000 & 1.000 & 1.000 & 1.000 \\
	\bottomrule
    \end{tabularx}
}

\section{Growth, return, and income growth}
\label{sec:GrowthReturn}

\subsection{Firm growth}
\label{sec:Growth}

The statistical distribution of firm growth rates (henceforth FGD) in the data has been of considerable interest to scholars. \cite{Ashton1926} is the first to document that the growth of British textile businesses in the period $1884-1924$, measured by the number of spindles employed, was heavy-tailed. Nevertheless, the work of \cite{Gibrat1931}, assuming growth is Gaussian, remained the workhorse of firm growth modelling. ``Gibrat's law'' and the dominance of the assumption that growth is Normally distributed stem from a simple, theoretically appealing application of the Central Limit Theorem (CLT). If yearly growth is the result of many smaller i.i.d shocks with finite variance, then yearly growth converges to a Normal distribution. There is now a large literature empirically rejecting Gibrat's predictions. See review by \cite{Sutton1997}. I defer further discussion of the literature on firm growth and its various findings to the companion paper \cite{Parham2022c}, and concentrate henceforth on the data.

Two common ways of measuring growth are using (i) percents and (ii) log-point growth rates. \cite{Parham2022a} (Section 3.2) discusses measures of growth and shows the appropriate measure for firm growth is log-point growth, as percentage growth introduces a \emph{convexity bias}. Let $M_t$ denote one of the measures of firm size discussed above in Section~\ref{sec:Scale}. Firm growth is then defined as the difference in log size between subsequent periods
\begin{equation} \label{eq:dlog}
\text{dlog}(M_t) = \log(M_t) - \log(M_{t-1})
\end{equation}

The first six columns in Panel (a) of Table~\ref{tab:FGDDescMom} present the descriptive statistics of the various firm growth distributions (FGD). Mean growth is close to zero, as expected from a stationary distribution of size. Almost all measures exhibit negative skewness, and all measures have kurtosis significantly higher than 3, indicating heavy tails. Note that equity and total value growth (EQ and VL) adjust for capital dispensations (and infusions), making them proper measures of investment return, at the yearly horizon.

\RPprep{Growth - Descriptive statistics}{0}{0}{FGDDescMom}{%
    Panel (a) of this table presents the first four central moments of the firm growth distribution (FGD), based on six measures of size. Variable definitions are in Table~\ref{tab:DataDef}. Time- and scale-adjusted versions are marked with $\widetilde{XX}$, and the adjustment procedure is described in Appendix~\ref{sec:AppdxFE}. The panel also presents the persistence of each growth measure, using the \cite{ArellanoBover1995}/\cite{BlundellBond1998} panel estimator. Panel (b) presents the correlations between the growth measures, with the upper triangular presenting correlations between unadjusted measures (XX) and the lower triangular correlations between time- and scale-adjusted measures ($\widetilde{\text{XX}}$).
}
\RPtab{%
    \begin{tabularx}{\linewidth}{Frrrrrrrrrrrr}
    \toprule
	\multicolumn{13}{l}{\textit{Panel (a): Growth = dlog(XX) moments}}\\
	\midrule
	& EQ & VL & KP & KT & SL & XS & $\widetilde{\text{EQ}}$ & $\widetilde{\text{VL}}$ & $\widetilde{\text{KP}}$ & $\widetilde{\text{KT}}$ & $\widetilde{\text{SL}}$ & $\widetilde{\text{XS}}$\\
	\midrule
    $M_{1}$ & -0.02 &  0.02 &  0.01 &  0.01 &  0.01 &  0.02 & -0.01 &  0.02 &  0.00 &  0.01 &  0.01 &  0.01 \\
    $M_{2}$ &  0.57 &  0.34 &  0.33 &  0.24 &  0.30 &  0.37 &  0.54 &  0.30 &  0.31 &  0.23 &  0.27 &  0.34 \\
    $M_{3}$ & -0.71 & -0.41 & -0.36 &  0.31 & -0.54 & -0.04 & -0.79 & -0.52 & -2.76 & -0.34 & -2.46 & -1.64 \\
    $M_{4}$ &  8.84 &  13.3 &  27.2 &  16.5 &  35.6 &  28.3 &  8.56 &  10.8 &  64.4 &  23.1 &  53.5 &  32.3 \\ \\
    Pers.   & -0.05 & -0.08 &  0.16 &  0.16 &  0.07 & -0.17 & -0.01 & -0.03 &  0.15 &  0.16 &  0.11 & -0.13 \\
    s.e.    & .002 & .002 & .003 & .003 & .003 & .002 & .002 & .002 & .003 & .003 & .003 & .002 \\
    \end{tabularx}

    \begin{tabularx}{3.2in}{Frrrrrr}
	\\ \multicolumn{7}{l}{\textit{Panel (b): Growth correlations}}\\
	\midrule
	& EQ & VL & KP & KT & SL & XS \\
	\midrule
	EQ &  --- & .867 & .136 & .324 & .203 & .001 \\
	VL & .871 &  --- & .113 & .316 & .191 & .004 \\
	KP & .138 & .104 &  --- & .579 & .376 & .214 \\
	KT & .306 & .275 & .611 &  --- & .501 & .198 \\
	SL & .234 & .207 & .416 & .530 &  --- & .514 \\
	XS & .022 & .026 & .217 & .209 & .584 &  --- \\
    \end{tabularx}
}

Figure~\ref{fig:FGDfacts} presents graphical evidence on the FGD. Panel (a) presents a histogram of the growth in sales, denoted FGD(SL), along with two fitted Normal curves. The first curve is the standard MLE-fitted Normal (i.e., fitted by minimizing a squared-error metric), while the second is instead fitted by minimizing least absolute deviations (LAD). When heavy tails are involved, the two minimization objectives yield significantly different results, as squared-errors give more weight to fitting the tails, while LAD give more weight to fitting the center-mass (at the expense of the tails). The deviation from Normality is striking, as can be seen in the MLE-fitted Normal's q-q plot in Panel (b) and the LAD-fitted Normal's q-q in Panel (c). The LAD-fitted Normal shows better fit around the center, but exhibits even stronger deviation around the tails. Panel (d) presents the q-q vs. the DLN, showing an excellent fit. Similar results hold when using the other size measures.

\RPprep{Growth - Stylized facts}{0}{0}{FGDfacts}{%
    This figure presents stylized facts of the firm growth distribution (FGD). Panel (a) presents the FGD(SL), with MLE- and LAD-fitted Normal distributions, and panels (b) and (c) present the respective q-q plots. Panel (d) presents the q-q plot of FGD(SL) vs. the DLN. Panel (e) presents the FGD time trend by plotting the $\{25,50,75\}^{th}$ percentiles of the FGD per year. Panel (f) plots binscatters of the median and IQR of FGD(SL) per 49 scale bins, ignoring the top and bottom 1\% of the data. Panel (g) presents the (10,25,50,75,90)$^{th}$ percentiles of FGD(SL) for the 49 scale bins. Panel (h) presents the time- and scale-adjusted FGD($\widetilde{\text{SL}}$), overlaid with an MLE-fitted DLN, and Panel (i) presents the respective q-q plot.
}
\RPfig{%
	\begin{tabular}{ccc} 
		\subfigure[FGD(SL) w/ Normals ]{\includegraphics[width=2in]{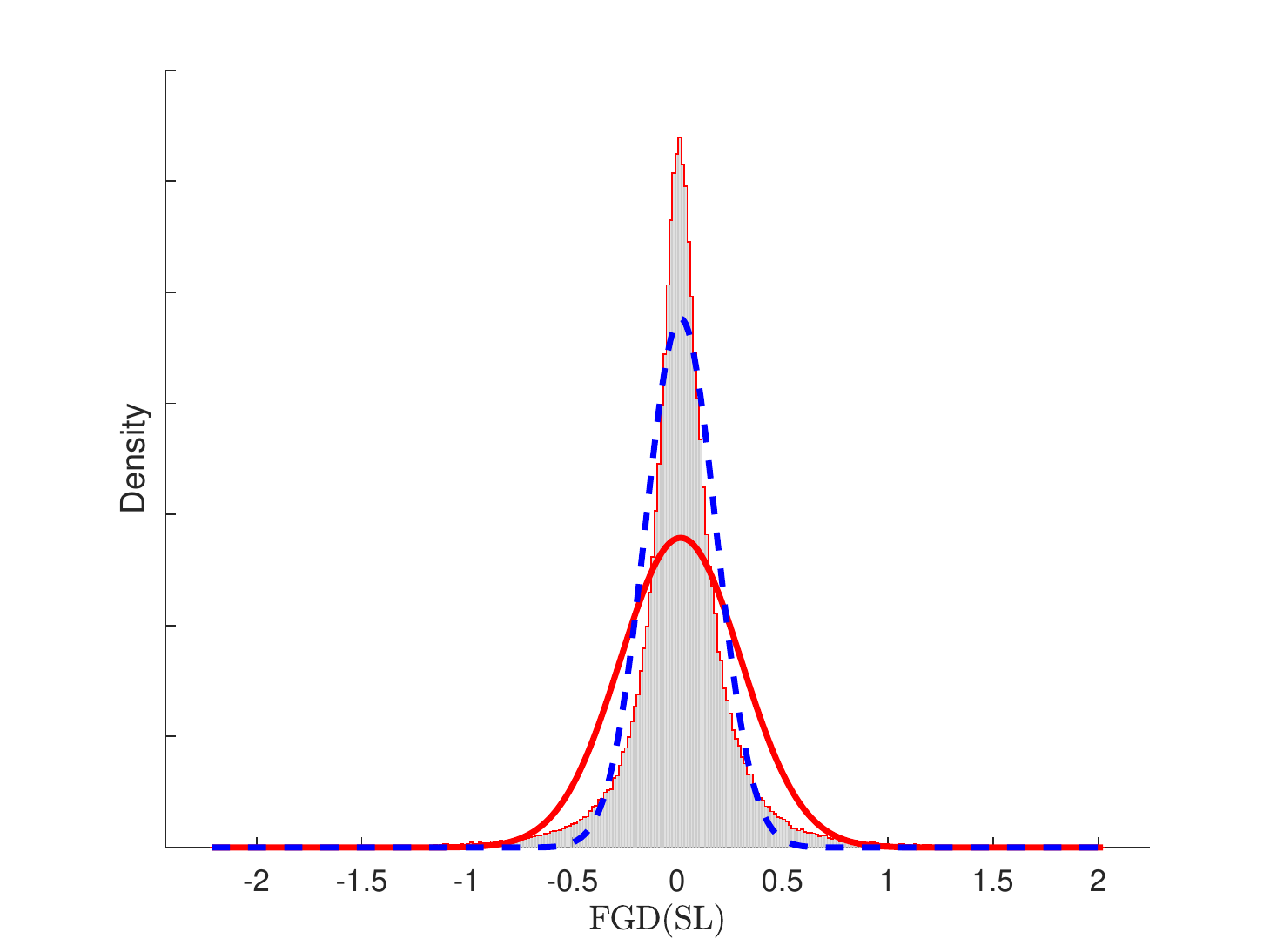}} & \subfigure[q-q FGD(SL) vs. Normal ]{\includegraphics[width=2in]{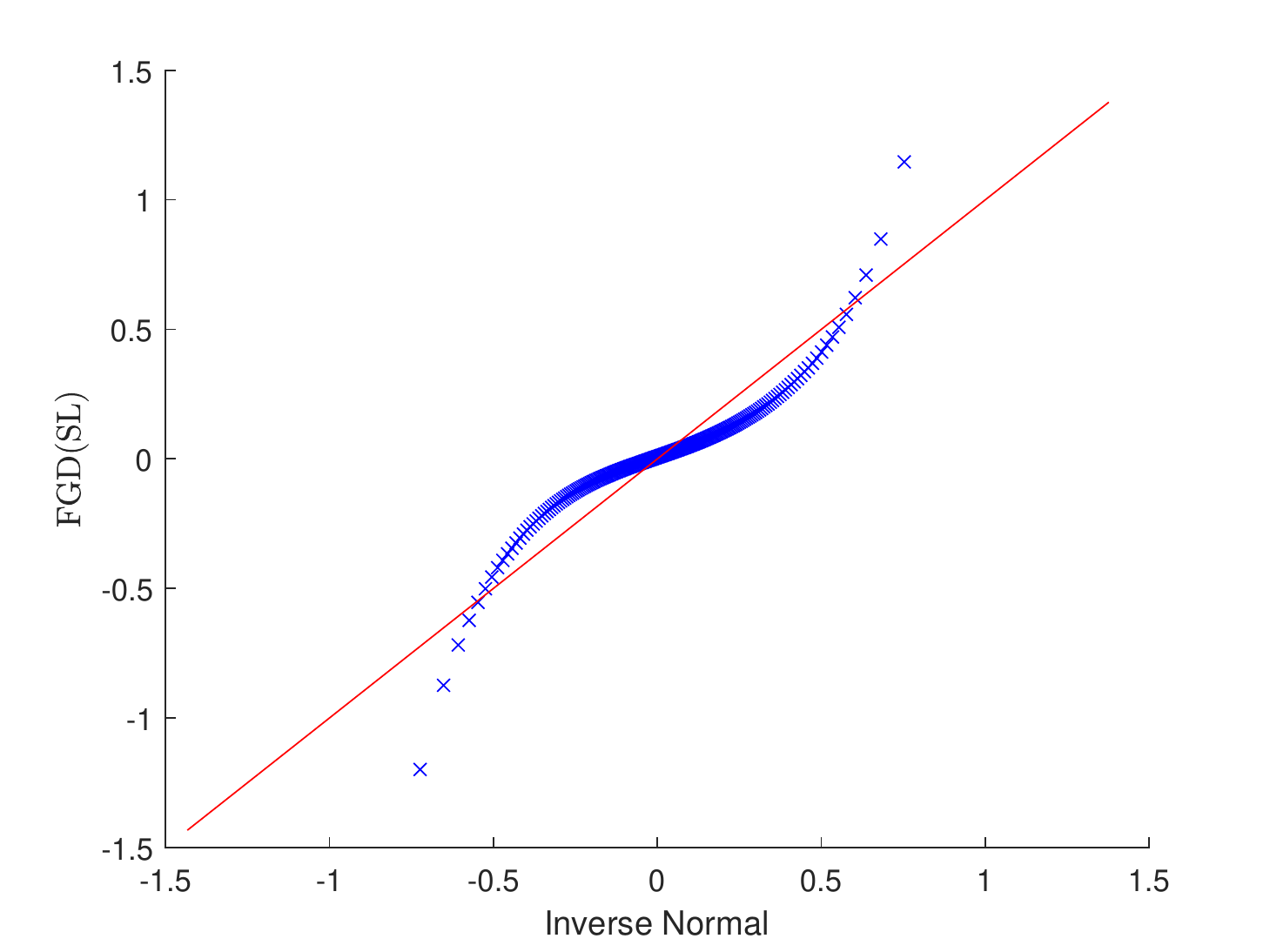}} &
		\subfigure[q-q FGD(SL) vs. Normal ]{\includegraphics[width=2in]{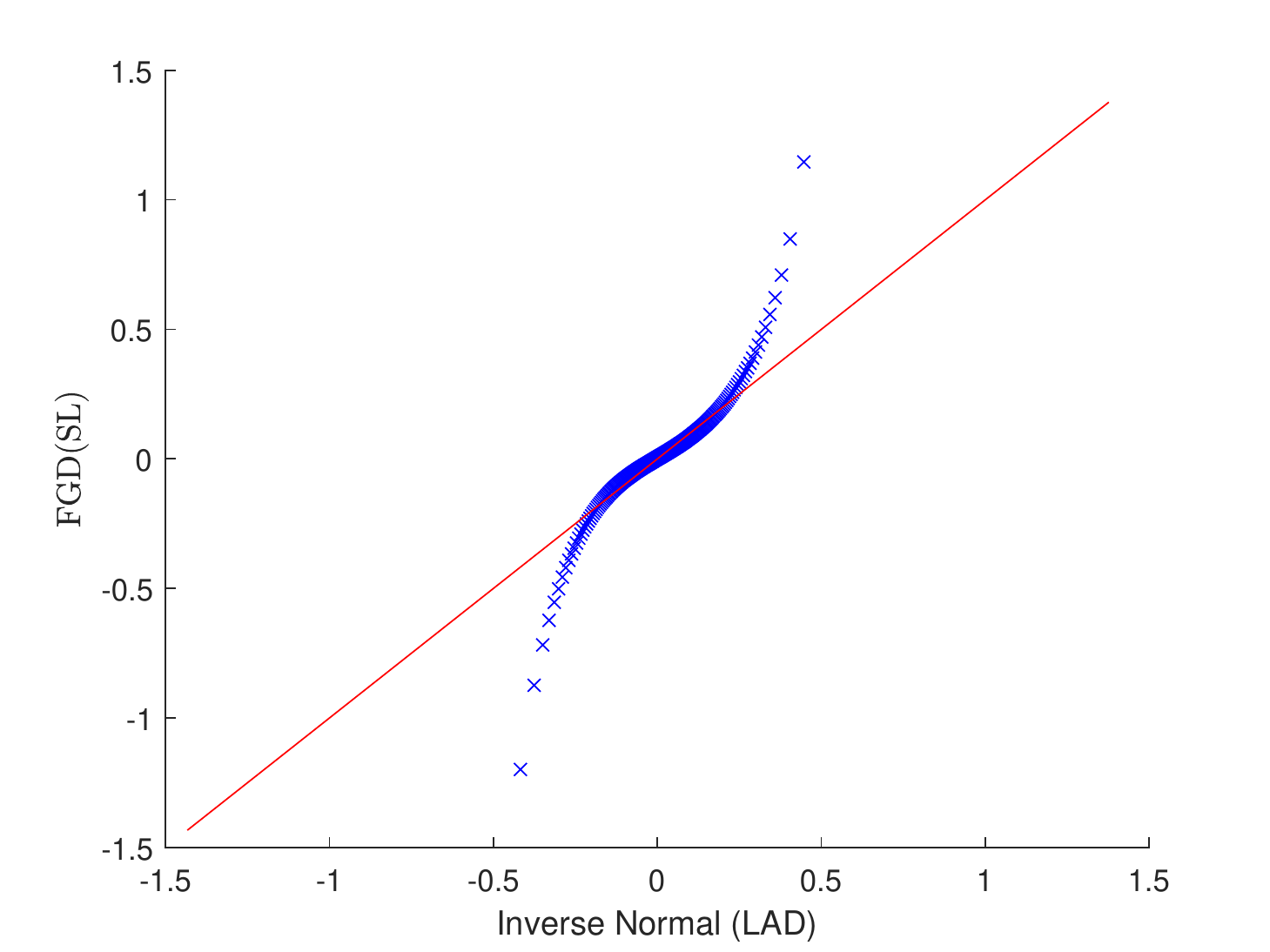}} \\ \\
		\subfigure[q-q FGD(SL) vs. DLN ]{\includegraphics[width=2in]{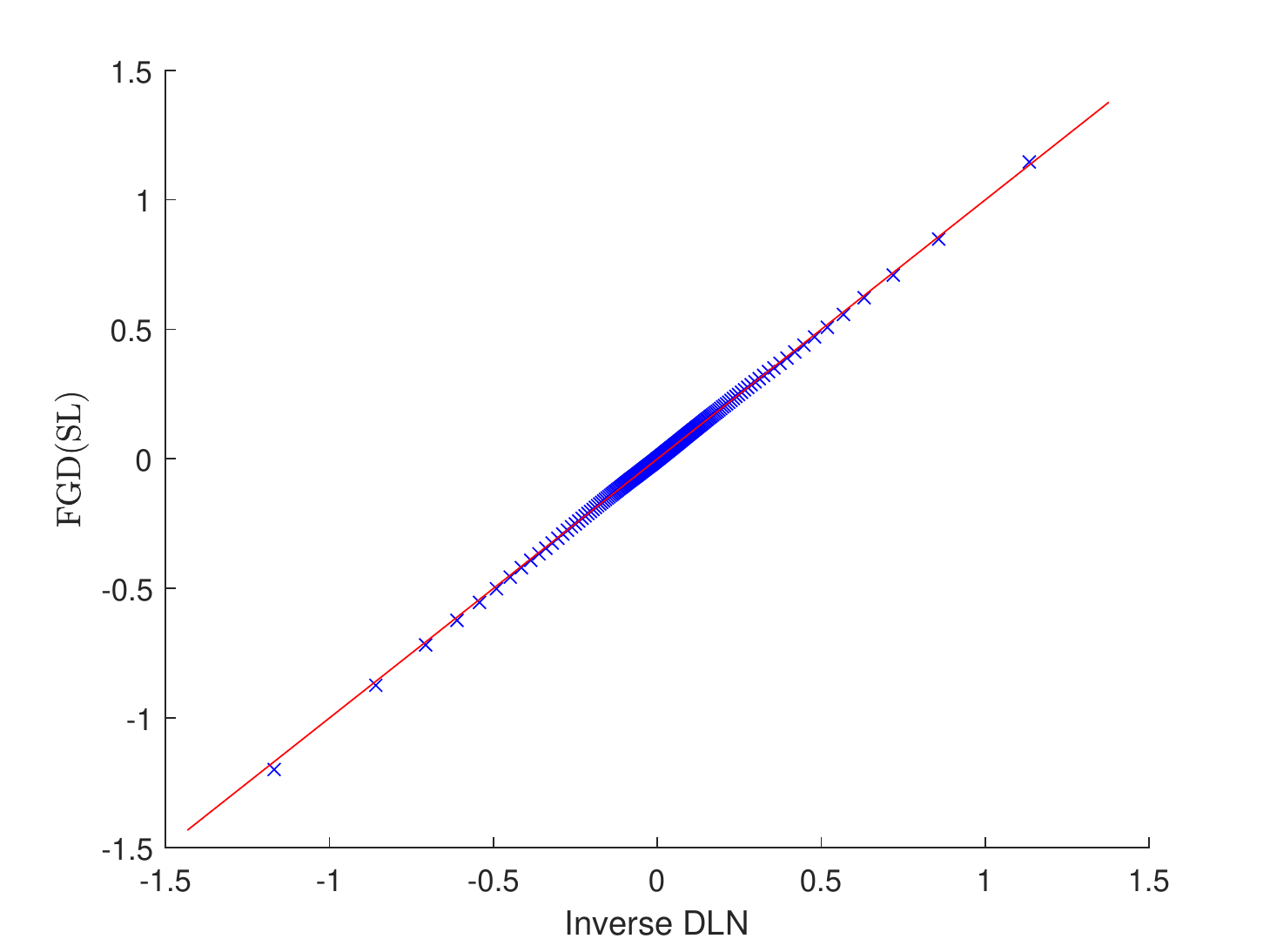}} &
		\subfigure[Time trend of FGD ]{\includegraphics[width=2in]{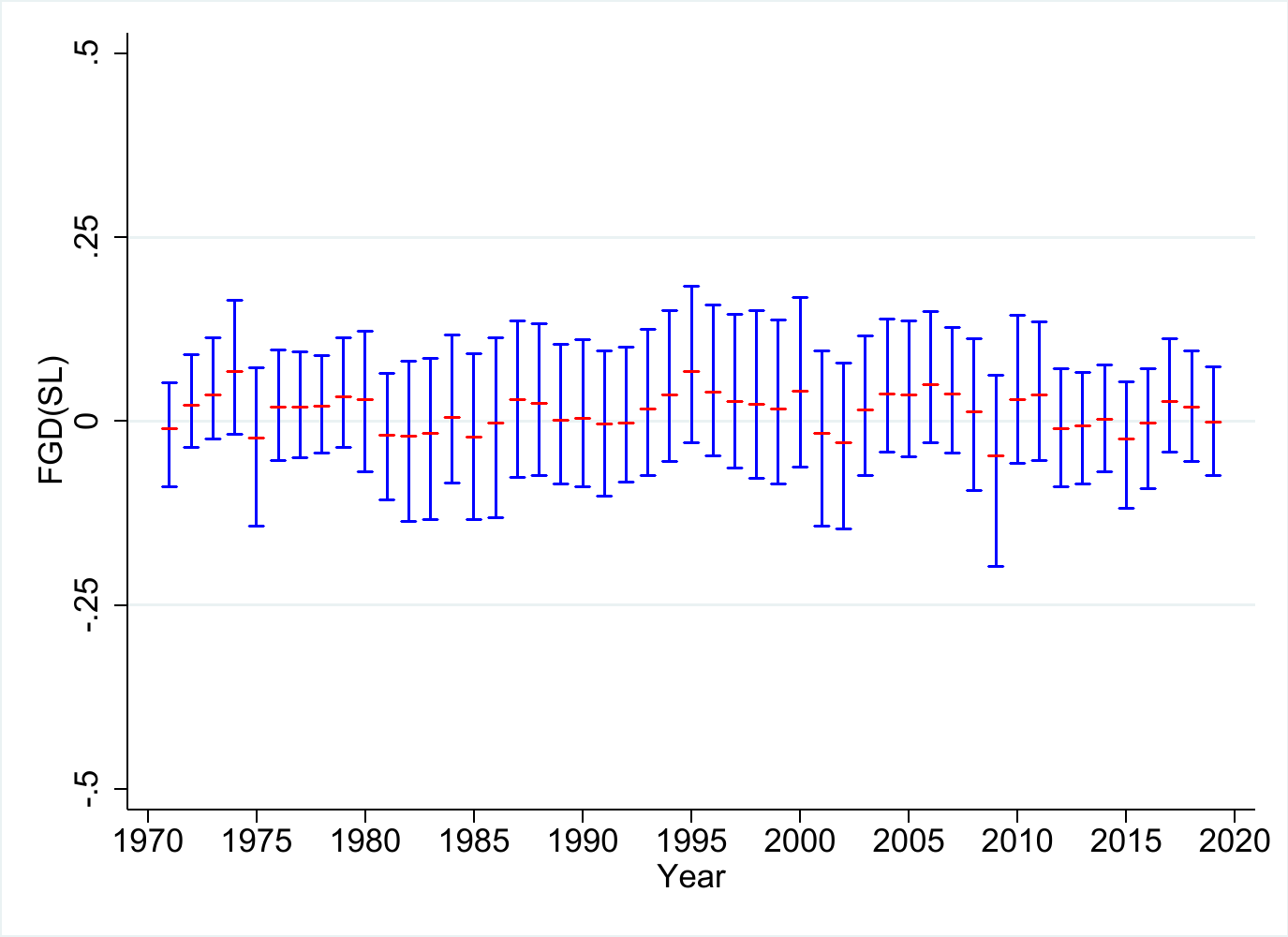}} &
		\subfigure[FGD(SL) Median+IQR ]{\includegraphics[width=2in]{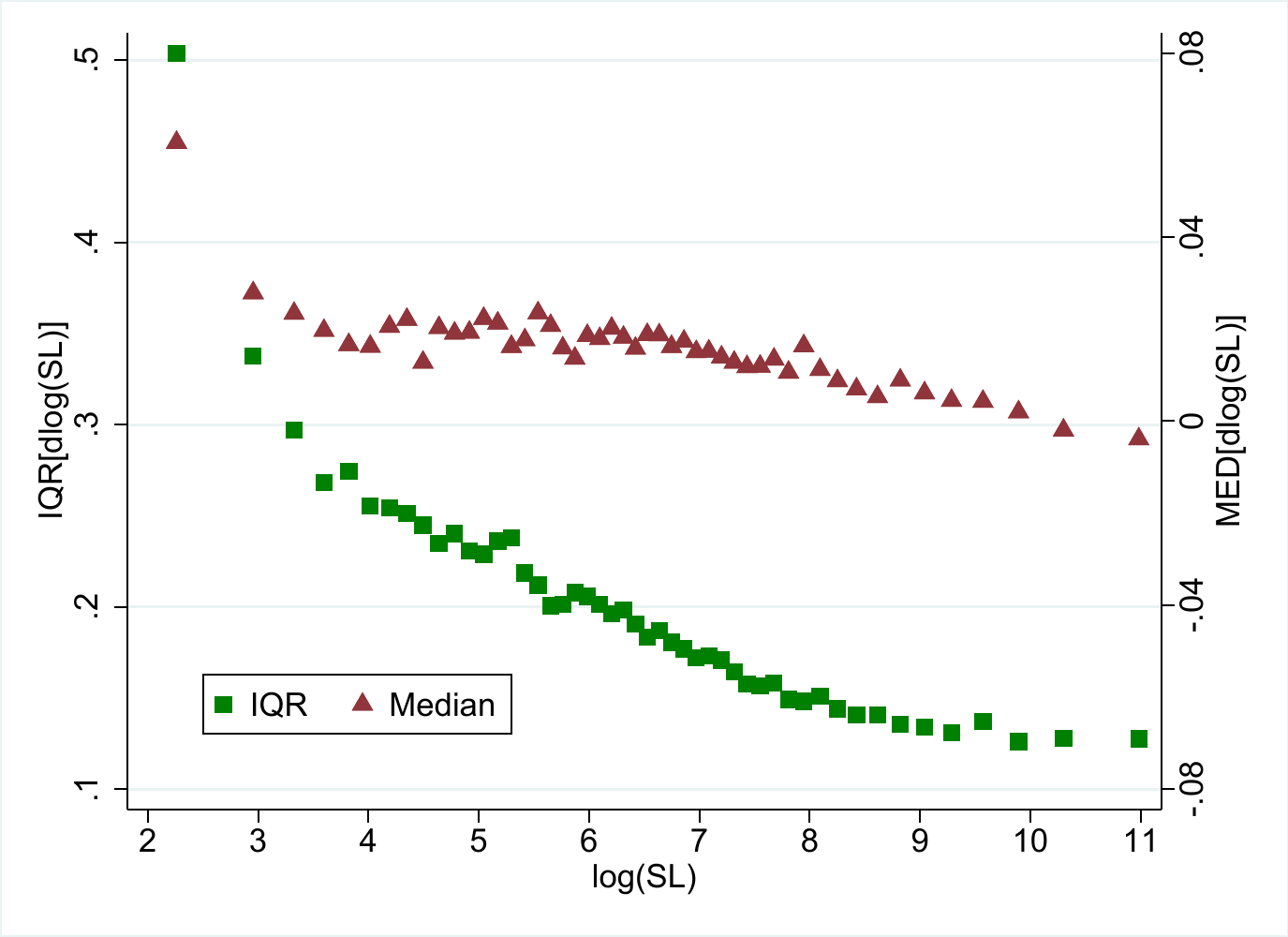}} \\ \\
		\subfigure[FGD(SL) by scale ]{\includegraphics[width=2in]{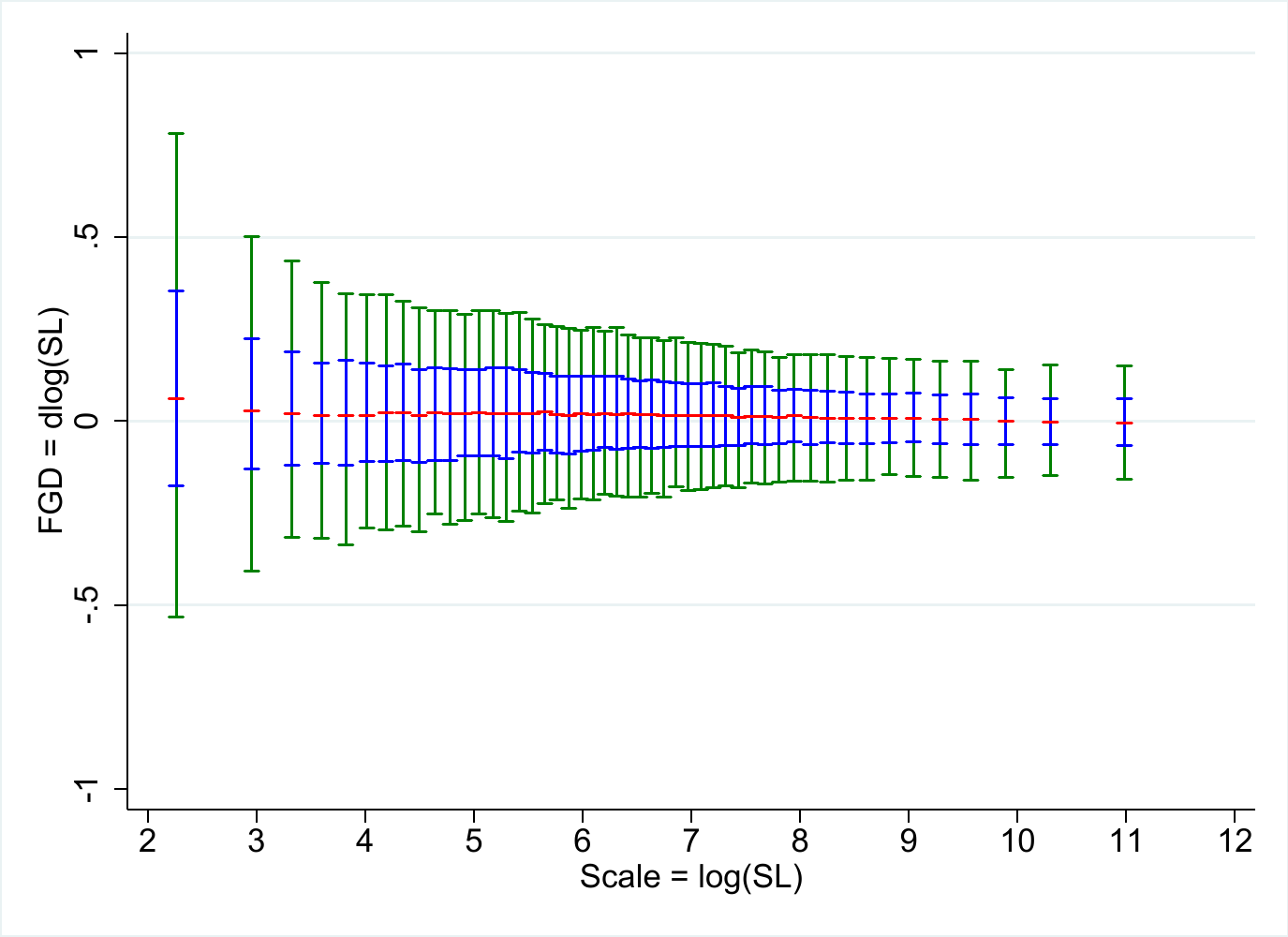}} &
		\subfigure[FGD($\widetilde{\text{SL}}$) w/ DLN ]{\includegraphics[width=2in]{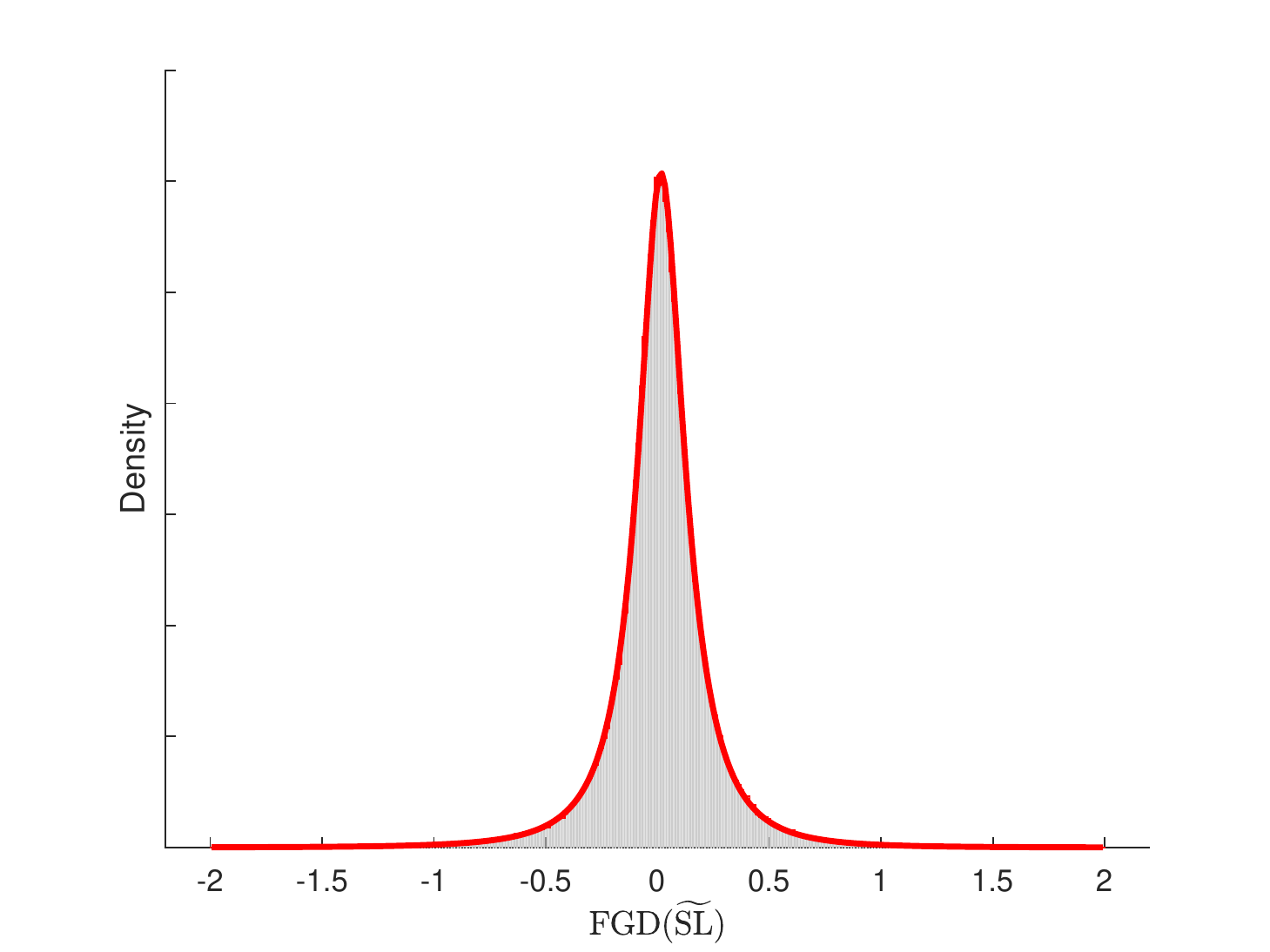}} &
		\subfigure[q-q FGD($\widetilde{\text{SL}}$) vs. DLN ]{\includegraphics[width=2in]{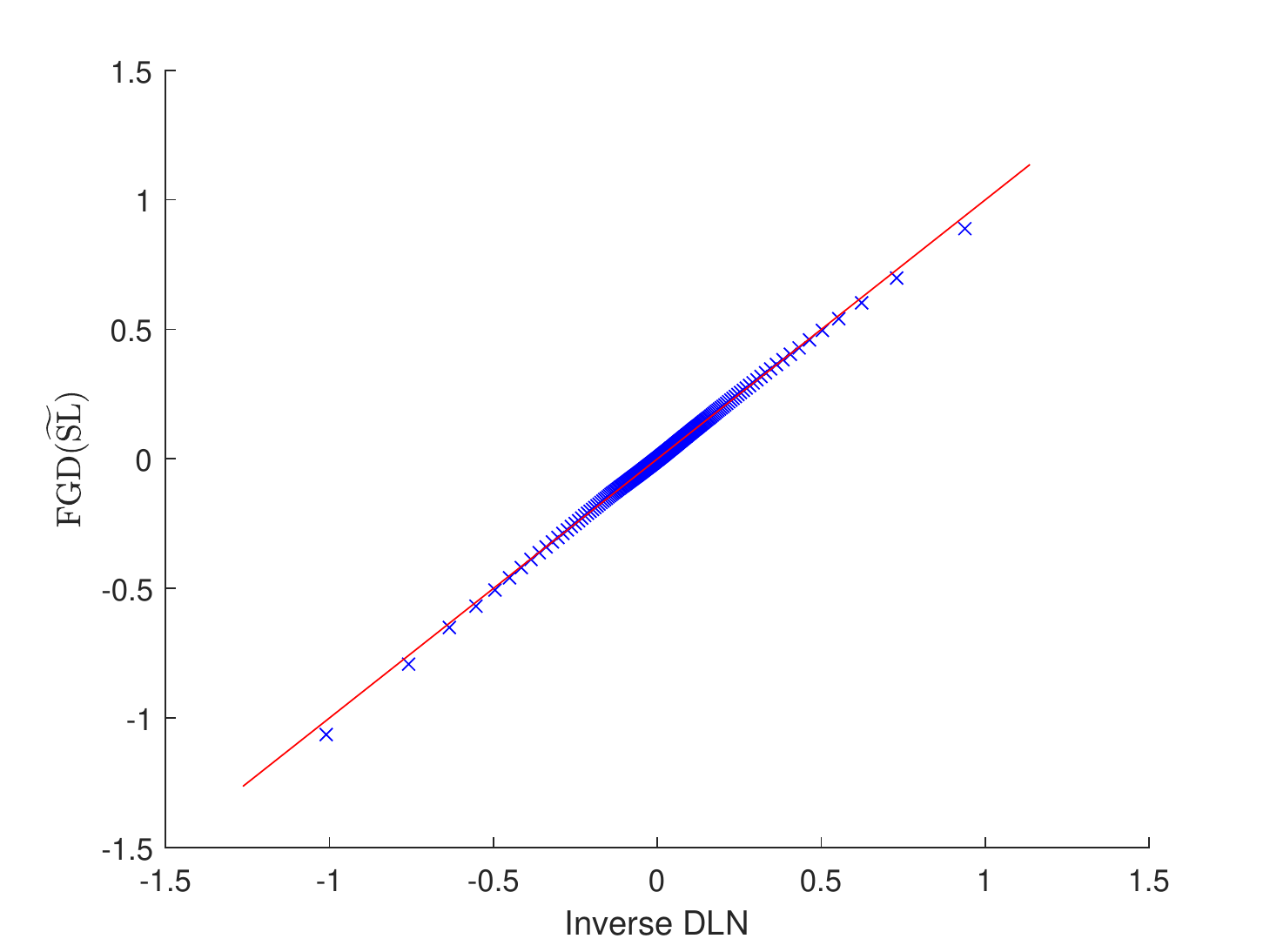}} \\ \\
	\end{tabular}
}

Patterns related to the business-cycle are evident in Panel (e), which presents the time-trends of the FGD(SL) (see e.g. the years 1975, 2001, 2008). A cleaner look at the FGD hence requires adjustments (by taking year fixed effects), similar to those previously employed for firm scale. Panel (f) uses \emph{binscatter} plots to presents two other systematic patterns in the FGD.\footnote{For an econometric review of binscatter methods, see \cite{CattaneoEtAl2021}.} The FGD presents moderately decreasing growth level with scale, and substantial decreasing growth dispersion with scale (i.e., conditional heteroscedasticity of growth rates). These two patterns are not new, and were first documented by \cite{Mansfield1962} with later empirical support by \cite{Hall1987} and \cite{StanleyEtAl1996}. In Panel (f), I again split the data into $49$ equal bins and calculate the bin's median growth rate: $\text{Median}[\text{dlog}(SL)]$ and the bin's dispersion of growth: $\text{IQR}[\text{dlog}(SL)]$. The panel then plots these values w.r.t. firm scale. Panel (g) provides a second view of this phenomenon, by plotting the $(10,25,50,75,90)^{th}$ percentiles of FSD(SL) for each bin.

The scale-dependant patterns in panels (f) and (g) raise concerns that observed FGD heavy tails might be an aggregation artifact. In the words of \cite{Fama1965}, \begin{quote}``Perhaps the most popular approach to explaining long-tailed distributions has been to hypothesize that the distribution [...] is actually a mixture of several normal distributions with possibly the same mean, but substantially different variances.''\end{quote} I hence turn to analyzing the decreasing dispersion of growth rates more closely.

The six panels of Figure~\ref{fig:FGDDisp} present direct evidence of this conditional heteroskedasticity, for each of the six firm size measures previously used $M=\{EQ, VL, KP, KT, SL, XS\}$. The figure plots $\log(\text{IQR}[\text{dlog}(M)])$ per bin as a function of $\log(M)$, along with fitted regression lines. For all six measures, the R$^2$ of the regressions are high ($>0.93$), the coefficients are similar in magnitude ($-0.13\pm 0.02$), and are highly significant (t-stat $\approx$ 40). Results are similar with 98 bins and when using standard deviation rather than IQR.

\RPprep{Growth - Decreasing dispersion}{0}{0}{FGDDisp}{%
    This figure presents the dependence of firm growth dispersion on firm scale. For each of the six firm size measures $XX=\{EQ,VL,KP,KT,SL,XS\}$, the sample is first split into $49$ equal bins, based on firm scale, $\log(XX)$, truncating the top and bottom 1\% of the FSD, such that each bin contains 2\% of the observations. For each bin, the corresponding panel presents the log of each bin's growth dispersion $\log(\text{IQR}[dlog(XX)])$ as a function of scale $\log(XX)$, along with fitted regression lines.
}
\RPfig{%
	\begin{tabular}{ccc}
		\subfigure[FGD(EQ)] {\includegraphics[width=2in]{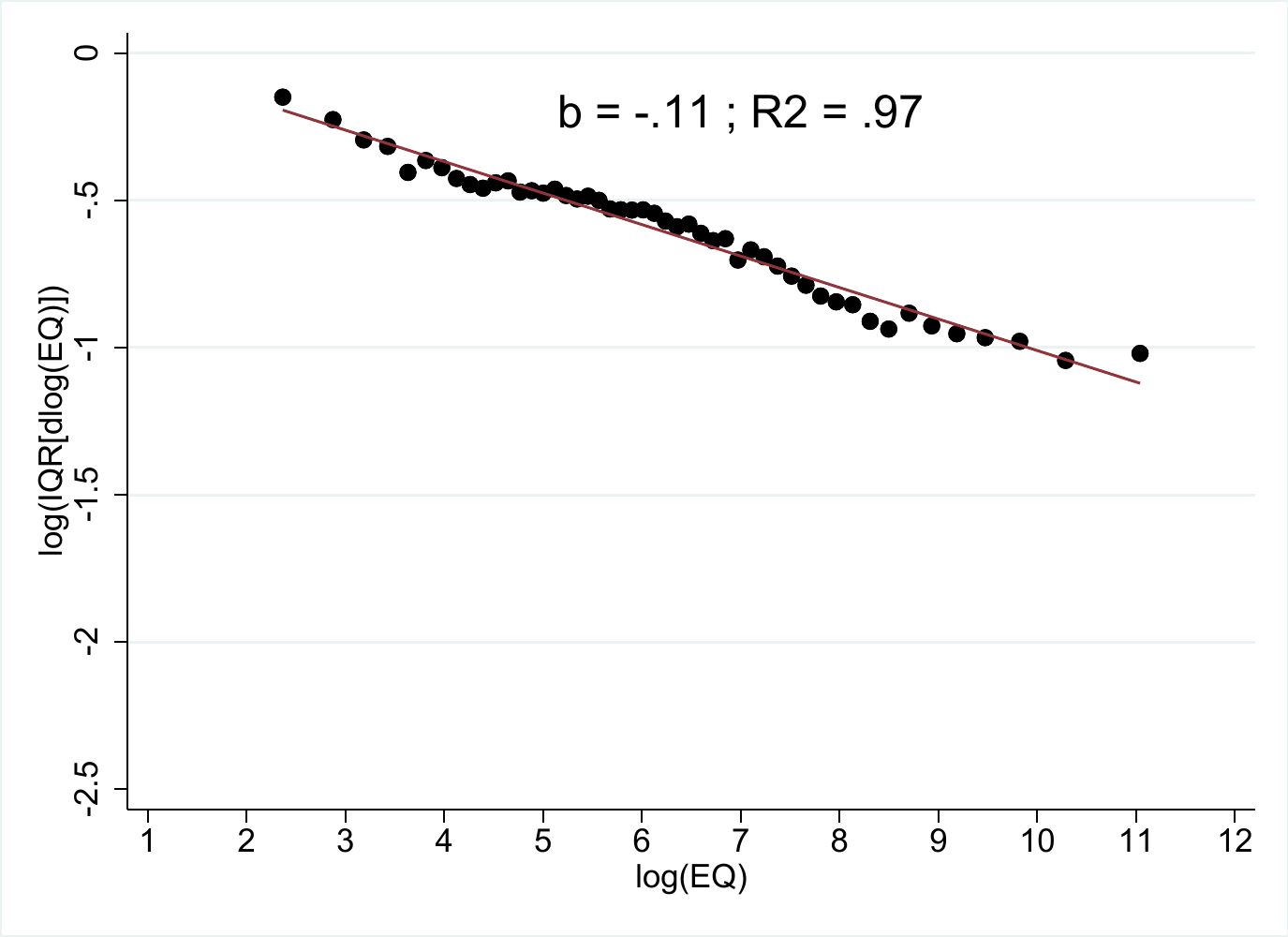}}
		& \subfigure[FGD(VL)]
		{\includegraphics[width=2in]{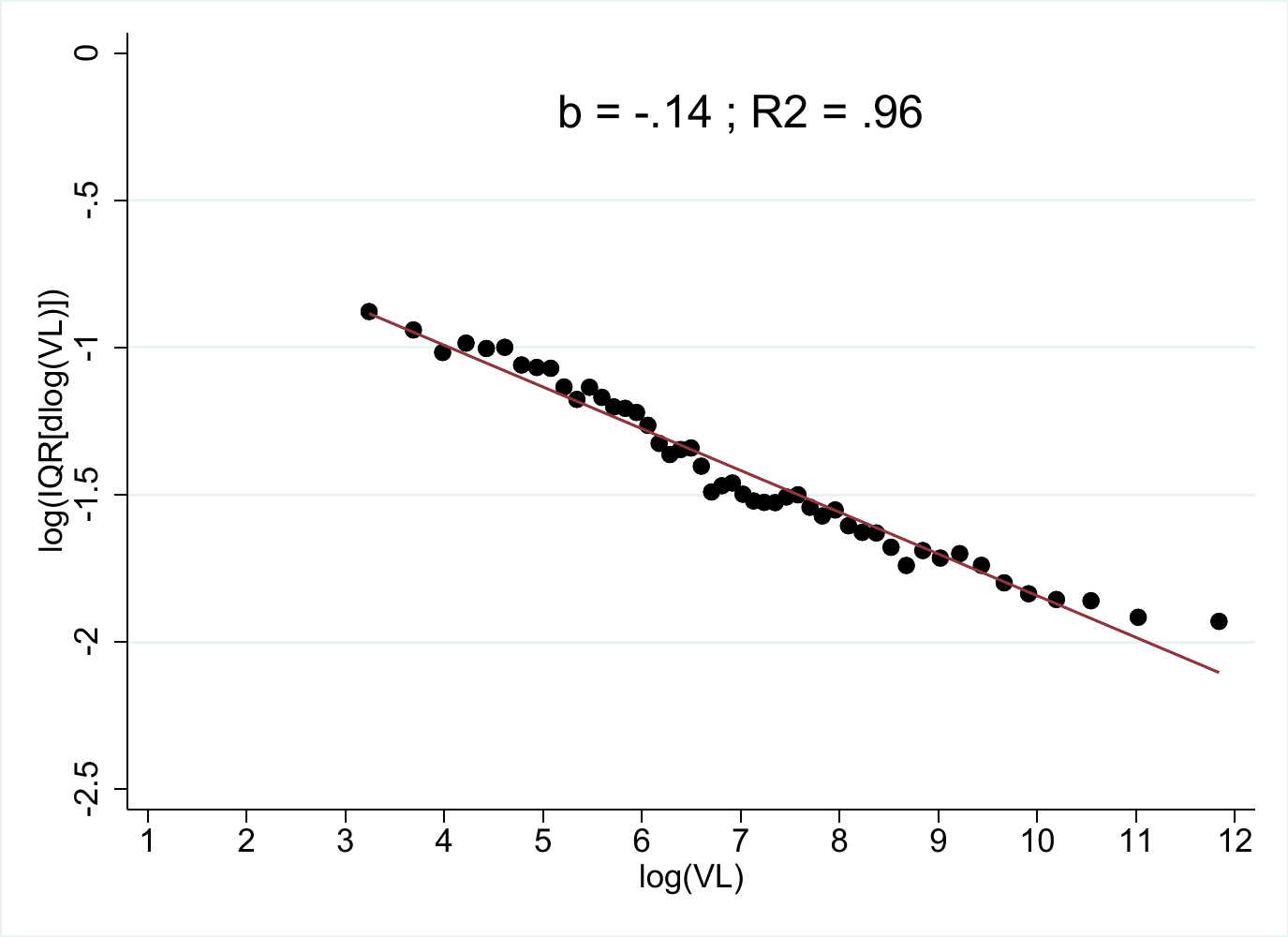}}
		& \subfigure[FGD(KP)]
		{\includegraphics[width=2in]{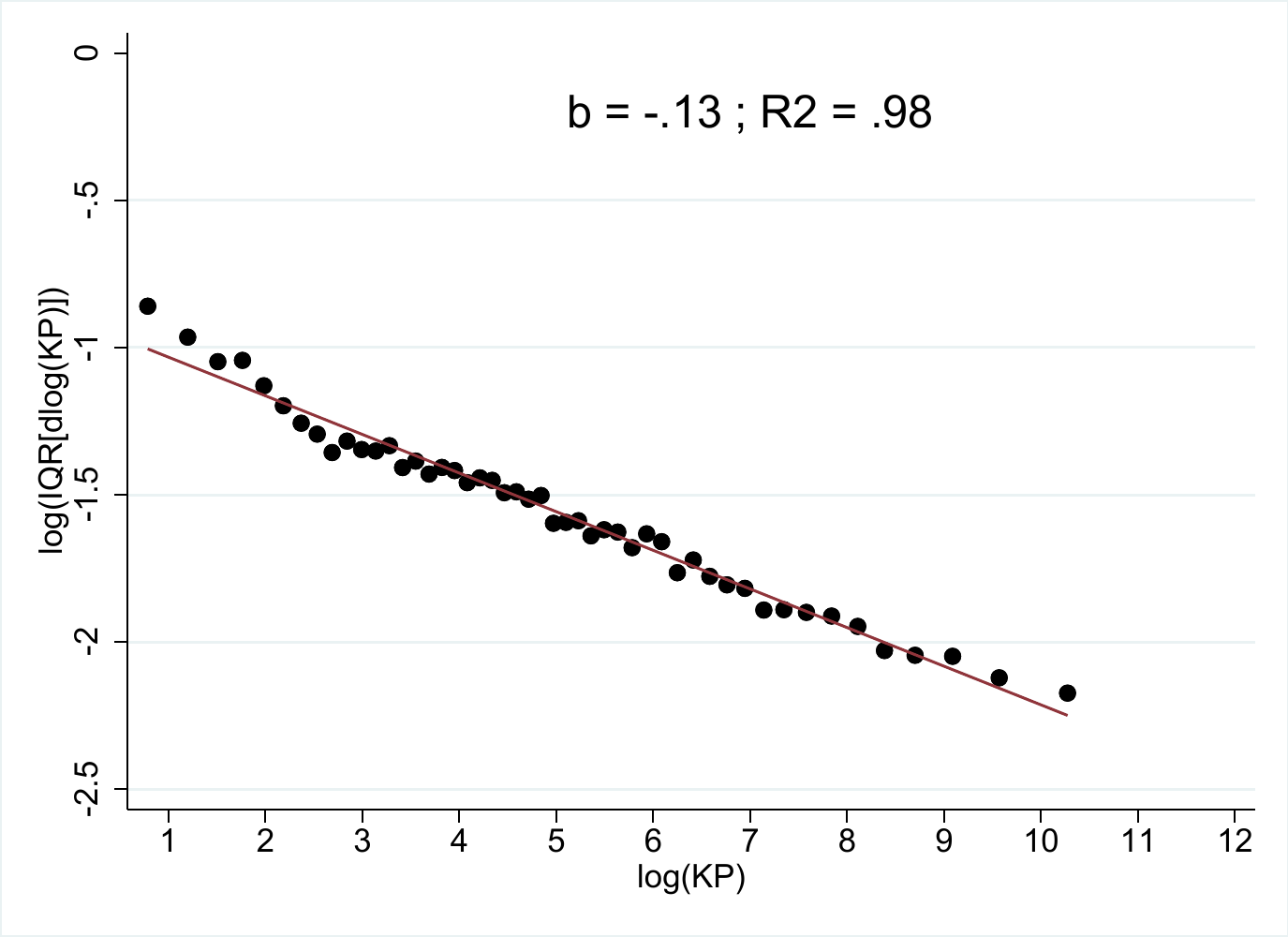}} \\ \\
		\subfigure[FGD(KT)] {\includegraphics[width=2in]{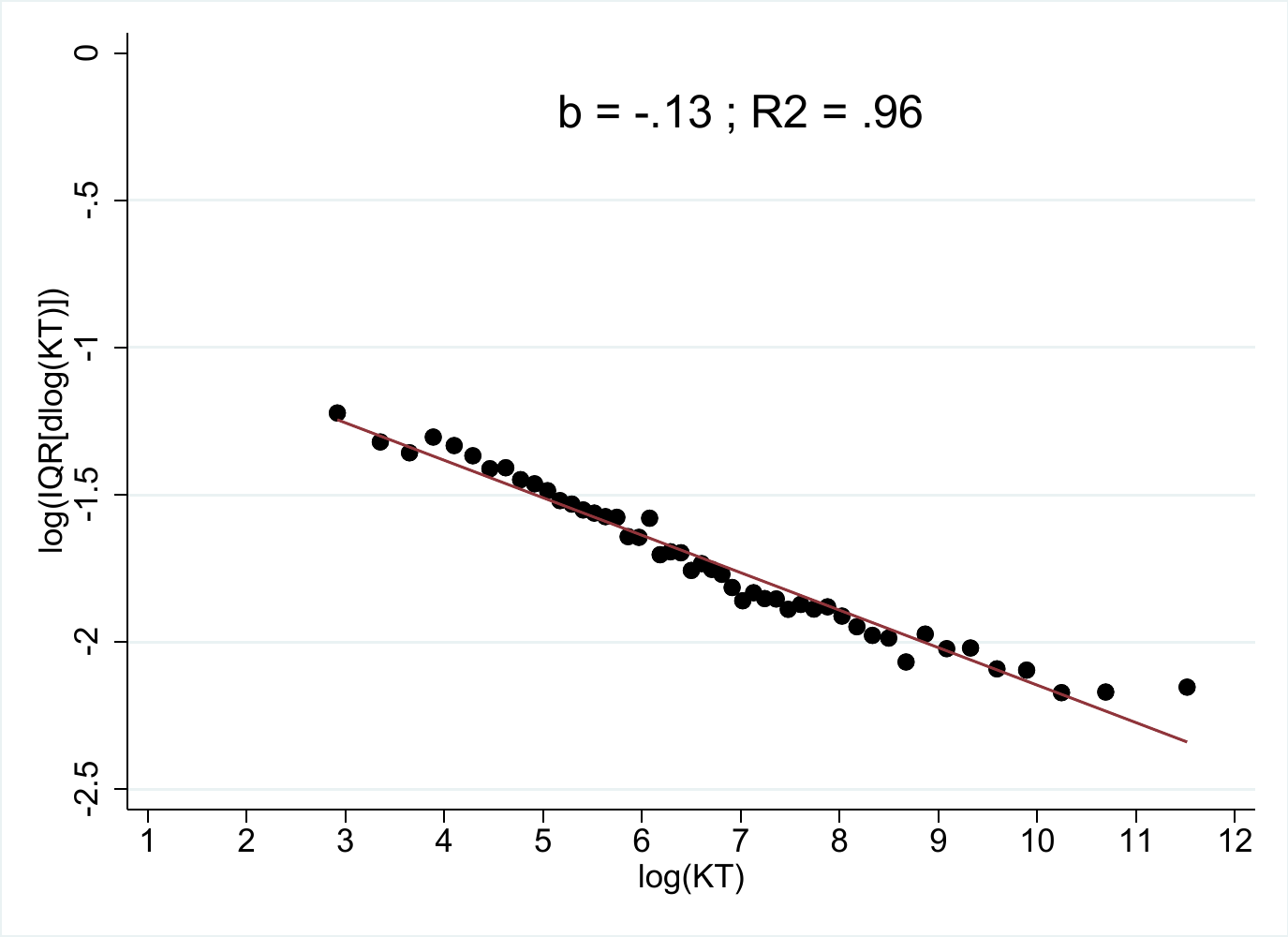}}
		& \subfigure[FGD(SL)]
		{\includegraphics[width=2in]{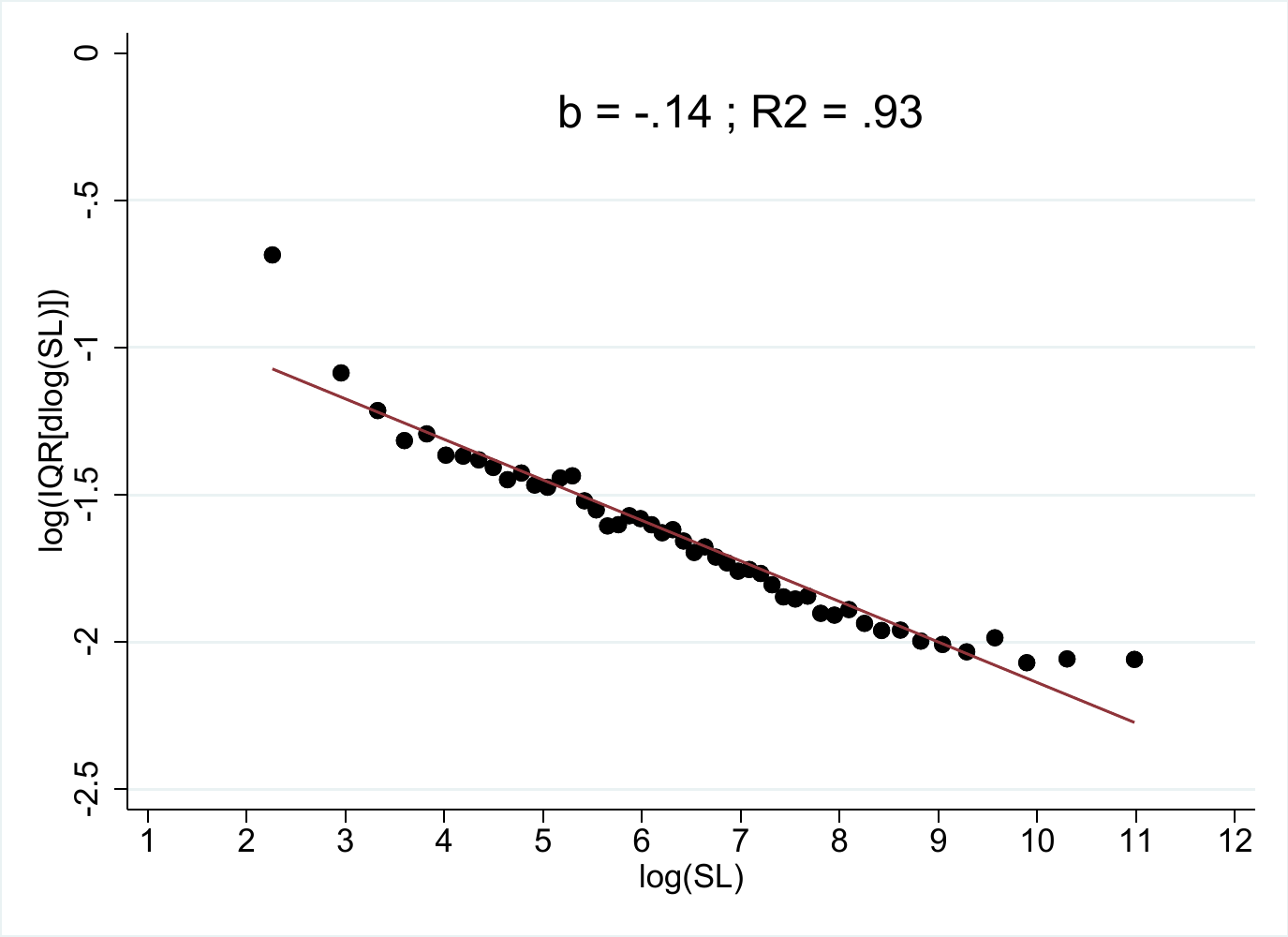}}
		& \subfigure[FGD(XS)]
		{\includegraphics[width=2in]{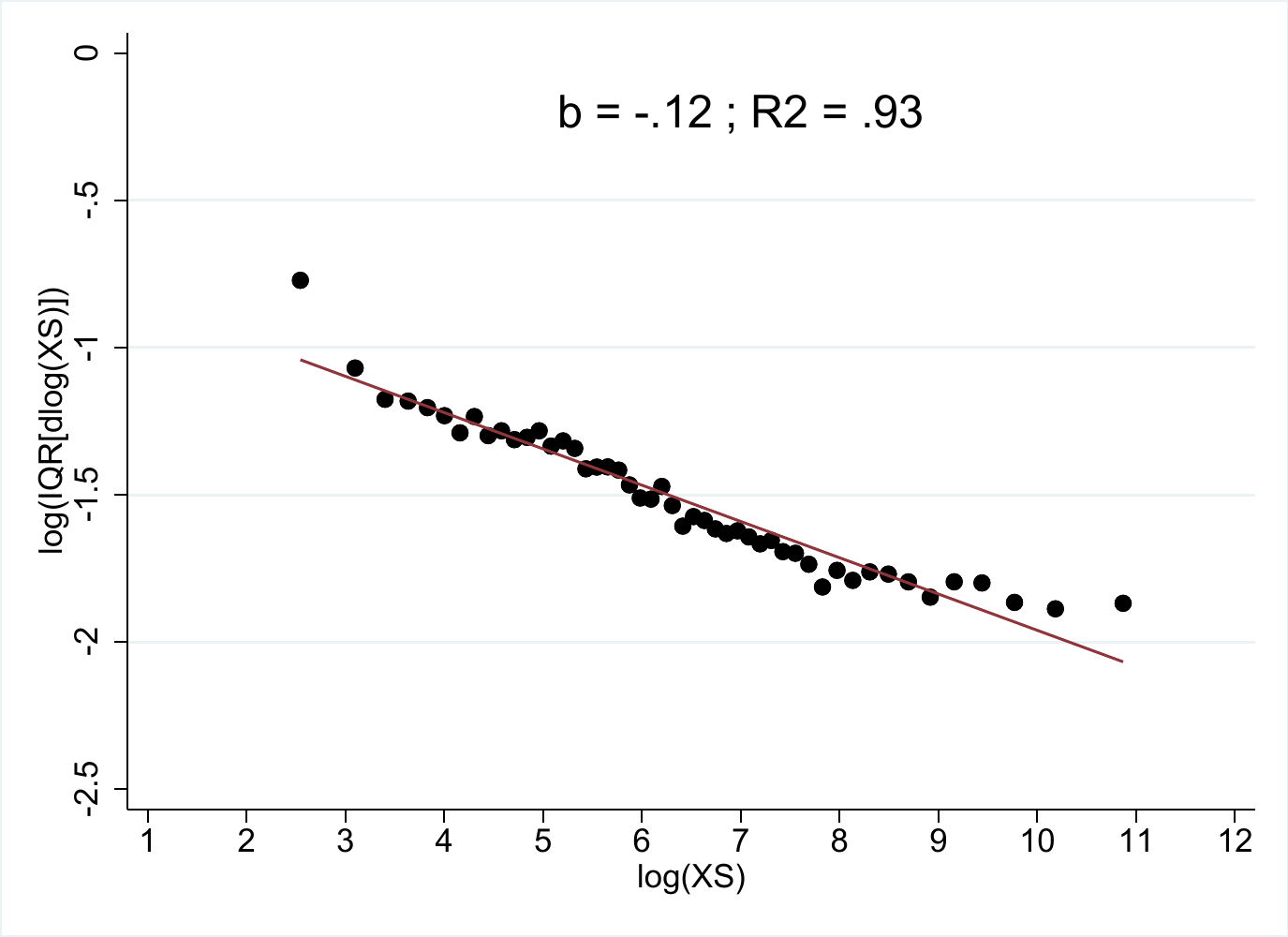}} \\ \\
	\end{tabular}
}

These results, and especially the high R$^2$ values, are strong indication of a \emph{mechanical} relation between scale and growth dispersion. They are also intuitively clear upon inspection. The likelihood that a firm with \$$100$M in sales will increase sales to \$$200$M in a single year is intuitively higher than the likelihood a firm with \$$100$B in sales will double its sales in a single year. The predominant explanation of this mechanical relation, e.g. \cite{Sutton2002}, is simple and derived from the logic of modern portfolio theory.

To see that, assume that a firm is a bundle of projects. If project growth rates are not fully correlated within the firm, then a ``portfolio'' of such projects (i.e., the firm) will have growth dispersion that is decreasing with the number of projects in the portfolio, similar to the diversification effect in a portfolio of stocks. A straightforward application of portfolio theory implies a linear relation between $\log(SD[\text{dlog}(M)])$ and $\log(M)$, and further implies that if projects are of the same size and have uncorrelated growth rates, then the slopes in Figure~\ref{fig:FGDDisp} should be $-0.5$. As discussed by \cite{Sutton2002}, that the slopes in Figure~\ref{fig:FGDDisp} are higher than $-0.5$ is not surprising. A positive correlation between the growth rates of projects, which is likely for projects of the same firm, or an equally likely deviation from the assumption that all projects are of the same size, will lead to less ``diversification'' and higher slope. I conclude:

\begin{observation}
    \label{ob:5}
    FGD dispersion is systematically decreasing with scale.
\end{observation}

To offset this mechanical relation, I use the systematic nature of the decreasing dispersion apparent in Figure~\ref{fig:FGDDisp} to adjust the location and dispersion of firm growth rates as a function of their scale. I.e., I take both time and scale fixed-effects from the entire FGD distribution (see Appendix~\ref{sec:AppdxFE} for details). This is similar to the strategy used by \cite{Barro1991} to adjust the heteroskedasticity of GDP growth based on country size.

The last six columns in Panel (a) of Table~\ref{tab:FGDDescMom} present the descriptive statistics of the adjusted versions of the FGDs. Skewness increases significantly in the adjusted versions, and kurtosis remains high and in most cases even increases. This implies the heavy-tails are not a scale-aggregation artifact. Panel (h) of Figure~\ref{fig:FGDfacts} presents the time- and scale-adjusted sales growth distribution, denoted FGD($\widetilde{\text{SL}}$), overlaid with a fitted DLN distribution. The fit is again remarkable, as can also be seen in Panel (i), presenting the q-q plot of FGD($\widetilde{\text{SL}}$) vs. the DLN.

The excellent fit of the DLN is confirmed by formal distributional tests reported in Table~\ref{tab:FGDdist}. I again concentrate on the Stable, Laplace, and DLN distributions. The Stable distribution has been the workhorse candidate for modelling firm growth since the early works of \cite{Mandelbrot1961} and \cite{Fama1963}, while the Laplace has later been proposed as a better alternative by \cite{StanleyEtAl1996} and the following literature. The tests generally reject the Stable and Laplace on all growth measures (at the 5\% confidence level). The DLN is not rejected for any raw or adjusted growth measure, with the exception of growth in physical capital. The relative likelihood tests are unequivocal - the DLN is strongly favored as a statistical model of growth, even for the rejected KP. I conclude:

\begin{observation}
    \label{ob:6}
    The FGD is distributed DLN.
\end{observation}

\RPprep{Growth - Distributional tests}{1}{0}{FGDdist}{%
    This table presents results of tests of distribution equality for the FGD, based on the scale measures described in Table~\ref{tab:DataDef}, as well as their time- and scale-adjusted versions, denoted $\widetilde{XX}$. K-S is a Kolmogorov–Smirnov test; C-2 is a binned $\chi^2$ test with 50 bins; A-D is an Anderson-Darling test. Panels (a)-(c) report the test statistics and their p-values for the Stable, Laplace, and DLN, respectively. Panel (d) reports the relative likelihoods for each distribution.
}
\RPtab{%
    \begin{tabularx}{\linewidth}{Frrrrrrrrrrrr}
    \toprule
	& EQ & VL & KP$^{a}$ & KT$^{a}$ & SL & XS & $\widetilde{\text{EQ}}$ & $\widetilde{\text{VL}}$ & $\widetilde{\text{KP}}^{a}$ & $\widetilde{\text{KT}}^{a}$ & $\widetilde{\text{SL}}$ & $\widetilde{\text{XS}}$\\ 
	\midrule
	\\ \multicolumn{13}{l}{\textit{Panel (a): FGD vs. Stable}}\\
	\midrule
    K-S   & 0.013 & 0.017 & 0.016 & 0.011 & 0.012 & 0.009 & 0.012 & 0.019 & 0.021 & 0.012 & 0.013 & 0.011 \\
    p-val & 0.050 & 0.040 & 0.042 & 0.055 & 0.052 & 0.067 & 0.053 & 0.036 & 0.034 & 0.052 & 0.050 & 0.056 \\
    C-2   & 115   & 324   & 107   & 91    & 88    & 67    & 80    & 230   & 193   & 88    & 80    & 59    \\
    p-val & 0.036 & 0.020 & 0.037 & 0.040 & 0.040 & 0.046 & 0.042 & 0.025 & 0.027 & 0.041 & 0.042 & 0.049 \\ 
    A-D   & 4.37  & 10.60 & 6.73  & 3.79  & 4.04  & 2.77  & 3.34  & 9.36  & 14.93 & 3.70  & 4.31  & 3.50  \\
    p-val & 0.045 & 0.032 & 0.039 & 0.047 & 0.046 & 0.052 & 0.049 & 0.034 & 0.028 & 0.048 & 0.045 & 0.048 \\ 
    \\ \multicolumn{13}{l}{\textit{Panel (b): FGD vs. Laplace}}\\
	\midrule
    K-S   & 0.009 & 0.040 & 0.046 & 0.027 & 0.041 & 0.047 & 0.016 & 0.023 & 0.041 & 0.019 & 0.034 & 0.043 \\
    p-val & 0.063 & 0.018 & 0.015 & 0.027 & 0.017 & 0.014 & 0.043 & 0.032 & 0.017 & 0.036 & 0.022 & 0.016 \\
    C-2   & 52    & 368   & 509   & 290   & 415   & 657   & 98    & 121   & 320   & 137   & 225   & 436   \\
    p-val & 0.052 & 0.019 & 0.015 & 0.022 & 0.017 & 0.012 & 0.038 & 0.035 & 0.020 & 0.033 & 0.025 & 0.016 \\ 
    A-D   & 1.69  & 23.97 & 37.45 & 18.72 & 33.79 & 51.45 & 4.43  & 8.21  & 23.19 & 8.19  & 17.01 & 33.23 \\
    p-val & 0.061 & 0.023 & 0.018 & 0.025 & 0.019 & 0.014 & 0.045 & 0.036 & 0.023 & 0.036 & 0.027 & 0.019 \\ 
    \\ \multicolumn{13}{l}{\textit{Panel (c): FGD vs. DLN}}\\
	\midrule
    K-S   & 0.004 & 0.008 & 0.015 & 0.006 & 0.003 & 0.001 & 0.001 & 0.005 & 0.018 & 0.005 & 0.005 & 0.003 \\
    p-val & 0.129 & 0.069 & 0.045 & 0.087 & 0.155 & 0.748 & 0.819 & 0.106 & 0.039 & 0.096 & 0.096 & 0.165 \\
    C-2   & 9     & 39    & 105   & 17    & 7     & 5     & 5     & 14    & 167   & 12    & 11    & 5     \\
    p-val & 0.133 & 0.059 & 0.037 & 0.089 & 0.171 & 0.430 & 0.438 & 0.100 & 0.030 & 0.110 & 0.115 & 0.270 \\ 
    A-D   & 0.23  & 1.08  & 4.47  & 0.50  & 0.15  & 0.06  & 0.02  & 0.46  & 7.77  & 0.47  & 0.39  & 0.11  \\
    p-val & 0.112 & 0.070 & 0.045 & 0.089 & 0.130 & 0.184 & 0.349 & 0.091 & 0.037 & 0.090 & 0.095 & 0.144 \\ 
    \\ \multicolumn{13}{l}{\textit{Panel (d): Distribution comparison}}\\
	\midrule
	\multicolumn{2}{l}{AIC R.L.:} \\
    Stable  & 0.000 & 0.000 & 0.000 & 0.000 & 0.000 & 0.000 & 0.000 & 0.000 & 0.000 & 0.000 & 0.000 & 0.000 \\
    Laplace & 0.000 & 0.000 & 0.000 & 0.000 & 0.000 & 0.000 & 0.000 & 0.000 & 0.000 & 0.000 & 0.000 & 0.000 \\
    DLN     & 1.000 & 1.000 & 1.000 & 1.000 & 1.000 & 1.000 & 1.000 & 1.000 & 1.000 & 1.000 & 1.000 & 1.000 \\
	\multicolumn{2}{l}{BIC R.L.:} \\
    Stable  & 0.000 & 0.000 & 0.000 & 0.000 & 0.000 & 0.000 & 0.000 & 0.000 & 0.000 & 0.000 & 0.000 & 0.000 \\
    Laplace & 0.000 & 0.000 & 0.000 & 0.000 & 0.000 & 0.000 & 0.000 & 0.000 & 0.000 & 0.000 & 0.000 & 0.000 \\
    DLN     & 1.000 & 1.000 & 1.000 & 1.000 & 1.000 & 1.000 & 1.000 & 1.000 & 1.000 & 1.000 & 1.000 & 1.000 \\
	\bottomrule
    \end{tabularx}
    \begin{flushleft}
    $^a$ Using the \{Non-Bank\} data subset \\
    \end{flushleft}
}

While a full economic discussion of \emph{why} the FGD distributes DLN is left to the companion paper \cite{Parham2022c}, a simple intuitive explanation can be gleaned by thinking of flow vs. stock of value. If the flow of economic value (e.g., firm income) is distributed DLN per Observation~\ref{ob:4}, and the stock of economic value (e.g., firm capital) is a running sum of flows, then the growth in the stock will tend to have the same distribution as the flow.

Turning to time-series aspects of firm growth, the last two rows in Panel (a) of Table~\ref{tab:FGDDescMom} present the persistence of the raw and adjusted growth measures. Total and equity value growth (i.e., total and equity returns) exhibit mild reversal at the yearly frequency, while capital and sales growth exhibit moderate momentum, with persistence in the range $0.11$ to $0.16$. Expenses, both raw and adjusted, exhibit moderate reversal, with persistence of $-0.17$ and $-0.13$ for raw and adjusted expense growth, respectively.

The different growth measures are considerably less correlated than the scale measures, as can be seen in Panel~(b) of Table~\ref{tab:FGDDescMom}. Equity and total value growth are strongly correlated, which is not surprising. Sales growth is moderately correlated with both capital and expense growth. The most surprising finding in Panel (b) is that expense growth is \emph{not} negatively correlated with equity or total value growth, but rather is effectively uncorrelated with them.

\subsection{A closer look at equity returns}
\label{sec:Return}

One measure of firm growth stands tall above all others in financial economics --- equity returns, or FGD(EQ). Note that the buy and hold equity return at the yearly frequency, in log-points, is
\begin{equation} \label{eq:adj_dlog}
\text{dlog}(EQ_t) = \log(EQ_t + DE_t) - \log(EQ_{t-1})
\end{equation}
or simply equity growth adjusted for capital dispensations to/from equity holders. A second reason to focus on equity growth, besides its  disproportionate importance, is the availability of higher-frequency data. While growth in sales can be computed yearly or sometimes quarterly, growth in equity value for publicly traded firms can be calculated at nearly any desired frequency. Does the observation that equity returns distribute DLN hold at higher frequencies, based on a different data source, and for excess returns as well?

This closer look begins with daily stock returns for the entire CRSP universe from 1/1970 to 12/2019, totalling $83$M raw observations and $63$M non-missing observations for $30$K firms. I use those to calculate dividend and split adjusted daily, monthly and yearly returns, in log-point terms. I also calculate excess returns relative to the Fama-French 3-factor (FF3) model, using factor returns from Ken French's data library and betas based on 12-month rolling windows of daily returns.

One challenge with analyzing the distribution of high frequency return data is that the effect of discrete prices on returns can no longer be neglected. Prior to 4/2001, stock prices were generally quoted as fractions in base 16 (e.g., 3 and 7/16 dollars per share). Even post the 2001 ``decimalization,'' prices are quoted to two decimal places (e.g., \$ 3.44 per share). This means returns are not continuous --- a share beginning trading at 3+7/16 can remain at 3+7/16 (have 0\% return), go up at least to 3+8/16 (have 1.82\% return) or go down at least to 3+6/16 (have -1.82\% return). Return values in the range $(-1.82,1.82)\backslash\{0\}$ are unobtainable. This causes the high-frequency return distributions to be ``saw-toothed'' and eliminates any ability to cogently analyze their distributional form. To overcome this issue, I add a tiny amount of uniform noise to stock prices prior to calculating returns. In essence, I assume a quoted price of 3+7/16 means the actual price is uniformly distributed between $3 + 7/16 \pm 1/32$. Similarly, a quoted price of 3.44 means the actual price is uniformly distributed between $3.44 \pm 0.005$. This procedure unrolls the effect of discrete prices and prevents ``bunching'' at the high-frequency return distributions.

Table~\ref{tab:FGDEQDescMom} presents the descriptive statistics of the various equity return flavors. The twelve flavors are \{yearly,monthly,daily\}X\{raw,excess\}X\{unadjusted,adjusted\} with the adjusted flavors adjusting for time and scale fixed effects as before. All returns are considerably negatively skewed and exhibit heavy tails, as can be seen by the third and fourth moments. Hence, table~\ref{tab:FGDEQDescMom} reports the median and IQR in additions to the first four central moments of each distribution. Both mean and median of the excess return distributions are negative, at all horizons, which is surprising as intuition would imply excess returns should be zero on average.

\RPprep{Equity return - Descriptive statistics}{0}{1}{FGDEQDescMom}{%
    This table presents the first four central moments, as well as the median and IQR, of the firm equity return distribution FGD(EQ). Twelve flavors are presented: \{yearly,monthly,daily\}X\{raw,excess\}X\{unadjusted,adjusted\}. The returns are split and dividend adjusted, at the relevant horizon, with excess returns calculated relative the the Fama-French 3-factor model. The data are the CRSP universe 1970-2019. Time- and scale-adjusted versions are marked with $\widetilde{XXX}$, and the adjustment procedure is described in Appendix~\ref{sec:AppdxFE}.
}
\RPtab{%
    \begin{tabularx}{\linewidth}{Frrrrrrrrrrrr}
    \toprule
	& \multicolumn{4}{c}{Yearly} & \multicolumn{4}{c}{Monthly} & \multicolumn{4}{c}{Daily} \\
	& Raw & FF3 & $\widetilde{\text{Raw}}$ & $\widetilde{\text{FF3}}$ & Raw & FF3 & $\widetilde{\text{Raw}}$ & $\widetilde{\text{FF3}}$ & Raw & FF3 & $\widetilde{\text{Raw}}$ & $\widetilde{\text{FF3}}$ \\
	\midrule
    $M_{1}$ & -0.0072 & -0.1018 & 0.0131  & -0.0899 & -0.0004 & -0.0084 & 0.0008  & -0.0089 & -0.0001 & -0.0005 & 0.0004  & -0.0003  \\
    $M_{2}$ & 0.6227  & 0.6058  & 0.5763  & 0.5395  & 0.1658  & 0.1578  & 0.1513  & 0.1404  & 0.0471  & 0.0461  & 0.0387  & 0.0362   \\
    $M_{3}$ & -2.6784 & -2.5665 & -2.0654 & -1.9818 & -0.7209 & -0.5199 & -0.8319 & -0.9166 & -0.2377 & -0.2086 & -0.4708 & -0.5924  \\
    $M_{4}$ & 92      & 105     & 33      & 32      & 26      & 35      & 33      & 35      & 356     & 390     & 296     & 307      \\
    MED     & 0.0701  & -0.0341 & 0.0674  & -0.0363 & 0.0053  & -0.0035 & 0.0037  & -0.0048 & 0.0000  & -0.0005 & 0.0000  & -0.0005  \\
    IQR     & 0.5140  & 0.4695  & 0.4997  & 0.4627  & 0.1277  & 0.1140  & 0.1277  & 0.1155  & 0.0260  & 0.0237  & 0.0267  & 0.0245   \\
    \end{tabularx}
}

This surprising fact is both explained and escalated by the graphical evidence in Figure~\ref{fig:FGDEQfacts}. Panel (a) presents median and IQR binscatters for yearly raw returns, with Panel (b) presenting them for monthly excess returns and Panel (c) for daily excess returns. In all cases, we observe an \emph{increase} in median return with scale (measured by log lagged equity value). This increase is striking as it seemingly goes against the well-known ``size effect'' in the asset pricing literature, and holds for raw and excess return at each of the frequencies considered. Note however that the bins in Panels (a)-(c) of Figure~\ref{fig:FGDEQfacts} are not tradable portfolios, so these results should be considered with care. Return dispersion is decreasing with scale, as with all other growth measures. While surprising, the data suggest the following observation:

\begin{observation}
    \label{ob:7}
    The FGD(EQ) location and dispersion are increasing and decreasing with scale, respectively.
\end{observation}

\RPprep{Equity return - Stylized facts}{0}{0}{FGDEQfacts}{%
    This figure presents stylized facts of the equity return distribution FGD(EQ). Panel (a) presents median and IQR binscatters for yearly raw unadjusted returns per 49 scale bins, ignoring the top and bottom 1\% of the data, with scale defined as lagged log(EQ).Panels (b) and (c) present binscatters for monthly and daily excess unadjusted returns relative to FF3. Panels (d)-(f) present the histograms for the respective flavors, overlaid with fitted DLN distributions, and Panels (g)-(i) present the respective q-q plots vs. the DLN.
}
\RPfig{%
	\begin{tabular}{ccc} 
		\subfigure[Raw (Y) Median+IQR ]{\includegraphics[width=2in]{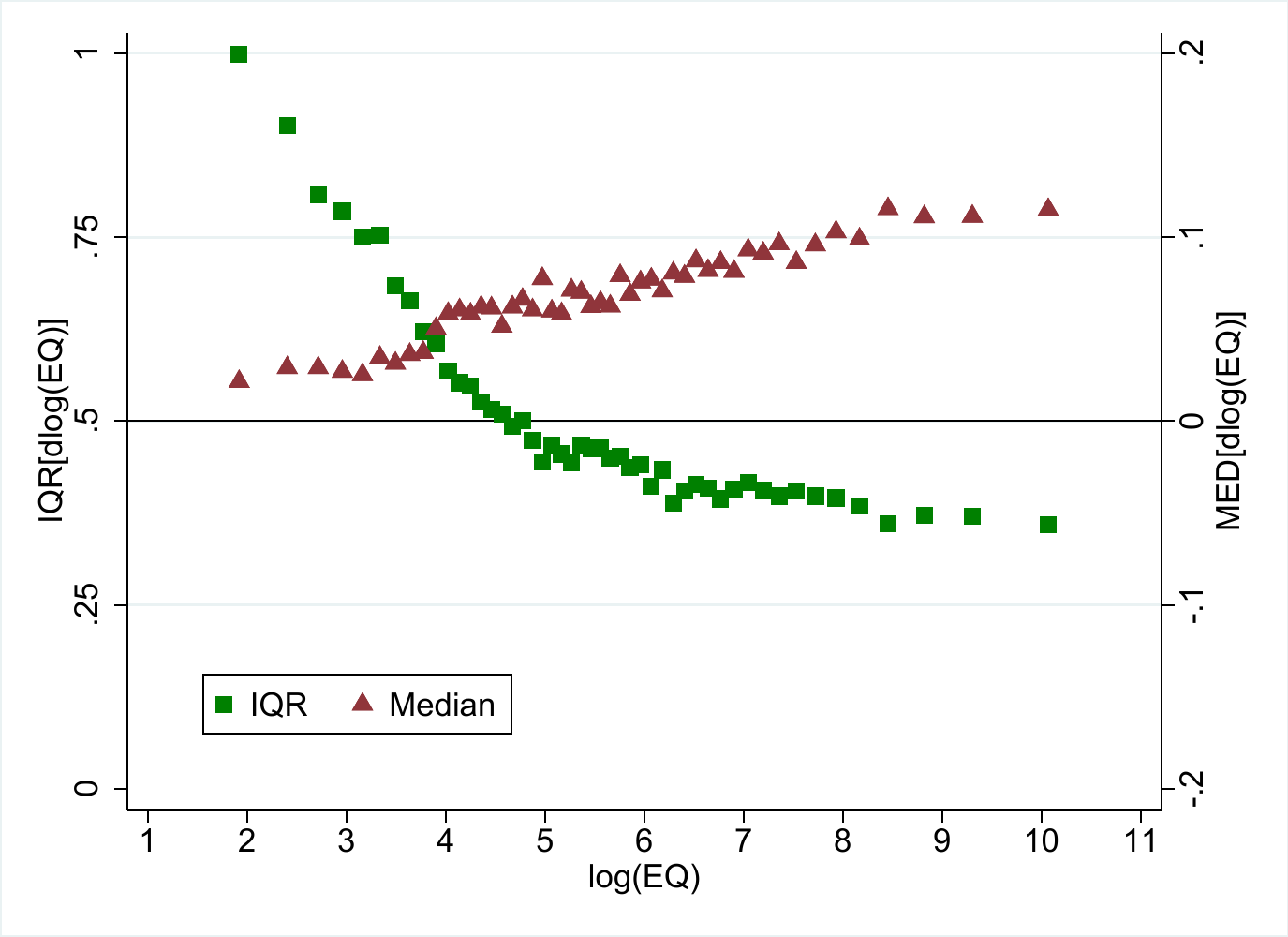}} & \subfigure[FF3 (M) Median+IQR ]{\includegraphics[width=2in]{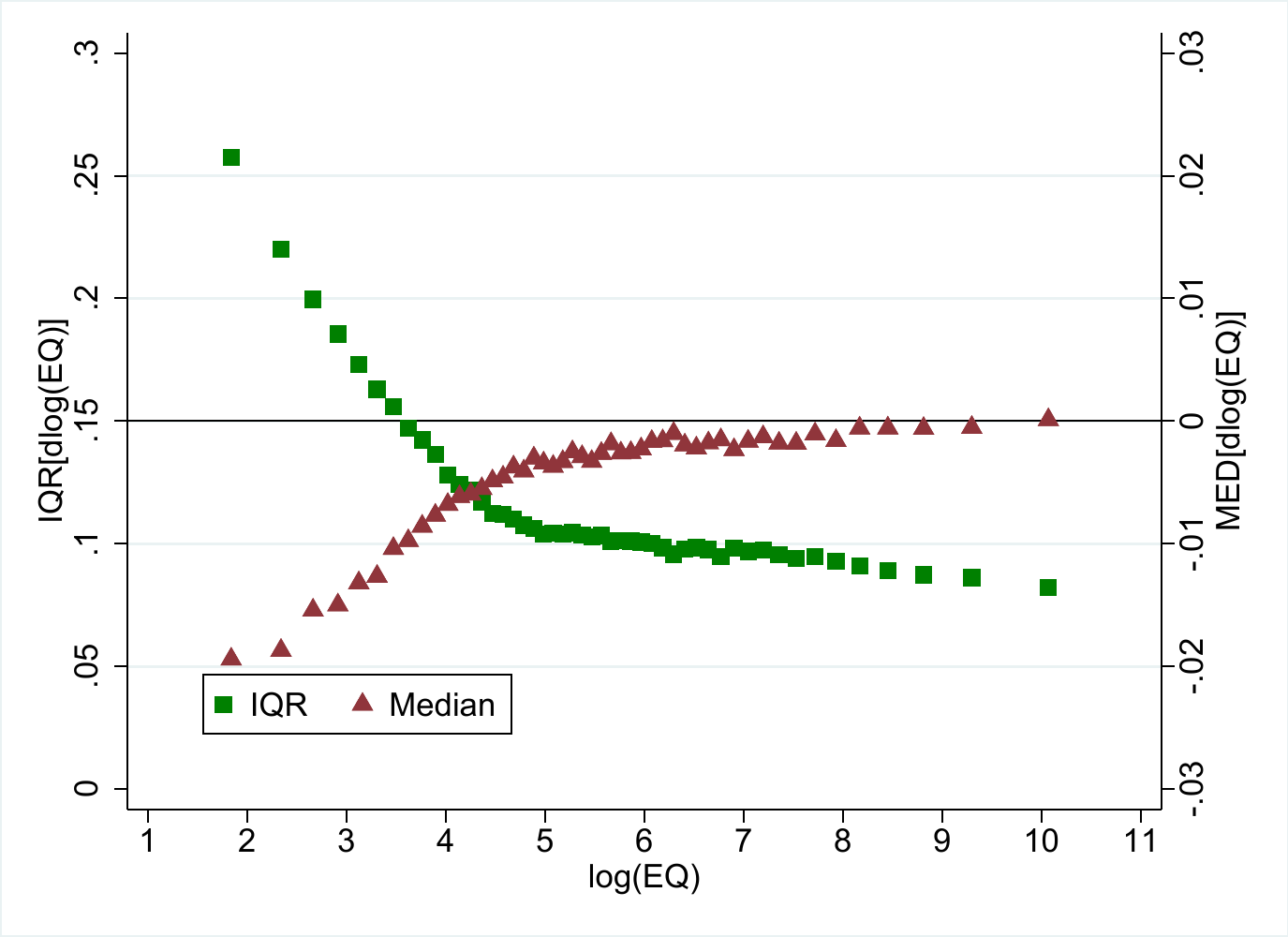}} &
		\subfigure[FF3 (D) Median+IQR ]{\includegraphics[width=2in]{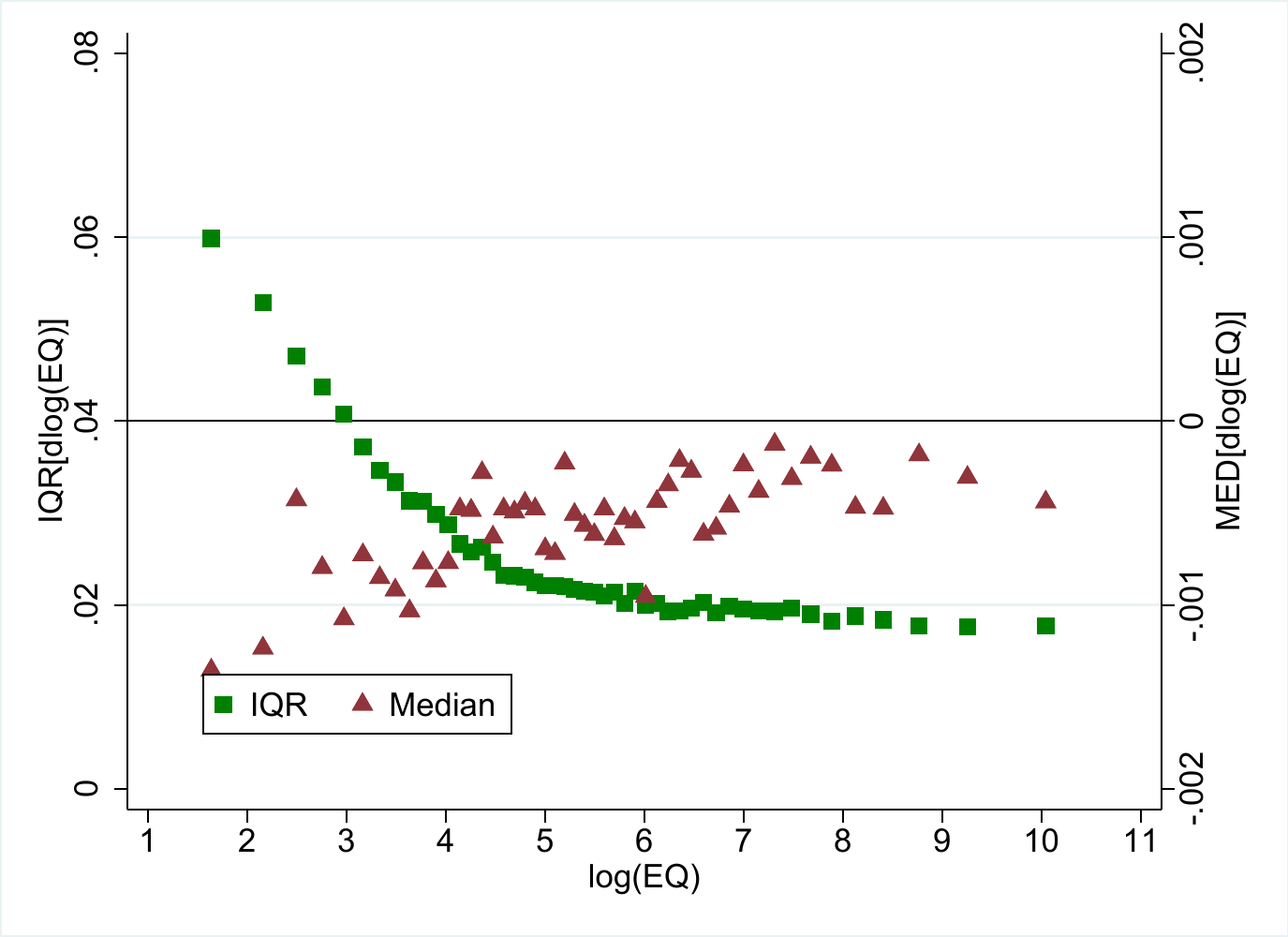}} \\ \\
		\subfigure[Raw (Y) w/ DLN ]{\includegraphics[width=2in]{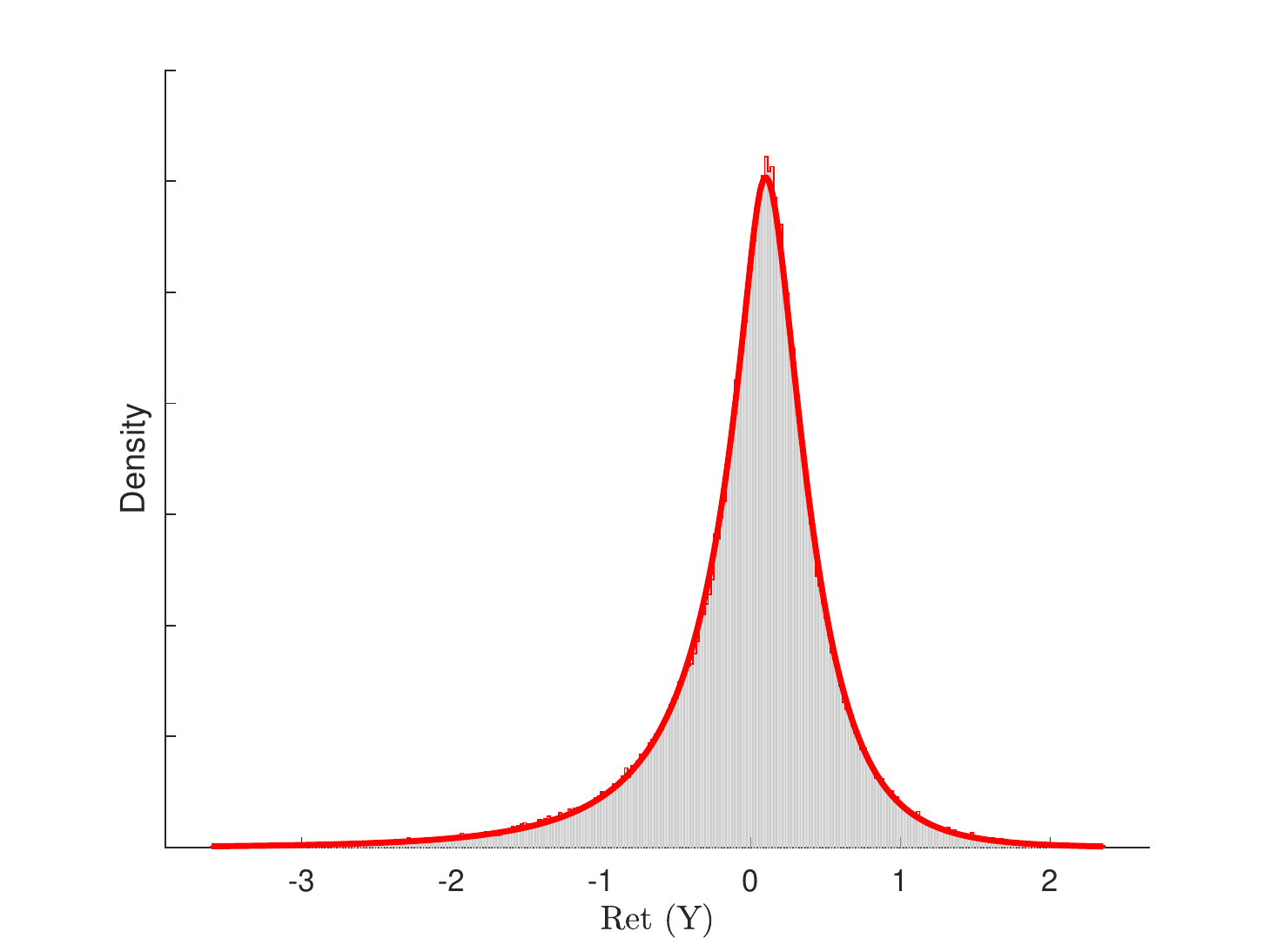}} &
		\subfigure[FF3 (M) w/ DLN ]{\includegraphics[width=2in]{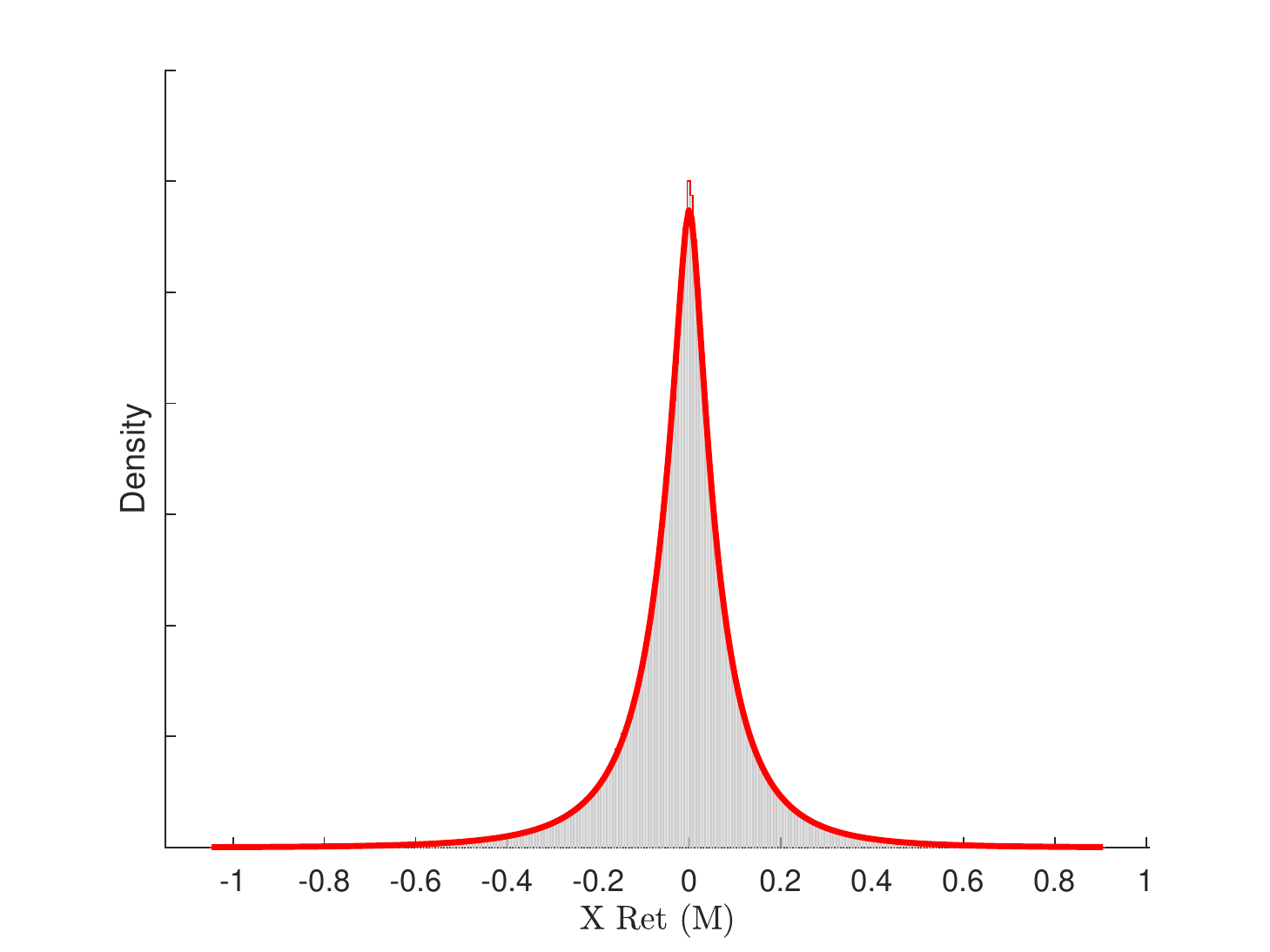}} &
		\subfigure[FF3 (D) w/ DLN ]{\includegraphics[width=2in]{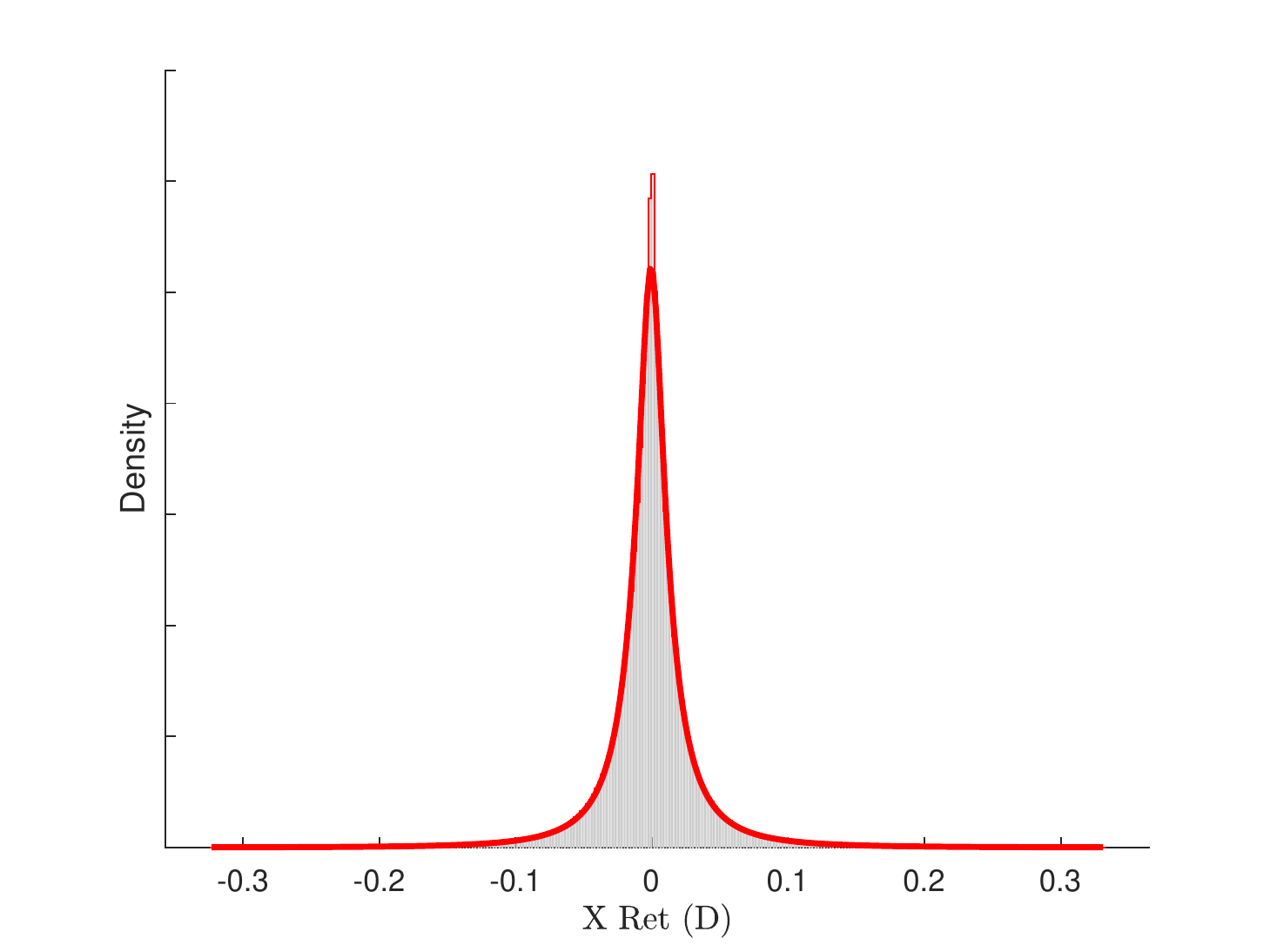}} \\ \\
		\subfigure[q-q Raw (Y) vs. DLN ]{\includegraphics[width=2in]{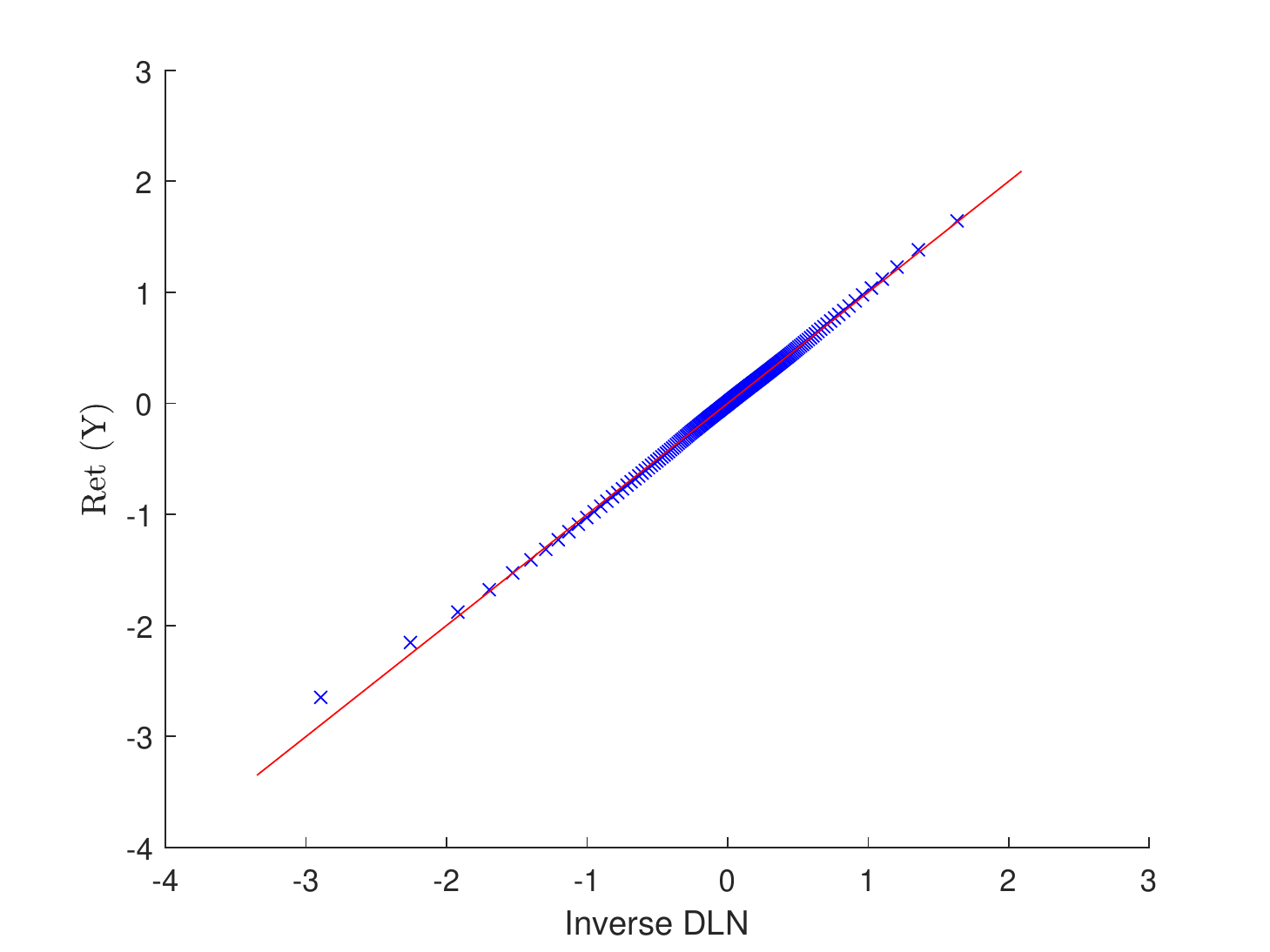}} &
		\subfigure[q-q FF3 (M) vs. DLN ]{\includegraphics[width=2in]{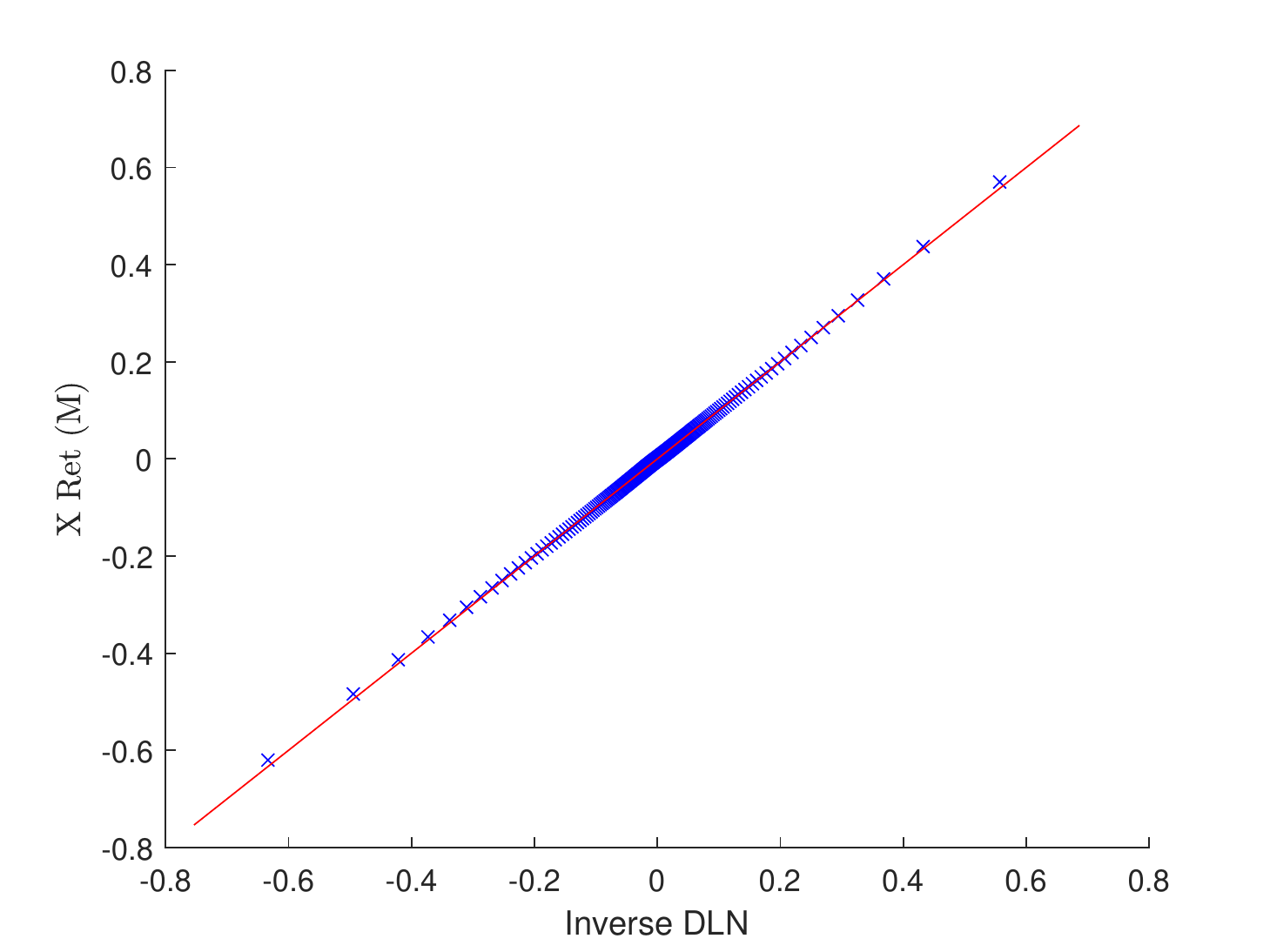}} &
		\subfigure[q-q FF3 (D) vs. DLN ]{\includegraphics[width=2in]{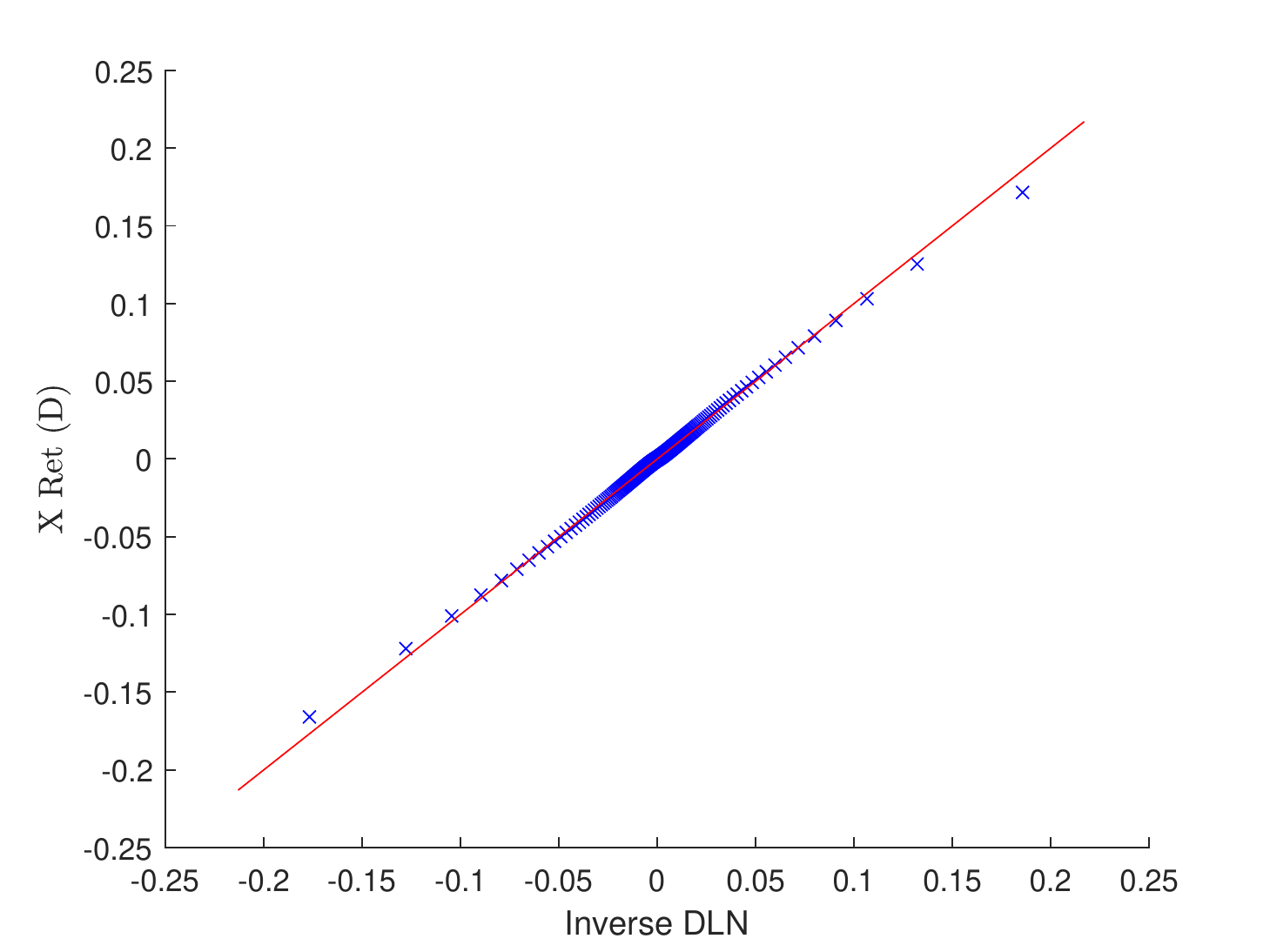}} \\ \\
	\end{tabular}
}

Next, Panels (d)-(f) present the distribution histograms for the respective flavors considered above. All histograms are overlaid with fitted DLN distributions, with remarkable fit. Panels (g)-(i) present the respective q-q plots vs. the DLN, again exhibiting excellent fit. The fit is likewise confirmed by the formal distributional tests reported in Table~\ref{tab:FGDEQdist}, which handily reject the Stable and Laplace in favor of the DLN. I conclude:

\begin{observation}
    \label{ob:8}
    The FGD(EQ) is distributed DLN.
\end{observation}

\RPprep{Equity return - Distributional tests}{1}{0}{FGDEQdist}{%
    This table presents results of tests of distribution equality for the following flavors of the FGD(EQ): \{yearly,monthly,daily\}X\{raw,excess\}X\{unadjusted,adjusted\}. Excess returns relative the the Fama-French 3-factor model are denoted FF3. Time- and scale-adjusted versions are marked with $\widetilde{XXX}$. K-S is a Kolmogorov–Smirnov test; C-2 is a binned $\chi^2$ test with 50 bins; A-D is an Anderson-Darling test. Panels (a)-(c) report the test statistics and their p-values for the Stable, Laplace, and DLN, respectively. Panel (d) reports the relative likelihoods for each distribution.
}
\RPtab{%
    \begin{tabularx}{\linewidth}{Frrrrrrrrrrrr}
    \toprule
	& \multicolumn{4}{c}{Yearly} & \multicolumn{4}{c}{Monthly} & \multicolumn{4}{c}{Daily} \\
	& Raw & FF3 & $\widetilde{\text{Raw}}$ & $\widetilde{\text{FF3}}$ & Raw & FF3 & $\widetilde{\text{Raw}}$ & $\widetilde{\text{FF3}}$ & Raw & FF3 & $\widetilde{\text{Raw}}$ & $\widetilde{\text{FF3}}$ \\
	\midrule
    \\ \multicolumn{13}{l}{\textit{Panel (a): FGD(EQ) vs. Stable}}\\
	\midrule
    K-S   & 0.017 & 0.017 & 0.015 & 0.015 & 0.013 & 0.015 & 0.011 & 0.013 & 0.021 & 0.014 & 0.014 & 0.015  \\
    p-val & 0.040 & 0.041 & 0.044 & 0.044 & 0.049 & 0.045 & 0.056 & 0.050 & 0.035 & 0.046 & 0.047 & 0.043  \\
    C-2   & 236   & 258   & 126   & 180   & 173   & 195   & 108   & 154   & 359   & 201   & 178   & 204    \\
    p-val & 0.024 & 0.023 & 0.034 & 0.028 & 0.029 & 0.027 & 0.037 & 0.031 & 0.019 & 0.027 & 0.029 & 0.027  \\
    A-D   & 8.54  & 9.12  & 4.88  & 6.74  & 6.07  & 6.67  & 4.18  & 5.75  & 8.32  & 6.16  & 5.77  & 6.70   \\
    p-val & 0.035 & 0.034 & 0.043 & 0.039 & 0.040 & 0.039 & 0.046 & 0.041 & 0.036 & 0.040 & 0.041 & 0.039  \\
    \\ \multicolumn{13}{l}{\textit{Panel (b): FGD(EQ) vs. Laplace}}\\
	\midrule
    K-S   & 0.027 & 0.030 & 0.016 & 0.015 & 0.023 & 0.026 & 0.013 & 0.013 & 0.045 & 0.051 & 0.025 & 0.027  \\
    p-val & 0.028 & 0.024 & 0.042 & 0.044 & 0.032 & 0.028 & 0.048 & 0.048 & 0.015 & 0.012 & 0.029 & 0.027  \\
    C-2   & 134   & 203   & 72    & 55    & 163   & 244   & 62    & 58    & 714   & 761   & 159   & 182    \\
    p-val & 0.033 & 0.027 & 0.044 & 0.050 & 0.030 & 0.024 & 0.048 & 0.049 & 0.011 & 0.010 & 0.030 & 0.028  \\
    A-D   & 9.21  & 14.66 & 4.00  & 3.92  & 12.62 & 19.38 & 4.52  & 5.27  & 50.80 & 62.69 & 15.26 & 17.40  \\
    p-val & 0.034 & 0.028 & 0.046 & 0.047 & 0.030 & 0.025 & 0.044 & 0.042 & 0.014 & 0.012 & 0.028 & 0.026  \\
    \\ \multicolumn{13}{l}{\textit{Panel (c): FGD(EQ) vs. DLN}}\\
	\midrule
    K-S   & 0.004 & 0.004 & 0.002 & 0.003 & 0.004 & 0.003 & 0.002 & 0.003 & 0.046 & 0.013 & 0.006 & 0.007  \\
    p-val & 0.113 & 0.115 & 0.256 & 0.166 & 0.115 & 0.148 & 0.277 & 0.148 & 0.015 & 0.050 & 0.084 & 0.080  \\
    C-2   & 10    & 12    & 3     & 6     & 8     & 5     & 2     & 9     & 729   & 60    & 18    & 21     \\
    p-val & 0.123 & 0.111 & 0.962 & 0.199 & 0.152 & 0.353 & 1.000 & 0.136 & 0.011 & 0.048 & 0.086 & 0.078  \\
    A-D   & 0.38  & 0.42  & 0.03  & 0.13  & 0.13  & 0.10  & 0.05  & 0.19  & 37.20 & 1.74  & 0.28  & 0.47   \\
    p-val & 0.096 & 0.093 & 0.297 & 0.138 & 0.137 & 0.148 & 0.199 & 0.120 & 0.018 & 0.061 & 0.105 & 0.090  \\
\\ \multicolumn{13}{l}{\textit{Panel (d): Distribution comparison}}\\
	\midrule
	\multicolumn{2}{l}{AIC R.L.:} \\
    Stable  & 0.000 & 0.000 & 0.000 & 0.000 & 0.000 & 0.000 & 0.000 & 0.000 & 0.000 & 0.000 & 0.000 & 0.000 \\
    Laplace & 0.000 & 0.000 & 0.000 & 0.000 & 0.000 & 0.000 & 0.000 & 0.000 & 0.000 & 0.000 & 0.000 & 0.000 \\
    DLN     & 1.000 & 1.000 & 1.000 & 1.000 & 1.000 & 1.000 & 1.000 & 1.000 & 1.000 & 1.000 & 1.000 & 1.000 \\
	\multicolumn{2}{l}{BIC R.L.:} \\
    Stable  & 0.000 & 0.000 & 0.000 & 0.000 & 0.000 & 0.000 & 0.000 & 0.000 & 0.000 & 0.000 & 0.000 & 0.000 \\
    Laplace & 0.000 & 0.000 & 0.000 & 0.000 & 0.000 & 0.000 & 0.000 & 0.000 & 0.000 & 0.000 & 0.000 & 0.000 \\
    DLN     & 1.000 & 1.000 & 1.000 & 1.000 & 1.000 & 1.000 & 1.000 & 1.000 & 1.000 & 1.000 & 1.000 & 1.000 \\
	\bottomrule
    \end{tabularx}
}

\subsection{Income growth}
\label{sec:IGrowth}

The last growth distribution I analyze is the firm income growth distribution (FIGD). Measuring income growth is complicated by the fact income is distributed DLN as per Observation~\ref{ob:4}. Incomes are sometimes negative, and the question ``what is the growth in income for a firm that lost \$100M last year and earned \$80M this year'' is not generally well-formed. \cite{Parham2022a} tackles the measurement of growth in DLN variables, showing that for a DLN RV $W_{t} = Y^{p}_{t} - Y^{n}_{t}$, the growth in $W$ from $t$ to $t+1$ is given by
\begin{equation} \label{eq:DLNGROWTH}
\frac{Y^{p}_{t}\cdot\text{dlog}\left(Y^{p}_{t+1}\right) - Y^{n}_{t}\cdot\text{dlog}\left(Y^{n}_{t+1}\right)}{\left|Y^{p}_{t} - Y^{n}_{t}\right|}
\end{equation}

With a coherent growth measure in hand, we can turn to Table~\ref{tab:FIGDDescMom} which presents the descriptive statistics of income growth for the four income measures of Section~\ref{sec:Income}, as well as their adjusted flavors. Panel (a) presents the moments for all valid observations, while Panels (b) and (c) separately consider income growth when the initial income is positive and negative, respectively. All FIGD flavors exhibit heavy tails (high kurtosis) and some have high skewness, so the panels report the medians as well as means. The difference in median between the three panels is instructive.

\RPprep{Income growth - Descriptive statistics}{0}{0}{FIGDDescMom}{%
    Panel (a) of this table presents the first four central moments and the median of the firm income growth distribution (FIGD), based on the four measures of income from Table~\ref{tab:FIDdist} and the DLN growth measure of \cite{Parham2022a}. Time- and scale-adjusted versions (relative to total capital KT) are marked with $\widetilde{XX}$, and the adjustment procedure is described in Appendix~\ref{sec:AppdxFE}. Panels (b) and (c) condition on the sign of beginning-of-period values, such that Panel (b) considers the distribution of income growth from initial positive income values and Panel (c) on income growth from initial negative income values.
}
\RPtab{%
    \begin{tabularx}{\linewidth}{Frrrrrrrr}
    \toprule
	& CF & CA & DI & DE & $\widetilde{\text{CF}}$ & $\widetilde{\text{CA}}$ & $\widetilde{\text{DI}}$ & $\widetilde{\text{DE}}$\\ 
	\midrule
    \\ \multicolumn{9}{l}{\textit{Panel (a): All observations}}\\
	\midrule
    $M_{1}$ & 0.1910  & 0.0429  & 0.3102  & 0.1328  & 0.1956  & 0.0463  & 0.2857  & 0.1231   \\
    $M_{2}$ & 1.4549  & 0.6568  & 2.9249  & 1.9885  & 1.4661  & 0.6096  & 2.9196  & 2.1924   \\
    $M_{3}$ & 1.5010  & 0.0776  & 1.3826  & -0.3515 & 1.8638  & 0.4423  & 1.3225  & -0.4057  \\
    $M_{4}$ & 12.77   & 10.20   & 12.23   & 17.20   & 14.22   & 10.39   & 12.18   & 17.61    \\
    Med     & 0.0384  & 0.0351  & -0.0642 & 0.0219  & 0.0395  & 0.0348  & -0.0781 & 0.0199   \\
    \\ \multicolumn{9}{l}{\textit{Panel (b): Conditional on beginning positive}}\\
	\midrule
    $M_{1}$ & 0.0093  & -0.0051 & -0.2018 & -0.0130 & 0.0240  & 0.0067  & -0.2197 & -0.0292  \\
    $M_{2}$ & 1.0958  & 0.5207  & 2.4838  & 1.5073  & 1.1252  & 0.5038  & 2.4857  & 1.7336   \\
    $M_{3}$ & 0.7635  & -0.8674 & 1.4885  & 0.6271  & 1.0839  & -0.3578 & 1.4130  & 0.0673   \\
    $M_{4}$ & 12.06   & 9.94    & 14.16   & 16.56   & 12.45   & 8.83    & 14.03   & 17.14    \\
    Med     & -0.0073 & 0.0248  & -0.4012 & -0.0406 & -0.0064 & 0.0234  & -0.4158 & -0.0618  \\
    \\ \multicolumn{9}{l}{\textit{Panel (c): Conditional on beginning negative}}\\
	\midrule
    $M_{1}$ & 1.2514  & 0.5799  & 1.0553  & 0.4629  & 1.2126  & 0.5006  & 1.0204  & 0.4897   \\
    $M_{2}$ & 2.9827  & 1.7058  & 3.3746  & 3.2869  & 3.0355  & 1.5580  & 3.3585  & 3.4147   \\
    $M_{3}$ & 1.8312  & 0.3717  & 1.2318  & -1.5929 & 2.3865  & 1.1673  & 1.1591  & -1.1858  \\
    $M_{4}$ & 13.42   & 8.02    & 11.74   & 18.86   & 15.75   & 10.60   & 11.57   & 19.07    \\
    Med     & 0.9469  & 0.4485  & 0.8686  & 0.7920  & 0.6945  & 0.3143  & 0.8366  & 0.6382   \\
    \end{tabularx}
}

In Panel (a), median income growth rates are low and positive, as expected, with the exception of growth in total dispensations FIGD(DI). When concentrating on positive beginning-of-period values in Panel (b), median FIGD(DI) is strongly negative, indicating a reversal in total dispensations to firm owners. Other income measures have median growth values relatively close to zero in that case as well. Panel (c) presents a starkly different situation: when beginning of period income is negative, we observe strong positive median growth in income during the period --- firms with negative income revert quickly and strongly.

Figure~\ref{fig:FIGDfacts} provides further evidence on income growth. Panel (a) presents the time-trends of the FIGD(CF), exhibiting similar business-cycle movements as before. Panel (b) presents the FIGD(CF) overlaid with a fitted DLN distribution, exhibiting good fit, which is corroborated by panel (c) presenting the q-q plot of the fit. Panel (d) presents the dependence of the FIGD(CF) on scale (measured as log firm value), showing stable median growth close to zero and decreasing dispersion with scale. The decreasing dispersion of cash-flow growth fits the patterns observed in Figure~\ref{fig:FGDDisp}, including a similar decreasing dispersion coefficient, as seen in panel (e).

\RPprep{Income growth - Stylized facts}{0}{0}{FIGDfacts}{%
    This figure presents stylized facts of the firm income growth distribution (FIGD). Panel (a) presents the FIGD time trend by plotting the $\{25,50,75\}^{th}$ percentiles of FIGD(CF) per year. Panel (b) presents the FIGD(CF) histogram overlaid with an MLE-fitted DLN, and panel (c) presents the respective q-q plot. Panel (d) presents the (25,50,75)$^{th}$ percentiles of FIGD(CF) for the 49 scale bins, and panel (e) plots the binscatter of their dispersion. Panel (f) plots the percentiles of FIGD(DI), and panel (g) plots its histogram, fitted with a DLN. Panels (h) and (i) again present the FIGD(DI) histogram, but split the sample to growth from positive values and growth from negative values, respectively.
}
\RPfig{%
	\begin{tabular}{ccc} 
		\subfigure[Time trend of FIGD ]{\includegraphics[width=2in]{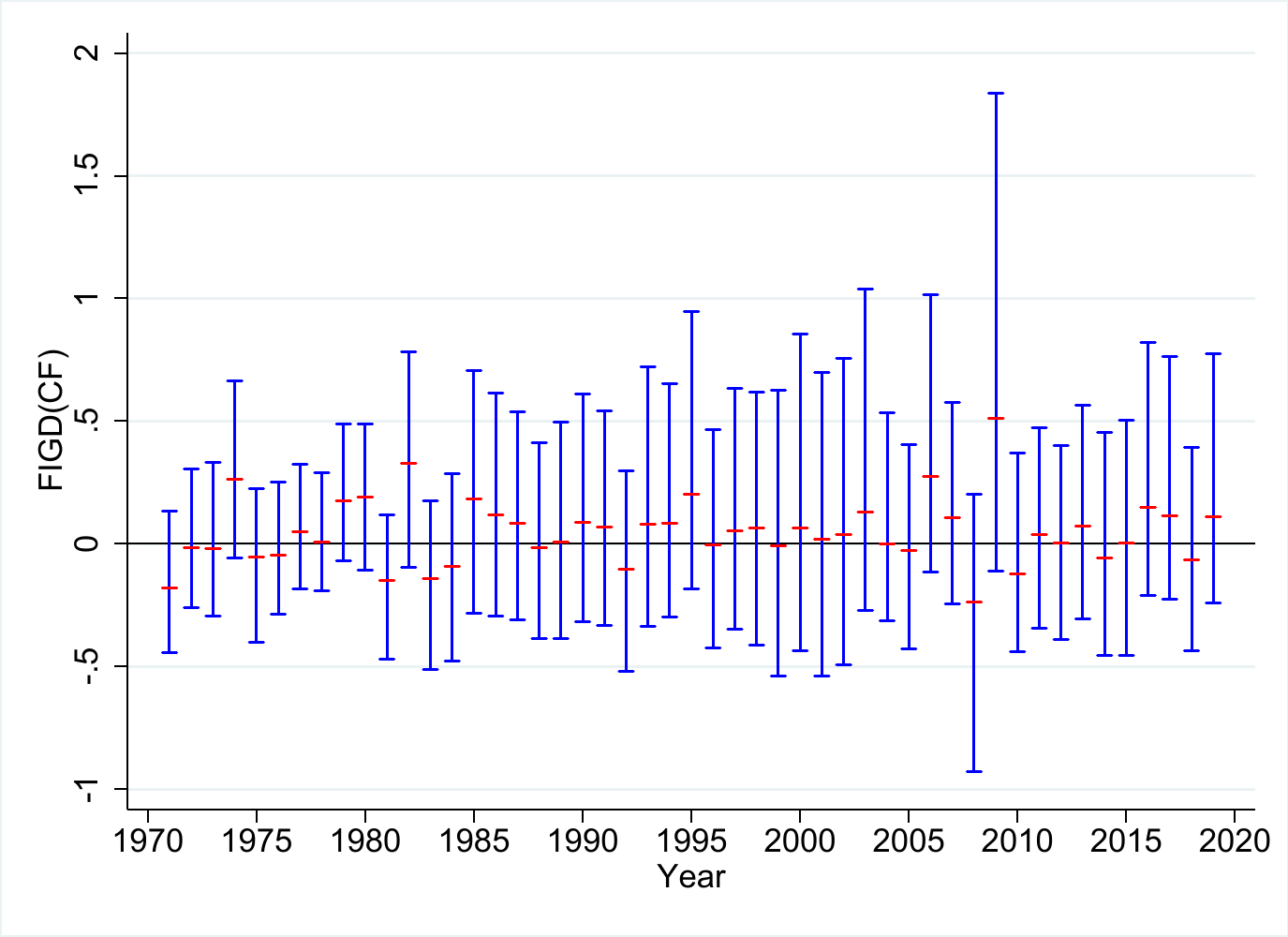}} & \subfigure[FIGD(CF) w/ DLN ]{\includegraphics[width=2in]{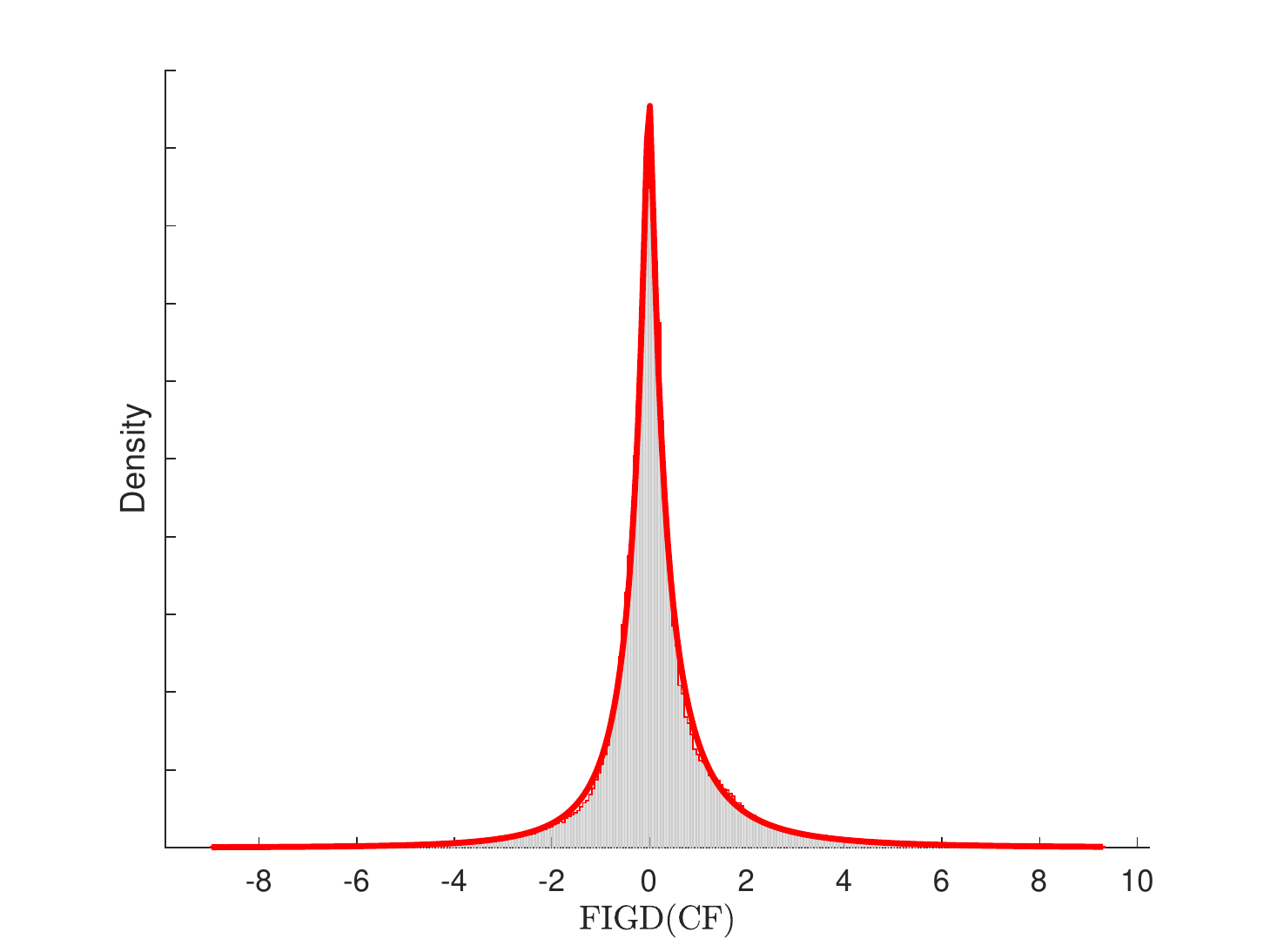}} &
		\subfigure[q-q FIGD(CF) vs. DLN ]{\includegraphics[width=2in]{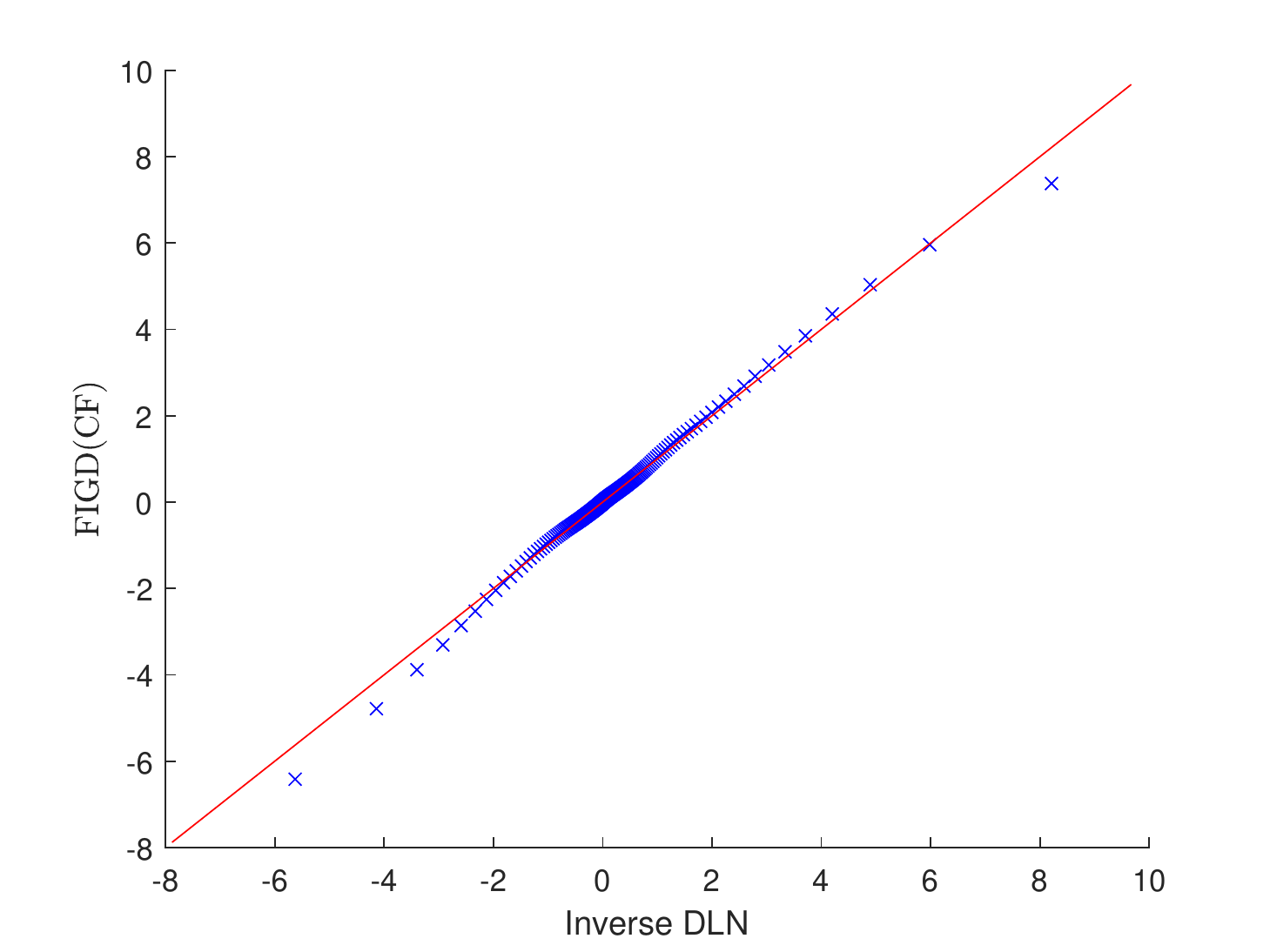}} \\ \\
		\subfigure[FIGD(CF) by scale ]{\includegraphics[width=2in]{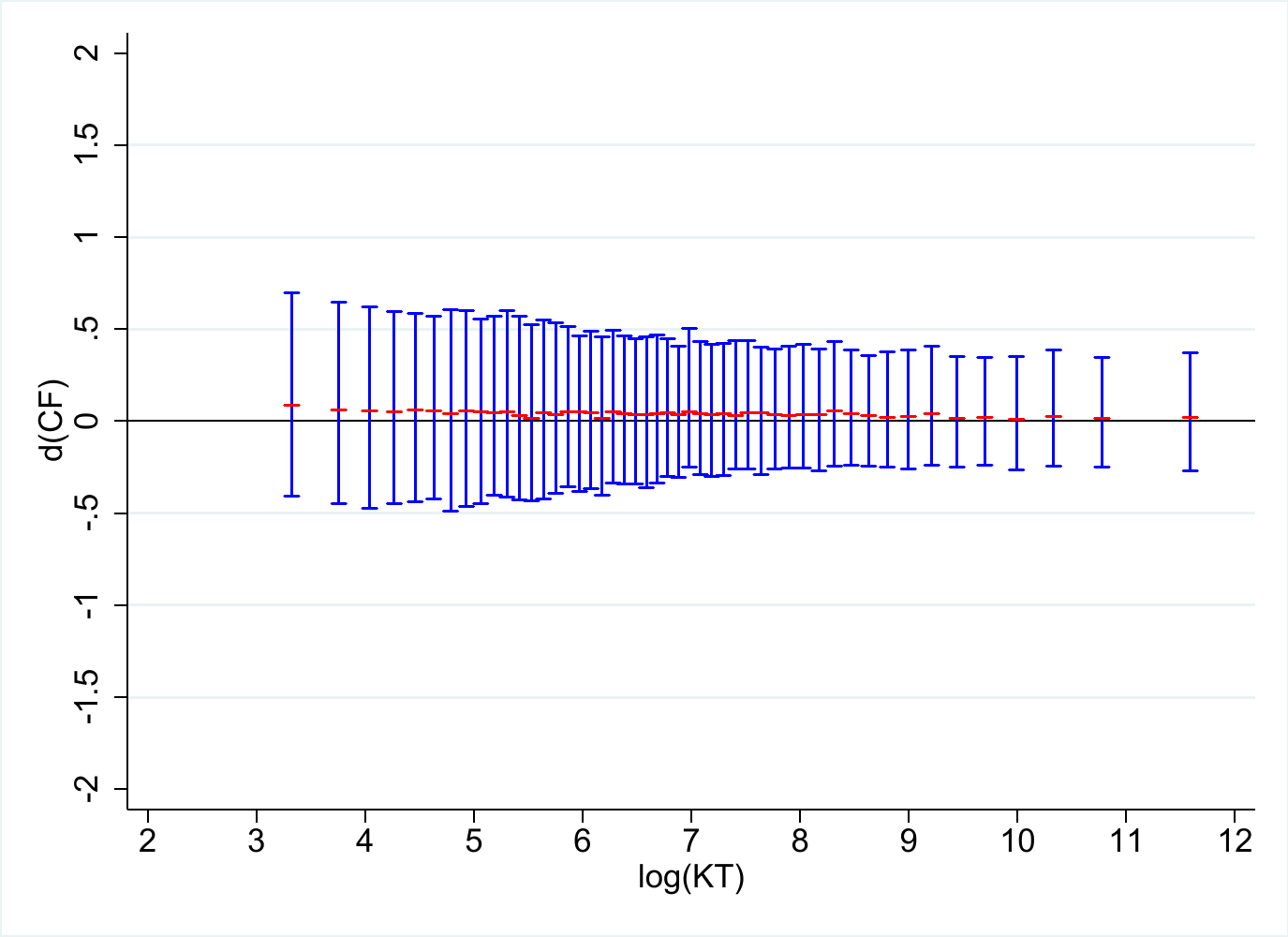}} &
		\subfigure[FIGD(CF) dispersion ]{\includegraphics[width=2in]{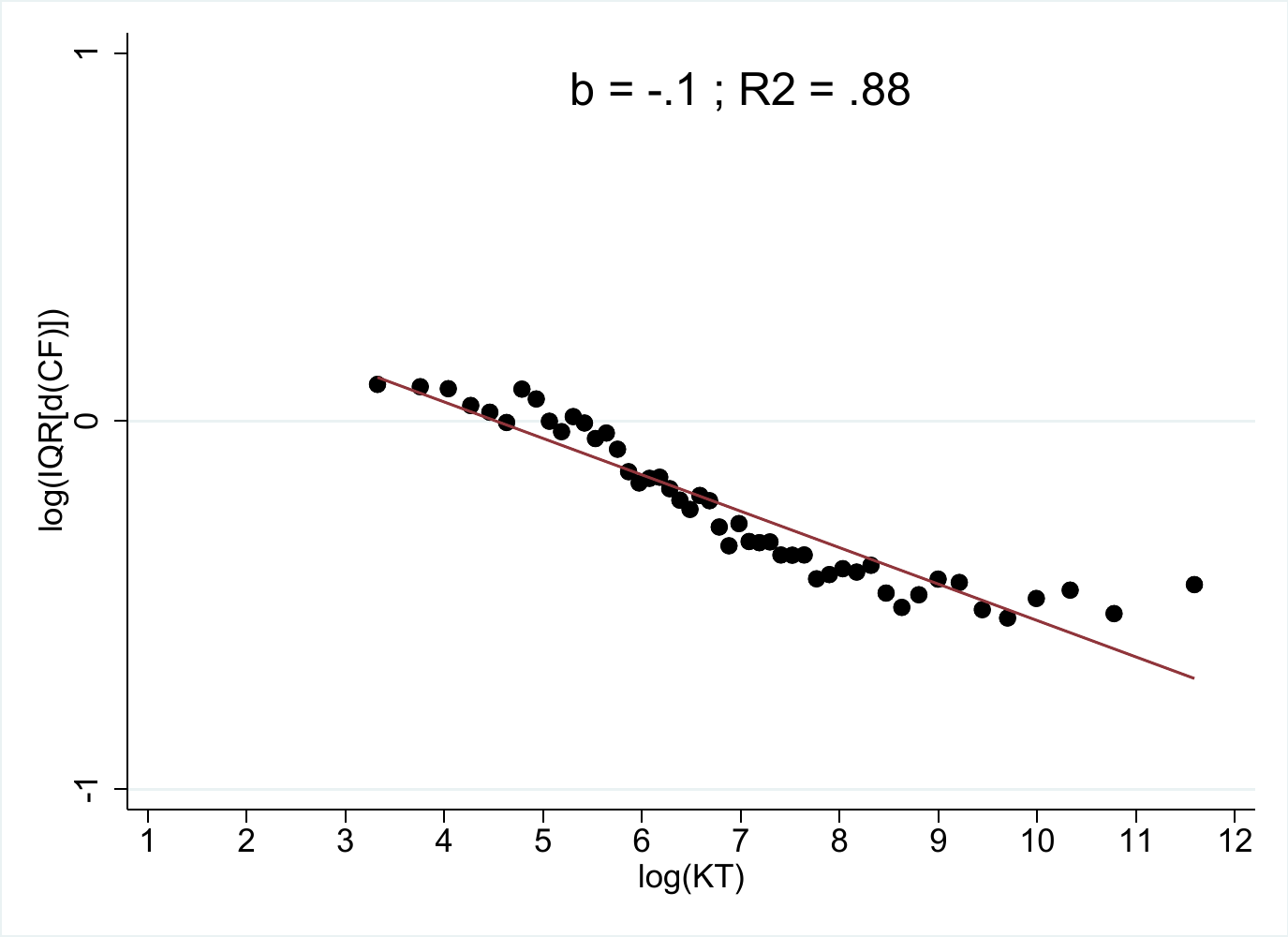}} &
		\subfigure[FIGD(DI) by scale ]{\includegraphics[width=2in]{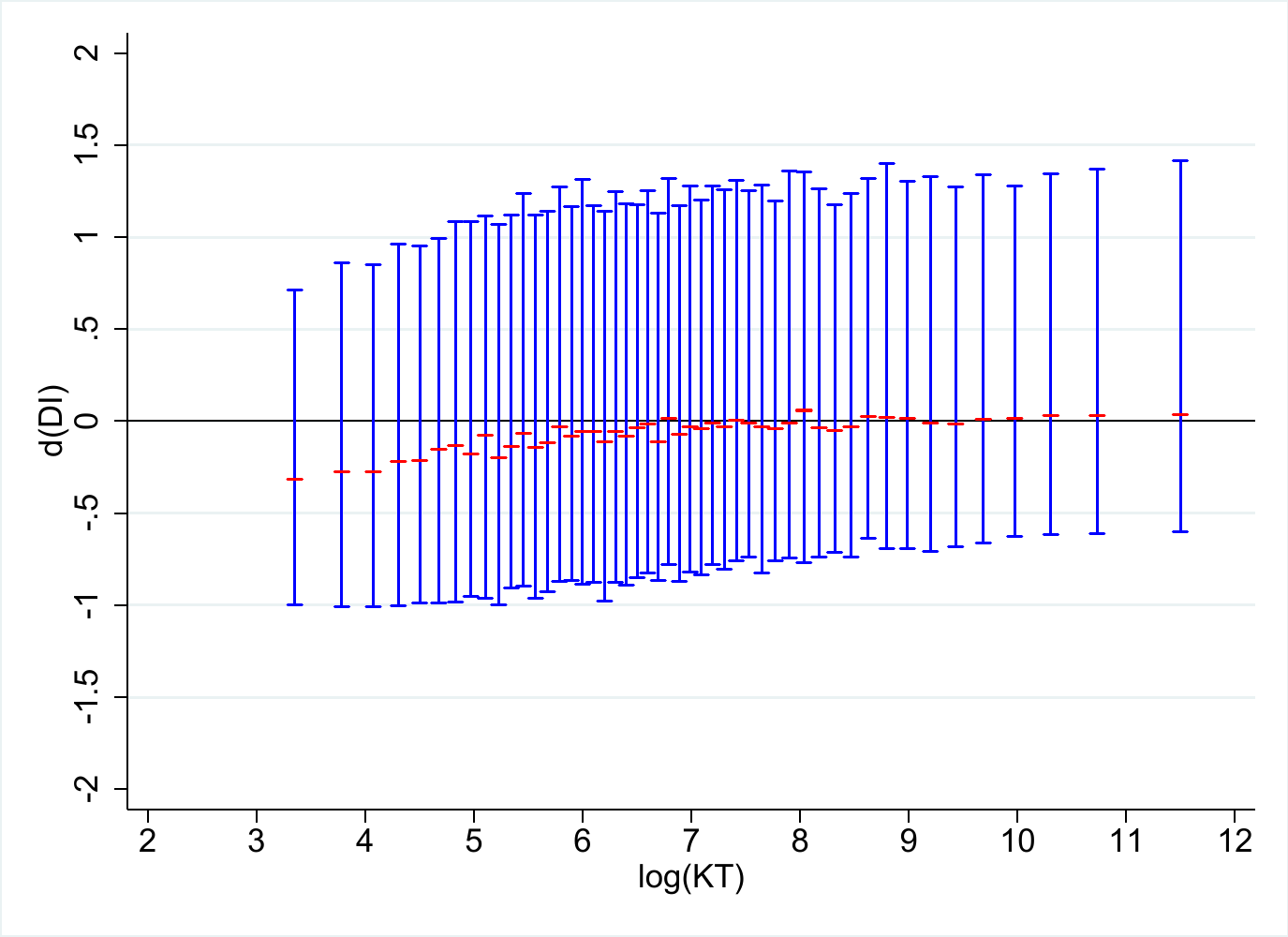}} \\ \\
		\subfigure[FIGD(DI) w/ DLN ]{\includegraphics[width=2in]{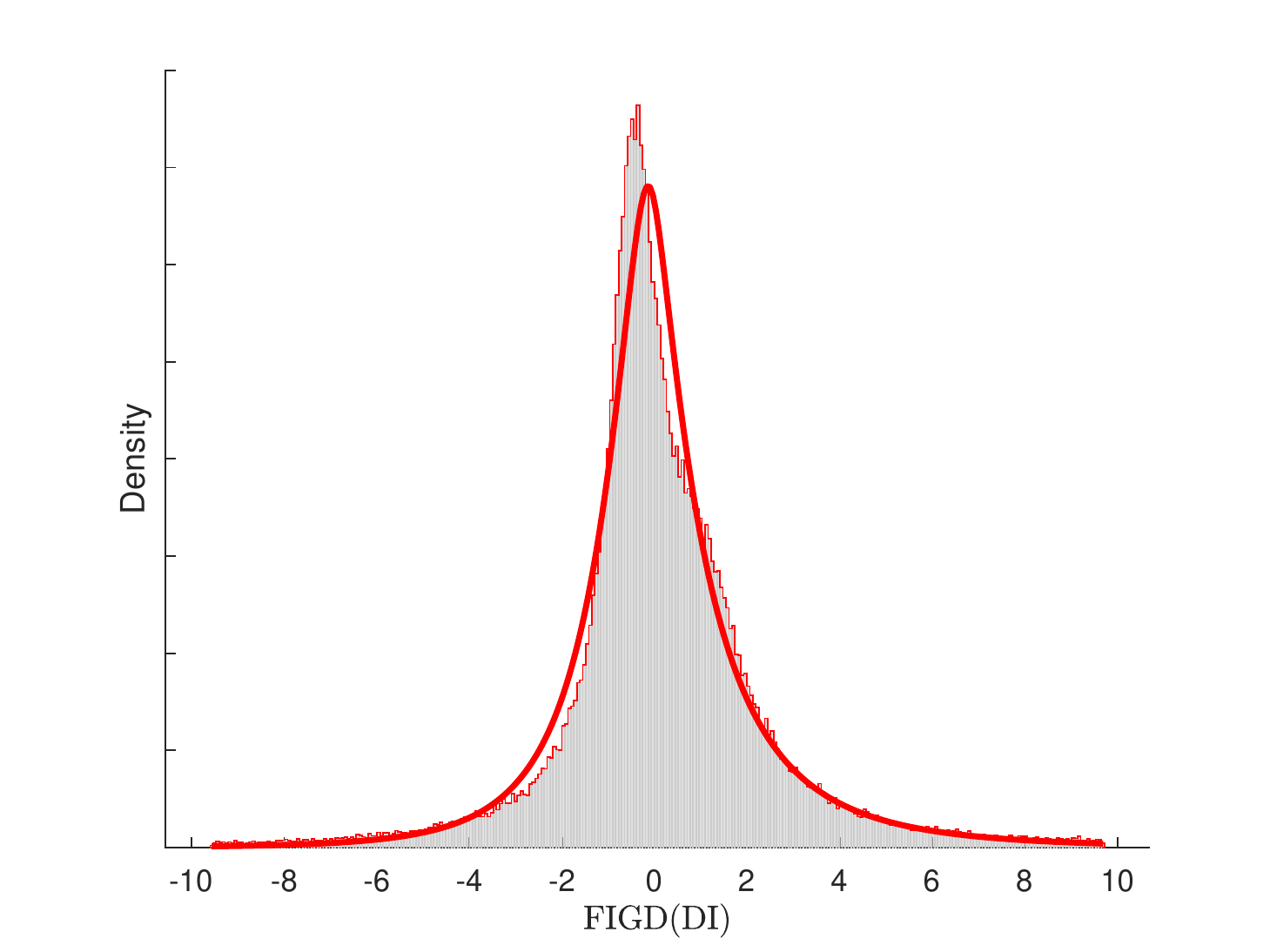}} &
		\subfigure[FIGD($\widetilde{\text{DI}})^{+}$ w/ DLN ]{\includegraphics[width=2in]{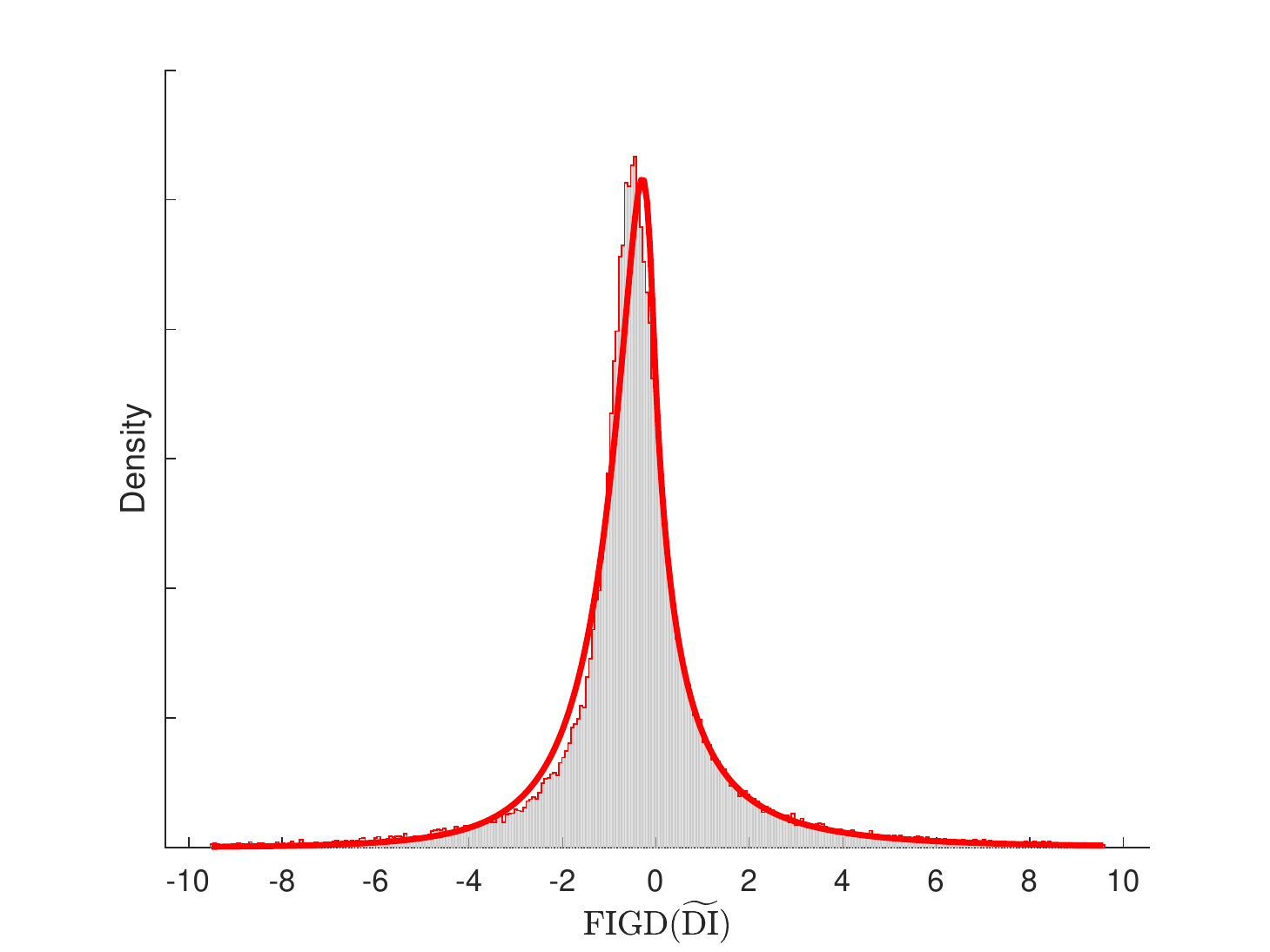}} &
		\subfigure[FIGD($\widetilde{\text{DI}})^{-}$ w/ DLN ]{\includegraphics[width=2in]{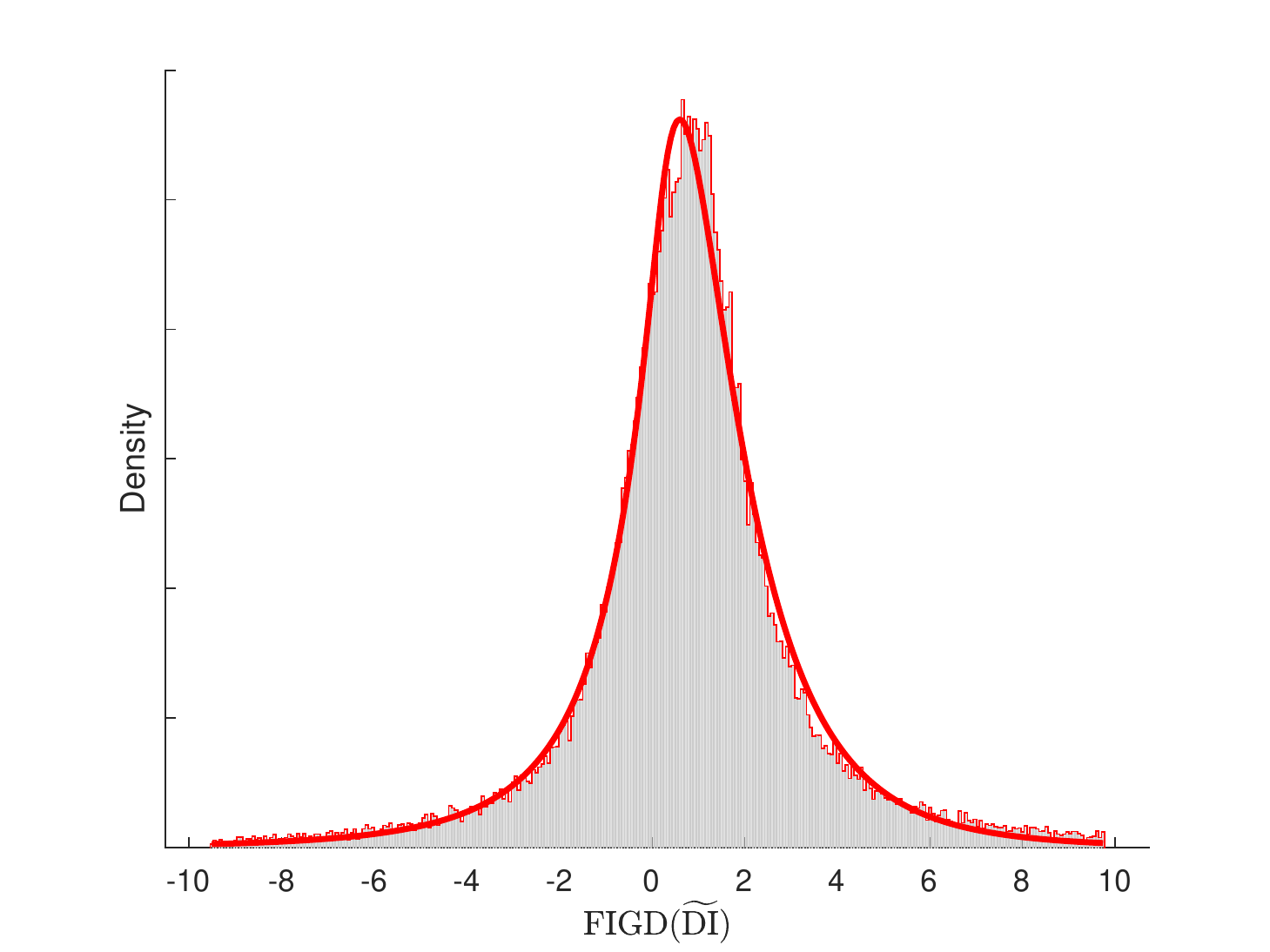}} \\ \\
	\end{tabular}
}

The same is however not true for firm total capital dispensations. Panel (f) presents the dependence of FIGD(DI) on scale, exhibiting increasing median and constant dispersion. Small firms have negative median capital dispensation growth, meaning they are prone to ``death spirals''. Panel (g) presents the FIGD(DI) overlaid with fitted DLN and exhibiting a fairly poor fit. The fit somewhat improves if we separately consider the growth from negative vs. from positive values, presented in panels (h) and (i), respectively. The difference in the location of both distributions, presented in panels (b) and (c) of Table~\ref{tab:FIGDDescMom}, can be observed in the last two panels of Figure~\ref{fig:FGDDisp}.

Distributional tests for the FIGD are reported in Table~\ref{tab:FIGDdist}. The Stable and Laplace distributions are again generally rejected. DLN is not rejected for FIGD(CF) and FIGD(CA), but unlike in previous tests, it is rejected for FIGD(DI) and FIGD(DE). The relative likelihood tests nevertheless continues to strongly favor the DLN over the Stable and Laplace for all flavors of the FIGD, as before. I conclude:

\begin{observation}
    \label{ob:9}
    The FIGD(CF) has decreasing dispersion with scale and is distributed DLN.
\end{observation}

\RPprep{Income growth - Distributional tests}{1}{0}{FIGDdist}{%
    This table presents results of tests of distribution equality for the FIGD, based on the income measures described in Table~\ref{tab:DataDef}, as well as their time- and scale-adjusted versions (relative to total capital KT), denoted $\widetilde{XX}$. K-S is a Kolmogorov–Smirnov test; C-2 is a binned $\chi^2$ test with 50 bins; A-D is an Anderson-Darling test. Panels (a)-(c) report the test statistics and their p-values for the Stable, Laplace, and DLN, respectively. Panel (d) reports the relative likelihoods for each distribution. The raw variables (columns 1-4) use the \{Good\} data subset. The scale adjusted versions use the \{Non-Bank\} subset.
}
\RPtab{%
    \begin{tabularx}{\linewidth}{Frrrrrrrr}
    \toprule
	& CF & CA & DI & DE & $\widetilde{\text{CF}}$ & $\widetilde{\text{CA}}$ & $\widetilde{\text{DI}}$ & $\widetilde{\text{DE}}$\\ 
	\midrule
    \\ \multicolumn{9}{l}{\textit{Panel (a): FIGD vs. Stable}}\\
	\midrule
    K-S   & 0.066 & 0.008 & 0.031 & 0.045  & 0.014 & 0.007 & 0.030 & 0.020  \\
    p-val & 0.008 & 0.070 & 0.024 & 0.015  & 0.046 & 0.079 & 0.024 & 0.036  \\
    C-2   & 486   & 59    & 342   & $>$999 & 120   & 42    & 344   & 183    \\
    p-val & 0.015 & 0.049 & 0.019 & 0.006  & 0.035 & 0.057 & 0.019 & 0.028  \\
    A-D   & 63.91 & 2.96  & 13.03 & 24.09  & 5.86  & 2.12  & 13.48 & 4.72   \\
    p-val & 0.012 & 0.051 & 0.030 & 0.022  & 0.041 & 0.057 & 0.029 & 0.044  \\
    \\ \multicolumn{9}{l}{\textit{Panel (a): FIGD vs. Laplace}}\\
	\midrule
    K-S   & 0.078  & 0.082  & 0.042 & 0.062  & 0.073  & 0.070  & 0.043 & 0.045  \\
    p-val & 0.004  & 0.004  & 0.017 & 0.009  & 0.006  & 0.006  & 0.016 & 0.015  \\
    C-2   & $>$999 & $>$999 & 555   & $>$999 & $>$999 & $>$999 & 570   & $>$999 \\
    p-val & 0.001  & 0.000  & 0.014 & 0.000  & 0.002  & 0.003  & 0.013 & 0.005  \\
    A-D   & 116.2  & 165.1  & 22.81 & 85.91  & 105.8  & 116.9  & 23.26 & 64.98  \\
    p-val & 0.006  & 0.003  & 0.023 & 0.009  & 0.007  & 0.006  & 0.023 & 0.012  \\
    \\ \multicolumn{9}{l}{\textit{Panel (c): FIGD vs. DLN}}\\
	\midrule
    K-S   & 0.009 & 0.005 & 0.024 & 0.033 & 0.007 & 0.005 & 0.026 & 0.015  \\
    p-val & 0.065 & 0.094 & 0.031 & 0.022 & 0.074 & 0.095 & 0.028 & 0.044  \\
    C-2   & 88    & 18    & 327   & 648   & 59    & 25    & 366   & 218    \\
    p-val & 0.040 & 0.085 & 0.020 & 0.012 & 0.049 & 0.073 & 0.019 & 0.026  \\
    A-D   & 1.85  & 0.54  & 10.63 & 12.12 & 1.11  & 0.69  & 12.03 & 3.78   \\
    p-val & 0.059 & 0.086 & 0.032 & 0.031 & 0.070 & 0.080 & 0.031 & 0.047  \\
    \\ \multicolumn{9}{l}{\textit{Panel (d): Distribution comparison}}\\
	\midrule
	\multicolumn{2}{l}{AIC R.L.:} \\
    Stable  & 0.000 & 0.000 & 0.000 & 0.000 & 0.000 & 0.000 & 0.000 & 0.000 \\
    Laplace & 0.000 & 0.000 & 0.000 & 0.000 & 0.000 & 0.000 & 0.000 & 0.000 \\
    DLN     & 1.000 & 1.000 & 1.000 & 1.000 & 1.000 & 1.000 & 1.000 & 1.000 \\
	\multicolumn{2}{l}{BIC R.L.:} \\
    Stable  & 0.000 & 0.000 & 0.000 & 0.000 & 0.000 & 0.000 & 0.000 & 0.000 \\
    Laplace & 0.000 & 0.000 & 0.000 & 0.000 & 0.000 & 0.000 & 0.000 & 0.000 \\
    DLN     & 1.000 & 1.000 & 1.000 & 1.000 & 1.000 & 1.000 & 1.000 & 1.000 \\
	\bottomrule
    \end{tabularx}
}

\section{Investment, ratios, and dynamism}
\label{sec:InvestDyn}

\subsection{Physical investment and depreciation}
\label{sec:PhysInvestment}

Firm scale and growth patterns are inexorably related to firm investment --- the means of growth and capital accumulation. I next review the data on the firm investment distribution (FND), composed of four panels: total investment, measured as changes in total capital net of depreciation (IT); physical investment, measured as changes in physical capital net of depreciation (IP); physical investment, measured in the traditional way as capital expenditures minus capital sales (IA); and depreciation as reported by firms (DP). Depreciation, while not an instance of investment, is nevertheless closely related to it and is hence included in the analysis. 

Table~\ref{tab:FNDDescMom} reports descriptive statistics of the FND. For the three investment types, which are sometimes negative, the table presents the statistics separately for positive and negative investment, in asinh scale. Low skewness and kurtosis close to 3 on both sides indicates approximate (log-)Normality on each side, an observation which also extends to the (always positive) depreciation panel. Similar to firm income, the 1-lag persistence on the positive side is relatively high, while on the negative side persistence is low or negative. Firms with positive investment persist while firms with negative investment revert. Depreciation is highly persistent, as expected from an accounting variable largely out of the firm's control. More than a third of total investment observations are negative - implying an analysis of investment patterns which ignores negative investments is fairly incomplete. Investment intensities are skewed and heavy-tailed, but exhibit no significant scaling behavior. The same is not true for depreciation intensity (relative to physical capital KP), which is decreasing with scale.

\RPprep{Investment - Descriptive statistics}{0}{0}{FNDDescMom}{%
    The first four columns of this table present statistics for the positive side of the FND (incl. depreciation DP), while the next three concentrate on the negative side and the last four on the entire investment intensity distribution (i.e., investment divided by total capital KT, and DP/KP). Variable definitions are in Table~\ref{tab:DataDef}. The positive and negative sides are in asinh scale. $M_1$ to $M_4$ are the first four central moments. Persistence for the positives and negatives is the 1-lag pooled auto-correlation, while for intensities it is the \cite{ArellanoBover1995}/\cite{BlundellBond1998} panel estimate. \%Ob is the percent of total observations in the positive/negative side, and may not sum to 100\% due to zero-valued obs. The scaling coefficient is the slope in a quantile regression of (asinh) income or of intensity on scale. All values are based on the \{Non-Bank\} data subset.
}
\RPtab{%
    \begin{tabularx}{\linewidth}{Frrrrrrrrrrr}
    \toprule
	& IT$^{+}$ & IP$^{+}$ & IA$^{+}$ & DP & IT$^{-}$ & IP$^{-}$ & IA$^{-}$ & $\widetilde{\text{IT}}$ & $\widetilde{\text{IP}}$ & $\widetilde{\text{IA}}$ & $\widetilde{\text{DP}}$\\
	\midrule
    $M_{1}$       & 4.99 & 4.04 & 4.02 & 3.16 & -4.28 & -3.15 & -2.45 & 0.09 & 0.07 & 0.08 & 0.26  \\
    $M_{2}$       & 2.10 & 2.19 & 2.24 & 2.11 & 2.05  & 2.25  & 1.78  & 0.36 & 0.16 & 0.12 & 0.32  \\
    $M_{3}$       & 0.15 & 0.33 & 0.34 & 0.37 & -0.35 & -0.57 & -0.61 & 3.96 & 6.99 & 7.39 & 7.74  \\
    $M_{4}$       & 2.89 & 2.68 & 2.63 & 2.89 & 2.96  & 2.67  & 2.84  & 35   & 126  & 136  & 123   \\
    Pers.         & 0.24 & 0.51 & 0.87 & 0.99 & 0.08  & -0.06 & -0.23 & 0.10 & 0.14 & 0.26 & 0.07  \\
    \%Ob          & 0.63 & 0.85 & 0.96 & /    & 0.37  & 0.15  & 0.03  & /    & /    & /    & /     \\
    $\beta_{scl}$ & 0.89 & 0.98 & 1.03 & 0.86 & -0.86 & -0.94 & -0.77 & 0.00 & 0.00 & 0.00 & -0.02 \\
    \end{tabularx}
}

Figure~\ref{fig:FNDfacts} provides graphical evidence on FND patterns. Panels (a)+(d) present the total investment distribution FND(IT) overlaid with MLE-fitted DLN and its q-q plot. Panel (b)+(e) and (c)+(f) repeat for physical investment FND(IP) and for total investment intensity FND($\widetilde{\text{IT}})$, respectively. Investment intensity is measured relative to total capital KT. The FND exhibits excellent fit to the DLN. Panel (g) presents the dependence of FND(IT) on scale, with investment showing a similar pattern to the ``hollow middle'' of the income distribution in panel (g) of Figure~\ref{fig:FIDfacts}. Panel (h) then presents the dependence of investment intensity FND($\widetilde{\text{IT}})$ on firm scale, exhibiting decreasing dispersion of intensity.

\RPprep{Investment - Stylized facts}{0}{0}{FNDfacts}{%
    This figure presents stylized facts of the firm investment distribution (FND). Panels (a)+(d) present the total investment distribution FND(IT) overlaid with MLE-fitted DLN and its q-q plot. Panel (b)+(e) and (c)+(f) repeat for physical investment FND(IP) and for total investment intensity FND($\widetilde{\text{IT}})$, respectively. Panel (g) presents the dependence of the FND on firm scale, by presenting the (10,25,50,75,90)$^{th}$ percentiles of FND(IT) conditional on the sign of IT, for 49 KT scale bins. Panel (h) presents the dependence of investment intensity FND($\widetilde{\text{IT}})$ on scale. Panel (i) presents the dependence of physical depreciation intensity, DP/KP, on physical capital scale KP. All figures are based on the \{Non-Bank\} data subset.
}
\RPfig{%
	\begin{tabular}{ccc} 
		\subfigure[FND(IT) w/ DLN ]{\includegraphics[width=2in]{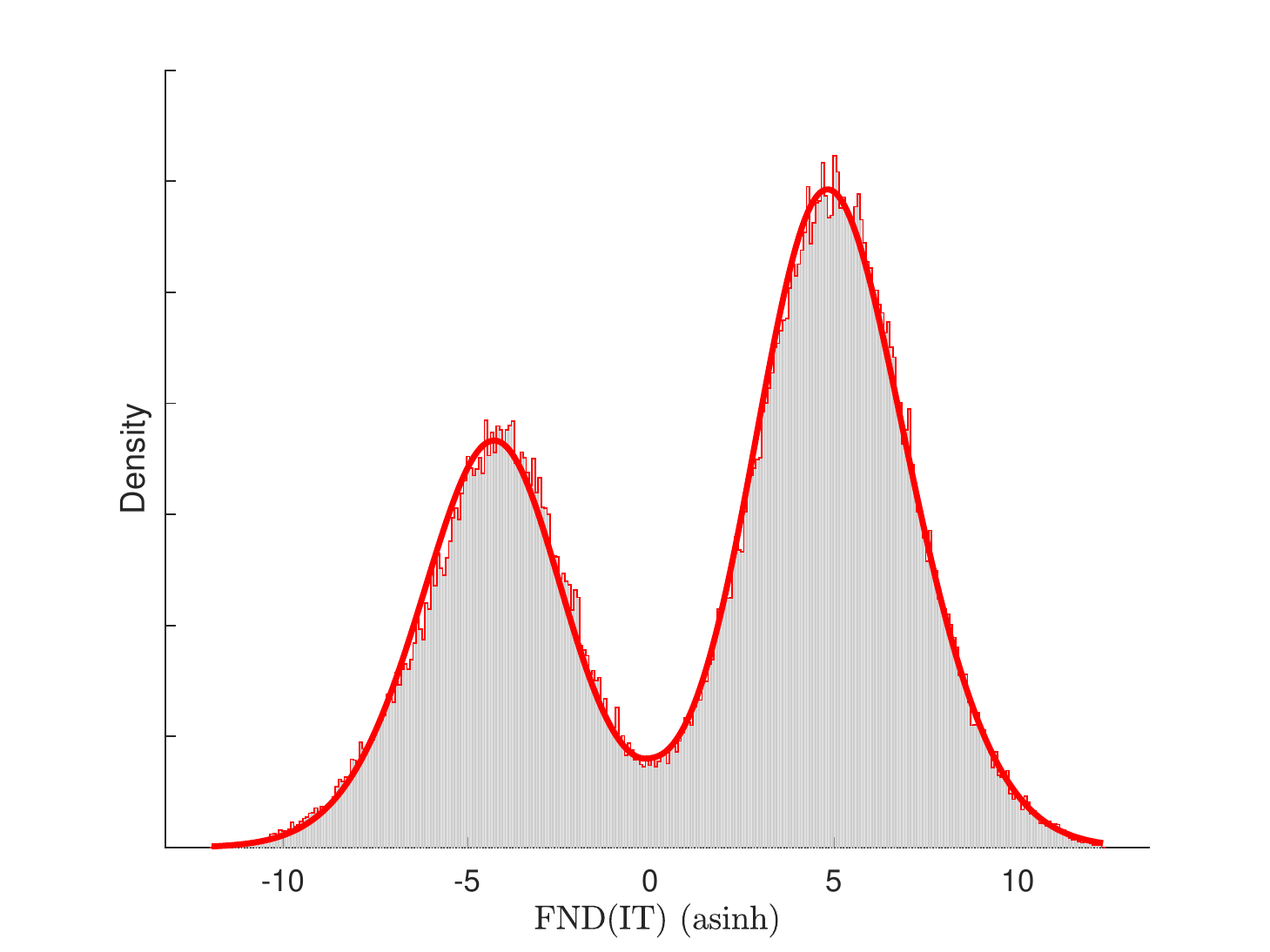}} &     \subfigure[FND(IP) w/ DLN ]{\includegraphics[width=2in]{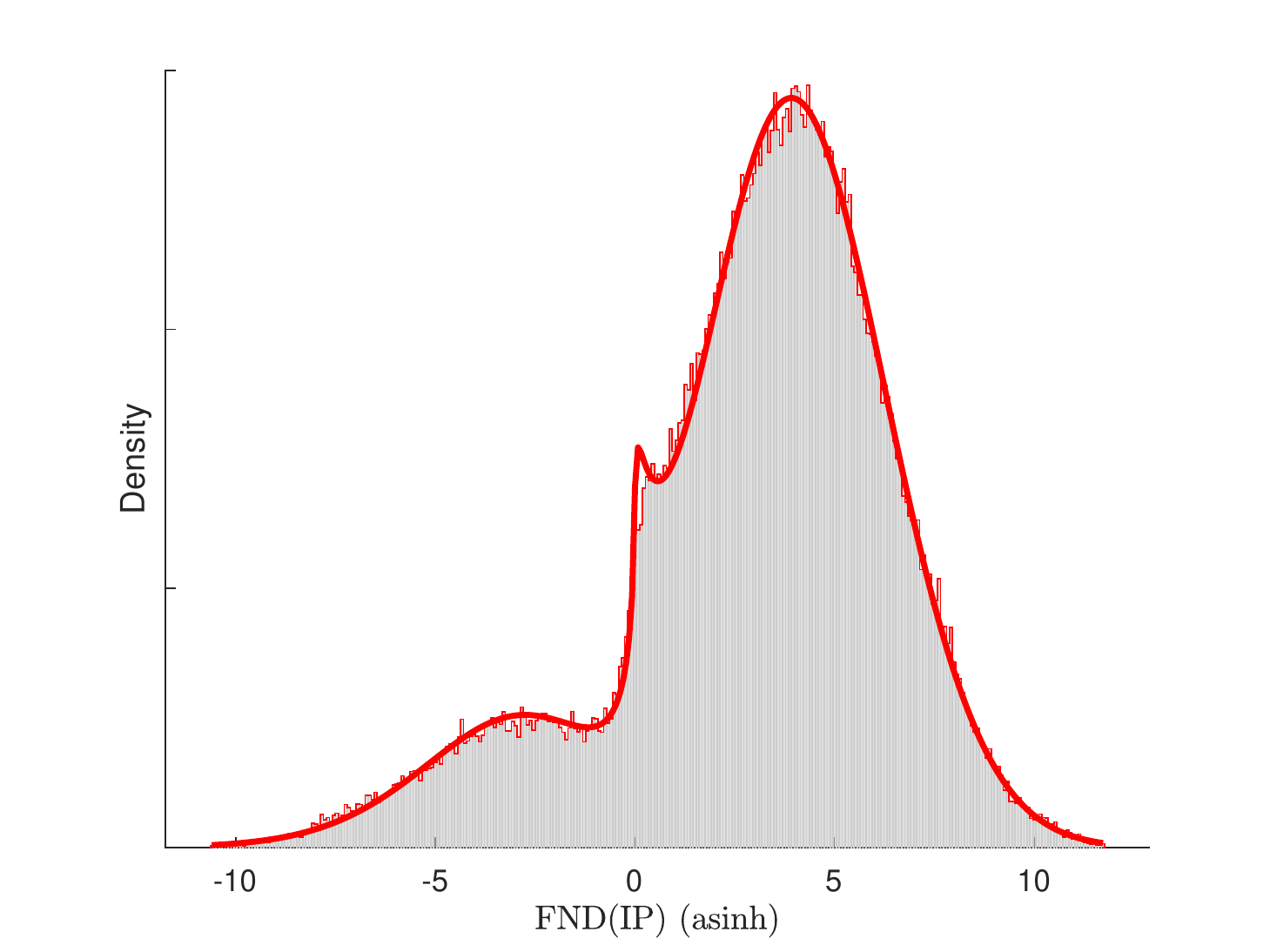}} &
		\subfigure[FND($\widetilde{\text{IT}}$) w/ DLN ]{\includegraphics[width=2in]{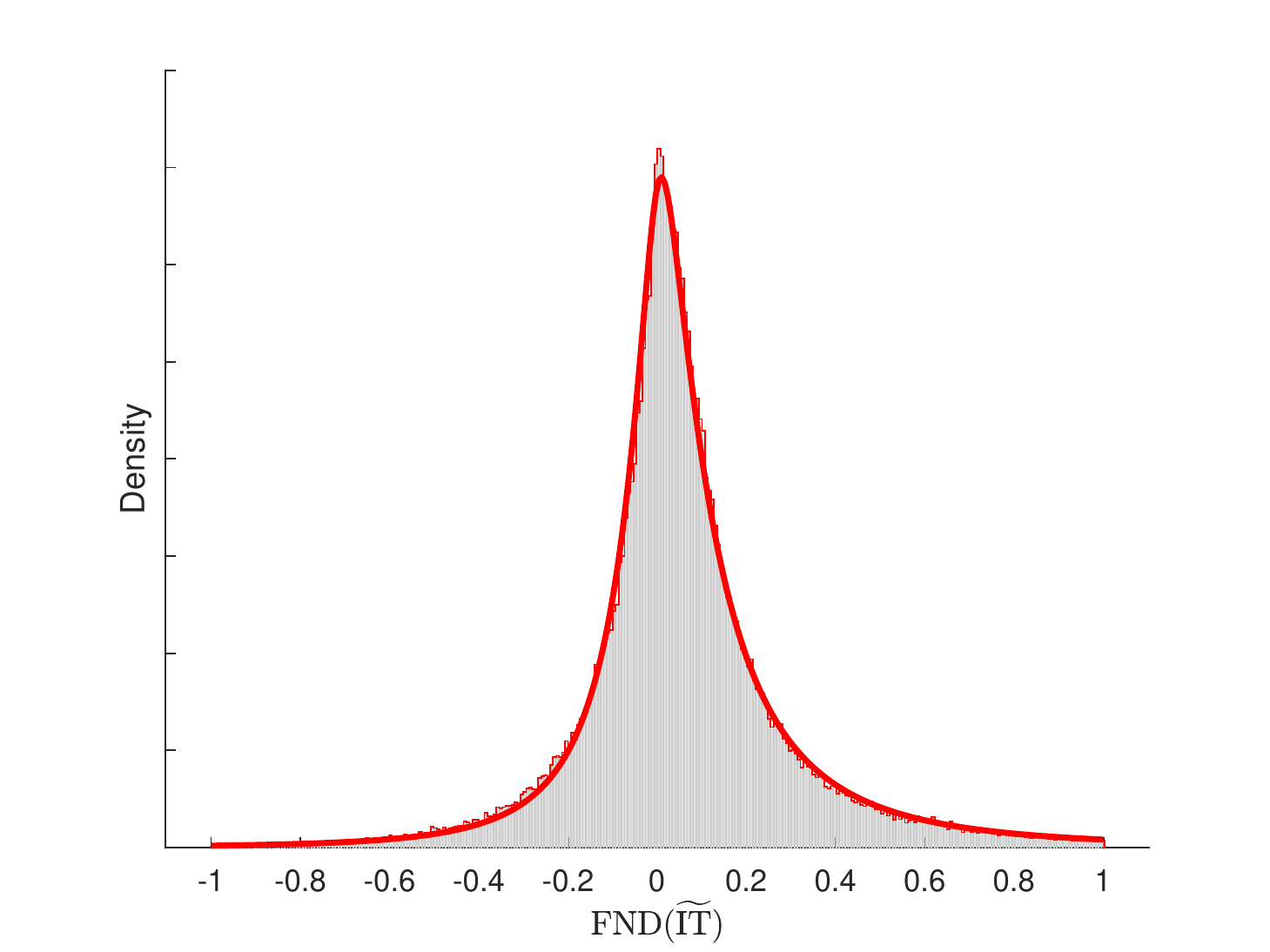}} \\ \\
		\subfigure[q-q FND(IT) vs. DLN ]{\includegraphics[width=2in]{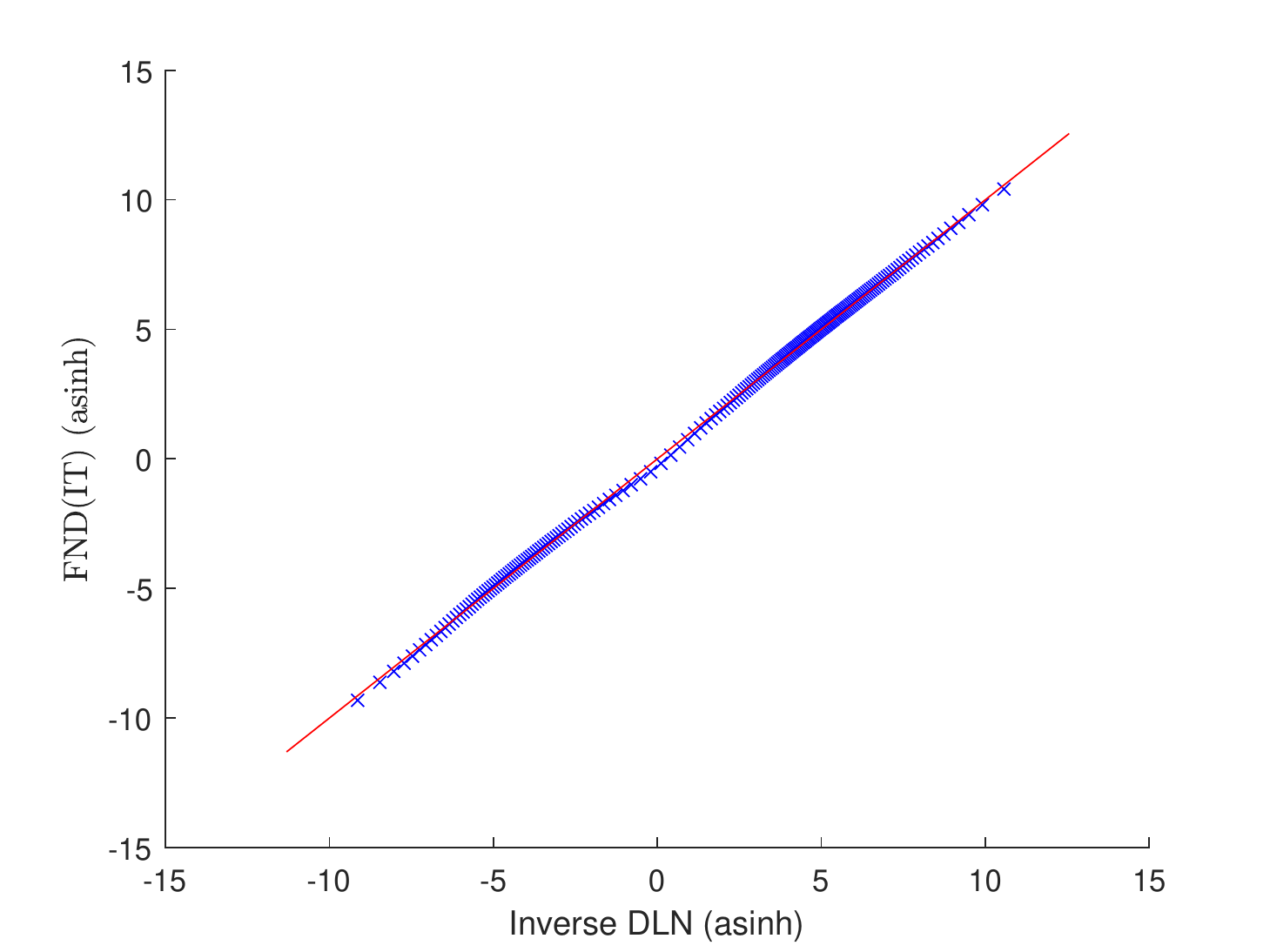}} &
		\subfigure[q-q FND(IP) vs. DLN ]{\includegraphics[width=2in]{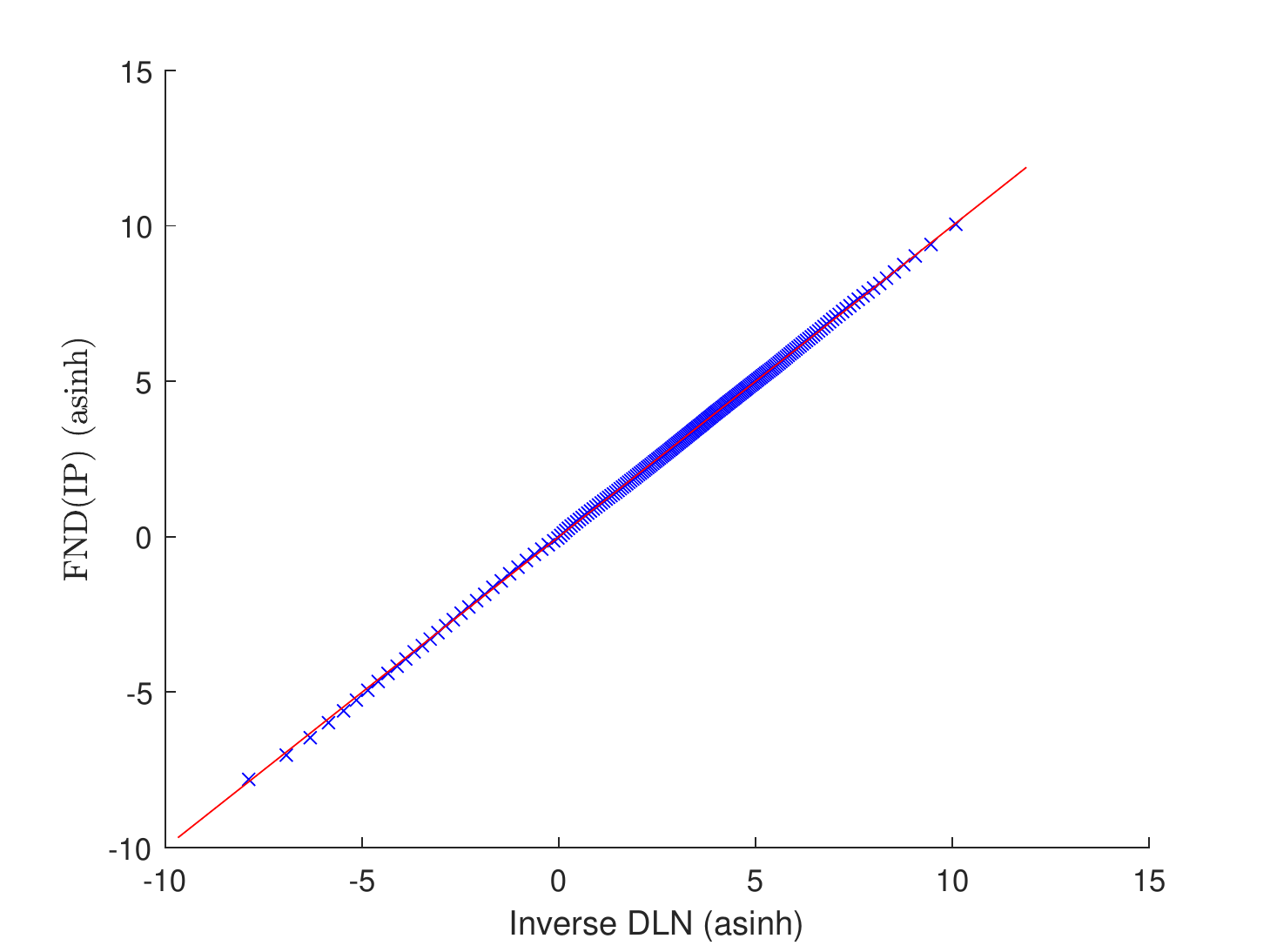}} &
		\subfigure[q-q FND($\widetilde{\text{IT}}$) vs. DLN ]{\includegraphics[width=2in]{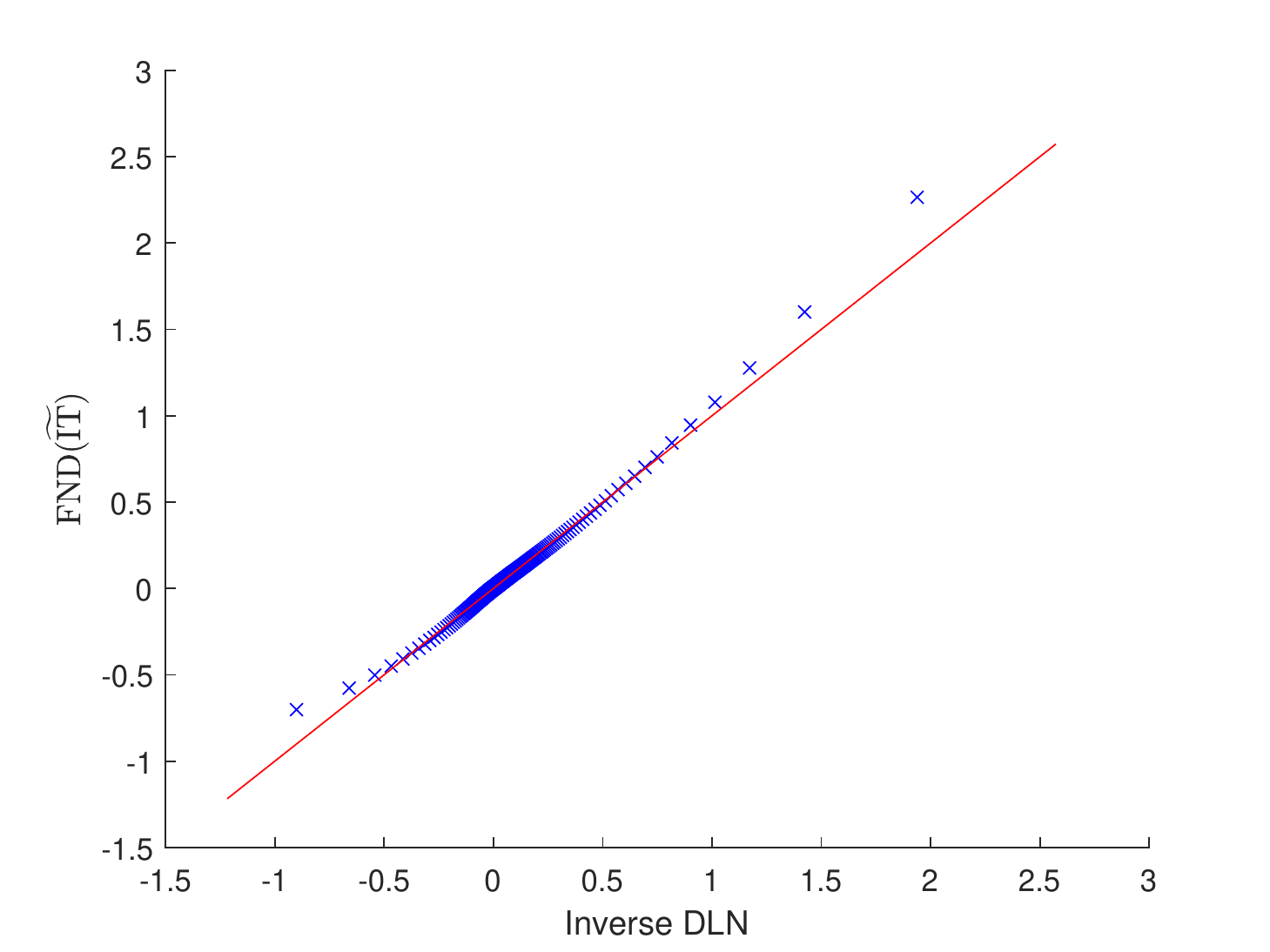}} \\ \\
		\subfigure[FND(IT) by scale ]{\includegraphics[width=2in]{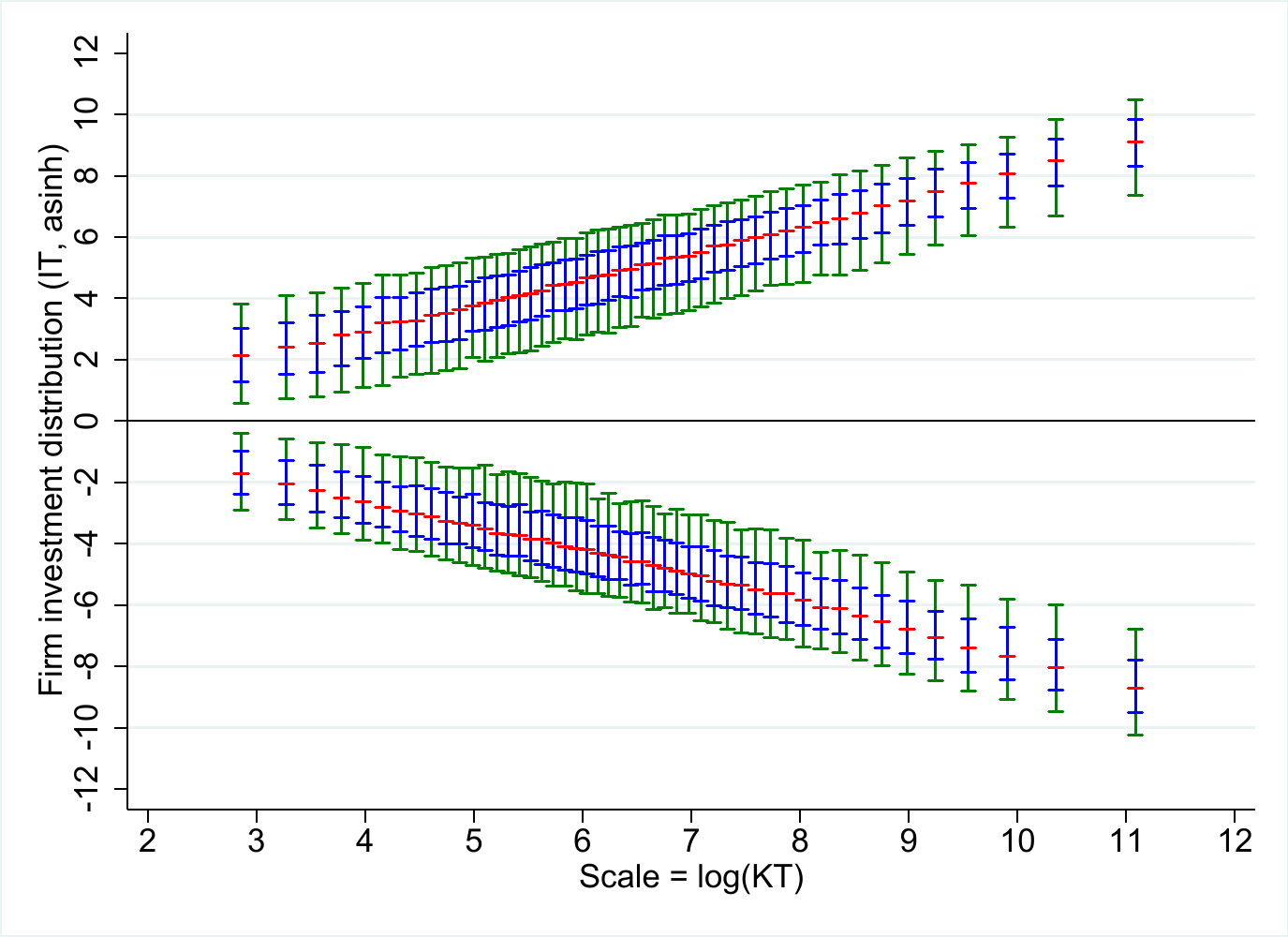}} &
		\subfigure[FND($\widetilde{\text{IT}}$) by scale ]{\includegraphics[width=2in]{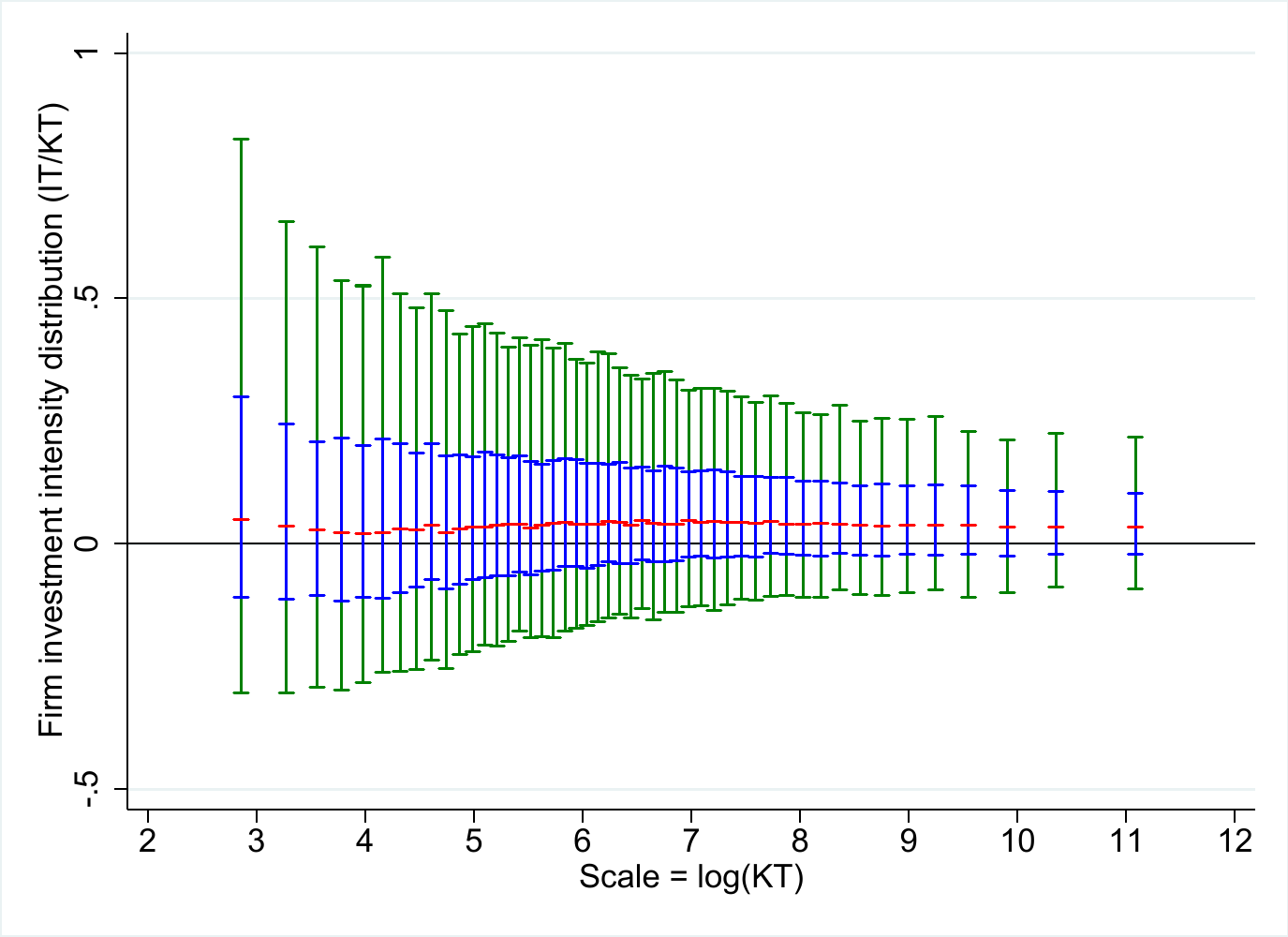}} &
		\subfigure[DP/KP by scale ]{\includegraphics[width=2in]{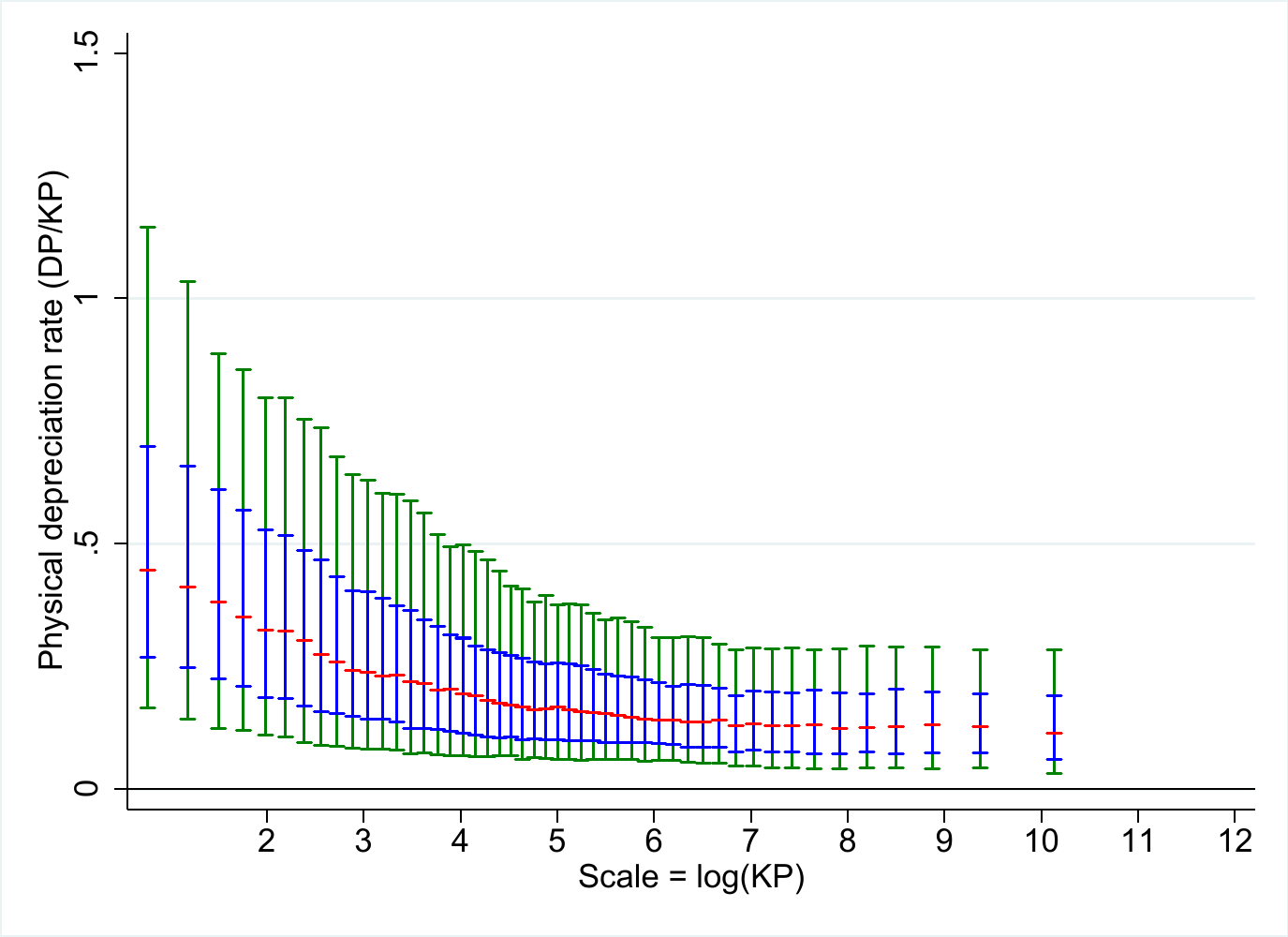}} \\ \\
	\end{tabular}
}

Panel (i) then presents the dependence of physical depreciation intensity, DP/KP, on physical capital scale. The importance of DP/KP is that it measures the rate of physical depreciation, often denoted $\delta_K$ in theoretical work. Panel (i) shows that depreciation rates are decreasing with scale. Small firms have physical depreciation rates as high as 40\%, while large firms have depreciation rates around 10\%. The same pattern, though more muted, is observed when one considers the relation of total depreciation intensity, DP/KT, on total capital scale. I conclude:

\begin{observation}
    \label{ob:10}
    Physical depreciation rates are decreasing with scale.
\end{observation}

Once again, the distributional tests confirm that the FND distributes DLN, as can be seen in Table~\ref{tab:FNDdist}. The differences are especially stark here, with the Stable and Laplace distributions strongly rejected. I conclude:

\begin{observation}
    \label{ob:11}
    The FND is distributed DLN.
\end{observation}

\RPprep{Investment - Distributional tests}{1}{0}{FNDdist}{%
    This table presents results of tests of distribution equality for the FND, based on the investment measures described in Table~\ref{tab:DataDef}, as well as their intensities relative to total capital KT, denoted $\widetilde{XX}$. K-S is a Kolmogorov–Smirnov test; C-2 is a binned $\chi^2$ test with 50 bins; A-D is an Anderson-Darling test. Panels (a)-(c) report the test statistics and their p-values for the Stable, Laplace, and DLN, respectively. Panel (d) reports the relative likelihoods for each distribution. All values are based on the \{Non-Bank\} data subset.
}
\RPtab{%
    \begin{tabularx}{\linewidth}{Frrrrrr}
    \toprule
	& IT & IP & IA & $\widetilde{\text{IT}}$ & $\widetilde{\text{IP}}$ & $\widetilde{\text{IA}}$ \\ 
	\midrule
    \\ \multicolumn{7}{l}{\textit{Panel (a): FND vs. Stable}}\\
	\midrule
    K-S   & 0.026 & 0.048 & 0.067  & 0.020 & 0.120 & 0.068  \\
    p-val & 0.028 & 0.014 & 0.007  & 0.036 & 0.000 & 0.007  \\
    C-2   & 380   & 728   & $>$999 & 193 & $>$999  & $>$999 \\
    p-val & 0.018 & 0.011 & 0.005  & 0.027 & 0.002 & 0.007  \\
    A-D   & 13.75 & 56.04 & 112.0  & 12.25 & 188.1 & 76.00  \\
    p-val & 0.029 & 0.013 & 0.007  & 0.031 & 0.002 & 0.010  \\
    \\ \multicolumn{7}{l}{\textit{Panel (a): FND vs. Laplace}}\\
	\midrule
    K-S   & 0.279  & 0.350  & 0.444  & 0.079 & 0.081  & 0.064  \\
    p-val & 0.000  & 0.000  & 0.000  & 0.004 & 0.004  & 0.008  \\
    C-2   & $>$999 & $>$999 & $>$999 & 924   & $>$999 & $>$999 \\
    p-val & 0.000  & 0.000  & 0.000  & 0.008 & 0.002  & 0.007  \\
    A-D   & $>$999 & $>$999 & $>$999 & 93.14 & 145.9  & 95.42  \\
    p-val & 0.000  & 0.000  & 0.000  & 0.008 & 0.004  & 0.008  \\
    \\ \multicolumn{7}{l}{\textit{Panel (c): FND vs. DLN}}\\
	\midrule
    K-S   & 0.006 & 0.004 & 0.007 & 0.008 & 0.006 & 0.013  \\
    p-val & 0.086 & 0.127 & 0.074 & 0.071 & 0.082 & 0.049  \\
    C-2   & 27    & 14    & 45    & 42    & 29    & 82     \\
    p-val & 0.070 & 0.099 & 0.055 & 0.057 & 0.068 & 0.042  \\
    A-D   & 0.84  & 0.18  & 1.41  & 1.84  & 1.25  & 3.72   \\
    p-val & 0.076 & 0.121 & 0.065 & 0.059 & 0.067 & 0.047  \\
    \\ \multicolumn{7}{l}{\textit{Panel (d): Distribution comparison}}\\
	\midrule
	\multicolumn{2}{l}{AIC R.L.:} \\
    Stable  & 0.000 & 0.000 & 0.000 & 0.000 & 0.000 & 0.000 \\
    Laplace & 0.000 & 0.000 & 0.000 & 0.000 & 0.000 & 0.000 \\
    DLN     & 1.000 & 1.000 & 1.000 & 1.000 & 1.000 & 1.000 \\
	\multicolumn{2}{l}{BIC R.L.:} \\
    Stable  & 0.000 & 0.000 & 0.000 & 0.000 & 0.000 & 0.000 \\
    Laplace & 0.000 & 0.000 & 0.000 & 0.000 & 0.000 & 0.000 \\
    DLN     & 1.000 & 1.000 & 1.000 & 1.000 & 1.000 & 1.000 \\
	\bottomrule
    \end{tabularx}
}

\subsection{R\&D investment}
\label{sec:KnowInvestment}

A special type of investment considered separately is R\&D investment. A major reason is that the stock of R\&D, denoted \emph{knowledge capital}, is unobservable. We hence lack measures of knowledge depreciation, cannot construct investment series based on changes in capital adjusted for depreciation (as we did for physical capital), and observe only strictly positive R\&D expenditure values.

Panels (a)-(c) of Figure~\ref{fig:FRDfacts} present the distributions of (logs of) R\&D investment, R\&D investment intensity relative to total capital KT, and R\&D investment intensity relative to firm sales SL, respectively. Panels (a)-(c) are also overlaid with fitted skew-Normal distributions, though the fit appears poor. Panel (c) exhibits especially striking deviation from skew-Normality, unlike any we encountered so far.

\RPprep{R\&D - Stylized facts}{0}{0}{FRDfacts}{%
    This figure presents stylized facts of firm R\&D investment. Panel (a) presents the (log) R\&D investment distribution overlaid with MLE-fitted skew-Normal. Panels (b) and (c) repeat for (log) R\&D intensity relative to KT and SL, respectively. Panels (d) and (e) present the distribution of (log) R\&D intensity by KT scale, and its binned median and skewness. Panel (f) repeats panel (e) , but for (log) R\&D intensity relative to SL scale. Panels (g) and (h) present the distribution of (log) physical investment intensity IP by KT scale, and its binned median and skewness, but limits the sample to R\&D performing firms. Panel (i) repeats panel (h) for non-R\&D performing firms. All figures are based on the \{Non-Bank\} data subset.
}
\RPfig{%
	\begin{tabular}{ccc} 
		\subfigure[log(RD) w/ SN ]{\includegraphics[width=2in]{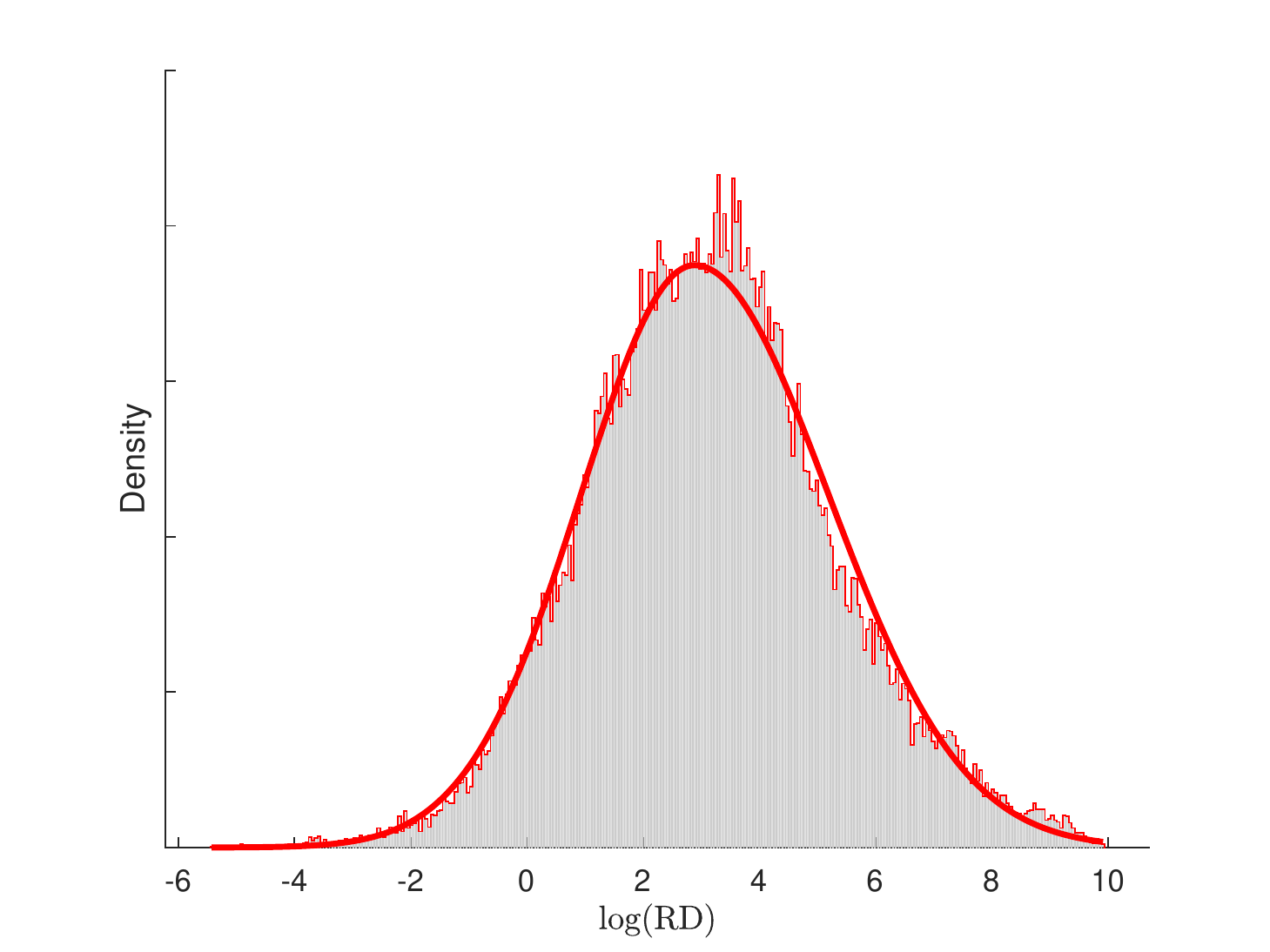}} &     \subfigure[log(RD/KT) w/ SN ]{\includegraphics[width=2in]{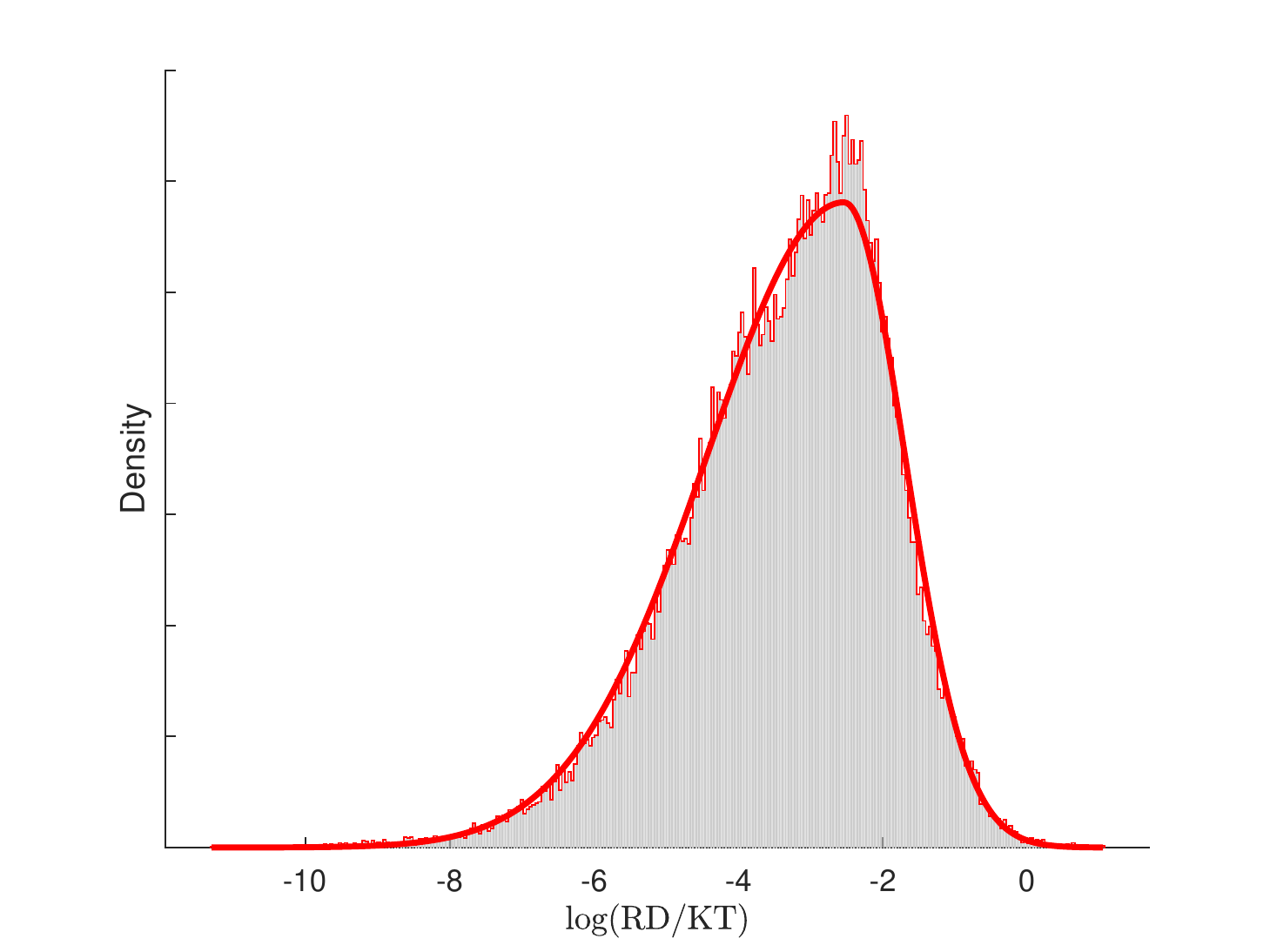}} &
		\subfigure[log(RD/SL) w/ SN ]{\includegraphics[width=2in]{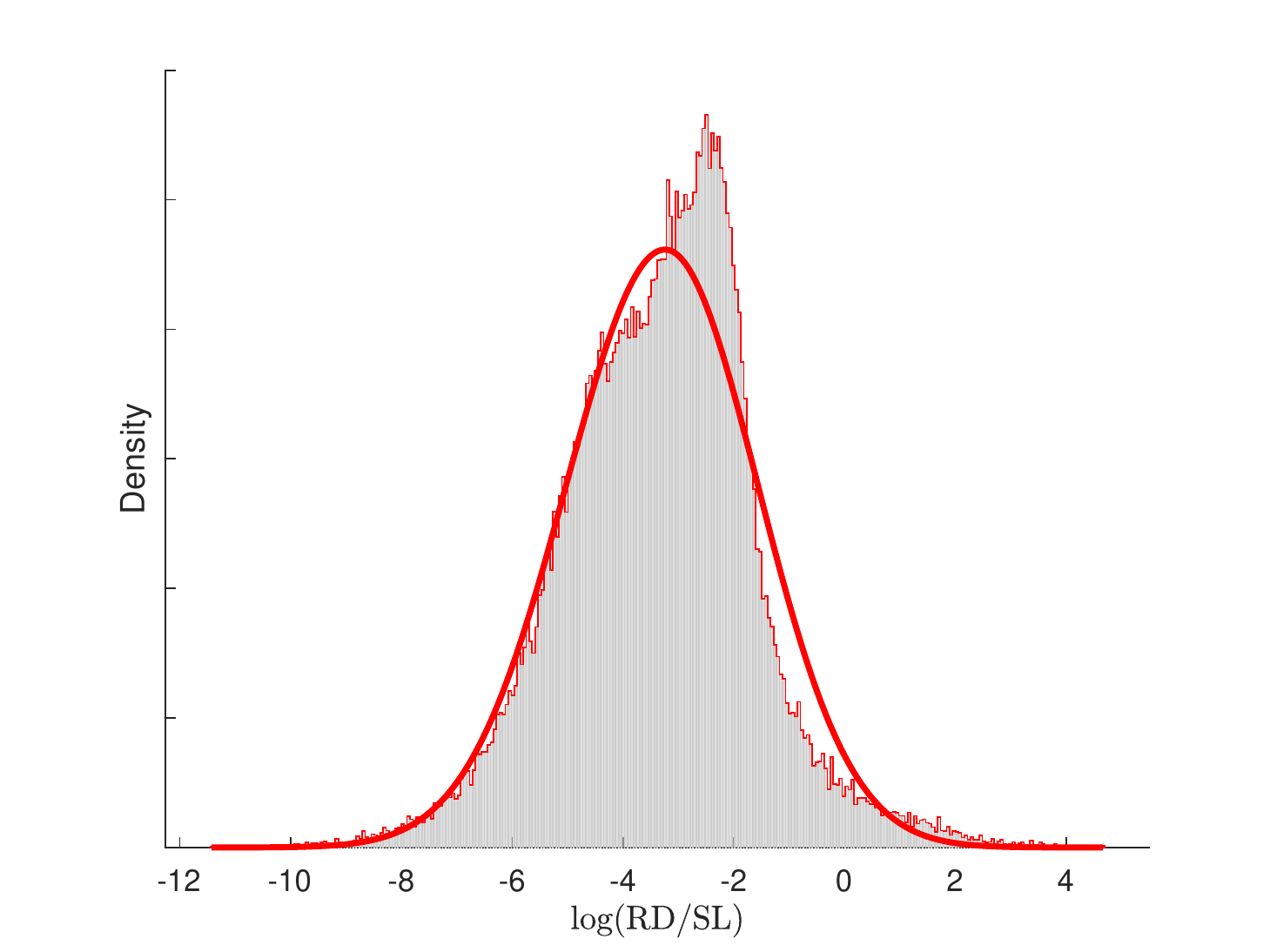}} \\ \\
		\subfigure[log(RD/KT) by scale ]{\includegraphics[width=2in]{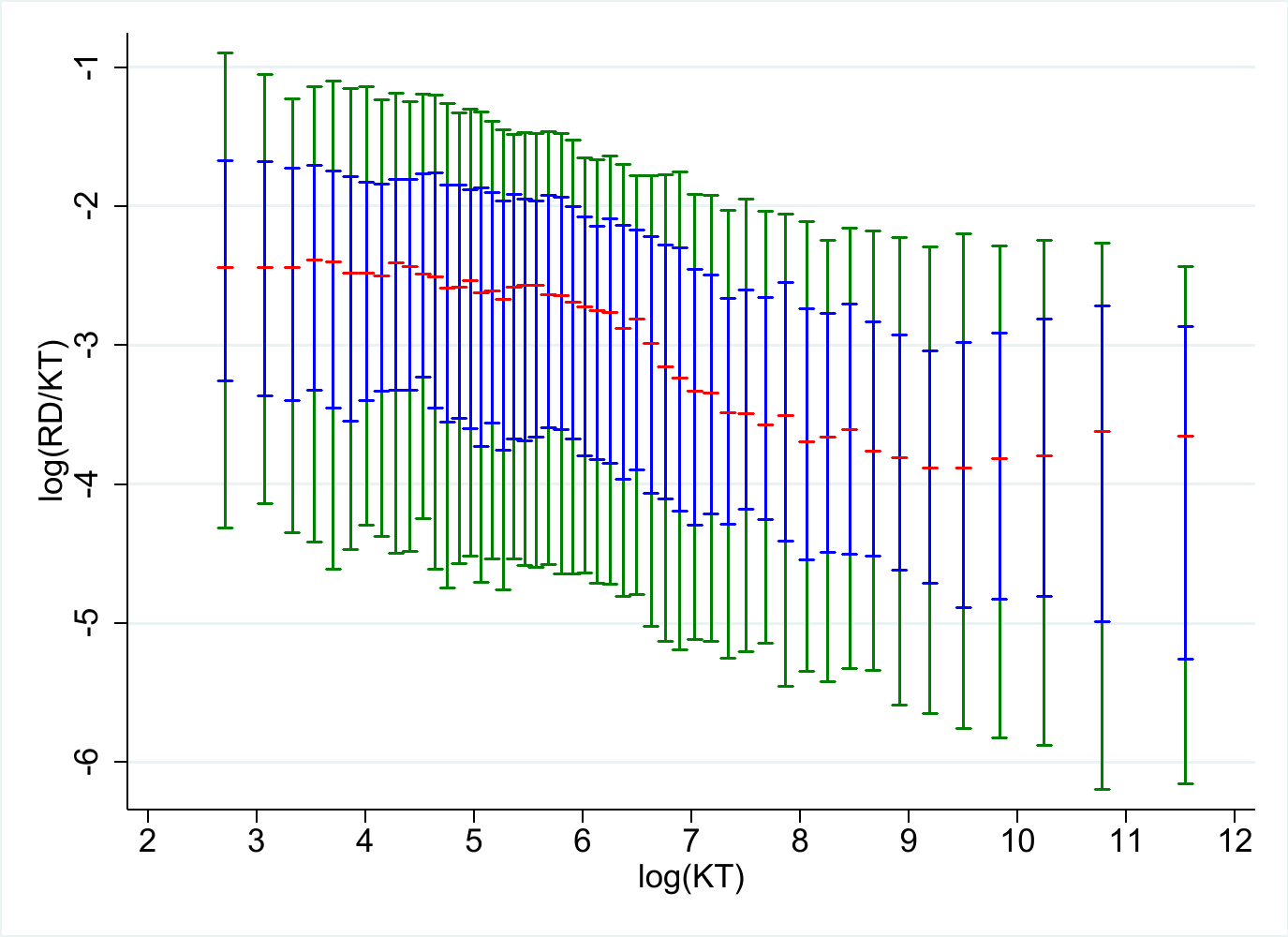}} &
		\subfigure[log(RD/KT) Med+Skew ]{\includegraphics[width=2in]{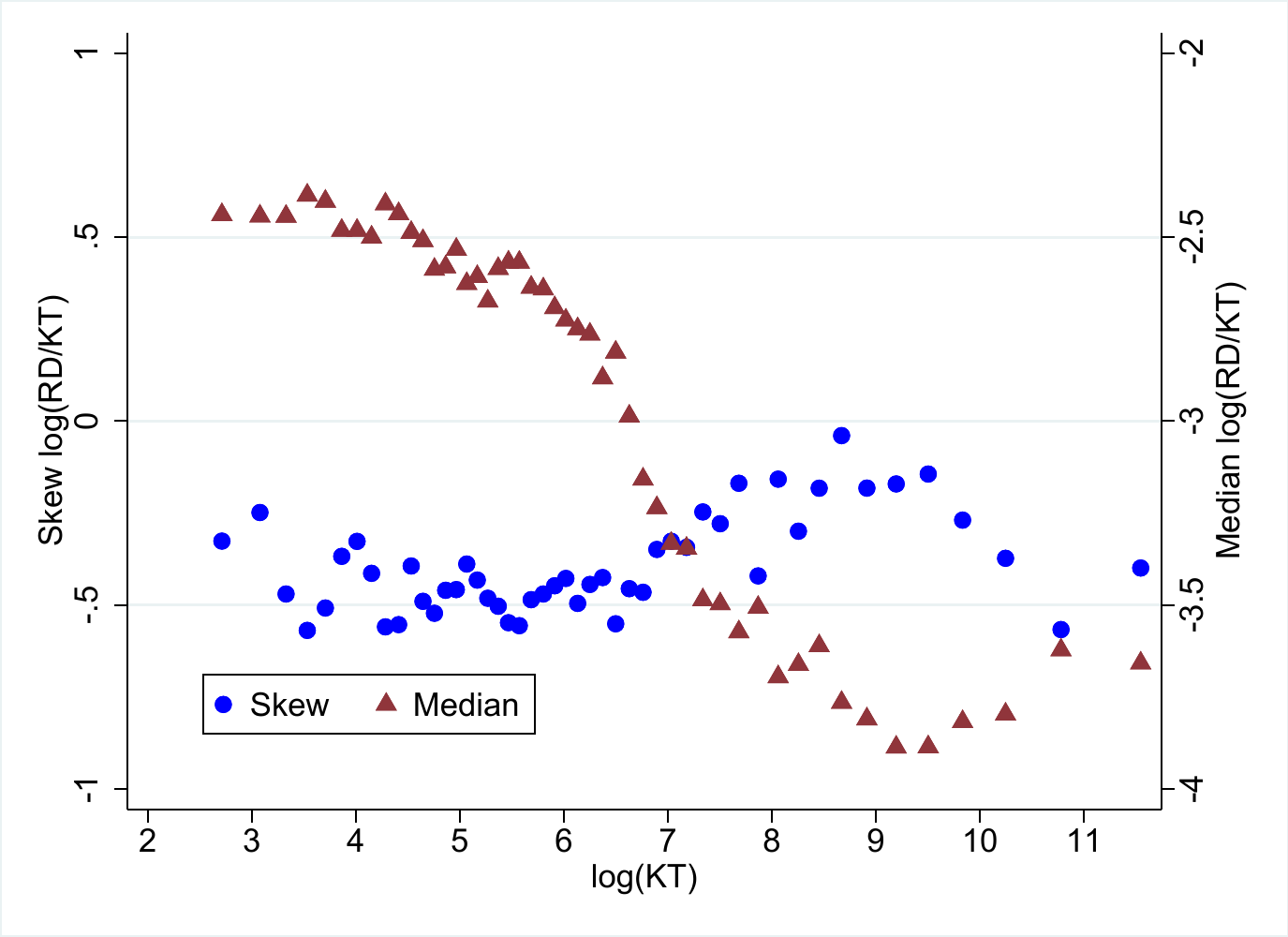}} &
		\subfigure[log(RD/SL) Med+Skew ]{\includegraphics[width=2in]{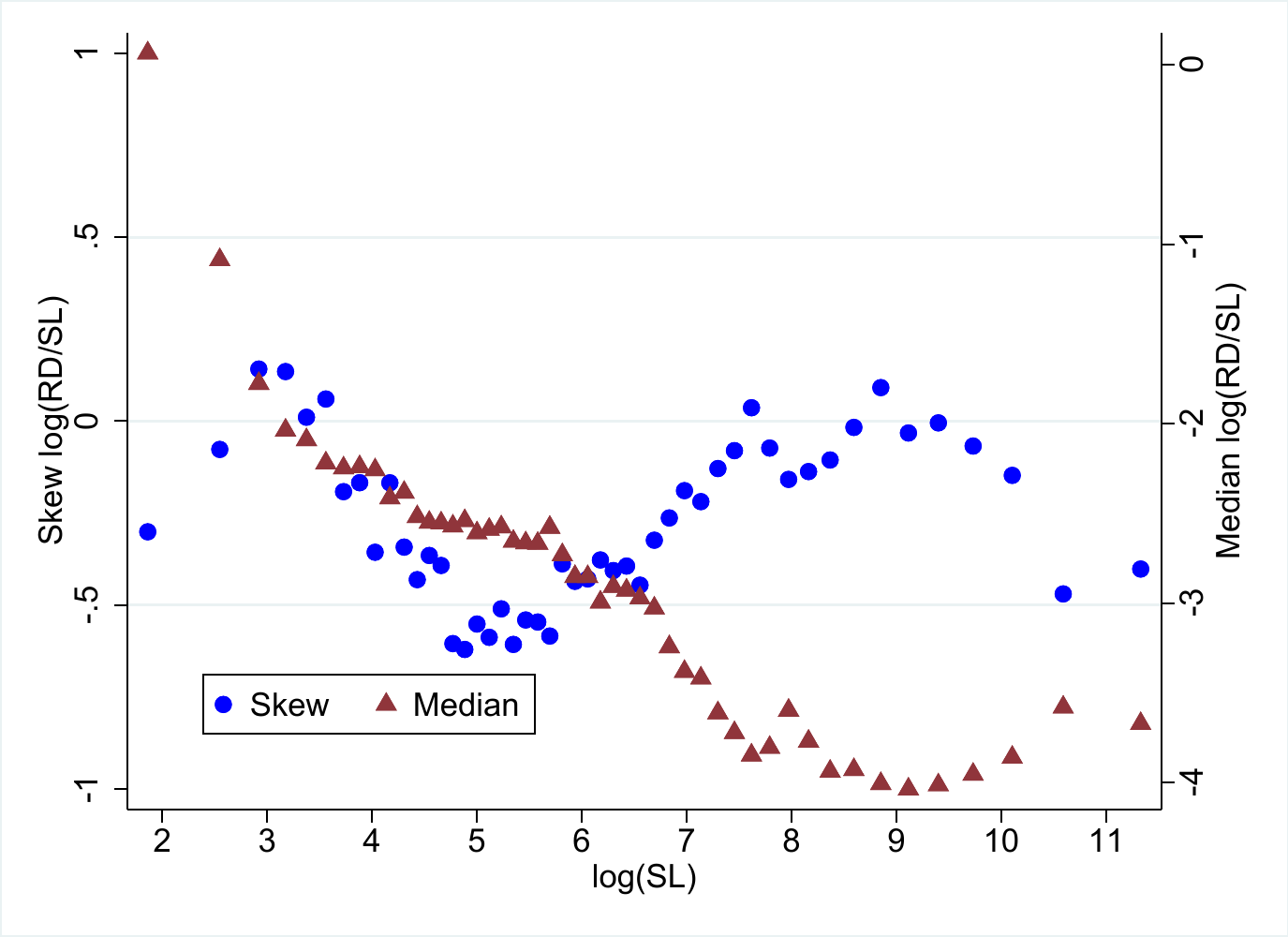}} \\ \\
		\subfigure[log(IP/KT) by scale ]{\includegraphics[width=2in]{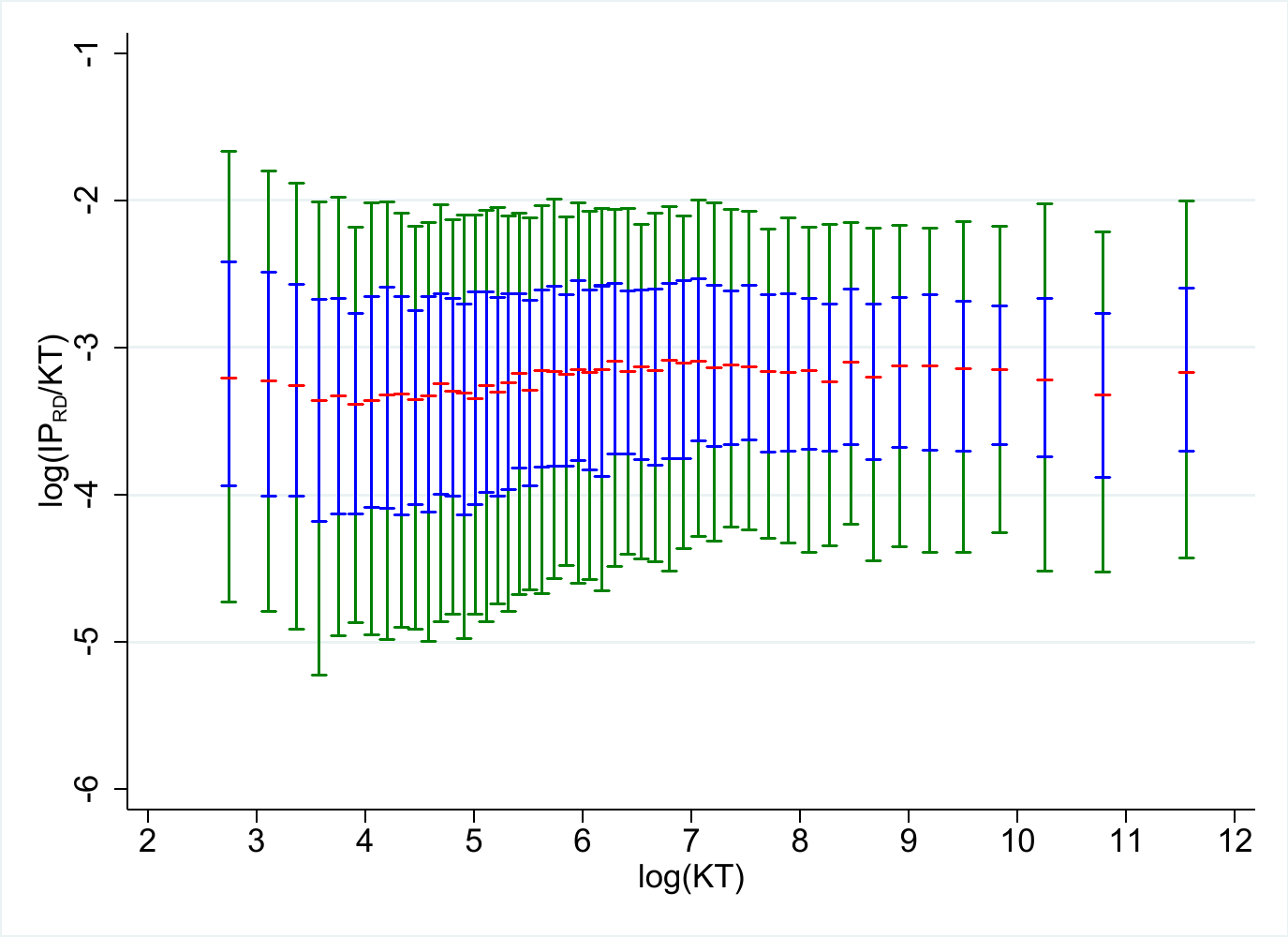}} &
		\subfigure[log(IP/KT) Med+Skew ]{\includegraphics[width=2in]{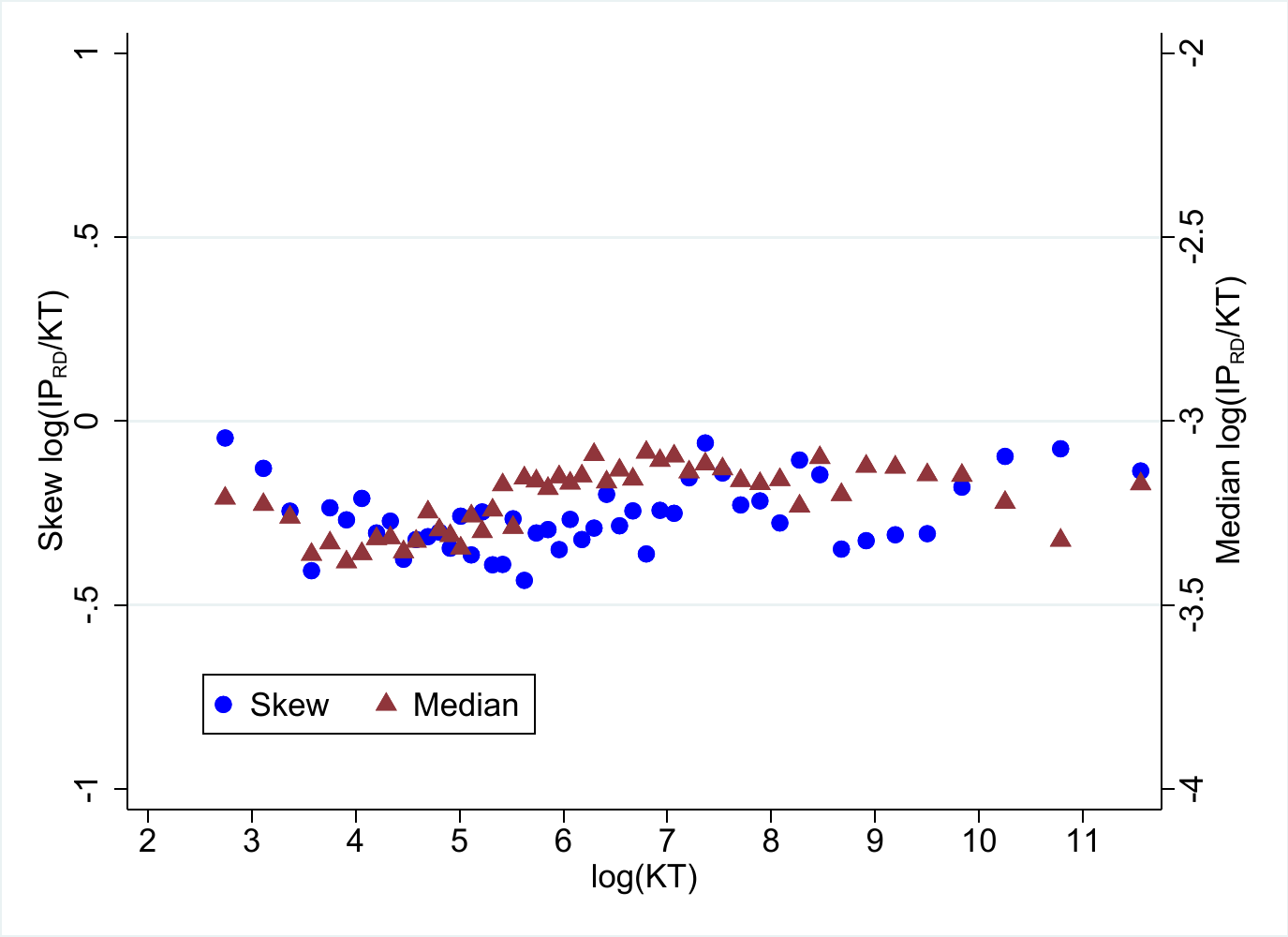}} &
		\subfigure[log(IP/KT) Med+Skew ]{\includegraphics[width=2in]{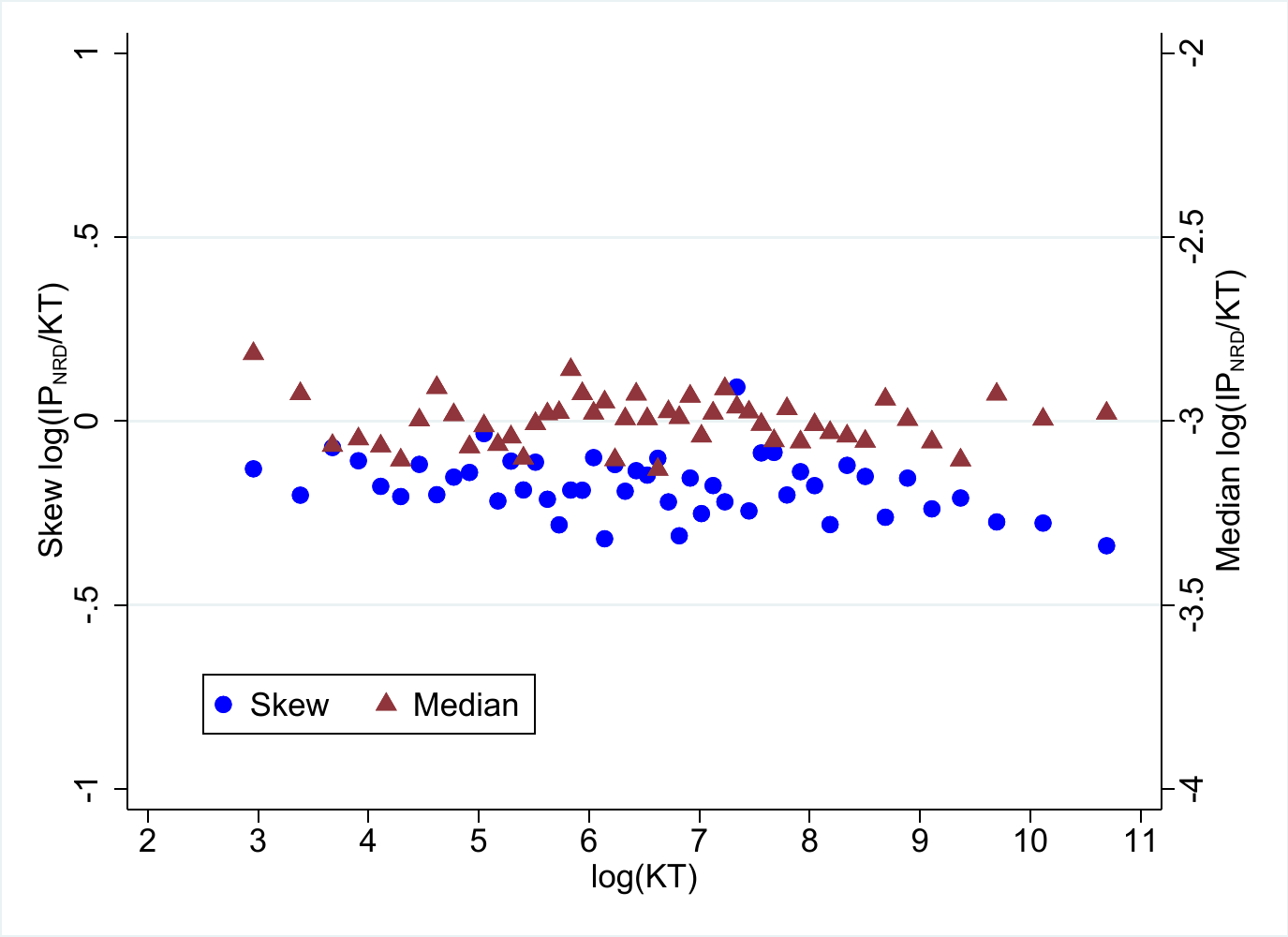}} \\ \\
	\end{tabular}
}

To explore these patterns, panel (d) plots the distribution of (log) R\&D intensity per scale bin, which exhibits significant decrease in median intensity with scale. The decrease is not smooth, but rather we observe a steep decline between scales 6 and 8. This pattern is clear in panels (e) and (f), which present the per-bin median and skewness of (log) R\&D intensity, relative to total capital KT and to sales SL, respectively. A second fact apparent in panels (e) and (f) is the strong negative relation between the median and skewness of (log) R\&D intensity, especially for non-tiny firms (sales or capital above \$100M, or scale 4.6).

As panels (g)-(i) show, these patterns are fairly unique to the R\&D intensity distribution, and are not evident at (log) physical investment intensity IP/KT. Panels (g) and (h) present the distribution of log(IP/KT), and its binned median and skewness, for R\&D performing firms. We in fact observe a slight increase in intensity with scale. Panel (h) presents the binned median and skewness of log(IP/KT) for the non-R\&D performing firms in the sample, again showing fairly constant intensity.

The relation between (log) investment intensity's median and skewness is further reviewed in Table~\ref{tab:FRDDescMom}, along with other descriptive statistics. Panel (a) shows both R\&D and physical (log) intensities share similar mean and s.d., but R\&D intensity is significantly more persistent than physical intensity. R\&D has a strong decreasing median with scale, as seen in Figure~\ref{fig:FRDfacts}. The dependence of median (log) investment intensity on scale is generally positive for R\&D-performing firms and generally negative for non-R\&D-performing firms.

\RPprep{R\&D - Descriptive statistics}{0}{0}{FRDDescMom}{%
    This table presents moments of (log) R\&D and physical investment intensities. The first two columns describe R\&D intensities relative to KT and SL. The next four describe physical intensities for R\&D-performing firms, and the last four repeat for non-R\&D-performing firms. Only positive investments are considered. Panel (a) describes the first four moments of each distribution, its persistence of investment based on the \cite{ArellanoBover1995}/\cite{BlundellBond1998} panel estimator, and its coefficient of decreasing median with scale. Panels (b)-(d) reports results of skewness-median regressions, for all firms, small firms ($<\$100$M in KT or SL), and large firms ($>\$100$M), respectively.
}
\RPtab{%
    \begin{tabularx}{\linewidth}{Frrrrrrrrrr}
    \toprule
    & & & \multicolumn{4}{c}{R\&D-performing} & \multicolumn{4}{c}{non-R\&D-performing} \\
	log of & $\frac{RD}{KT}$ & $\frac{RD}{SL}$ & $\frac{IP}{KT}$ & $\frac{IP}{SL}$ & $\frac{IA}{KT}$ & $\frac{IA}{SL}$ & $\frac{IP}{KT}$ & $\frac{IP}{SL}$ & $\frac{IA}{KT}$ & $\frac{IA}{SL}$\\
	\midrule
    \\ \multicolumn{11}{l}{\textit{Panel (a): Descriptive moments}}\\
	\midrule
    $M_{1}$       & -3.171 & -3.022 & -3.301 & -3.120 & -3.399 & -3.224 & -3.074 & -3.154 & -3.143 & -3.184 \\
    $M_{2}$       & 1.413  & 1.730  & 1.106 & 1.237 & 1.063 & 1.136 & 1.280 & 1.580 & 1.201 & 1.501  		\\
    $M_{3}$       & -0.709 & 0.092  & -0.957 & -0.207 & -0.737 & 0.133 & -0.631 & 0.033 & -0.681 & 0.115  	\\
    $M_{4}$       & 3.791  & 4.039  & 6.349 & 5.801 & 5.464 & 5.642 & 4.825 & 3.978 & 4.917 & 3.862  		\\
    Pers.         & 0.706  & 0.673  & 0.217 & 0.212 & 0.381 & 0.352 & 0.158 & 0.191 & 0.386 & 0.390  		\\
    $\beta_{scl}$ & -0.239 & -0.339 & 0.019 & -0.030 & 0.080 & 0.031 & -0.004 & -0.124 & 0.058 & -0.055  	\\
    \\ \multicolumn{11}{l}{\textit{Panel (b): Binned skewness-median regressions - All firms}}\\
	\midrule
    $\beta_{med}$ & -0.173 & -0.057 & 0.126 & 0.070 & 0.101 & 0.081 & -0.200 & 0.009 & -0.810 & -0.163  	\\
    s.e.          & 0.027 & 0.038 & 0.157 & 0.066 & 0.067 & 0.076 & 0.174 & 0.060 & 0.086 & 0.087  			\\
	sig			  & *** & ~ & ~ & ~ & ~ & ~ & ~ & ~ & *** &   												\\
    R$^2$     & 0.462 & 0.024 & -0.008 & 0.002 & 0.026 & 0.003 & 0.007 & -0.021 & 0.647 & 0.050  		\\
    \\ \multicolumn{11}{l}{\textit{Panel (c): Binned skewness-median regressions - small firms}}\\
	\midrule
    $\beta_{med}$ & -1.193 & 0.044 & 1.387 & 0.049 & -0.419 & -0.063 & -0.058 & -0.215 & -0.130 & -0.275  	\\
    s.e.          & 0.737 & 0.080 & 0.387 & 0.067 & 0.347 & 0.060 & 0.208 & 0.028 & 0.243 & 0.047  			\\
	sig			  & ~ & ~ & *** & ~ & ~ & ~ & ~ & *** & ~ & ***  											\\
    R$^2$     & 0.128 & -0.057 & 0.542 & -0.039 & 0.044 & 0.008 & -0.152 & 0.904 & -0.113 & 0.844  		\\
    \\ \multicolumn{11}{l}{\textit{Panel (d): Binned skewness-median regressions - large firms}}\\
	\midrule
    $\beta_{med}$ & -0.214 & -0.371 & 0.019 & -0.797 & 0.285 & 0.638 & -0.286 & 0.291 & -1.066 & -0.567  	\\
    s.e.          & 0.030 & 0.027 & 0.221 & 0.280 & 0.091 & 0.147 & 0.223 & 0.158 & 0.113 & 0.249  			\\
	sig			  & *** & *** & ~ & ** & ** & *** & ~ & ~ & *** & *  										\\
    R$^2$     & 0.585 & 0.851 & -0.028 & 0.169 & 0.193 & 0.337 & 0.016 & 0.055 & 0.687 & 0.092  		\\
\end{tabularx}
}

We observe similar relations when considering the skewness-median relation in Panels (b)-(d). The relation is negative for R\&D intensity and for physical intensity at non-R\&D-performing firms, while it is generally positive for physical intensity at R\&D-performing firms. I conclude:

\begin{observation}
    \label{ob:12}
    R\&D intensity is non-linearly decreasing with scale.
\end{observation}

\begin{observation}
    \label{ob:13}
    R\&D intensity's median is negatively correlated with its skewness.
\end{observation}

\begin{observation}
    \label{ob:14}
    Physical intensity at (non) R\&D-performing firms exhibits patterns (similar) opposite those of R\&D intensity.
\end{observation}

\subsection{Ratios}
\label{sec:Ratios}

The non-linear decrease in R\&D intensity reported at Observation~\ref{ob:12} appears to indicate that small firms are somehow systematically and discontinuously different from large firms. This section considers some important firm ratios and yields a similar conclusion.

Figure~\ref{fig:RATfacts} reports the dependence of several important firm ratios on firm scale. Panel (a) presents the medians of debt, equity, and total payout ratios, defined as DD/DB, DE/EQ, and DI/VL, respectively. The panel shows a stable debt payout ratio of about 4\% across the firm scale distribution, indicating a stable cost of debt for firms of all sizes. The equity payout ratio exhibits remarkably different patterns. The median equity payout ratio of small firms is 0, while the median equity payout ratio of large firms is increasing with firm scale. The total payout ratio is fairly uniformly increasing with firm scale, as predicted by, e.g., \cite{ModiglianiMiller1958}. Panel (b) complements the picture by presenting the median debt-to-equity ratio of firms across the scale distribution, with larger firms linearly tending to operate with more debt. The probable cause of these facts is presented in Panel (c), which reports the median and mean net profitability (cashflow-to-value ratio) across the firm scale distribution. Small firms, below the median scale in the data ($\log(KT)=6.5$) tend to not produce enough cashflows to afford equity payouts, whereas large firms do.

Panels (d)-(f) present further evidence of a systematic difference between small and large firms. Panel (d) presents the mean and median of the firm expense (in)efficiency ratio XS/SL, as a function of firm scale. The median shows a smooth efficiency gain w/ scale (and an R$^2$=0.98), while the mean shows the average small firm loses money by having an efficiency ratio>1. Panel (e) presents a transformed version of the data in panel (d), $-\log$(XS/SL). We can see how negative efficiency skewness for small firms becomes positive efficiency skewness for large firms, with the crossing happening around median firm scale. Panel (f) complements the picture by presenting the mean and median of a second efficiency ratio - the sales to capital ratio SL/KT. Again, we can see clear differences between small and large firms - small firms are frugal on capital, while larger firms are more capital intensive.

These data, along with the R\&D data from the previous section, are commensurate with a view of the firm scale distribution in which firms below the median scale are ``laboratories'' and firms above the median scale are ``production factories''. But as the focus of the current paper is collecting and analyzing stylized facts, this insight is left for future work. I conclude:

\begin{observation}
    \label{ob:15}
    Small and large firms (below and above the median scale) are systematically different from each other.
\end{observation}

\begin{observation}
    \label{ob:16}
    Median debt-to-equity, firm efficiency, and total payout ratio grow linearly with scale.
\end{observation}

\RPprep{Ratios - Stylized facts}{0}{0}{RATfacts}{%
    This figure presents stylized facts of important firm ratios as a function of firm scale. Panel (a) presents the binned medians of debt, equity, and total payout ratios (DD/DB, DE/EQ, and DI/VL). Panel (b) presents the binned median debt-to-value ratio (DB/VL), and Panel (c) presents the binned median and mean of firm profitability (CF/VL). Panels (d)-(f) present the mean of median of three efficiency indices: expense (in)efficiency XS/SL, its transform -$\log$(XS/SL), and sales to capital efficiency SL/KT. All figures are based on the \{Non-Bank\} data subset.
}
\RPfig{%
	\begin{tabular}{ccc} 
		\subfigure[Payout ratios by scale  ]{\includegraphics[width=2in]{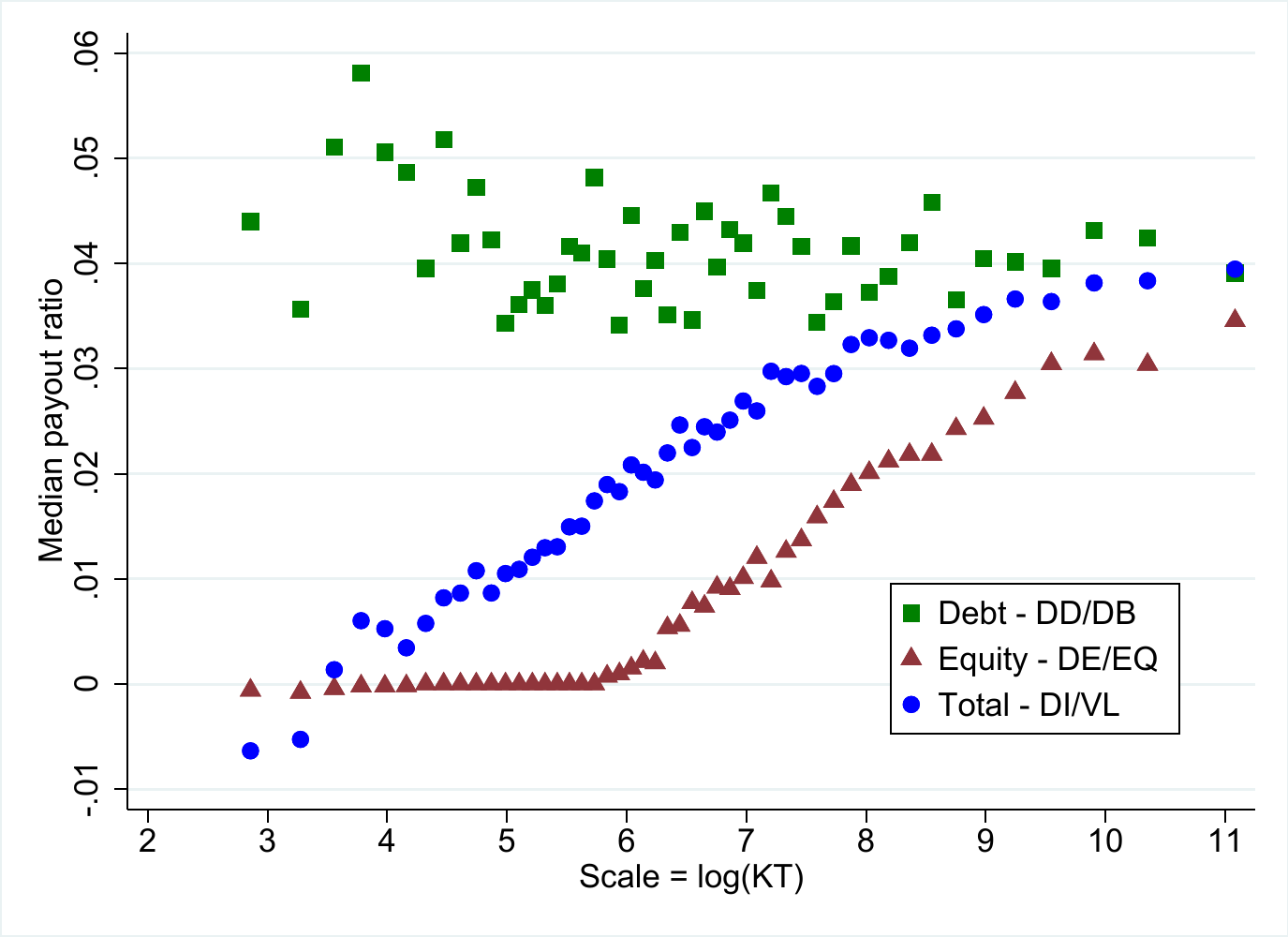}} &     \subfigure[Debt-to-equity by scale ]{\includegraphics[width=2in]{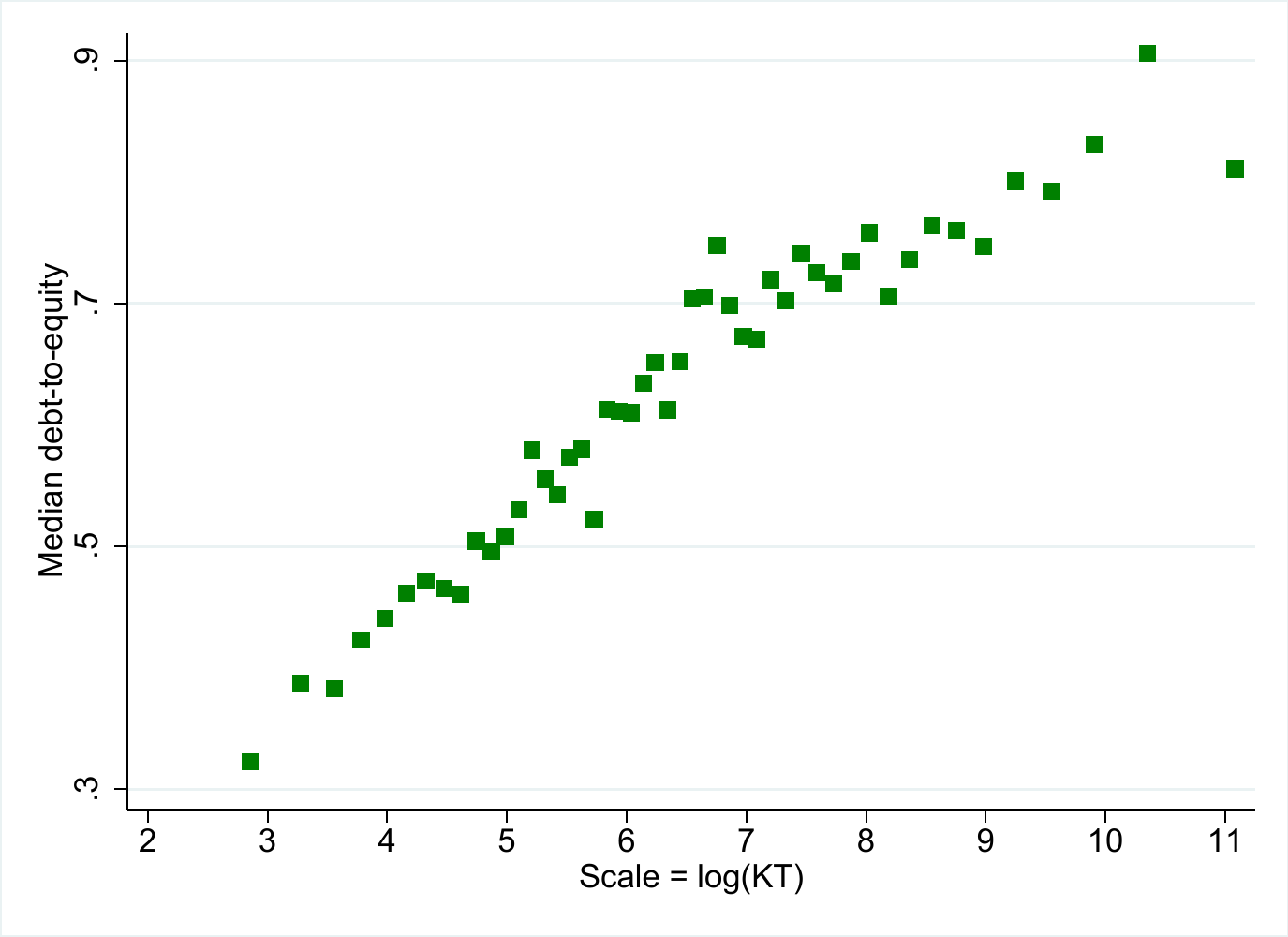}} &
		\subfigure[Cashflow-to-value by scale ]{\includegraphics[width=2in]{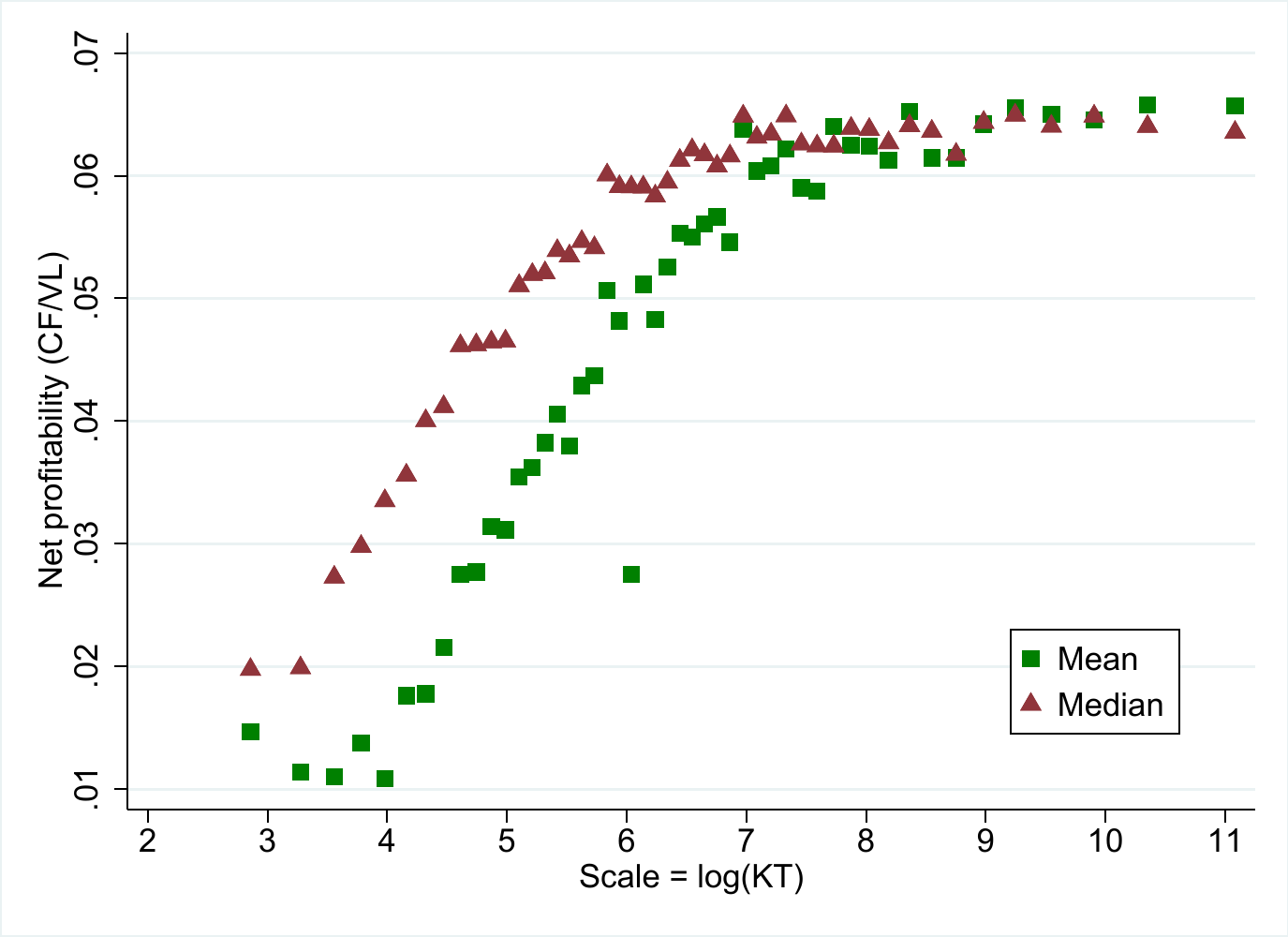}} \\ \\
		\subfigure[Expense eff. by scale ]{\includegraphics[width=2in]{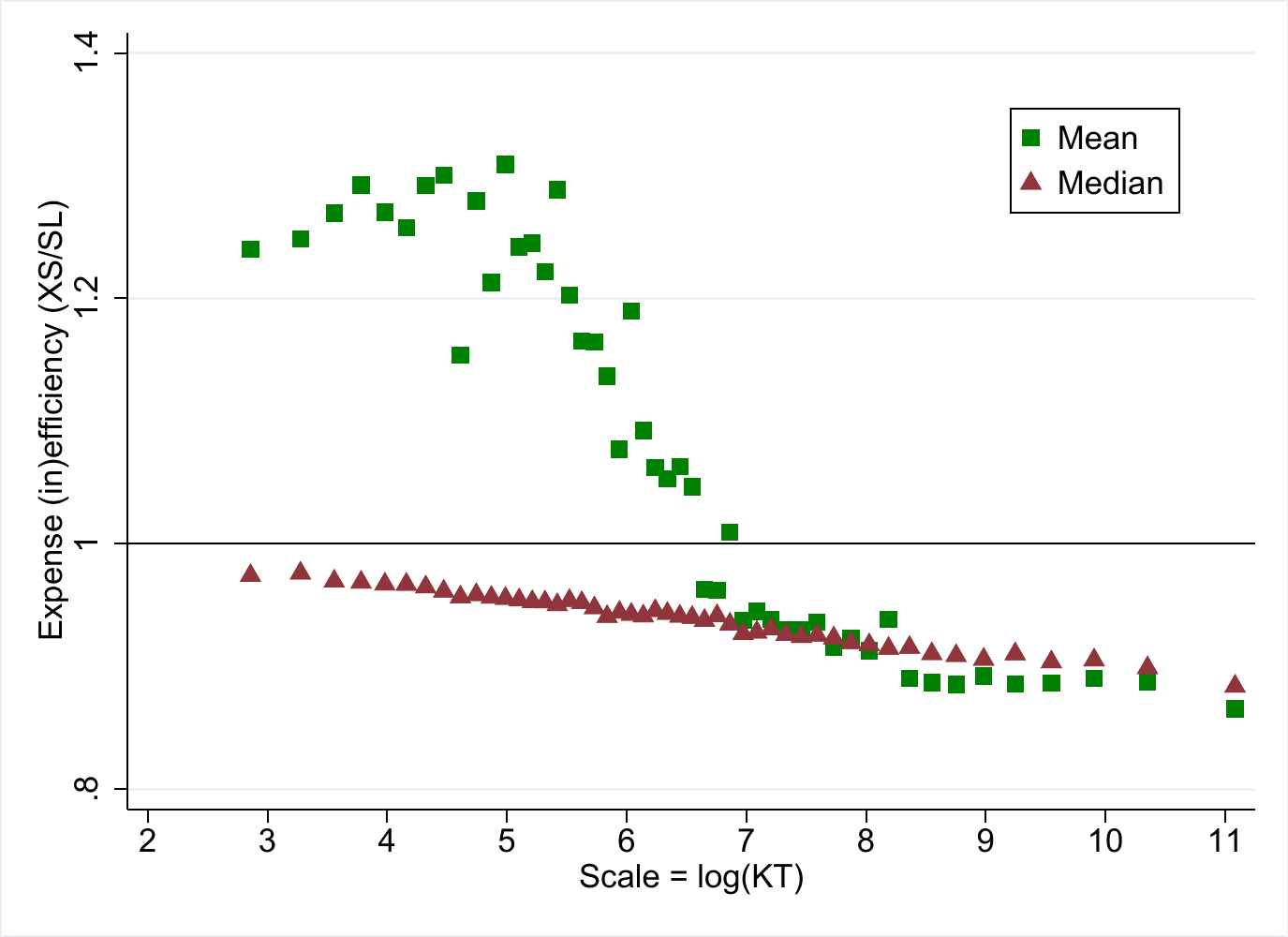}} &
		\subfigure[log eff. by scale ]{\includegraphics[width=2in]{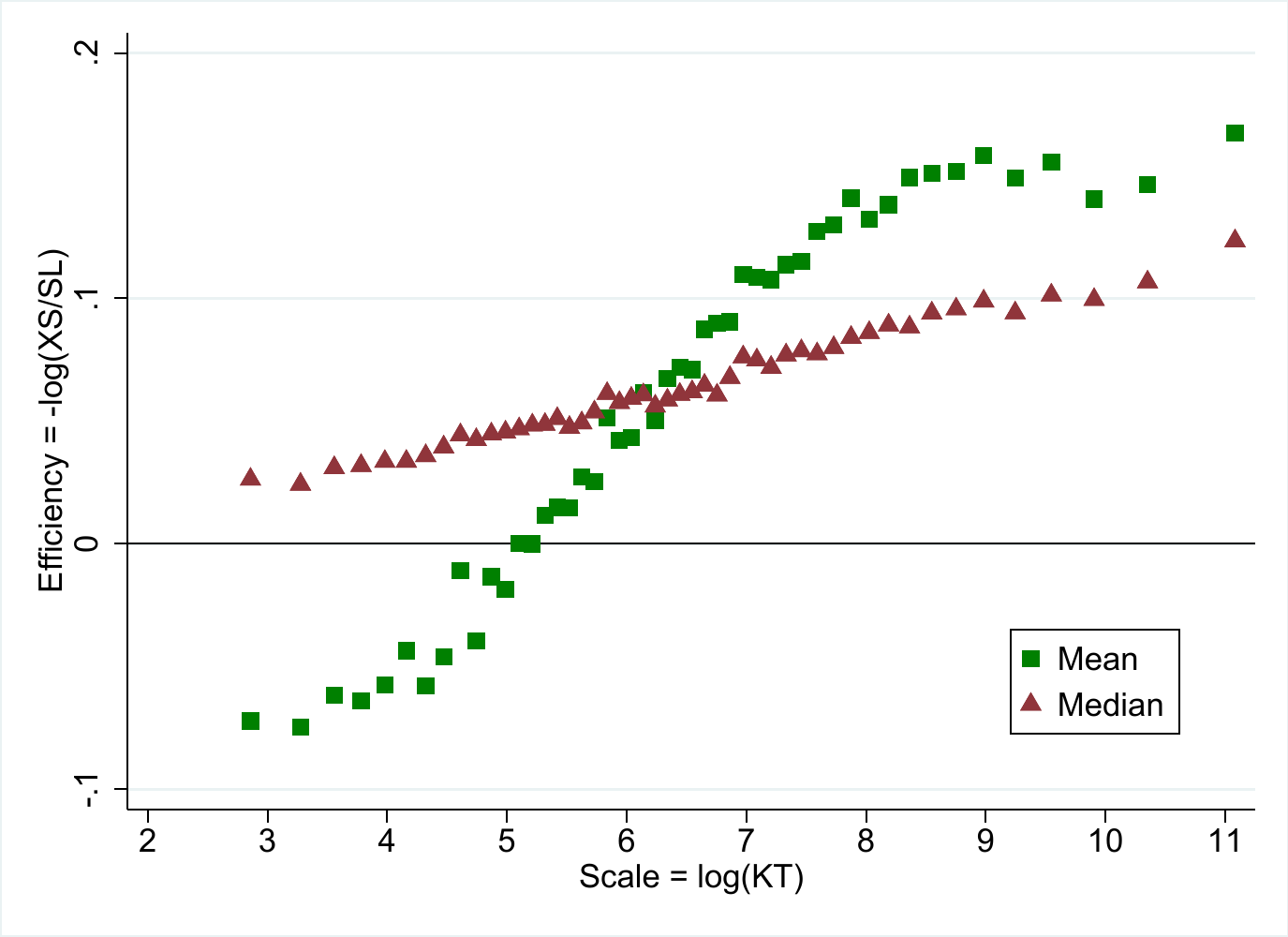}} &
		\subfigure[Sales-to-capital by scale ]{\includegraphics[width=2in]{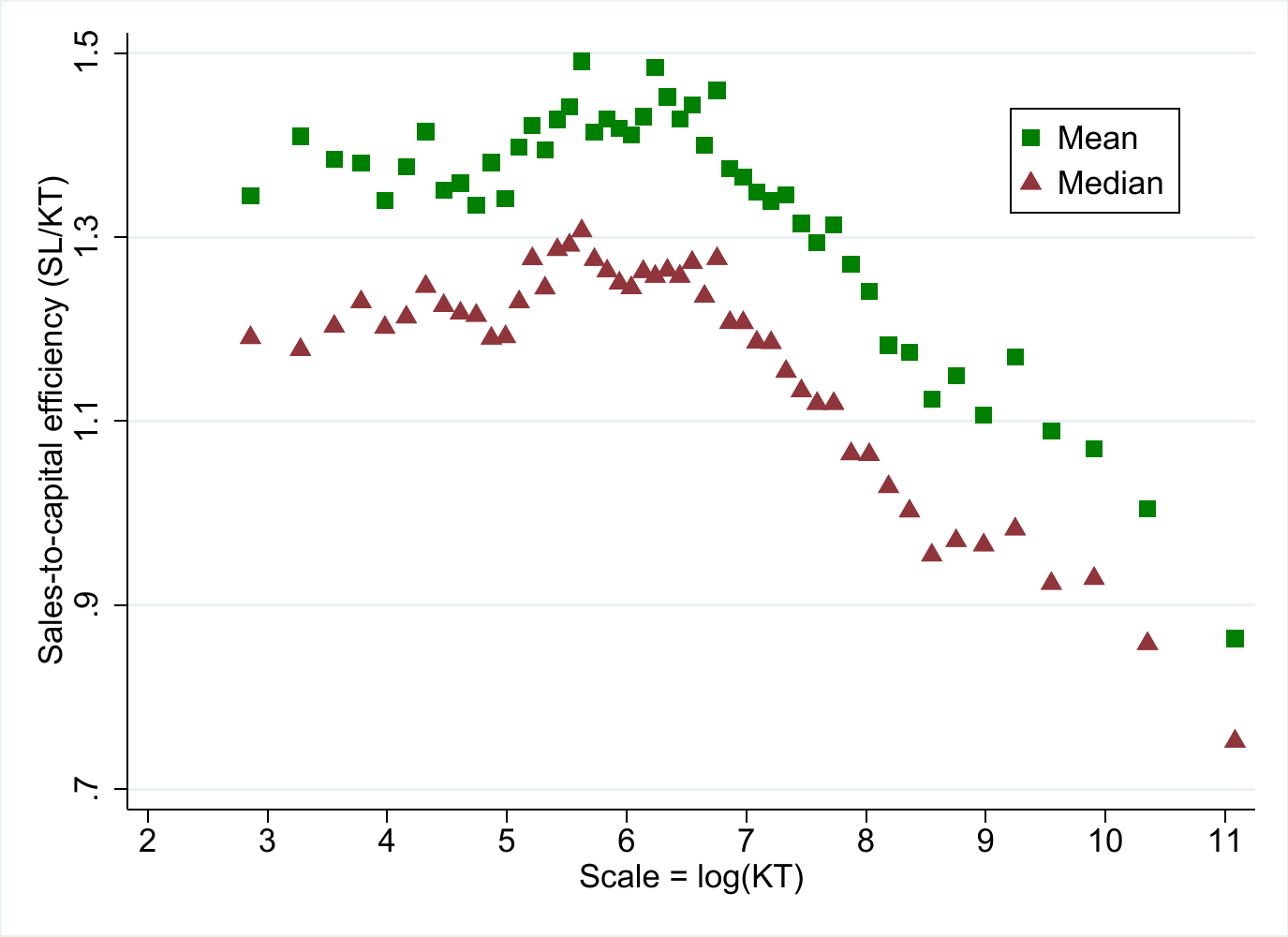}} \\ \\
	\end{tabular}
}

\subsection{Dynamism}
\label{sec:Dynamism}

The last firm distribution feature I review is dynamism, or the entry, exit, and long-term scaling dynamics of firms. Figure~\ref{fig:DYNfacts} presents the relevant facts for the US public-firm data. Panel (a) begins by presenting the long-term scale of firms conditional on their current scale. For each scale bin, the panel graphs the $\{1,10,25,50,75,90,99\}^{th}$ percentiles of the 10-year ahead scale distribution of firms in the bin. The medians align along the 45-degree line almost perfectly, the 10-90 range is approximately one scale up and down, and even the 1-99 range is only about two scales in any direction. Small firms remain small, large firms remain large, and there is little mixing even at long horizons. I conclude:

\begin{observation}
    \label{ob:17}
    Firm scale is highly persistent even in the long-term.
\end{observation}

\RPprep{Dynamism - Stylized facts}{0}{0}{DYNfacts}{%
    This figure presents stylized facts of firm dynamism. Panel (a) graphs the $\{1,10,25,50,75,90,99\}^{th}$ percentiles of the 10-year ahead KT scale distribution per current scale bin. Panel (b) graphs percentiles of the age distribution by scale, and panel (c) percentiles of the scale distribution by age, along with a fitted median regression line. Panels (d) and (e) plot exit and entry probability by scale bin, respectively, and include exponential-fit trend lines. Panel (f) plots exit probability by age, along with a linear trend line. Panel (g) plots the (log) number of observations by age in the entire sample, along with the fitted trend line. Panel (g) presents the distribution of scale at entry and at exit, and panel (h) plots the probability of entry and exit by year, with the horizontal line the unconditional exit probability. All figures are based on the \{Non-Bank\} data subset.
}
\RPfig{%
	\begin{tabular}{ccc} 
		\subfigure[10-year forward scale ]{\includegraphics[width=2in]{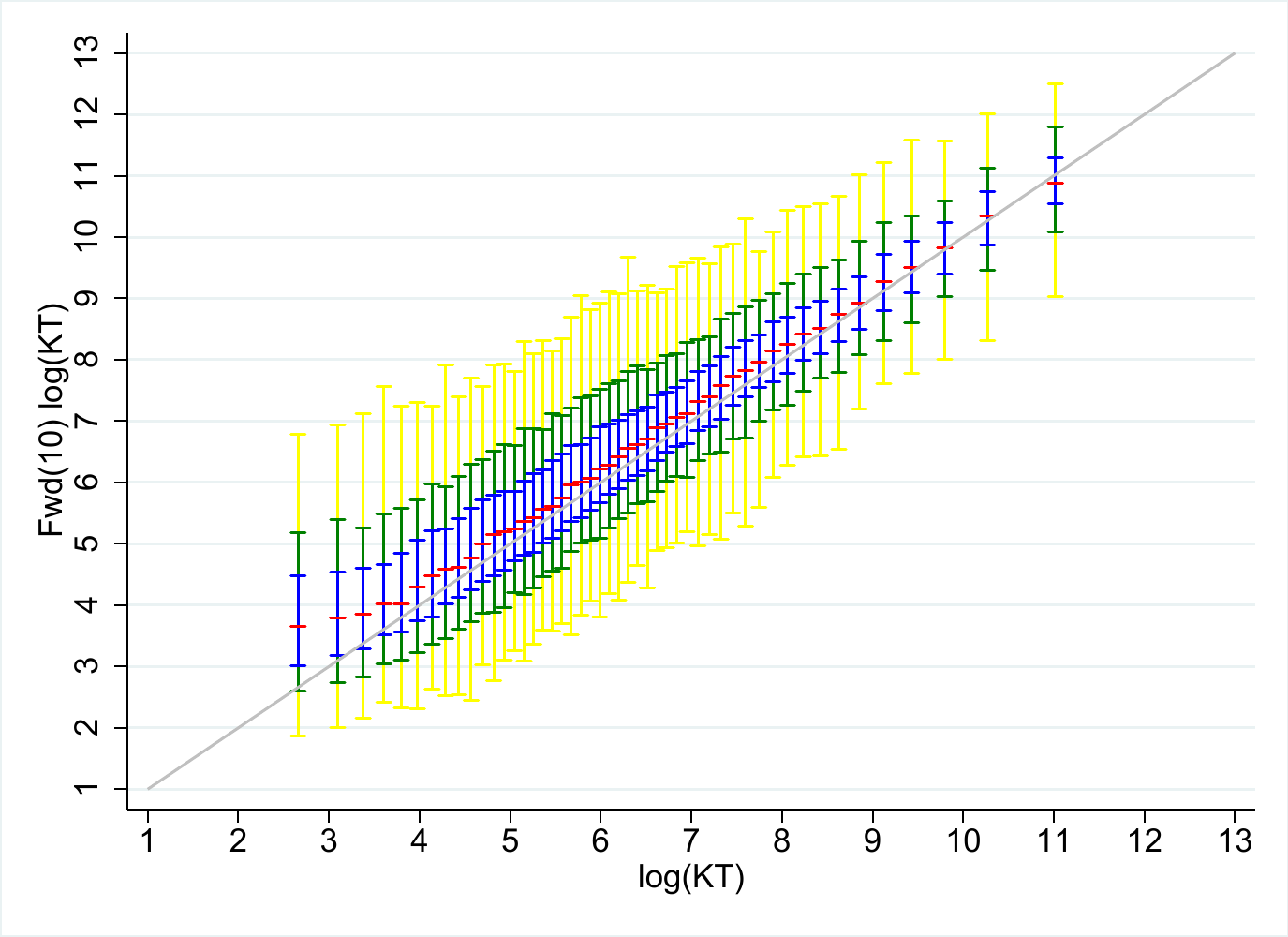}} &     \subfigure[Age dist. by scale ]{\includegraphics[width=2in]{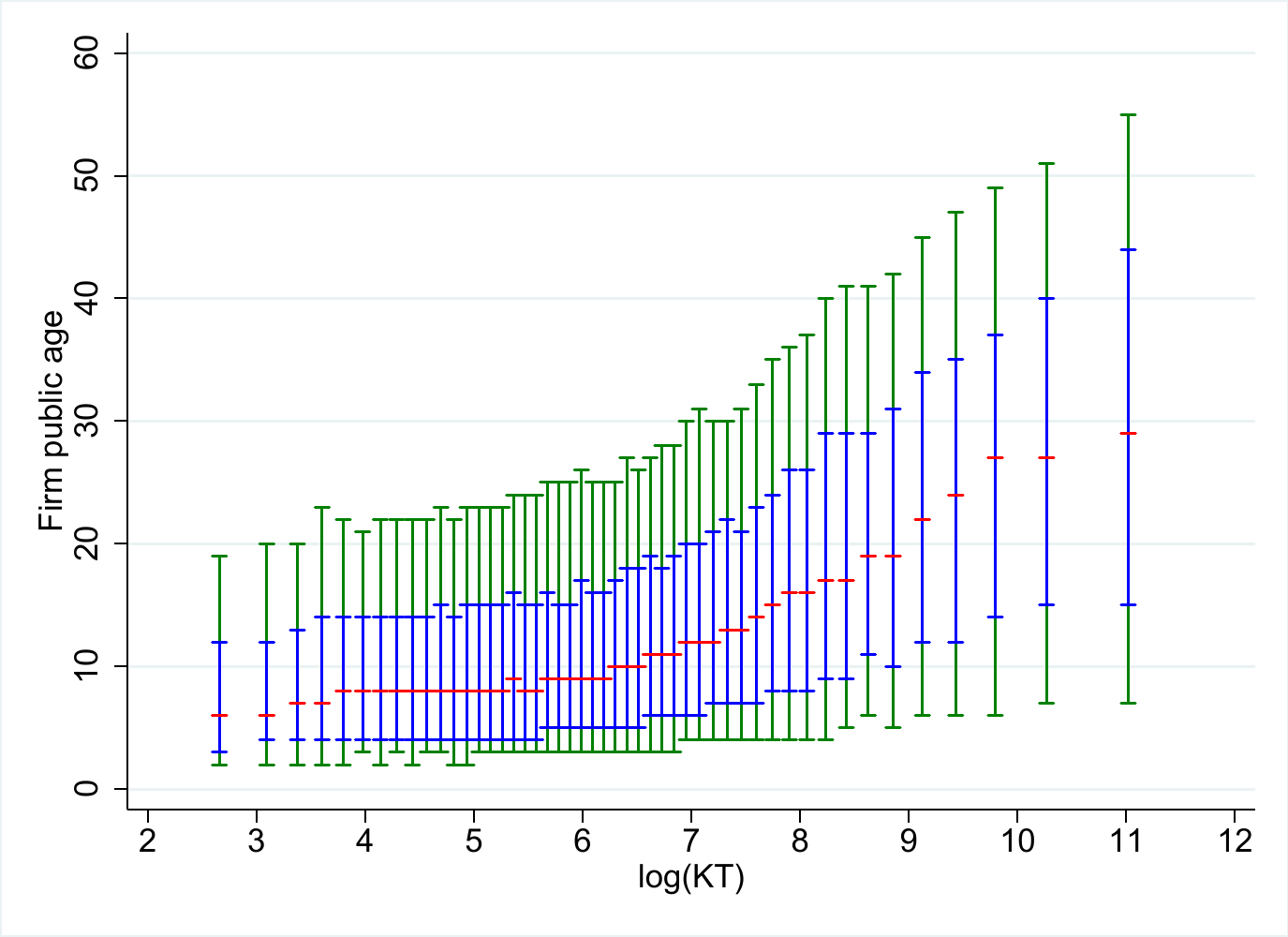}} &
		\subfigure[Scale dist. by age ]{\includegraphics[width=2in]{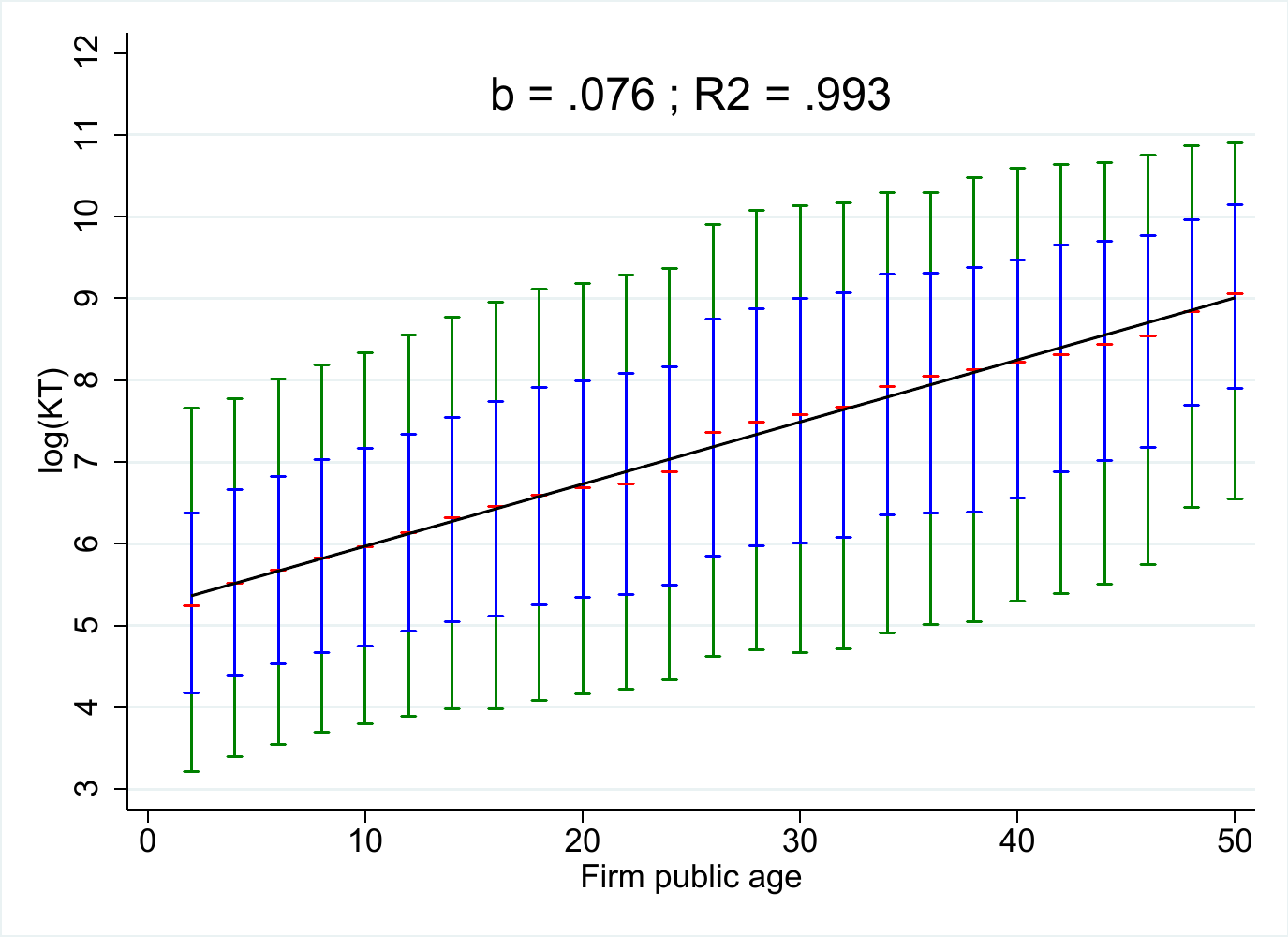}} \\ \\
		\subfigure[Exit prob. by scale ]{\includegraphics[width=2in]{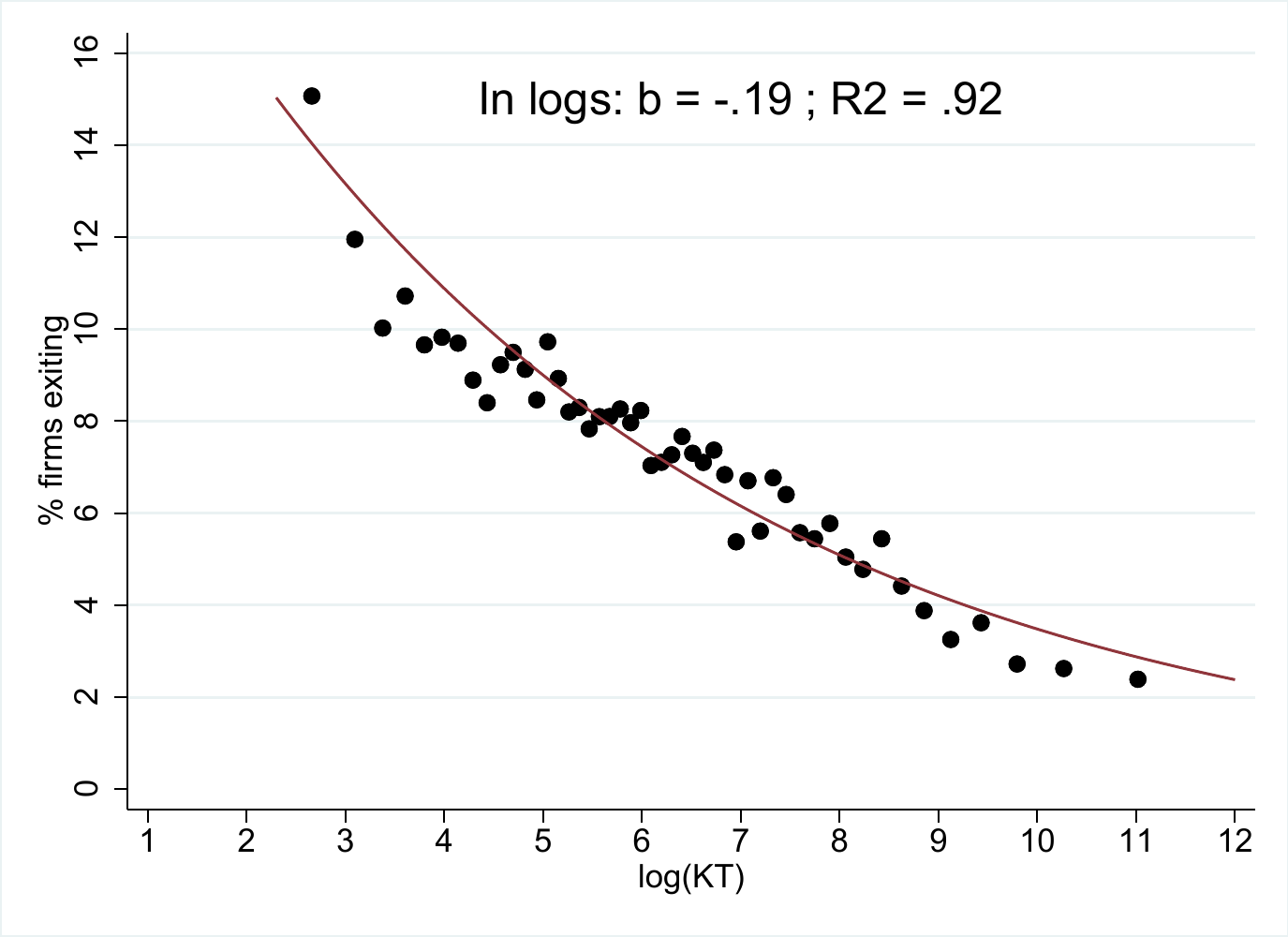}} &
		\subfigure[Entry prob. by scale ]{\includegraphics[width=2in]{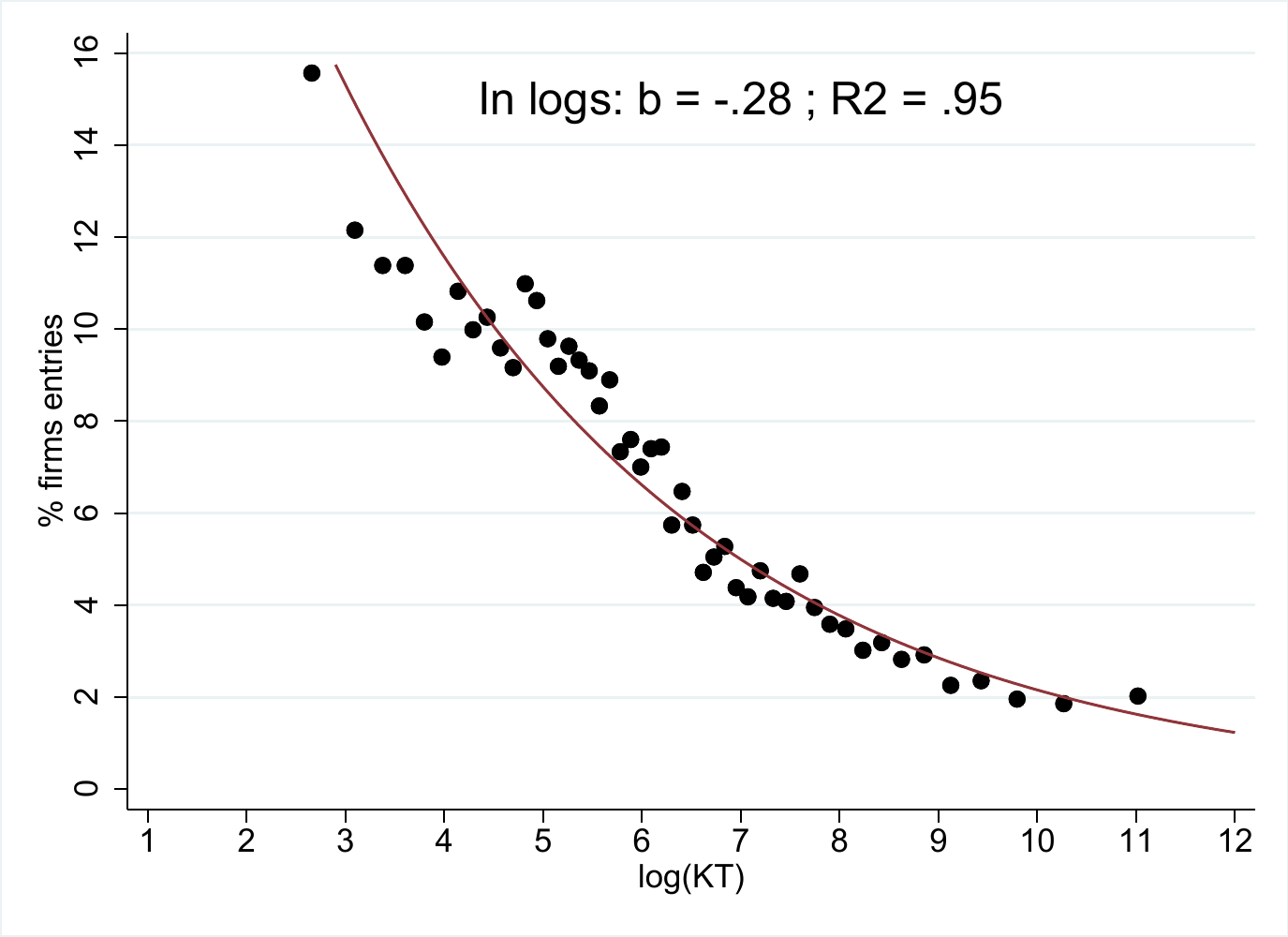}} &
		\subfigure[Exit prob. by age ]{\includegraphics[width=2in]{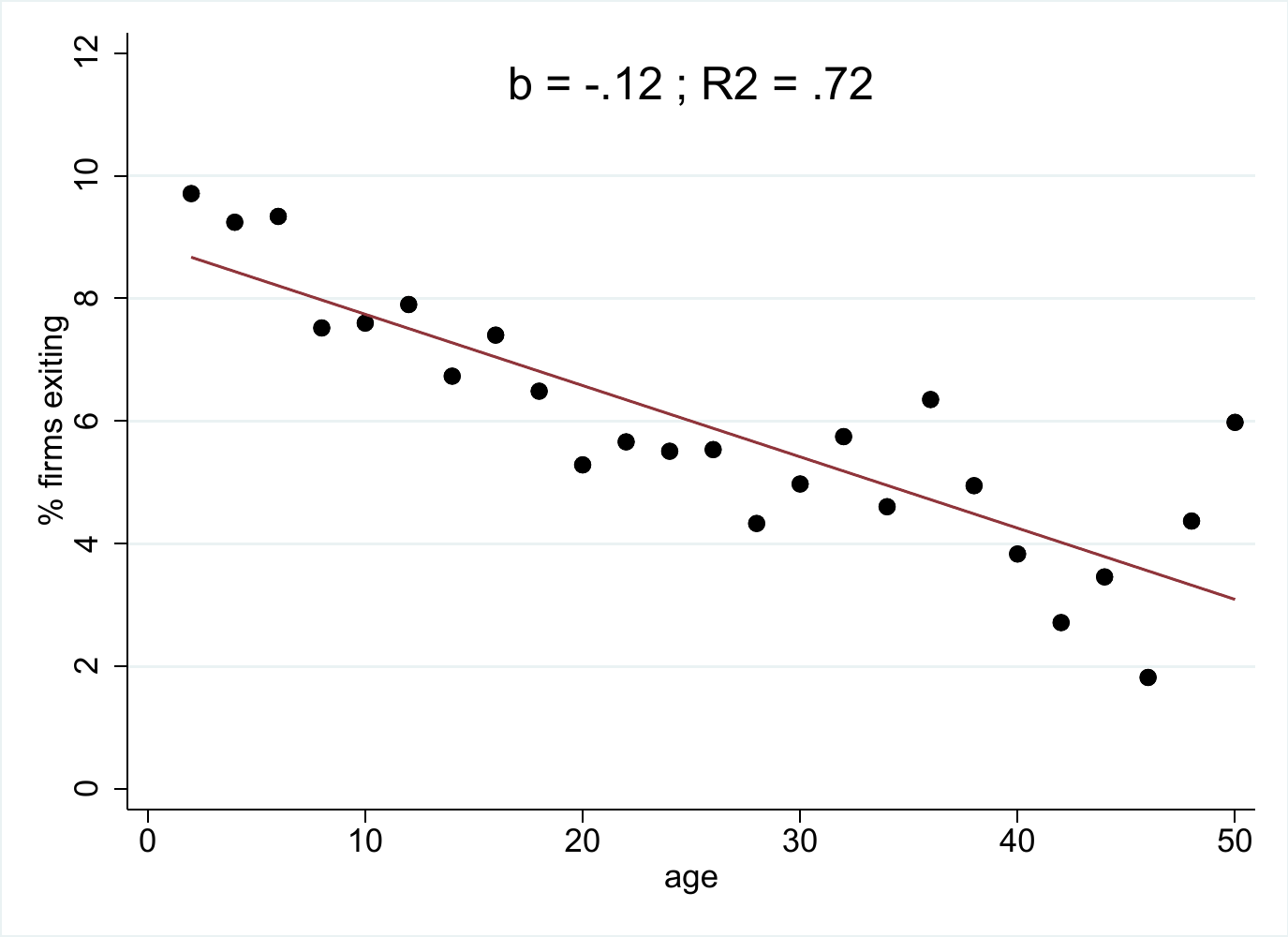}} \\ \\
		\subfigure[log(\#obs) by age ]{\includegraphics[width=2in]{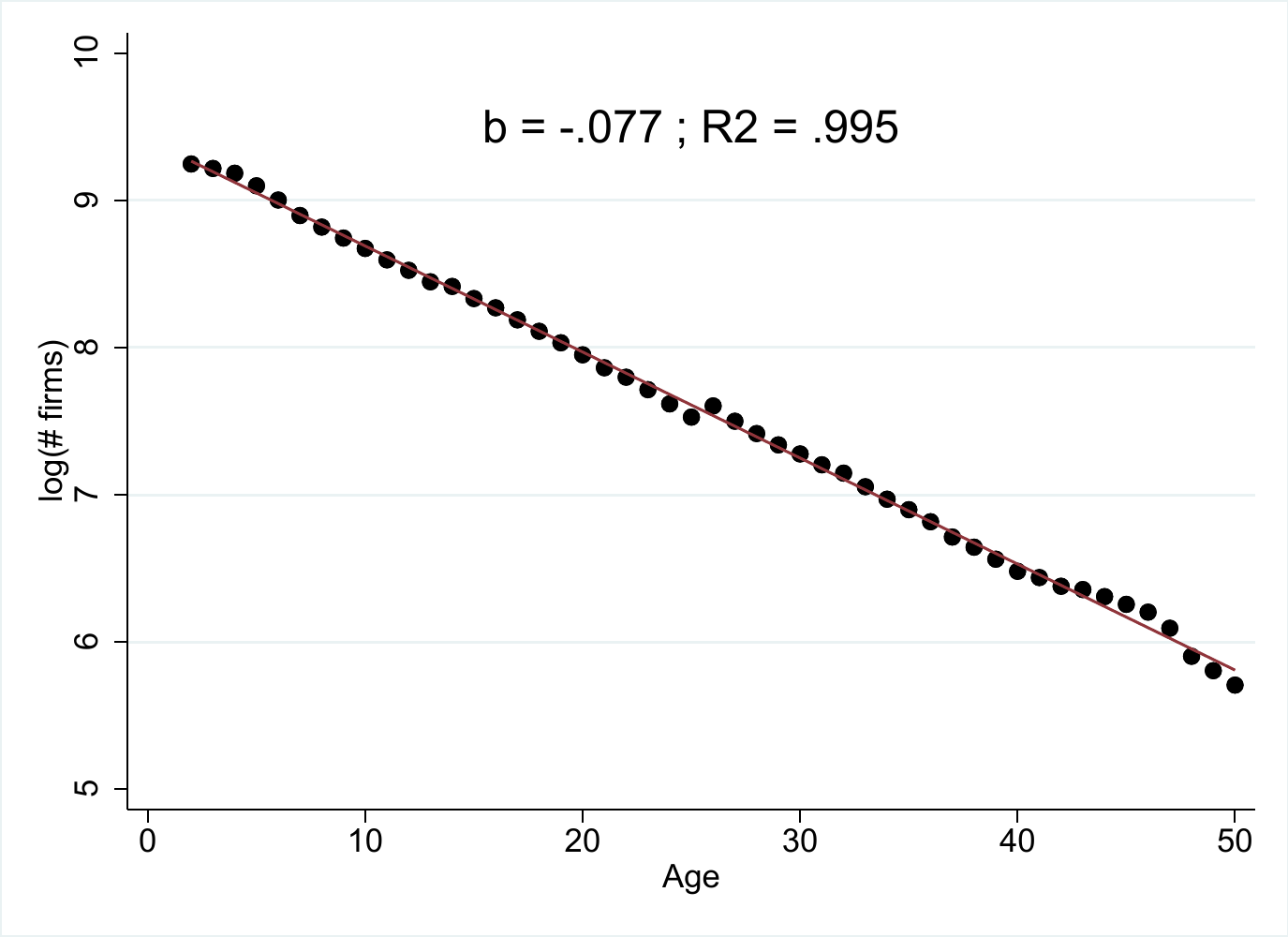}} &
		\subfigure[Scale at entry/exit ]{\includegraphics[width=2in]{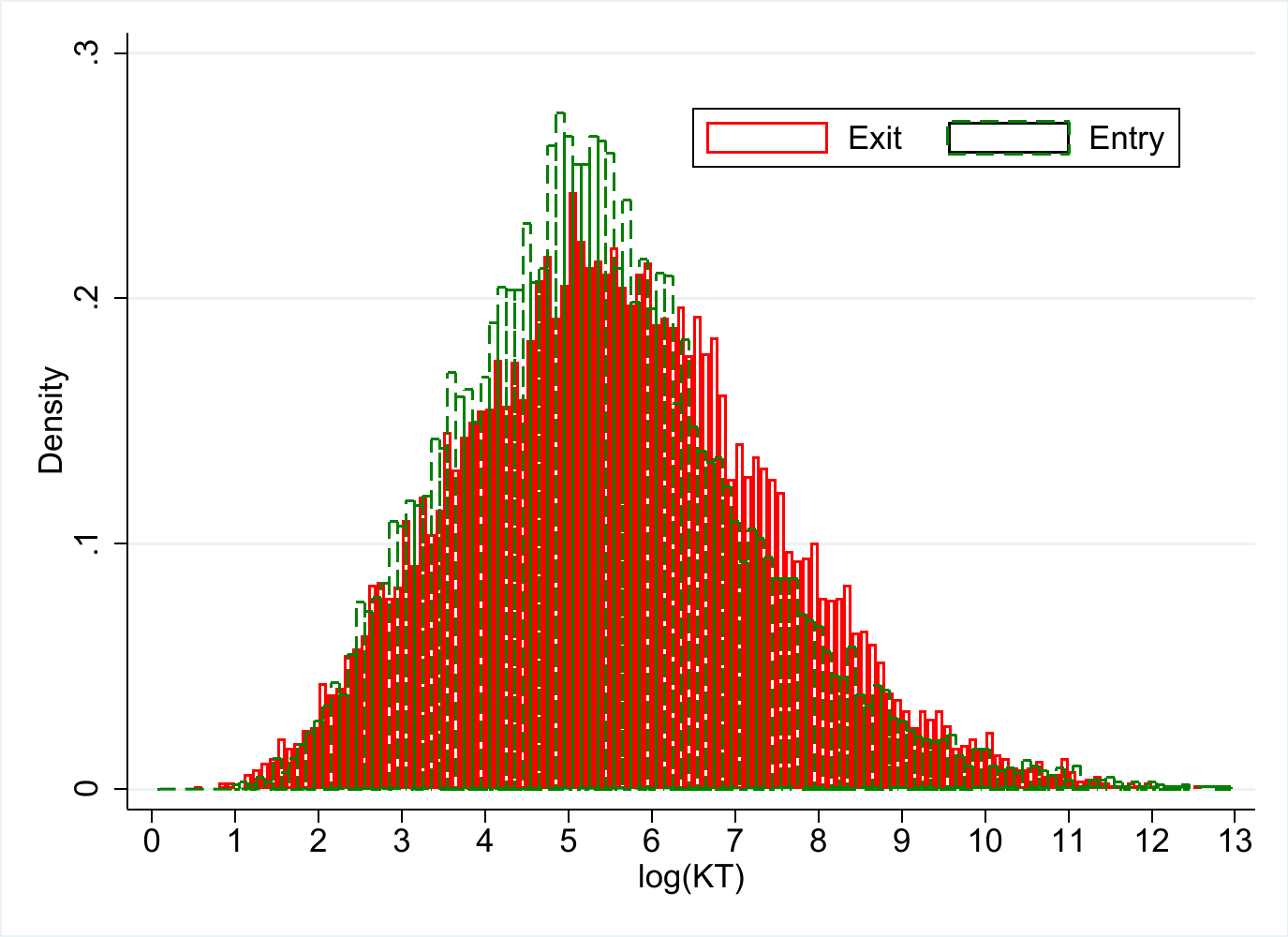}} &
		\subfigure[Entry/exit by year ]{\includegraphics[width=2in]{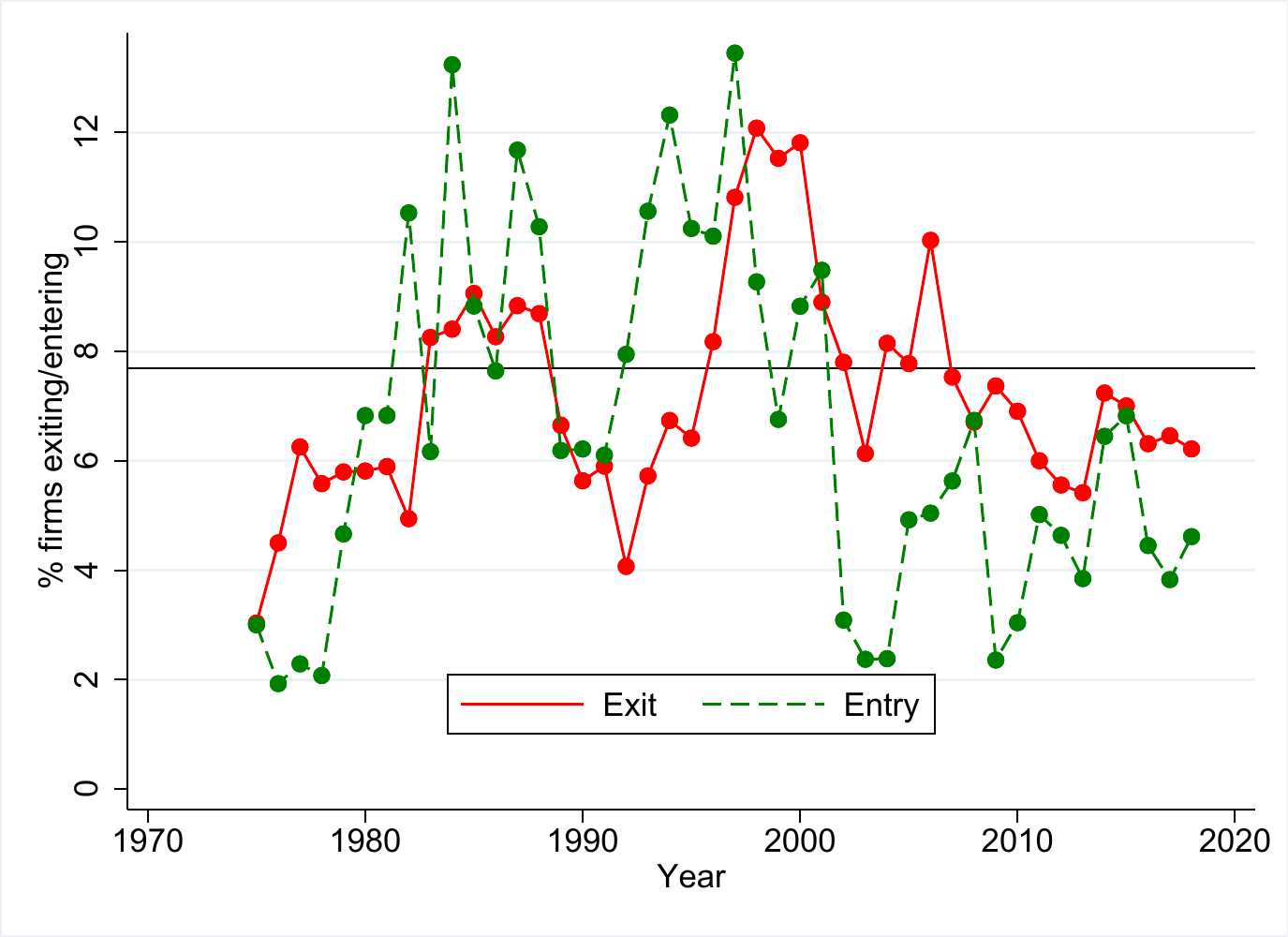}} \\ \\
	\end{tabular}
}

Panels (b) and (c) of Figure~\ref{fig:DYNfacts} present the relation between scale and public firm age. Public age is defined as the number of years since the firm appeared in the Compustat data-set, and is unrelated to the length of period the firm was private before becoming public. When considering the distribution of ages per scale in panel (b), we see the median public age of smaller firms is lower than 10 while the median age of the largest firms is closer to 30. Nevertheless, there is much dispersion in ages at the largest firm bin, with the 10$^{th}$ percentile of the age distribution below 10 years and the 90$^{th}$ percentile above 50. This relation is much crisper when one ``flips'' the figure and considers the scale distribution as a function of public age. Panel (c) exhibits a strong linear relation between public age and median scale, with median firm scale growing by $0.076$ for every year of survival. These panels cannot however ascertain whether firms that survive grow or whether large firms survive.

Panels (d)+(e) attend to this question, by considering the probability of exit and of entry by firm scale, respectively. Exit probability is strongly decreasing in scale, with about 15\% probability of exiting the sample every year for the smallest scale bin and about 2\% for the largest. The probability of entry by scale is similar, though has a faster decline (in logs). The uni-variate relation between exit probability and age depicted in Panel (f) appears to be an artifact of the relation between age and scale and the relation between scale and exit probability. The longevity of the largest firms has to do more with the fact that large firms have a lower hazard rate than the fact small firms take long to become large. This fact is also ascertained when one considers double-binning observations by scale and age, and regressing exit probability within bin on age and scale (unreported). Scale is overwhelmingly the better predictor of exit probability, and the residual from regressing exit probability on scale is not explained by age, while the residual of regressing exit probability on age is strongly explained by scale.

A second way to observe the fact age has limited impact on exit probability is using Panel (g). The panel plots the (log) number of firms in the pooled data, by firm age. The slope is uniform by age, and firms have an approximate exit probability of 7.4\% per year, which is also the pooled unconditional exit probability in the data. I conclude:

\begin{observation}
    \label{ob:18}
    Exit probability is strongly decreasing with scale, but there's only a weak relation between age and exit probability.
\end{observation}

Panel (h) is a different presentation of the data in panels (d)+(e) --- it presents the distribution of scale at entry and at exit, showing that both exit and entry happen throughout the FSD, but entering firms are somewhat smaller than exiting firms. Finally, Panel (i) presents the time-structure of entry/exit probability throughout the sample period, with the 80's and 90's generally showing much higher level of dynamism relative to the 70's, 00's and 10's.

\section{Summary}
\label{sec:SUM}

This paper analyzed the data on US public firms in the 50 year period 1970-2019. Firm distributions that were explored are the scale distribution FSD, income distribution FID, growth distribution FGD, equity return distribution FGE(EQ), income growth distribution FIGD, investment distribution FND, as well as firm R\&D investment, firm ratios, and firm dynamism. Special attention was given to scale effects and distributional forms.
\newline \newline
\noindent The following 18 observations were made:
\begin{enumerate}
    \item The FSD is distributed skew-Normal.
    \item Value and capital scales are highly persistent. Sales and particularly expenses somewhat less so.
    \item FSD measures are cointegrated.
    \item The FID is distributed DLN.
    \item FGD dispersion is systematically decreasing with scale.
    \item The FGD is distributed DLN.
    \item The FGD(EQ) location and dispersion are increasing and decreasing with scale, respectively.
    \item The FGD(EQ) is distributed DLN.
    \item The FIGD(CF) has decreasing dispersion with scale and is distributed DLN.
    \item Physical depreciation rates are decreasing with scale.
    \item The FND is distributed DLN.
    \item R\&D intensity is non-linearly decreasing with scale.
    \item R\&D intensity's median is negatively correlated with its skewness.
    \item Physical intensity at (non) R\&D-performing firms exhibits patterns (similar) opposite those of R\&D intensity.
    \item Small and large firms (below and above the median scale) are systematically different from each other.
    \item Median debt-to-equity, firm efficiency, and total payout ratio grow linearly with scale.
    \item Firm scale is highly persistent even in the long-term.
    \item Exit probability is strongly decreasing with scale, but there's only a weak relation between age and exit probability. 
\end{enumerate}
While some of the observations were previously documented in the literature, most are (to the best of my knowledge) new, namely: 4, 6-17. Three main findings are that the DLN distribution describes firm outcomes surprisingly well, that decreasing dispersion with scale is rampant in the data, and that small and large firms are systematically different from each other, possibly relating to their different R\&D activity profiles.

Observation 7 implies the relation between firm size and return in the data is actually opposite that posited by \cite{FamaFrench1992} and summarized by the "size factor", though further inquiry using tradeable portfolios is merited. Observation 8 similarly has significant asset pricing implications, especially given the large literature suggesting possible empirical distributions to describe asset returns for use in practical applications such as options pricing.

Similarly, observations 17, 18, and the entire dynamism discussion in Section~\ref{sec:Dynamism} imply that "scale is destiny", and firms seldom change their scale significantly through their life once becoming public. The process of growing from scale 0 to (approximately) the target scale for a given firm mostly happens while the firm is private. Furthermore, the relation between firm survival, size, and age is more complex, and it appears that the longevity of the largest firms has to do more with the fact that large firms have a lower hazard rate than the fact small firms take long to become large, thus casting doubt on the learning-by-doing mechanism.

Finally, the ubiquitous finding that the firm is ruled by the DLN distribution begs for a clear model explaining how it arises in firm dynamics. The follow-on paper \cite{Parham2022c} replaces the traditional ``Cobb-Douglas'' with a DLN production function in an otherwise standard q-theory model of the firm, and shows that the relevant stylized facts emerge naturally within the model.

\clearpage
\bibliographystyle{JFE}
\bibliography{MyLibrary}

\appendix

\renewcommand\thefigure{\thesection.\arabic{figure}}
\renewcommand\thetable{\thesection.\arabic{table}}
\renewcommand\theequation{\thesection.\arabic{equation}}
\setcounter{figure}{0}
\setcounter{table}{0}
\setcounter{equation}{0}

\clearpage
\section{Distributional fixed-effects}
\label{sec:AppdxFE}

This section describes the use of distributional fixed effects. The dimensions of fixed effects I am concerned with are \{time and scale\}X\{location and dispersion\} fixed effects. 

\subsection{Time fixed-effects}

The core idea is simply to standardize the yearly distributions prior to pooling them together. E.g., the transformation
\begin{equation} \label{eq:Adjust1}
T_{1}(X_{i,t}) =\frac{X_{i,t} - \text{Mean}[X_{*,t}]}{\text{Std}[X_{*,t}]}
\end{equation}
will standardize the distribution of X per year by deflating the location and scale of each observation within the year by the yearly location and scale. The pooled deflated observations are then not subject to the aggregation critique, as it relates to yearly location and dispersion changes.

A better transformation reflates each yearly distribution by the full-sample location and dispersion. Thus, the pooled adjusted sample maintains its location and dispersion, by using
\begin{equation} \label{eq:Adjust2}
T_{2}(X_{i,t}) =\frac{X_{i,t} - \text{Mean}[X_{*,t}]}{\text{Std}[X_{*,t}]}\text{Std}[X_{*,*}] + \text{Mean}[X_{*,*}]
\end{equation}

A second improvement is using median and IQR as measures of location and scale rather than mean and standard deviation. This is useful when using the transformations to adjust heavy-tailed distributions, in which the mean and s.d. are significantly less robust than the median and IQR. The transformation~\ref{eq:Adjust2} then becomes
\begin{equation} \label{eq:Adjust3}
T_{3}(X_{i,t}) =\frac{X_{i,t} - \text{Median}[X_{*,t}]}{\text{IQR}[X_{*,t}]}\text{IQR}[X_{*,*}] + \text{Median}[X_{*,*}]
\end{equation}

Note that the same transformation can also be specified in size (rather than scale) terms, i.e., if $Y = \text{exp}(X)$, one could equivalently specify the transformation~\ref{eq:Adjust3} as
\begin{equation} \label{eq:Adjust4}
T_{4}(Y_{i,t}) = \text{exp}\left(\left(\log\left(Y_{i,t}\right) - \text{Median}[\log(Y_{*,t})]\right)\frac{\text{IQR}[\log(Y_{*,*})]}{\text{IQR}[\log(Y_{*,t})]} + \text{Median}[\log(Y_{*,*})] \right)
\end{equation}

While the form of the transformation in~\ref{eq:Adjust4} appears to needlessly complicate the transformation in~\ref{eq:Adjust3}, the benefit of doing so becomes clear when one considers deflating doubly-exponential $Y$ variates (e.g., firm income), by the time-trend of the positive part of $Y$. E.g., if $Y \sim \text{DLN}$, we cannot use the transformations in~\ref{eq:Adjust3} or~\ref{eq:Adjust4}. A slight adjustment to~\ref{eq:Adjust4} is required,
\begin{equation} \label{eq:Adjust5}
T_{5}(Y_{i,t}) = \text{sign}\left(Y_{i,t}\right)\text{exp}\left(\left(\log\left(\left|Y_{i,t}\right|\right) - \text{Median}[\log(Y^{+}_{*,t})]\right)\frac{\text{IQR}[\log(Y^{+}_{*,*})]}{\text{IQR}[\log(Y^{+}_{*,t})]} + \text{Median}[\log(Y^{+}_{*,*})] \right)
\end{equation}
In which $\text{sign}(Y)$ is the sign step function, with a value of +1 for positive values and -1 for negative values, and $Y^+_{*,t}$ denotes all positive Y values in time $t$.

The transformation in~\ref{eq:Adjust5} adjusts both positive and negative values, in log dimensions, and maintains their sign, but does so based on the location and dispersion of positive values. For the DLN distributions we encounter, upward of 80\% of observations are positive. This allows us to avoid the dubious act of adjusting positive and negative values by different factors.

\subsection{Scale fixed-effects}

As discussed in Section~\ref{sec:Growth}, most firms measures exhibit scale-dependant heterogeneity in both location and dispersion. A straightforward adjustment approach would be to adjust each observation based on its scale bin. I.e., if $k$ indexes bin membership based on some measure of scale, as in the binscatter plots of Figure~\ref{fig:FGDDisp}, then one could adjust for scale effects by using the transformation
\begin{equation} \label{eq:Adjust6}
T_{6}(X_{i,t,k}) =\frac{X_{i,t,k} - \text{Median}[X_{*,*,k}]}{\text{IQR}[X_{*,*,k}]}\text{IQR}[X_{*,*,*}] + \text{Median}[X_{*,*,*}]
\end{equation}

A better approach, less subject to bias due to the discrete nature of the bins, is to simply adjust each value based on the fitted regression lines between the bins, in the appropriate binscatter. Letting $M_{i,t}$ denote a measure of firm size used as the basis for scale-normalization, define the following set of transformations:
\begin{equation} \label{eq:Adjust7}
\begin{split}
T_{7}^{loc}(M_{i,t-1}) & = \beta_0^{loc} + \beta_1^{loc}\cdot\log(M_{i,t-1}) \\
T_{7}^{dis}(M_{i,t-1}) & = \exp\left(\beta_0^{dis} + \beta_1^{dis}\cdot\log(M_{i,t-1})\right) \\
T_{7}(X_{i,t},M_{i,t-1}) & = \frac{X_{i,t} - T_{7}^{loc}(M_{i,t-1})} {T_{7}^{dis}(M_{i,t-1})}\cdot T_{7}^{dis}(\text{Median}[M_{*,*}]) + T_{7}^{loc}(\text{Median}[M_{*,*}]) \\
\end{split}
\end{equation}
in which $\beta_0^{loc}, \beta_1^{loc}$ are the intercept and slope from a quantile regression for the 50$^{th}$ percentile of $X_{i,t}$ on lagged scale $\log(M_{i,t-1})$, and $\beta_0^{dis}, \beta_1^{dis}$ are the intercept and slope of regressing $\log(\text{IQR}[X_{i,t}])$ on $\log(M_{i,t-1})$ (as in Figure~\ref{fig:FGDDisp}). The transformation $T_{7}$ in~\ref{eq:Adjust7} is conceptually equivalent to the transformation in~\ref{eq:Adjust6}, but is akin to assigning the observations to a continuum of bins.

\renewcommand\thefigure{\thesection.\arabic{figure}}
\renewcommand\thetable{\thesection.\arabic{table}}
\renewcommand\theequation{\thesection.\arabic{equation}}
\setcounter{figure}{0}
\setcounter{table}{0}
\setcounter{equation}{0}

\clearpage

\end{document}